\documentclass[prb,eqsecnum,twocolumn,groupedaddress,amsmath,floatfix]{revtex4}
\usepackage{graphicx}% Include figure files
\usepackage{bm}

\newcommand{\lambdah}{\hat{\lambda}}
\newcommand{\as}{a_s}

\newcommand{\vf}{v_{\rm F}}

\newcommand{\coupling}{\gamma}
\newcommand{\curH}{{\mathcal H}}
\newcommand{\phdag}{{\phantom{\dagger}}}
\newcommand{\phstar}{{\phantom{*}}}
\newcommand{\vol}{{V}}
\newcommand{\kf}{k_{\rm F}}
\newcommand{\pf}{p_{\rm F}}
\newcommand{\tkf}{\tilde{k}_{\rm F}}

\newcommand{\txi}{\varepsilon}%
\newcommand{\te}{E}
\newcommand{\dbmu}{\bar{h}}
\newcommand{\ef}{\epsilon_{\rm F}}

\newcommand{\curO}{{\mathcal O}}

\newcommand{\imag}{{\rm Im}}
\newcommand{\real}{{\rm Re}}

\newcommand{\width}{\Gamma_0}

\newcommand{\sfm}{SF$_{\rm M}$\,}

\newcommand{\bk}{{\bf k}}
\newcommand{\bq}{{\bf q}}
\newcommand{\bQ}{{\bf Q}}
\newcommand{\bp}{{\bf p}}

\newcommand{\grad}{{\bm{\nabla}}}

\newcommand{\Uh}{\hat{U}}

\newcommand{\bh}{\hat{b}}
\newcommand{\ch}{\hat{c}}
\newcommand{\alphah}{\hat{\alpha}}
\newcommand{\Bh}{\hat{B}}
\newcommand{\Qh}{\hat{Q}}
\newcommand{\Psih}{\hat{\Psi}}
\newcommand{\Qb}{\bar{Q}}
\newcommand{\Vb}{\bar{V}}
\newcommand{\Deltab}{\bar{\Delta}}
\newcommand{\hb}{\bar{h}}
\newcommand{\nb}{\bar{n}}
\newcommand{\mb}{\bar{m}}
\newcommand{\mub}{\bar{\mu}}
\newcommand{\Vt}{\tilde{V}}
\newcommand{\muh}{\hat{\mu}}
\newcommand{\hh}{\hat{h}}
\newcommand{\mg}{m}
\newcommand{\mgh}{\hat{m}}
\newcommand{\hc}{\hat{h}_c}
\newcommand{\hcone}{\hat{h}_{c1}}
\newcommand{\hctwo}{\hat{h}_{c2}}

\newcommand{\hfflo}{\hat{h}_{\rm FFLO}}

\newcommand{\Vh}{\hat{V}}
\newcommand{\Deltah}{\hat{\Delta}}
\newcommand{\deltah}{\hat{\delta}}

\newcommand{\br}{{\bf r}}

\newcommand{\be}{\begin{equation}}
\newcommand{\ee}{\end{equation}}
\newcommand{\bea}{\begin{eqnarray}}
\newcommand{\eea}{\end{eqnarray}}
\newcommand{\bse}{\begin{subequations}}
\newcommand{\ese}{\end{subequations}}

\newcommand{\chip}{\chi_{\rm P}}
\newcommand{\phaseshift}{\delta_{s}}
\newcommand{\fermiint}{\lambda}
\newcommand{\deltabcs}{\Delta_{\rm BCS}}
\newcommand{\deltahbcs}{\Deltah_{\rm BCS}}
\newcommand{\Hcz}{H_c^{\rm Z}}
\newcommand{\mubohr}{\mu_{\rm B}}
\input{epsf}

\begin{document}
\title{
BEC-BCS crossover, phase transitions and phase separation in 
polarized resonantly-paired superfluids
}
%-------------------------
\author{Daniel E.~Sheehy and Leo Radzihovsky}
\affiliation{
Department of Physics, 
University of Colorado, 
Boulder, CO, 80309}
\date{July 30, 2006}
%\maketitle
%--------------------------
\begin{abstract}
%--------------------------

We study resonantly-paired s-wave superfluidity in a degenerate gas of
two species (hyperfine states labeled by $\uparrow,\downarrow$) of
fermionic atoms when the numbers $N_{\uparrow}$ and  $N_{\downarrow}$
of the two species are {\it unequal\/}, i.e., the system
is \lq\lq polarized\rq\rq.  We find that  the continuous
crossover from the Bose-Einstein condensate (BEC) limit of
tightly-bound diatomic molecules to the Bardeen-Cooper-Schrieffer
(BCS) limit of weakly correlated Cooper pairs, studied extensively at
equal populations, is interrupted by a variety of distinct  phenomena
under an imposed population difference $\Delta N \equiv N_\uparrow -
N_\downarrow$.  Our findings are summarized by a \lq\lq
polarization\rq\rq\ ($\Delta N$) versus Feshbach-resonance  detuning
($\delta$) zero-temperature  phase diagram, which exhibits regions of
phase separation, a periodic FFLO superfluid, a polarized normal
Fermi gas and a polarized molecular superfluid consisting of a
molecular condensate and a fully polarized Fermi gas.  We describe
numerous experimental signatures of such phases and the transitions
between them, in particular focusing on their  spatial structure in
the inhomogeneous environment of an atomic trap.

%
%--------------------------
\end{abstract}
\maketitle
%--------------------------

\section{Introduction}
\label{SEC:intro}
\subsection{Background and Motivation}
%%%%%%%%%%%%%%%%%%%%%%%%%%%%%%%%%%%%%%%%%%%%%%%%%%%%%%%%%%%%%%%%%%%%%%%%%%%%%%%%%%%%%%%%%%%%%%%%%%%%%%%%%%%%%%%%%%
%
%%%%%%%%%%%%%%%%%%%%%%%%%%%%%%%%%%%%%%%%%%%%%%%%%%%%%%%%%%%%%%%%%%%%%%%%%%%%%%%%%%%%%%%%%%%%%%%%%%%%%%%%%%%%%%%%%%%

One of the most exciting recent developments in the study of
degenerate atomic gases  has been the 
observation~\cite{Regal,Zwierlein,Kinast,Bartenstein,Bourdel04,Chin,Greiner,Partridge05p,Zwierlein05p} of singlet paired
superfluidity of fermionic atoms  interacting via an s-wave Feshbach
resonance~\cite{Timmermans99,Dieckmann,Ohara,Regal03,Bourdel,Strecker,Duine04,Bartenstein05,Kohler}.

A crucial and novel feature of such experiments is the  {\it
tunability\/} of the position (detuning, $\delta$) of the Feshbach
resonance, set by the energy of  the diatomic molecular (\lq\lq
closed-channel\rq\rq) bound state relative to the open-channel atomic
continuum, which allows a degree of control over the fermion
interactions that is unprecedented in other (e.g., solid-state)
contexts.  As a function of the magnetic-field controlled detuning, $\delta$,
fermionic pairing is observed to undergo  the
Bose-Einstein condensate to Bardeen-Cooper-Schrieffer (BEC-BCS)  
crossover~\cite{Eagles,Leggett,Nozieres,sademelo,Timmermans01,Holland,
Ohashi,agr,Stajic,Levinsen,Gurarie} between the Fermi-surface
momentum-pairing BCS regime of strongly overlapping Cooper pairs (for
large positive detuning) to the coordinate-space pairing BEC regime of
dilute Bose-condensed diatomic molecules (for negative detuning).

Except for recent experiments~\cite{Zwierlein05,Partridge05,Zwierlein06,PartridgeComment,Zwierlein06p,Shin} that
followed our original work~\cite{shortpaper}, and a wave of recent
theoretical~\cite{Combescot01,Liu,Bedaque,Caldas,Mizushima,Carlson,Cohen,Castorina,
Sedrakian,Yang1,Pao,Son,Yang2,Dukelsky,YangSachdev,Pieri,Torma,Yi,Chevy,He,
Caldas06,DeSilva,Haque,SachdevYang,Bulgac06,HoZhai,LiuHu,Gu,Yang06,HuLiu,Iskin,
Imambekov,Jensen,Mannarelli,YiDuan,PaoYip,DeSilva06,Caldas06p,Chien,Koponen,Sedrakian06,
Parish,Chevy2,
Recati,He06,Gubbels,Martikainen}
activity, most work  has focused on the case of an  {\em equal} mixture
of two hyperfine states (forming a pseudo-spin 1/2 system), observed
to exhibit pseudo-spin singlet superfluidity near an s-wave Feshbach
resonance.  Here we present an extensive study of such systems for an
{\em unequal} number of atoms in the two pairing hyperfine states,
considerably extending results  and calculational details beyond those
reported in our recent Letter~\cite{shortpaper}.  Associating the two
pairing hyperfine states with up ($\uparrow$) and down ($\downarrow$)
projections of the pseudo-spin $1/2$, the
density difference $\delta n=n_\uparrow - n_\downarrow$ between the
two states is isomorphic to an imposed \lq\lq magnetization\rq\rq\
$\mg\equiv\delta n$ (an easily controllable experimental \lq\lq
knob\rq\rq), with the chemical potential difference
$\delta\mu=\mu_\uparrow-\mu_\downarrow$ corresponding to a purely Zeeman field
$h\equiv\delta\mu/2$.

This isomorphism makes a connection to a large body of work in a
related condensed-matter system,  namely BCS superconductors under an
applied Zeeman magnetic field, providing further motivation for our
study.  In contrast to a normal Fermi liquid that exhibits Pauli
paramagnetism, a  conventional homogeneous BCS state~\cite{Clogston,Sarma} at
zero temperature remains unmagnetized until, at a critical Zeeman
field $\Hcz$, it is  destroyed in a first-order transition to the
unpaired magnetized normal state.  A natural question is 
 whether there can be a
\lq\lq compromise\rq\rq\ state that exhibits both pairing {\it and\/}
nonzero magnetization.   One proposal for such a state dates back to
work of  Fulde and Ferrell (FF)~\cite{ff} and Larkin
and Ovchinnikov (LO)~\cite{lo}, and has been the subject of strong
interest for many
years~\cite{Aslamazov,Takada,Matsuo,Agterberg,Alford,Combescot,Bowers},  finding
putative realizations in  a variety of systems ranging from heavy-fermion
superconductors~\cite{CeCoIn5} to  dense quark
matter~\cite{Alford,Bowers}.  These so-called FFLO states,
theoretically predicted to be the ground state for a narrow range of
applied Zeeman field near the above-mentioned transition, are quite
unusual in that, while exhibiting off-diagonal long-range order (i.e., superfluidity),
 they spontaneously break rotational and translational
symmetry, forming a crystal of pairing order (i.e., 
a supersolid~\cite{AndreevLifshitz,Chester,Leggett70,KimChan}) 
with lattice vectors
$Q\approx k_{{\rm F}\uparrow} -k_{{\rm F}\downarrow}$, where $k_{{\rm
F}\sigma}$ is   the spin-$\sigma$ Fermi wavevector.

The observation of such magnetized superfluidity in condensed-matter
systems has been  elusive for  a variety of reasons, primarily because
an applied {\it physical\/} magnetic field $H$ couples not only to
spin polarization (i.e., the Zeeman effect) but also to the orbital motion
of {\it charged\/} electrons.  The latter coupling leads to
the Meissner effect, in which a charged superconductor expels an
externally applied magnetic field.  For type I superconductors, for
sufficiently large $H>H_c$, the energy cost of expelling the field
exceeds the condensation energy of the superconducting  state and the
system is driven normal via a first-order transition.   Because this
thermodynamic critical field $H_c  = \Delta \sqrt{4\pi N(\ef)}$ 
is much smaller
than~\cite{Footnote:FFLO} the critical (Clogston limit~\cite{Clogston}) Zeeman field 
$\Hcz = \Delta \sqrt{N(\ef)/\chi_P}$ (where $\chi_P$ is the
Pauli magnetic susceptibility), i.e., $H_c/\Hcz = \sqrt{4\pi \chi_P} \ll 1$,
 in condensed-matter systems the effects of a purely Zeeman field are
expected to be obscured by the orbital effects of the physical
applied magnetic field~\cite{Footnote:situation}.  The effects of
impurities~\cite{Aslamazov,Agterberg}, that are always present in condensed-matter systems, can
further complicate  the realization of FFLO and other magnetized
superfluid states (see, however, recent
work~\cite{CeCoIn5}).

In contrast, trapped atomic systems are  natural settings where
paired superfluidity  at a finite imposed magnetic moment,
$\Delta N = N_\uparrow - N_\downarrow$, can be experimentally studied
by preparing mixtures with different numbers $N_\uparrow,
N_\downarrow$ of two hyperfine-state species, with the (Legendre-conjugate)
chemical potential difference $\delta \mu = \mu_\uparrow -
\mu_\downarrow$ a realization of the corresponding effective (purely)
Zeeman field.

Beyond the aforementioned isomorphism in the weakly-paired BCS regime,
atomic gases interacting via  a Feshbach resonance allow studies of
magnetized paired superfluidity as a function of Feshbach resonance
detuning across the resonance and deep into the strongly-paired
molecular BEC regime, inaccessible in condensed-matter systems.

\subsection{Theoretical framework and its validity}
\label{SEC:tf}

The goal of our present work is to extend the study of  s-wave
paired resonant superfluidity to the case of an unequal number of the two
hyperfine-state species, $\Delta N/N = (N_\uparrow -
N_\downarrow)/N\neq 0$, namely, to calculate the phase diagram 
as a function of
detuning, $\delta$, and 
polarization, $\Delta N/N$ (or, equivalently, the chemical 
potential difference $\delta \mu$).

An appropriate microscopic model of such a Feshbach-resonantly
interacting fermion system is the so-called two-channel
model~\cite{Timmermans99,Holland,Ohashi,agr,Gurarie} that captures the
dynamics of atoms in the open channel, diatomic molecules in the
closed channel and the coupling between them, which  corresponds to the
decay of closed-channel diatomic molecules into two atoms in the open
channel.

As we will show, the model is characterized by a key dimensionless parameter
\bea
\gamma= \frac{\sqrt{8}}{\pi}\sqrt{\frac{\width}{\ef}},
\eea
determined by the ratio of the Feshbach resonance width $\width\approx 4m\mubohr^2 a_{bg}^2 B_w^2/\hbar^2$
(where $\mubohr$ is the Bohr magneton), to the
Fermi energy $\ef$, and describes the strength of the atom-molecule
coupling that can be extracted from the two-body scattering length 
observed~\cite{Regal03,Bartenstein05} to behave as 
\be
\as = a_{bg} \big( 1- \frac{B_w}{B-B_0}\big),
\label{eq:scatteringlengthformula}
\ee
as a function of the magnetic field $B$ near the resonance position $B_0$.
Equivalently, $\gamma$ is the ratio of the inter-atomic spacing
$n^{-1/3}$ (where $n$ is the atom density) to the effective range $r_0$ characterizing the energy dependence of
two-body scattering amplitude in the open channel. As was first emphasized by
Andreev, Gurarie and Radzihovsky~\cite{agr,Gurarie} in the
context of pairing in an unpolarized (symmetric) Fermi gas, in the
limit of a vanishingly narrow resonance, $\gamma\rightarrow 0$, the
two-channel model is exactly solvable by a mean-field
solution. Consequently, for a finite but narrow resonance $\gamma \ll
1$, the system admits a detailed {\em quantitative} analytical
description for an arbitrary value of detuning ({\it throughout\/} the
BEC-BCS crossover), with its accuracy controlled by a systematic
perturbative expansion in $\gamma$. 

This important observation also holds for asymmetric mixtures with an
arbitrary magnetization. Hence in the narrow Feshbach resonance limit
we can accurately study this system by a perturbative expansion in
$\gamma$ about a variational mean-field solution. To implement this we
compute the energy in a generalized BCS-like variational state, and
minimize  the energy over the variational parameters. Explicitly, the
variational state is parametrized by the bosonic condensate  order parameter 
$b_{\bQ}$, where $\bQ$ is a  center
of mass wavevector that allows for a periodically modulated 
condensate.  Our
choice of the variational ground state is sufficiently rich as to
include the normal state, the gapped singlet BCS state as well as its
molecular BEC cousin, a magnetized paired superfluid state with
gapless atomic excitations, a FFLO state~\cite{SingleQ,RajagopalComment}, and
inhomogeneous ground states that are a phase-separated coexistence of any two of the
above pure states.  Our variational ansatz is {\it not\/}, however,
general enough to allow for the very interesting possibility of a uniform
but anisotropic paired superfluid ground state, e.g., a nematic superfluid~\cite{Sedrakian,Yang2},
nor of more exotic multi-$\bQ$ FFLO superfluids~\cite{SingleQ,RajagopalComment}.

Most of our work focuses on the narrow resonance ($\gamma\ll 1$) limit,
studied within the two-channel model. Although experiments do not lie in
this regime, the analysis provides valuable and {\it quantitatively\/}
trustworthy predictions for the behavior of the system (throughout the 
phase diagram), at least in
this one nontrivial limit.
However, typical present-day
experiments~\cite{Regal,Zwierlein,Kinast,Bartenstein,Bourdel04,Chin,Greiner,Partridge05p,Zwierlein05p}
fall in a broad Feshbach resonance, $\gamma\simeq 10^1-10^4\gg 1$,
regime~\cite{noteBroad} (see Appendix~\ref{SEC:scatteringamplitude}).
 To make contact with these experiments we
also extend our results to the broad resonance limit, complementing
our two-channel analysis with an effective single-channel model.
 For  a broad resonance, we can ignore the
dispersion of the closed-channel molecular mode and integrate it out,
thereby reducing the two channel model to an effective single (open)
channel model with a tunable four-Fermi coupling related to the atomic s-wave 
scattering length.  Although in this broad resonance regime our
mean-field variational theory is not guaranteed to be {\em
quantitatively} accurate, we expect that it remains qualitatively
valid.
Indeed, we find a reassuring qualitative
consistency between these two approaches. 

Furthermore, to make detailed predictions for cold-atom experiments we
extend our bulk analysis to include a trap, $V_T(r)$. We do this
within the local density approximation (LDA). Much like the WKB
approximation, this corresponds to using expressions for the bulk
system, but with an effective local chemical potential
$\mu(r)=\mu-V_T(r)$ in place of $\mu$. The validity of the LDA
approximation relies on the smoothness of the trap potential, namely
that $V_T(r)$ varies slowly on the scale of the {\it longest\/} physical length
$\lambda$ (the Fermi wavelength, scattering length, effective range,
etc.) in the problem, i.e., $(\lambda/V_T(r)) dV_T(r)/dr\ll 1$. Its accuracy 
can be equivalently controlled by a small parameter that is the ratio 
of the single particle trap level spacing $\delta E$ to the
smallest characteristic energy $E_c$ of the studied phenomenon (e.g, the 
chemical potential, condensation energy, etc.), by requiring $\delta E/E_c \ll 1$.

\subsection{Outline}

The rest of the paper is organized as follows. In Sec.~\ref{SEC:Summary} we summarize
our main results.  In Sec.~\ref{SEC:Hamiltonian}, we review the standard two-channel and
one-channel models of fermions interacting via an s-wave Feshbach resonance.  Focusing on
a narrow resonance described by the two-channel model, we compute the
system's ground state energy in Sec.~\ref{SEC:MFT}. By minimizing it we map out
the detuning-polarization phase diagram. Focusing first on uniformly-paired
states (i.e., ignoring the FFLO state), we do this for the case of
positive detuning (the BCS and crossover regimes) in Sec.~\ref{SEC:BCS}, and for 
negative detuning (the BEC regime) in Sec.~\ref{SEC:BEC}, finding a variety of phases
and the transitions between them. In Sec.~\ref{SEC:FFLO}, we revisit the positive-detuning 
regime to study the periodically-paired FFLO state. In
Sec.~\ref{SEC:OneChannelModel}, we complement our analysis of the two-channel model with a
study of the one-channel model, finding an expected qualitative agreement.  To
make predictions that are relevant to cold-atom experiments, in Sec.~\ref{SEC:LDA}
we extend our bulk (uniform-system) analysis to that of a trap. We
conclude in Sec.~\ref{SEC:concludingremarks} with a discussion of our work in the context of
recent and future experiments and relate it to other theoretical studies 
that have recently appeared in the literature.  We relegate many technical details 
to a number of Appendices.

\section{Summary of principal results}
\label{SEC:Summary}

Having motivated our study and discussed its theoretical framework and validity,
below we present a summary of the main predictions of our work, thereby
allowing a reader not interested in derivations easy access to our
results. As discussed above, the quantitative validity of our calculations 
is guaranteed in the narrow resonance $\gamma\ll 1$ limit, with all expressions given to 
leading order in $\gamma$.  Our results (summarized by Figs.~\ref{fig:hphasetwo}-\ref{fig:ldabecintro})   
are naturally organized into a detuning-polarization phase diagram (see Figs.~\ref{fig:mphasetwo} and
\ref{fig:globalmagphasesingleintro}) and our presentation
logically splits into predictions for a bulk system and for a 
trapped system.

\subsection{Bulk system}
\label{SEC:SummaryUniform}
In this somewhat theoretically-minded presentation of our results, we
focus on the two-channel description, outlining the system's phenomenology
as a function of detuning, $\delta\simeq 2\mubohr (B-B_0)$ (with $B_0$ the magnetic field
at which the Feshbach resonance is tuned to zero energy), and chemical potential difference
$h$, or imposed polarization $\Delta N/N$.

\subsubsection{BCS regime: $\delta \gg 2 \ef$}

For large positive detuning, the closed-channel molecules are
energetically costly, and the atom density is dominated by
open-channel atoms, exhibiting a weak attraction mediated by virtual
closed-channel molecules. Consequently, for a weak chemical potential
difference (or Zeeman field) $h$ we find a standard s-wave singlet (non-polarized) BCS
superfluid ground state, that, as a function of detuning, undergoes a
BCS-BEC crossover that is identical to the well-studied crossover at 
vanishing $h$. Deep in the BCS regime, for fixed chemical
potential $\mu$, this singlet paired state becomes unstable when the
Zeeman field $h$ overwhelms the superconducting gap
$\deltabcs(\mu)$. For simplicity, for now ignoring the FFLO state, the
BCS superfluid then undergoes a first-order transition to the
partially-polarized normal Fermi gas state at
\bse
\bea
\label{hcsummary}
h_c(\mu,\delta) &=& \frac{\deltabcs(\mu)}{\sqrt{2}},
\\
 &=&
4\sqrt{2}{\rm e}^{-2} \mu{\rm e}^{
-\frac{\delta - 2\mu}{\gamma\sqrt{\ef\mu}}},
\label{eq:deltanoughtsummary}
\eea
\ese
consistent with established results first found by Clogston~\cite{Clogston} and Sarma~\cite{Sarma}.
 As is generically the case for first-order
transitions, thermodynamic quantities exhibit jump discontinuities
as $h$ crosses $h_c$. In particular, we find  the discontinuities
in the atomic, molecular and total densities are, respectively,
\bse
\bea
&&\hspace{-1.3cm}\Delta n_a \equiv n_{SF}^a - n_N^a \simeq \frac{c\Delta_{BCS}^2}{\sqrt{\mu}} \big(
1-\frac{1}{2}\ln \frac{\Delta_{BCS}}{8{\rm e}^{-2}\mu} 
\big)>0,
\\
&&\hspace{-1.3cm}\Delta n_m \equiv  n_{SF}^m - n_N^m\simeq  \frac{\Delta_{BCS}^2}{g^{2}}>0,
\\
&&\hspace{-1.3cm}\Delta n \equiv n_{SF} - n_N, \nonumber
 \\
&&\hspace{-.6cm}\simeq \frac{2\Delta_{BCS}^2}{g^2}+ \frac{c\Delta_{BCS}^2}{\sqrt{\mu}} \big(1-
\frac{1}{2}\ln \frac{\Delta_{BCS}}{8{\rm e}^{-2}\mu} 
\big)>0,
\eea
\ese
where $n_{SF}^a$ and $n_N^a$ are the free atomic densities in the superfluid and
normal states, respectively, $n_{SF}^m$ and $n_{N}^m$ are their molecular
analogs and $n$ is the total (whether free or paired) atom density.
Here, $c\equiv m^{3/2}/\sqrt{2}\pi^2 \hbar^3$ is defined by the three-dimensional density of states
$N(\epsilon) = c\sqrt{\epsilon}$.
 The
species imbalance (magnetization) $\mg=(N_\uparrow-N_\downarrow)/\vol$ also exhibits a
jump discontinuity
\be
\Delta \mg \equiv \mg_{SF} - \mg_N \simeq -2c h_c\sqrt{\mu}<0,
\ee
across $h_c$.

%-----------------------------
%
% fig%1
%
\begin{figure}[bth]
\vspace{3.2cm}
\hspace{-.1cm}
\centering
\setlength{\unitlength}{1mm}
\begin{picture}(40,80)(0,0)
\put(-55,55){\begin{picture}(0,0)(0,0)
\includegraphics{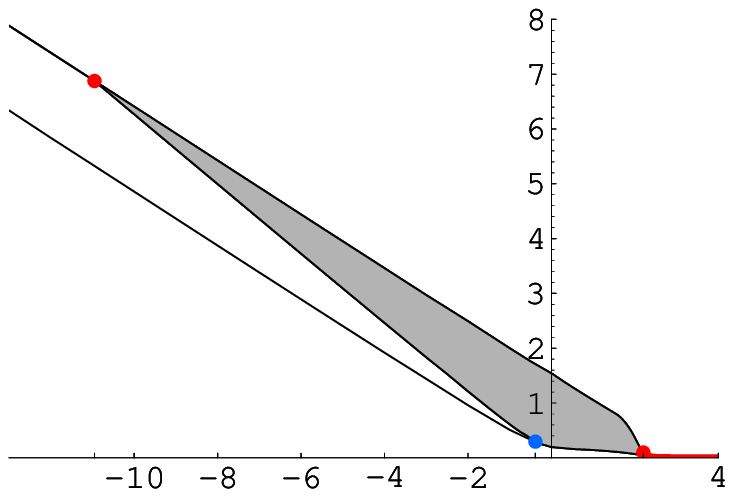}
\end{picture}}
\put(56,107) {(a)}
\put(37,112) {\large{$h/\ef$}}
\put(14.5,88) {$h_{c2}$}
\put(17,81) {$h_{c1}$}
\put(9,82) {$h_{m}$}
\put(20,100) {Spin-Polarized Normal (N)}
\put(59.5,60) {\large{$\frac{\delta}{\ef}$}}
%\put(-10,56.5) {\large{$\deltah_c$}}
\put(-9.5,57) {$\deltah_c$}
\put(38,57) {$\deltah_M$}
\put(50,57) {$\deltah_{*}$}
%\put(-8.5,89.5) {\vector(2,1){4}}
\put(-7.0,94) {SF$_{\rm M}$}
%\put(26.5,79.5) {\vector(-2,-1){6}}
\put(27.5,73) {PS}
\put(-17,63.5) {\large{BEC-BCS Superfluid(SF)}}
\put(55,66) {\vector(0,-1){5}}
\put(51,66.5) {FFLO}
\put(-60,0){\begin{picture}(0,0)(0,0)
\includegraphics{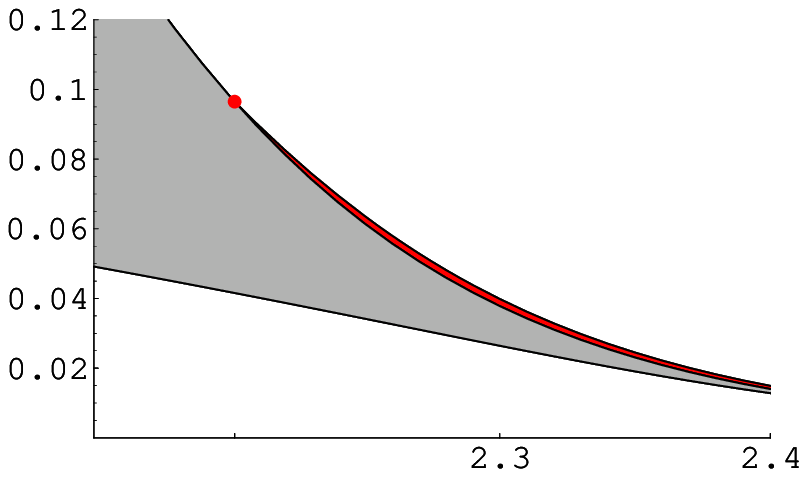}
\end{picture}}
\put(56,45) {(b)}
\put(-20,52) {\large{$h/\ef$}}
\put(59.5,5) {\large{$\frac{\delta}{\ef}$}}
\put(33.5,21.5) {\vector(-2,-1){4}}
\put(34.5,20.5) {\large{FFLO}}
\put(-12,31.5) {\large{PS}}
\put(0,9) {\large{SF}}
\put(-2,2.5) {$\deltah_*$}
\put(2,20.5) {$h_{c1}$}
\put(20,40) {\large{N}}
%\put(14,29.5)  {$h_{\rm FFLO}$}
%
\put(13,20)  {$h_{c2}$}
\put(15,21.5) {\vector(2,1){4}}
%
%--------------------------------
%
\put(18.5,30.5)  {$h_{\rm FFLO}$}
\put(18,31) {\vector(-2,-1){4}}
%--------------------------------
%
\end{picture}
\vspace{-.5cm}
\caption{(Color online)  (a) Global phase diagram as a function of Feshbach resonance 
detuning at fixed total atom number $N$ and chemical potential 
difference $h$ within the two-channel model, showing homogeneous superfluid (SF), magnetized 
superfluid (\sfm), FFLO and spin-polarized normal (N) phases as well as a regime of phase separation
(PS, shaded).
(b) Zoom-in of the positive-detuning BCS regime, showing
the regime of phase separation (gray) and FFLO state (red online). See 
Fig.~\ref{fig:hsingle} for  a similar phase diagram within
the one-channel model. $\deltah_*\equiv \delta_*/\ef$, 
$\deltah_M \equiv \delta_M/\ef$ and the 
tricritical point $\deltah_c\equiv \delta_c/\ef$ indicate
limiting detuning values for FFLO, \sfm and phase separated states, respectively.}
\label{fig:hphasetwo} 
%
%  phasediagram_twochannel.nb (both)
%
\end{figure}
%------------------------------

In the more experimentally-relevant (to cold atoms) fixed total atom
number $N$ ensemble, the density difference between the normal and
superfluid states makes it impossible for the normal state to satisfy
the imposed number constraint immediately above $h_{c1}(N,\delta) \equiv
h_c(\mu_{SF}(N),\delta)$. [Here, $\mu_{SF,N}$ are the superfluid (SF) and normal (N) state
chemical potentials corresponding to the imposed $N$.] Consequently, we
find that in the fixed number ensemble the transition at $h_c$ opens
up into a coexistence region $h_{c1}(\delta) < h < h_{c2}(\delta)$ in which the gas is an
inhomogeneous mixture of phase-separated superfluid and normal states,
with respective volume fractions $x(h,\delta)$ and $1-x(h,\delta)$ varying
according to 
\be
x(h,\delta) = \frac{n  - n_N[h,\mu_c(h)]}{n_{SF}[\delta,\mu_c(h)]
- n_N[h,\mu_c(h)]},
\ee
to satisfy the imposed total atom number constraint $n= N/\vol = x n_{SF} +
(1-x) n_N$. Here, $\mu_c(h)$ is the critical chemical potential as a function of $h$, implicitly given by
Eq.~(\ref{eq:deltanoughtsummary}), specifying the SF-N first-order phase boundary.
The coexistence region is bounded by $h_{c2}(\delta)$ from above,
corresponding to a Zeeman field below which a pure
partially-polarized normal Fermi gas phase cannot satisfy the imposed
number constraint while remaining the ground state.

A more careful analysis, that includes a periodically-paired FFLO
state in the variational ansatz, shows that, in fact, deep in the BCS
regime, for detuning $\delta > \delta_*\approx 2 \ef$, the first-order
transition is actually (from the singlet BCS state) into a magnetized
superfluid FFLO state. As illustrated in Fig.~\ref{fig:hphasetwo}b and consistent with
many deep-BCS studies\cite{ff,lo,Takada}, we find that this fragile state only
survives over a narrow sliver of Zeeman fields (or imposed
polarization),\cite{RajagopalComment}
undergoing a continuous\cite{MFTcomment,Landau,Brazovskii,Alexander} transition at $h_{\rm FFLO}(\delta)$ to
a partially-polarized normal Fermi gas ground state. Our work is an
extension of these earlier BCS studies to a varying detuning (controlling
the strength of the attractive interactions), with our
main result in this regime the location of the critical detuning
\be
\label{eq:deltastarintro}
\delta_* \approx \ef\big[2 - \frac{\gamma}{2} \ln (0.159 \gamma)\big],
\ee
accurately predicted for a narrow resonance ($\gamma \ll 1$), below which the FFLO
state is unstable for any $h$ or $\Delta N$. As seen in Fig.~\ref{fig:hphasetwo}b, 
above $\delta_*$, deep
in the BCS regime, the boundaries $h_{c1}$, $h_{c2}$, and $h_{\rm FFLO}$ display
exponential behavior [defining $\Delta_F \equiv \deltabcs(\ef)$]:
\bse
\bea
&&\hspace{-1cm}h_{c1}(\delta) \approx  \frac{1}{\sqrt{2}} \Delta_F 
\exp\big[- \frac{\delta^2\Delta_F^2}{8\gamma^2\ef^4}\big],\,\,\, \delta \gg 2\ef,
\label{eq:hconebcsintro}
\\
\label{eq:hctwobcsintro}
&&\hspace{-1cm}h_{c2}(\delta) \approx \frac{1}{\sqrt{2}} \Delta_F 
\exp\big[- \frac{\delta\Delta_F^2 }{16\gamma \ef^3}\big], \,\,\, \delta \gg 2\ef,
\\
&&\hspace{-1cm}h_{\rm FFLO}(\delta)\approx \eta\Delta_F  \exp\big[-\frac{\eta^2\delta\Delta_F^2 }{8\gamma \ef^3}\big],
 \,\,\, \delta \gg 2\ef,
\eea
\ese
never crossing with increasing detuning, the latter two boundaries asymptoting to a ratio
$h_{\rm FFLO}/h_{c2} = \eta \sqrt{2} \approx 1.066$ previously found by Fulde and
Ferrell~\cite{ff}.

\subsubsection{Crossover regime: $0 < \delta < 2\ef$}

As the detuning is lowered below approximately $2\ef$, it becomes favorable
(for low chemical potential difference $h$) to convert a finite
fraction of the Fermi sea (between $\mu\approx \delta/2$ and $\ef$) into
Bose-condensed molecules.  Since in this crossover regime the pairing
strength (proportional to the growing molecular density) is no longer
driven by  Fermi surface pairing and is therefore no longer
exponentially weak, the response to the chemical potential difference,
$h$, (or the imposed polarization, $\Delta N$) changes qualitatively from the
weakly-paired BCS regime discussed above. 

Consequently, the FFLO state, driven by mismatched Fermi-surface pairing, 
is no longer stable for $\delta < \delta_*$ [given by Eq.~(\ref{eq:deltastarintro})],
leading to a direct first-order transition between a paired superfluid
and a partially polarized normal Fermi gas (see Fig.~\ref{fig:hphasetwo}).

As above, for  imposed atom number the phase boundary splits
into a superfluid-normal coexistence region  bounded by $h_{c1}(\delta)$ and
$h_{c2}(\delta)$. The lower  boundary, $h_{c1}(\delta)$, is still determined by
the value of the gap, $\Delta$. However, because the gap is in turn set
by the molecular condensate $\Delta = g \sqrt{n_m(\delta)}$, that in this regime
is no longer exponentially small but grows as a
power-law with detuning reduced below $2\ef$, $h_{c1}(\delta)$ is also a
power-law in $2\ef-\delta$. We note, however, that while $h_{c1}(\delta)$ is
significantly larger than its exponentially small value in the BCS
regime, it nevertheless remains small compared to $\mu$ for a narrow resonance,
justifying a linear response (in $h$) approximation for the normal-state energy near $h_{c1}$.  

The upper boundary, $h_{c2}(\delta)$, (no longer in the linear
Pauli-paramagnetic regime, i.e., $h_{c2} \,{\not\!\ll} \,\mu$) is increased even 
more dramatically beyond that of Eq.~(\ref{eq:hctwobcsintro}), due to the
superfluid-normal density difference enhanced by a large molecular
density $n_m$ that here is a finite fraction of the total atom    
density. This considerably spreads the coexistence region for
$\delta$ below $2\ef$, with $h_{c1}(\delta)$ given by 
\bea
&&\hspace{-.75cm}h_{c1}(\delta)\simeq \ef\sqrt{\frac{\gamma}{3}} 
\sqrt{1- \big(\frac{\delta}{2\ef}\big)^{3/2}},\,\,\, \delta_M < \delta < 2\ef,
\label{hconecrossintro}
\eea
and the critical detuning $\delta_M$ given by 
\be
\delta_M \approx -1.24 \ef\sqrt{\gamma}.
\ee
Here, $h_{c2}(\delta)$ is implicitly given by 
\bea
&&\hspace{-.75cm}\ef^{3/2} \simeq \frac{1}{2}\Big[
\big(\frac{\delta}{2} + h_{c2}(\delta)\big)^{3/2} 
\\
&&\hspace{-.75cm}\qquad\qquad - \big(\frac{\delta}{2} - h_{c2}(\delta)\big)^{3/2} 
\Theta\big(\frac{\delta}{2} - h_{c2}(\delta)\big)\Big],\,\,\,  \nonumber
\delta < 2\ef, 
\eea
which is well-approximated by 
 \bea
&&\hspace{-.75cm}h_{c2}(\delta)\simeq  2^{2/3}\ef - \frac{\delta}{2}, \,\,\,\delta < 2^{2/3} \ef,
\label{hctwocrossintro}
\eea
over most of the range of interest.

\subsubsection{BEC regime: $\delta < 0$}

In the BEC regime, a new {\em uniform magnetized} superfluid (\sfm) phase appears, for
detunings below $\delta_M$,
when the population imbalance $\Delta N = N_{\uparrow} - N_{\downarrow}$ is imposed.
The \sfm ground state consists of closed-channel singlet molecules,
with the remaining unpaired atoms forming a fully-polarized Fermi sea
that carries the imposed magnetization. 

%-----------------------------%
% fig%2
%
\begin{figure}[bth]
\vspace{1.4cm}
\centering
\setlength{\unitlength}{1mm}
\begin{picture}(40,40)(0,0)
\put(-53,0){\begin{picture}(0,0)(0,0)
\includegraphics{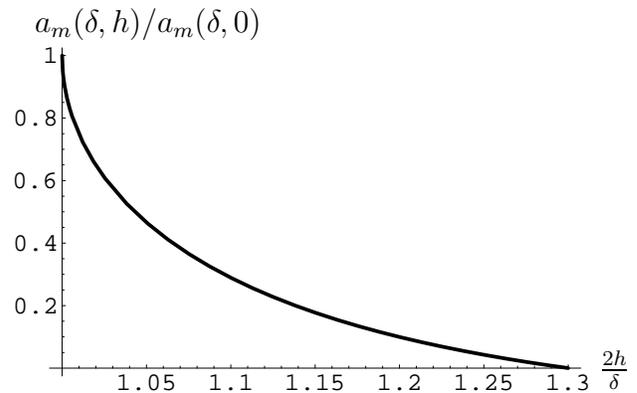}
\end{picture}}
\put(-19,50) {\large{$a_m(\delta,h)/a_m(\delta,0)$}}
\put(56,4) {\large{$\frac{2h}{\delta}$}}
\end{picture}
\vspace{-.5cm}
\caption{(Color online) Plot of molecular scattering length $a_m(\delta,h)$ [Eq.~(\ref{eq:amsum})], 
normalized to its value
at $h = 0$, as a function of $2h/|\delta|$.}
\label{molscatlenplot}
%
%  phasediagram_twochannel.nb 
%
\end{figure}
%------------------------------

 As illustrated in the phase diagram, Fig.~\ref{fig:hphasetwo}, for $\delta < \delta_M$ a 
spin-singlet (unpolarized) molecular (BEC) superfluid undergoes a 
second-order quantum phase transition at
\be
h_m(\delta) \approx |\delta|/2,\,\,\,\, \delta<\delta_M,
\ee
to the magnetized \sfm superfluid. The transition turns into a (sharp at low $T$) 
crossover at any finite temperature, since then
the magnetization is finite for arbitrarily small $h$, even below $h_m$. In
the narrow-resonance limit, the phase boundary $h_m(\delta)$ is determined
by the vanishing of the majority-species chemical potential
$\mu_\uparrow=h - |\delta|/2$. Physically, this corresponds to the
condition that the gain in the Zeeman energy $-h\Delta N$ compensates the
loss of the molecular binding energy $|\delta|/2$ per atom.

For intermediate negative detuning $\delta_c < \delta < \delta_M$, upon
increasing the Zeeman field $h$ the magnetized superfluid, \sfm,
undergoes a first-order transition at $h_c(\delta)$ [given by the implicit
Eq.~(\ref{eq:firstordercondition2}) in the main text].
In the more
experimentally-relevant fixed atom number $N$ ensemble this transition
at $h_c(\delta)$ opens up (as is standard for a first-order transition)
into a phase-separated region of coexistence between the \sfm and a fully-polarized 
normal atomic gas. This phase-separated regime is bounded by
\bea
\label{eq:hconebecintro}
h_{c1}(\delta) &\approx& 0.65 |\delta|,\,\,\,\,\,\,\,\delta_c<\delta<0,
\\
h_{c2}(\delta) &\approx& 2^{2/3}\ef+ \frac{|\delta|}{2},\,\,\,\,\,\,\delta<0,
\eea
that are continuations of the boundaries 
[Eqs.~(\ref{eq:hconebcsintro}), (\ref{eq:hctwobcsintro}), (\ref{hconecrossintro}) and (\ref{hctwocrossintro})]
found in the BCS and crossover
regimes above. We find that the \sfm-phase separation instability,
initiated at $h_{c1}$, is signaled by an enhanced compressibility and a
corresponding suppression (with increasing $h$) of the molecular
scattering length
\be
a_m \simeq \frac{\pi^2\ef \gamma^2}{16\sqrt{m} |\delta|^{3/2}}
F_4\Big(\frac{2h}{|\delta|}\Big),
\label{eq:amsum}
\ee
plotted in Fig.~\ref{molscatlenplot}. 
The function $F_4(x)$ is defined in Eq.~(\ref{eq:f}) of the main text.
In the narrow resonance limit, $\gamma\ll 1$, indeed 
$h_{c1}$ is determined by the vanishing of $a_m$, as given in Eq.~(\ref{eq:hconebecintro}) above.

As illustrated in Fig.~\ref{fig:hphasetwo}a, for sufficiently large negative detuning,
$\delta < \delta_c$, the $h_{c1}(\delta)$ and $h_{c2}(\delta)$ boundaries cross and the first-order
\sfm-N transition and the corresponding phase-coexistence region are
eliminated. The \sfm then undergoes a direct continuous transition at
$h_{c2}(\delta)$ into a fully-polarized normal state.

%-----------------------------%
% fig%3
%
\begin{figure}[bth]
\vspace{3.2cm}
\centering
\setlength{\unitlength}{1mm}
\begin{picture}(40,80)(0,0)
\put(-55,55){\begin{picture}(0,0)(0,0)
\includegraphics{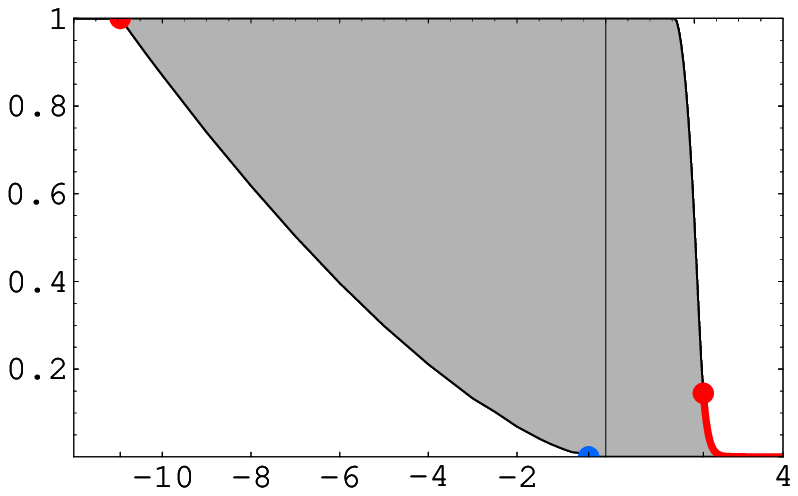}
\end{picture}}
\put(-16,107) {$\frac{\Delta N}{N}$}
\put(-14,100) {(a)}
\put(58,59) {\large{$\frac{\delta}{\ef}$}}
\put(-11.5,56.5) {$\deltah_c$}
\put(36,56.5) {$\deltah_M$}
\put(48,56.5) {$\deltah_*$}
\put(55,68.5) {\vector(0,-1){7}}
\put(51.0,69) {FFLO}
\put(50,95.5) {\large{N}}
\put(15,87.5) {\large{PS}}
\put(-6,72) {\large{SF$_{\rm M}$}}
\put(48,85) {$\scriptstyle\Delta N_{c2}$}
\put(15,75) {$\scriptstyle\Delta N_{c1}$}
\put(-55,0){\begin{picture}(0,0)(0,0)
\includegraphics{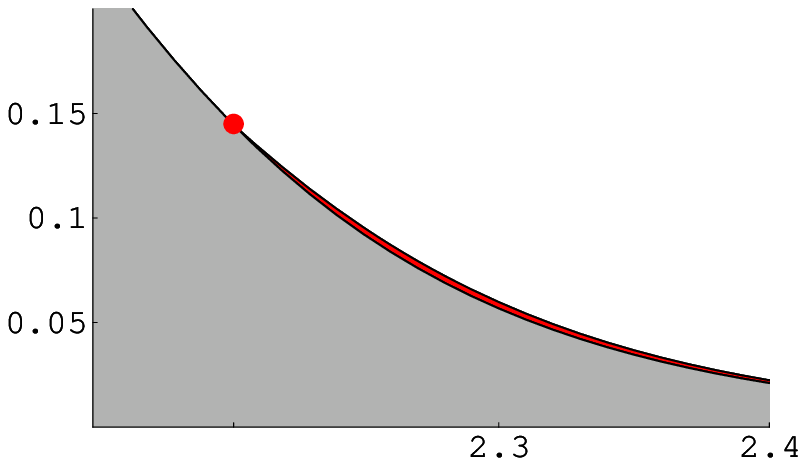}
\end{picture}}
\put(-15,52) {$\frac{\Delta N}{N}$}
\put(-13,45) {(b)}
\put(-0,3.5) {$\deltah_*$}
\put(60,6) {\large{$\frac{\delta}{\ef}$}}
\put(38.75,22.0) {\vector(-2,-1){7}}
\put(39.5,21.0) {\large{FFLO}}
\put(25.5,42.5) {\large{N}}
\put(-6,17) {\large{PS}}
\put(13,29) {$\scriptstyle\Delta N_{\rm FFLO}$}
\put(18,20) {$\scriptstyle\Delta N_{c2}$}
\end{picture}
\vspace{-.5cm}
\caption{(Color Online) Global phase diagram (a) as a function of Feshbach resonance detuning 
at fixed population difference 
$\frac{\Delta N}{N}=\frac{N_\uparrow - N_\downarrow}{N_\uparrow + N_\downarrow}$. The regime of
 phase separation (PS) is gray, and 
the FFLO regime (too thin to see on this scale) intervenes along a thin line at 
positive detuning, indicated with
 a thick dark (red online) line.
(b) Zoom in of the BCS regime at positive detuning showing the regime of phase separation and the 
FFLO regime.}
\label{fig:mphasetwo} 
%
%  phasediagram_twochannel.nb (both)
%
\end{figure}
%------------------------------

\begin{figure}[bth]
\vspace{2.5cm}\hspace{-1cm}
\centering
\setlength{\unitlength}{1mm}
\begin{picture}(40,40)(0,0)
\put(-55,5){\begin{picture}(0,0)(0,0)
\includegraphics{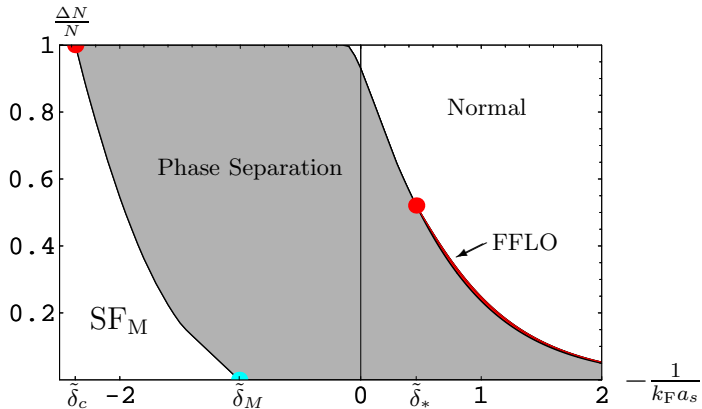}
\end{picture}}
\put(-16,57) {$\frac{\Delta N}{N}$}
\put(60,9) {\large{$-\frac{1}{\kf \as}$}}
\put(-14,6.8) {$\tilde{\delta}_c$}
\put(8,6.8) {$\tilde{\delta}_M$}
\put(31.5,6.8) {$\tilde{\delta}_*$}
\put(41.75,28.5) {\vector(-2,-1){4}}
\put(42.5,27.5) {FFLO}
\put(36.5,45.5) {Normal}
\put(-2,37.5) {Phase Separation}
\put(-11,17) {\large{SF$_{\rm M}$}}
\end{picture}
\vspace{-.5cm}
\caption{Polarization $\Delta N/N$ vs. detuning $\propto-\frac{1}{\kf \as}$ 
phase diagram of the one-channel model (appropriate for resonance width $\gamma\to \infty$)
within mean-field theory
 showing regimes of FFLO, superfluid (SF), magnetized superfluid (\sfm),
and  phase separation (PS). Note that here (in contrast to a narrow Feshbach resonance $\gamma \ll 1$, 
Fig.~\ref{fig:mphasetwo}) at unitarity, $(\kf \as)^{-1} = 0$, the boundary between N and PS is at 
$\Delta N/N<1$,
consistent with experiments~\cite{Zwierlein05}.}
\label{fig:globalmagphasesingleintro} 
%
%  magplotssingle.nb
%  
\end{figure}

It is now straightforward to convert the phase diagram and our other
predictions to the more experimentally-relevant ensemble of  fixed
total atom number $N$ and imposed atomic species difference
(polarization) $\Delta N=N_\uparrow-N_\downarrow$. As can be seen in
Fig.~\ref{fig:mphasetwo}, in this ensemble for positive detuning (BCS and crossover
regimes), the singlet BCS superfluid is confined to the detuning axis,
and is unstable to phase separation and coexistence with a normal
atomic gas for any imposed population imbalance. The upper boundary of
the coexistence region and the phase boundary of the FFLO phase are
then respectively given by
\bea
\frac{\Delta N_{c2}}{N} &\approx& \frac{3\Delta_F}{2\sqrt{2}\ef} \exp\big[- \frac{\delta \Delta_F^2 }{16\gamma\ef^3} \big],  
\\
\frac{\Delta N_{\rm FFLO}}{N} &\approx&   \frac{3\eta\Delta_F}{2\ef} 
\exp\big[- \frac{\eta^2\delta \Delta_F^2 }{8\gamma\ef^3} \big].  
\eea
As seen from the phase diagram these two boundaries cross at
$\delta_*$, thereby eliminating the FFLO state for lower detuning. For a narrow Feshbach resonance ($\gamma \ll 1$) at
even lower detuning (see Fig~\ref{fig:mphasetwo}(a))
\be
\delta_p \simeq 2^{2/3} \ef,
\ee
the normalized critical polarization $\Delta N_{c2}/N$ reaches unity according to
\be
\frac{\Delta N_{c2}}{N} \approx  1-\ef^{-\frac{3}{2}}(\delta - \delta_p)^{3/2}
\Theta(\delta- \delta_p).
\ee
This implies that for these lower detunings, $\delta < \delta_p$, only
a fully polarized normal Fermi gas can appear.  In contrast, as illustrated in 
Fig.~\ref{fig:globalmagphasesingleintro}, for a broad resonance (large $\gamma$) $\Delta N_{c2}/N<1$, consistent with 
experiments~\cite{Zwierlein05} that find $\Delta N_{c2}/N \approx 0.7$ near unitarity.  

For sufficiently negative detuning, $\delta < \delta_M$, the uniform magnetic
 superfluid state, \sfm, appears at low population
imbalance. For $\delta_c<\delta<\delta_M$ it becomes unstable to phase
separation and coexistence with the fully-polarized atomic Fermi gas for
$\Delta N > \Delta N_{c1}$, with
\be 
\frac{\Delta N_{c1}}{N}\approx 0.029 \Big( \frac{\delta}{\ef}  \Big)^{3/2}.
\ee
As this population imbalance is approached, the molecular
scattering length $a_m(\delta, \Delta N)$ drops precipitously 
and the corresponding molecular condensate Bogoliubov sound
velocity 
\be
 u(\delta,\Delta N)\simeq 
u_0(\delta)\sqrt{{\textstyle 1-\frac{\Delta N}{N}}}\sqrt{F_4\Big(1+{\textstyle\frac{2^{5/3}}{|\deltah|} 
 \big(\frac{\Delta N}{N}\big)^{2/3}}\Big)},
\ee
(where $\deltah \equiv \delta/\ef$) vanishes according to 
\be
 u\sim 0.38 u_0 \big(1-\frac{\Delta N_{c1}}{N}\big)^{1/2}
\big(1-\frac{\Delta N}{\Delta N_{c1}}
\big)^{1/2}, \,\, \Delta N \to \Delta N_{c1}. 
\ee
Finally, for $\delta < \delta_c$, the \sfm state is stable for any population
imbalance up to the fully polarized limit of $\Delta N =N$, where it is
identical to the fully polarized (single species) Fermi gas.

%-----------------------------
%
% fig%4
%
%------------------------------

%-----------------------------

\subsection{Trapped system}
\label{SEC:SummaryLDA}

To make a more direct connection with atomic-gas experiments we summarize
the extension of the above bulk results to the case of a trapped gas. This
is particularly simple to do in the case of a trap potential that
varies smoothly on the scale of the Fermi wavelength, i.e., when the Fermi
energy is much larger than the trap level spacing, the regime of most
atomic gas experiments. In this regime, the trap potential can be
easily taken into account via the local density approximation
(LDA). Namely, we treat the trapped system as a locally uniform one
(thereby taking advantage of our bulk results), but with a local
chemical potential given by $\mu(\br) = \mu - V_T(\br)$. The true chemical potential 
$\mu$
and chemical potential difference $h$ still appear and are determined by
constraints of the total atom number $N$ and species imbalance $\Delta
N$. For simplicity we consider an isotropic trap with $V_T(\br)=\frac{1}{2} m \Omega_T^2 r^2$, but
our results can be easily generalized to an arbitrary anisotropic
trap.

The phase transitions and coexistence discussed above are strikingly
accentuated by the trap. To see this, note that (within the LDA
approximation) the local  phenomenology of a trapped
cloud is that of the bulk one at an effective chemical potential
$\mu(r)$. Hence, a radial slice through a trapped cloud is an effective
chemical potential \lq\lq scan\rq\rq\ through the bulk-system phase diagram at 
fixed $\mu$ and $h$ (the latter displayed in the main text, Fig.~\ref{fig:phasediagramfixedmu}). 
Consequently, as we first predicted in our earlier
publication~\cite{shortpaper}, a trapped cloud consists of a combination of
shells of the various superfluid and normal phases, with the exact structure
determined by the value of detuning and population imbalance (or,
equivalently, the Zeeman field $h$). A critical value of the chemical potential
$\mu_c$ separating two phases in the bulk system translates, within LDA, into a critical
radius $r_c$, given by $\mu(r_c)=\mu_c$ (more generally a hypersurface),
that is the boundary between two corresponding phase shells in a trapped
gas.

%-----------------------------
%
% fig%5
%
\begin{figure}[bth]
\vspace{3.2cm}
\centering
\setlength{\unitlength}{1mm}
\begin{picture}(40,65)(0,0)
\put(-52,42){\begin{picture}(0,0)(0,0)
\includegraphics{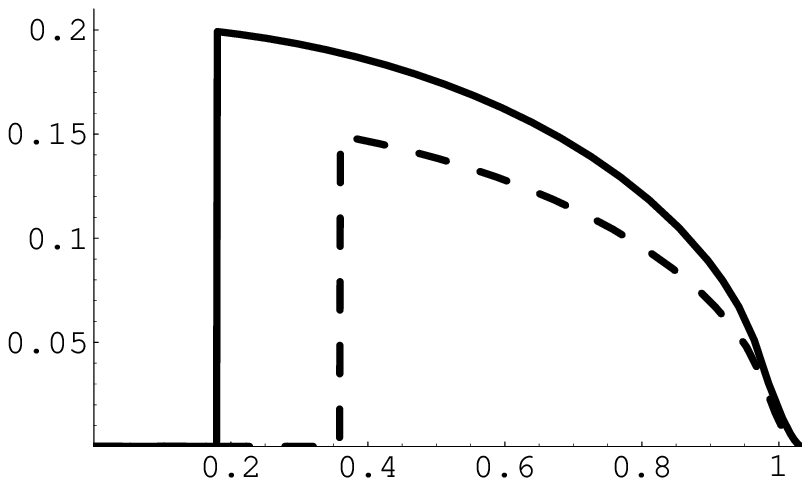}
\end{picture}}
\put(0,-3){\begin{picture}(0,0)(0,0)
\includegraphics{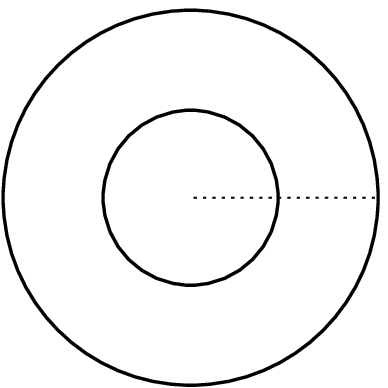}
\end{picture}}
\put(-18,92) {\large{(a)}}
\put(-18,37) {\large{(b)}}
\put(62,47) {\large{$\frac{r}{R_{0}}$}}
\put(-10,91) {\large{$\mg(r)/\mg_0$}}
\put(19,20) {\large{SF}}
\put(31.5,19.5) {\large{$r_c$}}
\put(42.75,19.0) {\large{$R_0$}}
\put(19.5,31.5) {\large{N}}
\end{picture}
%\vspace{-.5cm}
\caption{a) Local magnetization $\mg(r)$ radial profile
 confined to the normal outer shell of the cloud, $r>r_c$,
for coupling $(\kf |\as|)^{-1} = 1.5$ and 
$\frac{\Delta N}{N} = 0.15$ (dashed) and 
$\frac{\Delta N}{N} = 0.20$. (b) Sequence of shells, with increasing
radius, implied by the magnetization profiles in (a). }
\label{fig:magplotlda}
%
%   
% 25OctLDAnumbers2.nb
% shellfig1.fig
% 
\end{figure}
%------------------------------

More concretely, on the BCS side of the resonance (ignoring for
simplicity the narrow sliver of the FFLO state), the normal and
(singlet BCS-) SF phases translate into two shells of a trapped
cloud. Because the bulk $h_c(\mu)$ [see Eq.~(\ref{eq:deltanoughtsummary})] is an increasing
function of $\mu$ and $\mu(r)$ is a decreasing function of $r$, the SF phase
(that in the bulk appears at high chemical potential and low $h$) forms the cloud's
inner core of radius $r_c(\delta,h,N)$ set by $h_c(\mu(r_c)) = h$. The
normal phase forms the outer shell with inner radius
$r_c(\delta,h,N)$ 
 and outer shell radius
$R_0(\delta,h,N)$ determined by the vanishing of the normal phase chemical
potential, $\mu_N(R_0,\delta,h,N)=0$. The resulting magnetization density
profile $m(r)$, confined to (and thereby vividly imaging) the outer
normal shell, is illustrated in Fig.~\ref{fig:magplotlda}a.
Following our original prediction~\cite{shortpaper}, this shell structure has
recently been experimentally observed in Refs.~\onlinecite{Zwierlein05,Partridge05},
and subsequently calculated theoretically by a number of groups.\cite{Torma,Chevy,Haque,DeSilva}

%-----------------------------
%
% fig%6
%
\begin{figure}[bth]
\vspace{3.2cm}
\centering
\setlength{\unitlength}{1mm}
\begin{picture}(40,65)(0,0)
\put(-52,42){\begin{picture}(0,0)(0,0)
\includegraphics{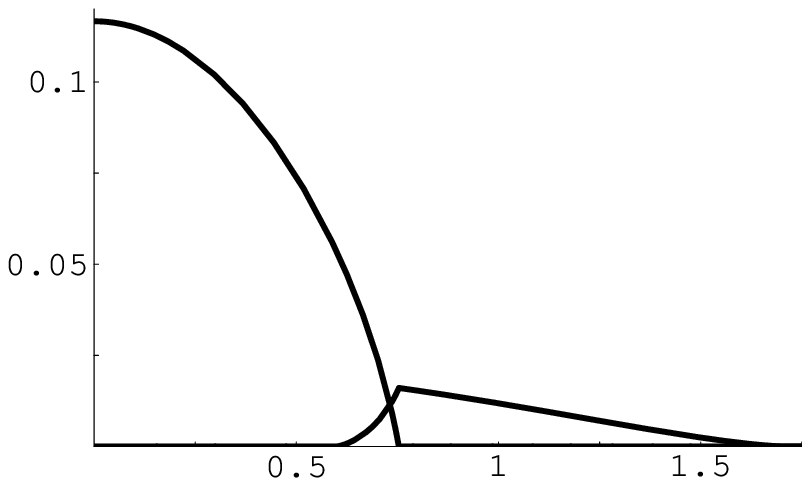}
\end{picture}}
\put(0,-3){\begin{picture}(0,0)(0,0)
\includegraphics{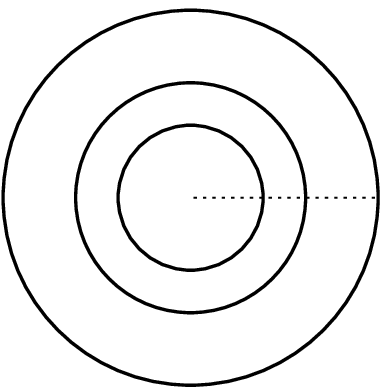}
\end{picture}}
\put(-18,92) {\large{(a)}}
\put(-18,37) {\large{(b)}}
\put(58,47) {\large{$\frac{r}{R_{TF0}}$}}
\put(-1,85) {\large{$\nb_m(r)$}}
\put(25,53) {\large{$\mb(r)$}}
\put(19,19) {\large{SF}}
\put(29.5,19.25) {$\scriptstyle R_{f1}$}
\put(34.25,19.25) {$\scriptstyle R_{TF}$}
\put(42.25,19.25) {$\scriptstyle R_{f2}$}
\put(18.5,27) {\large{\sfm}}
\put(19.5,34.5) {\large{N}}
\end{picture}
%\vspace{-.5cm}
\caption{a) Normalized molecular density $\nb_m(r)$ and normalized
magnetization $\mb(r)$ of a trapped fermion gas
as a function of radius $r$ (normalized 
to the radius of the unpolarized cloud, with parameters given in 
Sec.~\ref{SEC:ldabec}). b) Schematic of the sequence of shells, 
with increasing radius, implied by the curves in (a). }
\label{fig:ldabecintro}
%
%   
% 15June06.nb 
% shellfig.fig
% 
\end{figure}
%------------------------------

On the BEC side of the resonance, the appearance of an additional phase, the \sfm,
allows for the possibility of a triple-shell cloud structure
SF$\to$\sfm$\to$N with increasing radius at low population imbalance (small
$h$).  The inner ($R_{f1}$) and outer  ($R_{TF}$) boundaries of the \sfm shell occur where 
the population imbalance $m(r)$ becomes nonzero at the SF-\sfm boundary and where the 
molecular density $n_m(r)$ vanishes at the \sfm-N boundary, respectively. The radii of the
shells are given by the bulk critical fields $h_m(\mu(R_{f1}))=h$ and
$h_c(\mu(R_{TF}))=h$. The cloud's radial profile for $n_m(r)$ and $m(r)$ for this
case is illustrated in Fig.~\ref{fig:ldabecintro}. For larger population imbalance $R_{f1}$ is
driven to zero, resulting in a double-shell \sfm$\to$N structure. At even
higher $\Delta N$, $R_{TF}$ vanishes leading to a fully polarized, normal
cloud of radius $R_{f2}$.

Another interesting prediction of our work~\cite{shortpaper} is the possibility to
realize in resonantly-interacting degenerate Fermi gases the enigmatic
FFLO state~\cite{ff,lo}. In an idealized box-like trap the molecular occupation distribution
$n_\bq=\langle b^\dagger_\bq b_\bq^\phdag\rangle$ of this inhomogeneous 
superfluid exhibits finite
momentum pairing peaks set by Fermi surface mismatch $Q(\delta)\sim
k_{{\rm F}\uparrow} - k_{{\rm F}\downarrow}$
\be
Q(\delta)\approx\frac{\sqrt{2}\lambda \deltabcs(\ef)}{\hbar \vf},
\ee
(here $\lambda \simeq 1.200$) 
corresponding to reciprocal lattice vectors of this
supersolid~\cite{SingleQ}. This will translate into {\em spontaneous}
Bragg peaks (observable by projecting onto a molecular condensate~\cite{Regal})
appearing at $\hbar Q t/m$ in the cloud's shadow images
after expansion time $t$, akin to that exhibited by a superfluid
released from a lattice potential~\cite{GreinerOL}, where, in contrast to the
FFLO state, Bragg peaks are explicitly induced by the periodicity of
the optical potential.

However, realistic magnetic and optical traps produce a harmonic (more
generally, power-law, rather than box-like) confining potential, that
leads to a large density variation in the gas. Consequently, following
the above arguments, within LDA in a realistic trap the gas can only
exhibit a thin shell of FFLO phase, whose width $\delta r$ (among other parameters,
$N$, $\Delta N$, and $\delta$) is set by the inverse trap curvature and
proportional to the difference in critical chemical potentials for the
transition into the FFLO and normal phases. Clearly, to resolve the abovementioned 
Bragg peaks, the Bragg peak width $\hbar/\delta r$ must be much smaller than $Q$, or, 
equivalently, the FFLO shell width $\delta r$ must be larger than the FFLO period $2\pi/Q$.  
Since FFLO state is confined to a narrow sliver of the phase diagram and to be meaningful
(i.e., LDA reliable) the shell width must be much larger than the period $2\pi/Q$, its 
direct observation in density profiles may be difficult for present-day experiments.
Similar conclusions about the effect of the trap on the FFLO phase have been
found in recent work by Kinnunen et al.~\cite{Torma} that does not 
rely on the LDA approximation.

On the other hand, the identification of the FFLO state through less
direct probes maybe possible. For example, its spontaneous breaking of
orientational symmetry for an isotropic trap should be detectable in
noise experiments\cite{Altman,Greiner05,Lamacraft} sensitive to angle-dependence of pairing
correlations across a Fermi surface. 

Furthermore, gapless atomic excitations available in the FFLO and \sfm
states should be observable through Bragg spectroscopy and reflected
in thermodynamics such as power-law (rather than nearly activated
paired-superfluid) heat capacity. The latter should also exhibit a
latent heat peak across the first-order \sfm$\to$N, SF$\to$FFLO, FFLO$\to$N~\cite{MFTcomment},
and SF$\to$N phase 
transitions, and a nearly mean-field singularity across the continuous
SF$\to$\sfm transition.

Finally, we note that our above results (that are quantitatively
accurate at least in the narrow resonance limit) show no evidence of a
homogeneous but magnetized superfluid near the unitary limit $\as \to \infty$, nor in
the BCS and crossover regimes defined by positive detuning. We find
unambiguously that this magnetized superfluid  phase only appears on the BEC side for
sufficiently negative detuning, $\delta < \delta_M$, 
embodied in the \sfm ground state. 
We do, however, find that the corresponding inverse scattering length 
$(\kf a_{sM})^{-1} = -2\deltah_M/\pi \gamma$ shifts toward the unitary limit with increased
resonance width (saturating at $(\kf a_{sM})^{-1} \simeq 1$ within mean-field theory)
 and upon increasing temperature. Whether this is
sufficient or not to explain the putative existence of such a phase,
as claimed in the Rice experiments~\cite{Partridge05} (that were based on an apparent
observation, in the unitary limit, of a critical population imbalance for the
transition to phase separation), remains an open question.

%%%%%%%%%%%%%%%%%%%%%%%%%%%%%%%%%%%%%%%%%%%%%%%%%%%%%%%%%%%%%%%%%%%%%%%%%%%%%%%%%%%%%%%%%%%%%%%%%%%%%%%%%%%%%%
%%%%%%%%%%%%%%%%%%%%%%%%%%%%%%%%%%%%%%%%%%%%%%%%%%%%%%%%%%%%%%%%%%%%%%%%%%%%%%%%%%%%%%%%%%%%%%%%%%%%%%%%%%%%%%
%%%%%%%%%%%%%%%%%%%%%%%%%%%%%%%%%%%%%%%%%%%%%%%%%%%%%%%%%%%%%%%%%%%%%%%%%%%%%%%%%%%%%%%%%%%%%%%%%%%%%%%%%%%%%%

\section{Two-channel model of s-wave Feshbach resonance}
\label{SEC:Hamiltonian}

  The two-channel model of fermions interacting via an s-wave Feshbach resonance,
briefly discussed in Sec.~\ref{SEC:tf}, 
describes open-channel fermions ($\ch_{\bk\sigma}$) and closed-channel molecular bosons ($\bh_\bq$) coupled by 
molecule-atom interconversion~\cite{Timmermans99,Holland,Ohashi,agr,Rumer}.  It is 
characterized by the following model Hamiltonian:
\bea
\nonumber 
&&\hspace{-.9cm}\curH =  \sum_{\bk,\sigma} \epsilon_k  \ch_{\bk\sigma}^{\dagger} \ch_{\bk\sigma}^{\phdag}
 +\sum_\bq \big(\frac{\epsilon_q}{2} + \delta_0\big)\bh_\bq^\dagger \bh_\bq^\phdag
\\
\label{eq:bareham}
&&\hspace{-.9cm}\quad +\frac{g}{\sqrt{\vol}}\sum_{\bk,\bq}\Big( \bh^\dagger_\bq 
\ch_{\bk+\frac{\bq}{2}\downarrow}^{\phdag}\ch_{-\bk+\frac{\bq}{2}\uparrow}^{\phdag}
+ \ch_{-\bk+\frac{\bq}{2}\uparrow}^{\dagger}\ch_{\bk+\frac{\bq}{2}\downarrow}^{\dagger}\bh^\phdag_\bq
\Big),
\eea
where $\vol$ is the system volume (that we shall generally set to unity) $\epsilon_k \equiv k^2/2m$, 
$m$ is the atom mass and the \lq\lq bare\rq\rq\ detuning $\delta_0$
is related to the position of the Feshbach resonance  $\delta$ in
a way that we determine below. The molecule-atom interconversion term
is characterized by a coupling $g$ that measures the amplitude for the
decay of a closed-channel s-wave singlet diatomic molecule into a pair
of open-channel fermions. 

Although within some approximation the model $\curH$, Eq.~(\ref{eq:bareham}), can be
derived from a more microscopic starting point of purely fermionic
atoms interacting via a van der Waals potential and including exchange
and hyperfine interactions  (see e.g., Ref.~\onlinecite{Timmermans01}), the model's ultimate
justification is that, as we will show below, it reproduces the
Feshbach resonance phenomenology. Namely, for positive detuning $\delta>0$
the model exhibits a resonance at rest-energy $\delta$, whose width is
controlled by $g$, and exhibits a true molecular bound state at negative
detuning $\delta<0$.  Associated with the resonance the open-channel s-wave scattering
length $\as(\delta)$ diverges and changes sign as $-1/\delta$, when the
resonance at $\delta$ is tuned through zero energy.

\subsection{Scattering theory in the vacuum}

%-----------------------------
%
% fig%7
%
\begin{figure}[bth]
\vspace{1cm}
\centering
\setlength{\unitlength}{1mm}
\begin{picture}(40,50)(0,0)
%\put(-10,50){\begin{picture}(0,0)(0,0)
%\special{psfile=sfmpic1.eps vscale = 110 hscale=110}
%\end{picture}}
%
%
%
\put(-2,32){\begin{picture}(0,0)(0,0)
\includegraphics{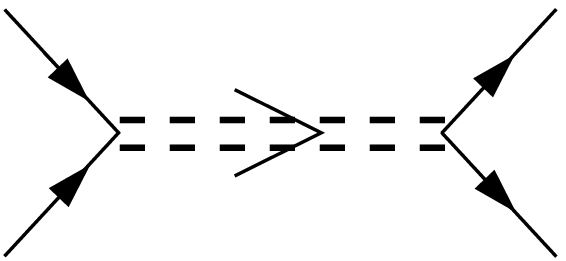}
\end{picture}}
\put(-20,10){\begin{picture}(0,0)(0,0)
\includegraphics{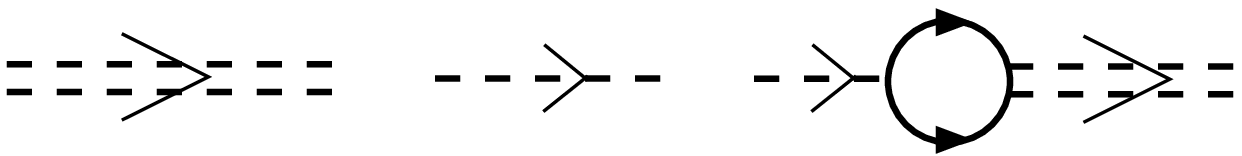}
\end{picture}}
\put(-20,55) {\large{$(a)$}}
\put(-20,25) {\large{$(b)$}}
\put(4.5,14.25) {$=$}
\put(25.5,14.25) {$+$}
\end{picture}
%\vspace{-.5cm}
\caption{(a) Feynman diagram corresponding to the atom scattering amplitude, with solid lines
indicating atoms and the single (double) dashed lines indicating the 
bare (full) molecular propagator. (b) A self-consistent
equation for the molecular propagator.}
\label{fig:scatteringdiagram} 
%
%  scatteringdiagram.fig
%
\end{figure}
%------------------------------

The above properties follow directly from the scattering amplitude $f_0(E)$,
that can can be easily computed exactly for two atom scattering in the
vacuum~\cite{agr,Gurarie}. The latter 
follows from the T-matrix $f_0(E)=-(m/2\pi\hbar^2) T(E)$ that, in a
many-body language, is given by the diagram in Fig.~\ref{fig:scatteringdiagram}a, with the
molecular propagator {\it exactly\/} given by the RPA-like geometric series of
fermion polarization bubbles illustrated in Fig.~\ref{fig:scatteringdiagram}b. With the details
of the calculations relegated to Appendix~\ref{SEC:scatteringamplitude} (see
also Refs.~\onlinecite{agr,Gurarie}), we find at low $E$
\be
f_0(E) = -\frac{\hbar}{\sqrt{m}}\frac{\sqrt{\width}}{E - \delta + i \sqrt{\width} \sqrt{E}},
\label{eq:scattampfin}
\ee
where $\delta$ is 
the renormalized (physical) detuning
and $\width$ is a parameter characterizing the width of the resonance, respectively given by 
\bea
\label{eq:deltaintro}
\delta &=& \delta_0 - g^2 \int \frac{d^3 p}{(2\pi\hbar)^3} \frac{m}{p^2},
\\
\width &\equiv& \frac{g^4 m^{3}}{16\pi^2\hbar^6}.
\label{Eq:width}
\eea
The integral in Eq.~(\ref{eq:deltaintro}) is cut off by the ultraviolet scale $\Lambda\simeq 2\pi/d$,
set by the inverse size $d$ of the closed-channel (molecular) bound state, so that
\be
\label{eq:deltaintrop}
\delta = \delta_0 - \frac{g^2m\Lambda}{2\pi^2\hbar^2},
\ee
giving the relation between the bare and physical detuning.  

The s-wave scattering length~\cite{LandauQM} $\as = -f_0(0)$ is then given by 
\be
\label{eq:as}
\as(\delta) = -\hbar
\sqrt{\frac{\width}{m}}
\frac{1}{\delta}.
\ee
Thus, to reproduce the experimentally observed dependence of the scattering 
length on the magnetic field $B$, Eq.~(\ref{eq:scatteringlengthformula}),
we take~\cite{Timmermans99} $\delta \approx
2\mu_B (B-B_0)$, representing the Zeeman energy splitting (at large 
two-atom separation) between the open and closed channels. The relevant
magnetic moment is naturally dominated by the two-atom {\it electronic\/} spin
state (hence the proportionality constant of the Bohr magneton $\mu_B$ above),
that is approximately a singlet for the closed channel and a triplet for the
open channel. Detailed multi-channel calculations (see, e.g., Ref.~\onlinecite{Kohler})
 allow a more accurate determination of
parameters when necessary. Equating $\as$ in Eq.~(\ref{eq:as}) 
with $\as^{\rm exp.}$,
Eq.~(\ref{eq:scatteringlengthformula}), allows us to determine the Feshbach resonance energy width $\width$ (or
equivalently the parameter $g$) arising in the model in terms of the 
experimentally measured \lq\lq width\rq\rq\ $B_w$ defined by the dependence of the scattering length
on $B$ in Eq.~(\ref{eq:scatteringlengthformula}):
\be
\width \approx \frac{4m\mubohr^2 a_{bg}^2 B_w^2}{\hbar^2}.
\ee

The bound states of the model are determined by the real negative-energy 
poles of $f_0(E)$, that can be shown to appear only at negative
detuning\cite{commentPhysicalBoundState}, $\delta < 0$. For small negative detuning, 
$|\delta| \ll \width$, the pole is at low energy, $E \ll \width$, so that $E$ in the
denominator of $f_0(E)$ can be neglected relative to $\sqrt{\width E}$. At more
negative detuning, $|\delta|\gg \width$, one can instead ignore the
$\sqrt{\width E}$ and the pole is simply given by the detuning $\delta$.
Together these limits give:
\bea
\label{sixpole}
\hspace{-0.7cm}
E_p(\delta)
&\approx&
\begin{cases}    -\frac{\delta^2}{\width}=-\frac{\hbar^2}{2m \as^2}, 
& \text{for $|\delta|\ll \width$,}\cr
 \delta, 
& \text{for $|\delta|\gg \width$,}\cr
\end{cases}
\eea
a standard result\cite{LandauQM} consistent with the observed phenomenology of
Feshbach resonances~\cite{Regal03,Bartenstein05}. The complete $E_p(\delta)$ interpolating
between these limiting expressions is given by
\bse
\label{eq:polepositionboth}
\bea
E_p(\delta) &=& -\frac{\width}{2} \Big[1+\frac{2|\delta|}{\width} - \sqrt{1+\frac{4|\delta|}{\width}}
\Big],\label{eq:polepositionpre}
\\
 &=& -\frac{2\hbar^2}{m r_0^2} \Big[1+\frac{|r_0|}{\as} - \sqrt{1+\frac{2|r_0|}{\as}} \Big],
\label{eq:poleposition}
\eea
\ese
obtained by solving the quadratic equation $1/f_0(E_p)=0$. In Eq.~(\ref{eq:poleposition}) we
expressed $E_p$ in terms of the scattering length $\as$ and the effective
range parameter $r_0= - 2\hbar/\sqrt{m \width}$. We note in passing that,
unlike the case of a non-resonant short-range potential where $r_0 > 0$
and measures the range of the potential (hence the name), here, for a
resonant interaction, $r_0 < 0$ is negative, with its
magnitude characterizing the closed-channel molecular lifetime
$\tau\sim \hbar/\sqrt{\width\delta}$ at positive detuning~\cite{agr,Gurarie}. In terms of $\as$
and $r_0$ the scattering amplitude takes the standard form\cite{LandauQM}
\be
f_0(k) =\frac{1}{-\as^{-1} + r_0 k^2/2 - ik},
\label{eq:fk}
\ee
where $k=\sqrt{2m_r E/\hbar^2}$ and $m_r = m/2$ is the two-atom reduced
mass.

%-----------------------------
%
% fig%8
%
\begin{figure}[bth]
\vspace{1.4cm}
\centering
\setlength{\unitlength}{1mm}
\begin{picture}(40,40)(0,0)
\put(-50,0){\begin{picture}(0,0)(0,0)
\includegraphics{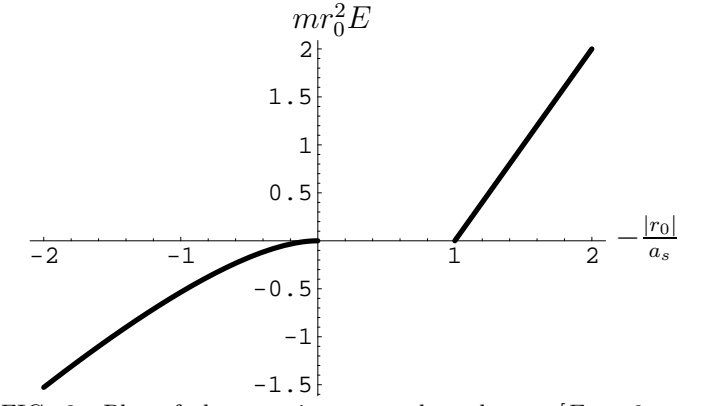}
\end{picture}}
\put(16,50) {\large{$mr_0^2 E$}}
\put(59,21.0) {\large{$-\frac{|r_0|}{\as}$}}
\end{picture}
\vspace{-.5cm}
\caption{Plot of the negative-energy bound-state [$E<0$, Eq.~(\ref{eq:poleposition})] and 
positive-energy resonance  [$E>0$, Eq.~(\ref{eq:resonanceposition})]  positions as
a function of  $-|r_0|/\as$.}
\label{fig:poleplot}
%
%  poleposition.nb 
%
\end{figure}
%------------------------------

As can be seen by analyzing Eq.~(\ref{eq:scattampfin}), the physical bound state
disappears~\cite{commentPhysicalBoundState} for positive $\delta$.
However, as illustrated in Fig.~\ref{fig:poleplot}, a positive energy resonance does not
appear until $\delta$ reaches the threshold value of $\delta_* = \width/2$,
corresponding to $|\as(\delta_*)| = |r_0|$. This absence of a
resonance for a range of positive detuning $0< \delta < \delta_*$ is a
property unique (due to the absence of a centrifugal barrier) to an s-wave
resonance, contrasting with finite angular momentum
resonances~\cite{Gurariepwave,Gurarie}.

For larger positive detuning, $\delta > \width/2$ ($|\as|<|r_0|$), a finite-width
resonance appears at a complex $E_p(\delta)$ given by 
\be
E_p(\delta) = E_r(\delta) -i\Gamma(\delta) ,
\ee
with
\bse
\bea
&&\hspace{-1cm}E_r(\delta) =  \delta - \frac{1}{2}\width = \frac{2\hbar^2}{m r_0^2} \big(\frac{|r_0|}{|\as|} -1\big),
\label{eq:resonanceposition}
\\
&&\hspace{-1cm}\Gamma(\delta)= 
 \frac{\width}{2}\sqrt{\frac{4\delta}{\width} -1}=\frac{2\hbar^2}{m r_0^2}\sqrt{2|r_0|/|\as| -1},
\eea
\ese
where the imaginary part $\Gamma$ measures the resonance width
(i.e., molecular decay rate is $\Gamma/\hbar$).

\subsection{Scattering at finite density: small parameter}

As already discussed in the Introduction, at finite density $n=\kf^3/3\pi^2$ the two-channel model
admits a dimensionless parameter $\gamma \propto 1/(\kf |r_0|)$ that is the ratio of
the average atom spacing $n^{-1/3}$ to the effective range.  The parameter $\gamma$ controls
a perturbative expansion (about an exactly solvable non-interacting $g=0$ limit) 
of any physical quantity and is given by 
\be
\gamma \equiv  \frac{g^2 N(\ef)}{\ef} = \frac{\sqrt{8}}{\pi} \sqrt{\frac{\width}{\ef}}
=\frac{g^2c}{\sqrt{\ef}} = \frac{8}{\pi} \frac{1}{\kf |r_0|},
\label{eq:gammadef}
\ee
related to 
the ratio of the Feshbach resonance width (controlled by the atom-molecule coupling $g$)
to the Fermi energy. 
The key observation is that $\gamma$ is independent of the scattering length $\as$ and detuning $\delta$,
and as such, once set small, remains small throughout the crossover, even for a Feshbach resonance 
tuned through zero. 
Hence, two-channel model predictions for a {\em narrow} Feshbach resonance, (defined by
$\gamma\ll 1$, i.e., width of the resonance much smaller than the Fermi
energy) are quantitatively accurate throughout the BEC-BCS crossover. 

This contrasts qualitatively with a one-channel model characterized by a dimensionless
gas parameter $n\as^3$ that diverges for a Feshbach resonance tuned to zero (i.e., \lq\lq on resonance\rq\rq) and 
therefore does not exhibit a small expansion parameter throughout the crossover. 

The dimensionless parameter $\gamma$ naturally emerges in a
perturbative expansion in atom-molecule coupling. More physically, 
it can also be deduced by estimating the ratio of the energy associated with
the atom-molecule Feshbach-resonance interaction to the kinetic energy,
i.e, the non-interacting part of the Hamiltonian Eq.~(\ref{eq:bareham}). To see this 
note that the atom-molecule coupling energy 
$E_{FR}$ per atom scales like 
\be
E_{FR} \sim g n^{1/2},
\ee
where we estimated the value of $\bh(\br)$ by $\bh\sim \sqrt{n}$. This interaction
energy is to be compared to the non-interacting part of the
Hamiltonian, i.e., the kinetic energy per atom
\be
E_0 \sim \epsilon_F,
\ee
with the square of the ratio 
\bea
\gamma&\sim& (E_{FR}/E_0)^2,
\\
&\sim& g^2 n/\ef^2,
\eea
giving the scale of the dimensionless parameter $\gamma$ in Eq.~(\ref{eq:gammadef}).

Another instructive way to estimate the interaction strength and to
derive the dimensionless coupling that controls perturbation theory is
to integrate out (in a coherent-state path-integral sense) the
closed-channel molecular field $b(\br)$ from the action. As $b(\br)$ couples to
atoms only linearly this can be done exactly by a simple Gaussian
integration. The resulting action only involves fermionic atoms that
interact via an effective four-Fermi {\it dispersive\/} vertex.  After
incorporating fermion-bubble self-energy corrections of the T-matrix
the latter is given by $T_k= (4\pi\hbar^2/m)f_\mu(k)\approx (4\pi \hbar^2/m)f_0(\kf)$, with a key factor
that is the finite-density analog of the scattering amplitude, $f_0(k)$,
Eq.~(\ref{eq:fk}). To gauge the strength of the molecule-mediated
interaction energy we compare the interaction per atom $(4\pi \hbar^2/m)f_0(k)n$ 
to the kinetic energy per atom $\ef$. Hence the dimensionless coupling
that is a measure of the atomic interaction is
\bea
\lambdah_k &\equiv&(4\pi \hbar^2/m)|f_0(\kf)| n/\ef, \\
&\sim & k_F |f_0(k_F)|,
\eea
dropping numerical prefactors.
At large detuning (i.e., deep in the BCS regime) $\lambdah_k\sim \kf |\as|\ll
1$ and the theory is perturbative in $\lambdah_k$.  However, as detuning is
reduced $|\as(\delta)|$ and $\lambdah_k(\delta)$ grow, and close to the resonance
$\as^{-1}$ may be neglected in the denominator of Eq.~(\ref{eq:fk}).  In this
regime, the coupling saturates at $\lambdah_k^\infty$:
\be
\lambdah_k^\infty\sim \frac{\kf}{|r_0 \kf^2/2 - i\kf|},
\ee
whose magnitude crucially depends on the dimensionless ratio $\gamma\propto
1/(k_F |r_0|)$, with 
\bea
\label{six}
\hspace{-0.7cm}
\lambdah_k^\infty
&\sim&
\begin{cases} \frac{1}{r_0\kf},
& \text{for $|r_0| \kf \gg 1$,}\cr
1,
& \text{for $|r_0| \kf \ll 1$.}\cr
\end{cases}
\eea
Hence, in contrast to two-particle vacuum scattering, in which  the cross
section diverges when the Feshbach resonance is tuned to zero energy, 
 at finite density, for
sufficiently large $\as$, the effective coupling $\lambdah_k$ ceases to grow
and saturates at $\lambdah_k^\infty$, with the saturation value depending on
whether this growth is cut off by the atom spacing $1/\kf$ or the effective
range $r_0$. The former case corresponds to a narrow resonance [$\gamma \propto (|r_0|\kf)^{-1} \ll 1$], 
with the
interaction remaining weak (and therefore perturbative) throughout
the BCS-BEC crossover, right through the strong-scattering $\frac{1}{\kf |\as|}=0$ point. In contrast, 
in the latter 
wide-resonance case [$\gamma \propto (|r_0|\kf)^{-1} \gg 1$], sufficiently
close to the unitary point $1/\as=0$ the effective coupling $\lambdah_k^\infty$, Eq.~(\ref{six}), grows to
$\curO(1)$ precluding  a perturbative expansion in atom
interaction near the unitary point.

\subsection{Relation to single-channel model}

In this latter broad-resonance limit, of relevance to most
experimentally-realized Feshbach resonances to date, the $r_0 k^2$ contribution to the
dispersion (arising from the molecular kinetic energy) of the
effective coupling $\lambdah_k$ can be neglected and one obtains an
effective single (open-) channel description. 

The reduction to a single-channel model in the broad resonance limit
can be executed in an operator formalism, with the
derivation becoming exact in the infinite Feshbach resonance width
($\gamma\rightarrow\infty$) limit\cite{Levinsen}.
The expression for the scattering length
\be
\frac{1}{\as} = - \frac{4\pi\hbar^2}{m g^2}\big(\delta_0 - \frac{g^2m\Lambda}{2\pi^2\hbar^2}
\big),
\label{eq:as2}
\ee
obtained from Eqs.~(\ref{Eq:width}), 
(\ref{eq:deltaintrop}), and (\ref{eq:as})
dictates that a proper transition to the broad resonance limit
corresponds to $g\to\infty$ while adjusting the bare detuning according to
\be
\delta_0 =  -\frac{g^2}{\fermiint},
\label{eq:connection}
\ee
such that the physical scattering length $\as$ remains fixed. This allows us to
trade the bare detuning $\delta_0$ and coupling $g$ for a new coupling $\fermiint$
that physically corresponds to a non-resonant attractive interaction
depth\cite{Gurarie} that can be used to tune the scattering length.
Inserting Eq.~(\ref{eq:connection}) into $\curH$, Eq.~(\ref{eq:bareham}), we obtain ($\vol = 1$):
\bea
\nonumber 
&&\hspace{-.75cm}\curH =  
\sum_{\bk,\sigma} \epsilon_k  \ch_{\bk\sigma}^{\dagger} \ch_{\bk\sigma}^{\phdag}
 +\sum_\bq \big(\frac{\epsilon_q}{2} - \frac{g^2}{\fermiint}\big)\bh_\bq^\dagger \bh_\bq^\phdag
\\
\label{eq:barehamnew}
&&\hspace{-.75cm}\quad +g\sum_{\bk,\bq}\Big( \bh^\dagger_\bq 
\ch_{\bk+\frac{\bq}{2}\downarrow}^{\phdag}\ch_{-\bk+\frac{\bq}{2}\uparrow}^{\phdag}
+ \ch_{-\bk+\frac{\bq}{2}\uparrow}^{\dagger}\ch_{\bk+\frac{\bq}{2}\downarrow}^{\dagger}\bh^\phdag_\bq
\Big),
\eea
The corresponding Heisenberg equation of motion governing the molecular
field $\bh_\bq$ dynamics is given by 
\bse
\bea
&&\hspace{-1.2cm} \dot{\bh}_\bq = -\frac{i}{\hbar} \big[
 \bh_\bq,\curH
\big],
\\
&&\hspace{-.8cm}\,=
-\frac{i}{\hbar}\Big[\big(\frac{\epsilon_q}{2} - \frac{g^2}{\fermiint}\big) \bh_\bq + g \sum_\bk
\ch_{\bk+\frac{\bq}{2}\downarrow}^{\phdag}\ch_{-\bk+\frac{\bq}{2}\uparrow}^{\phdag}
\Big].
\label{moleom}
\eea
\ese 
Now, in the large $g\to\infty$ limit (keeping $\fermiint$ fixed) the molecular
kinetic energy term $\propto \epsilon_q/2$ on the right and 
the  $\dot{\bh}_\bq$ term on the left are clearly subdominant,
reducing the Heisenberg equation to a simple constraint relation
\be
\bh_\bq = \frac{\fermiint}{g} \sum_k
\ch_{\bk+\frac{\bq}{2}\downarrow}^{\phdag}\ch_{-\bk+\frac{\bq}{2}\uparrow}^{\phdag}.  
\label{318}
\ee
Hence, we see that in the extreme broad-resonance limit the molecular
field's dynamics is \lq\lq slaved\rq\rq\ to that of pairs of atoms according to
Eq.~(\ref{318}). Substituting this constraint into the Hamiltonian,
Eq.~(\ref{eq:barehamnew}), allows us to eliminate the closed-channel molecular field in favor
of a purely open-channel atomic model with the Hamiltonian 
\be
\curH 
=   \sum_{\bk,\sigma}\epsilon_k \ch_{\bk\sigma}^\dagger  \ch_{\bk\sigma}^\phdag 
+ \fermiint\sum_{\bk\bq\bp} \ch_{\bk\uparrow}^\dagger \ch_{\bp\downarrow}^\dagger \ch_{\bk+\bq\downarrow}^\phdag
\ch_{\bp-\bq\uparrow}^\phdag,
\label{eq:singlechannelintro}
\ee
where we redefined the momenta in the interaction term to simplify
the final expression. 

For future reference, we note that the scattering length in the one-channel 
model, when expressed in terms of the coupling $\fermiint$ [by
combining Eqs.~(\ref{eq:deltaintro}) and   ~(\ref{eq:as})], is 
\bea
\frac{m}{4\pi \as\hbar^2} &=& \frac{1}{\fermiint} + \int \frac{d^3 k}{(2\pi\hbar)^3} \frac{1}{2\epsilon_k},
\label{eq:singlechannelintroas}
\eea
a well-known formula that can also be derived from the Hamiltonian Eq.~(\ref{eq:singlechannelintro}) directly.
 Evaluating the momentum
integral (cut off at $\Lambda$) gives the scattering length in the one-channel 
model
\be
\as(\fermiint) = \frac{\pi}{2\Lambda} \frac{\fermiint}{\fermiint + 2\pi^2\hbar^2/m\Lambda},
\ee
whose behavior is controlled by the new coupling $\fermiint$, with $\as(\fermiint)$
diverging at $\fermiint_0 = -2\pi^2\hbar^2/m \Lambda$.

A clear  advantage of the one-channel model is that, as shown above, it naturally emerges as the 
correct Hamiltonian
in the experimentally-relevant case of a wide resonance, $\gamma \gg 1$.  However, a 
notable disadvantage is that, in the most interesting
regime of a Feshbach resonance tuned to zero energy, its dimensionless gas 
parameter $\kf |\as|\to \infty$, precluding
a controlled perturbative calculation throughout the crossover. 
  Thus, in this manuscript we shall first compute using
the two-channel model that is under more stringent theoretical control, obtaining results that are 
{\it quantitatively\/}
 accurate for a narrow resonance and qualitatively accurate for a  wide resonance.  Then, with these 
well-controlled results as a guide, we shall re-derive the properties of the system within
(an uncontrolled) mean-field theory 
on the single-channel model, obtaining qualitative agreement with the results of the two-channel model.

\section{Two-channel model at finite atom density: Ground state energy}
\label{SEC:MFT}

Having introduced the two-channel model of fermionic atoms interacting via  an s-wave Feshbach 
resonance, we now study its ground state at finite atom density.
We work in the grand-canonical ensemble by introducing two chemical 
potentials 
\bse
\label{eq:mudefs}
\bea
\mu_\uparrow &=& \mu + h,
\\
\mu_\downarrow &=& \mu - h,
\eea
\ese
that tune the densities $n_\sigma$ of atoms in hyperfine spin states $\sigma = \uparrow, \downarrow$.  
Comprised of two opposite-spin fermions, the molecular density is controlled by
the sum  $2\mu = \mu_\uparrow + \mu_\downarrow$ of the two chemical potentials.  The appropriate grand-canonical
Hamiltonian is then given by
\be
\label{eq:onepre}
H= \curH - \sum_{\bk,\sigma} \mu_\sigma   \ch_{\bk\sigma}^{\dagger} \ch_{\bk\sigma}^{\phdag}
- 2\mu \sum_\bq \bh_\bq^\dagger \bh_\bq^\phdag.
\ee
Equivalently, it can be written in terms of the total chemical
potential $\mu$, that tunes the total (open and closed channel) atom
number $N$ and the chemical potential difference, $\delta \mu=
\mu_\uparrow-\mu_\downarrow$, or equivalently Zeeman field $h = \delta \mu/2$,
that tunes the polarization (difference in the two atom species)
$\Delta N = N_\uparrow-N_\downarrow$:
\bea
\label{eq:one}
H &=& \curH - \mu \hat{N} - h (\Delta \hat{N}),
\\
\hat{N} &\equiv& \sum_{\bk,\sigma}  \ch_{\bk\sigma}^{\dagger} \ch_{\bk\sigma}^{\phdag}
 +2 \sum_\bq \bh_\bq^\dagger \bh_\bq^\phdag,
\\
\Delta \hat{N}  &\equiv& 
\sum_{\bk} \big(\ch_{\bk\uparrow}^{\dagger} \ch_{\bk\uparrow}^{\phdag}- 
\ch_{\bk\downarrow}^{\dagger} \ch_{\bk\downarrow}^{\phdag}\big),
\label{eq:deltaNoperator}
\eea
with $\hat{N}$ the total atom number operator and $\Delta \hat{N} \equiv 
\hat{N}_\uparrow - \hat{N}_\downarrow$ the species asymmetry operator; their expectation values
are the imposed total atom number $N$ and polarization (magnetic moment) $\Delta N$.
 We shall also find it useful to 
define the magnetization via 
\be
\mg = (N_\uparrow - N_\downarrow)/\vol,
\label{eq:magfullygeneraldef}
\ee
 with $\vol$ the system volume.
For future reference, we note that the polarization is related to $\mg$ by
\be
\frac{\Delta N}{N} = \frac{\mg}{n} ,
\label{eq:deltanmrelation}
\ee
with $n = (N_\uparrow + N_\downarrow)/\vol$ the total atom density.

Combining Eq.~(\ref{eq:onepre}) with Eq.~(\ref{eq:bareham}) then yields the two-channel
model in the grand-canonical ensemble:
\bea
&& \hspace{-1cm}H =  \sum_{\bk,\sigma} (\epsilon_k -\mu_\sigma) \ch_{k\sigma}^{\dagger} \ch_{k\sigma}^{\phdag}
 + \sum_\bq\big(\frac{\epsilon_q}{2} + \delta_0-2\mu)\bh_\bq^\dagger \bh_\bq
\nonumber
\\
&& \hspace{-1cm} \quad +\frac{g}{\sqrt{V}}\sum_{\bk,\bq}\Big( \bh^\dagger_\bq
\ch_{\bk+\frac{\bq}{2}\downarrow}^{\phdag}\ch_{-\bk+\frac{\bq}{2}\uparrow}^{\phdag}
+ \ch_{-\bk+\frac{\bq}{2}\uparrow}^{\dagger}\ch_{\bk+\frac{\bq}{2}\downarrow}^{\dagger}\bh^\phdag_\bq
\Big),
\label{eq:meanfieldhamiltonianpre}
\eea
where henceforth we shall generally set $\vol = 1$.
We analyze this model, Eq.~(\ref{eq:meanfieldhamiltonianpre}), via a variational mean-field treatment
that, as discussed at length above, is quantitatively accurate for a
narrow Feshbach resonance with corrections controlled by powers of $\gamma\ll
1$. We parametrize the variational state by the value of a single
complex field $B_\bQ$, that dominates the expectation value of the
molecular operator $\bh_q$ according to $B_\bQ = \langle \bh_\bQ \rangle$.
Such a 
variational state captures a number of possible pure ground states
depending on $B_\bQ$, such as: (i) a normal Fermi gas, with $B_\bQ=0$, (ii) a
uniform (BEC-BCS) singlet superfluid, with $B_{Q=0}\neq 0$, (iii) a
periodically modulated FFLO superfluid, a supersolid, with
$B_{\bQ\neq0}\neq 0$, that in real space is characterized by a condensate~\cite{SingleQ}
$B(r)= B_\bQ e^{i \bQ\cdot  \br}$.  In addition, we will
admit the important possibility of an inhomogeneous coexistence of a
pair of these ground states, that, as we will see, arises over a large
portion of the detuning-polarization phase diagram  (Fig.~\ref{fig:mphasetwo}).

The grand-canonical Hamiltonian corresponding to this class of
variational states is given by
\bea
&& \hspace{-1cm}H =  \sum_{\bk,\sigma} (\epsilon_k -\mu_\sigma) \ch_{\bk\sigma}^{\dagger} \ch_{\bk\sigma}^{\phdag}
 + \big(\frac{\epsilon_Q}{2} + \delta_0-2\mu)|B_\bQ|^2
\nonumber
\\
&& \hspace{-1cm} \quad +g\sum_{\bk}\Big( B^*_\bQ 
\ch_{\bk+\frac{\bQ}{2}\downarrow}^{\phdag}\ch_{-\bk+\frac{\bQ}{2}\uparrow}^{\phdag}
+ \ch_{-\bk+\frac{\bQ}{2}\uparrow}^{\dagger}\ch_{\bk+\frac{\bQ}{2}\downarrow}^{\dagger}B_\bQ^\phstar
\Big).
\label{eq:meanfieldhamiltonian}
\eea
Its quadratic form in atomic operators,  $\ch_{\bk\sigma}^{\phdag}$,
$\ch_{\bk\sigma}^{\dagger}$, allows for an exact
treatment, leading to a ground state energy $E_G[B_\bQ]$, that we compute
below using two complementary approaches, via Green functions and via
a canonical transformation method. The subsequent minimization of
$E_G[B_\bQ]$ then unambiguously determines the phase behavior (i.e.~Figs.~\ref{fig:hphasetwo} and \ref{fig:mphasetwo})
as a function of detuning and Zeeman field, or equivalently, the polarization.

\subsection{Green-function approach}

In this subsection we compute the ground state energy of $H$, Eq.~(\ref{eq:meanfieldhamiltonian}), by
simply calculating the average 
\bse
\bea
&&\hspace{-1cm}E_G = \langle B_\bQ| H |B_\bQ\rangle,
\\
&&\hspace{-.4cm}= \big(\frac{\epsilon_Q}{2} + \delta_0-2\mu)|B_\bQ|^2 
+ \langle B_\bQ|H_f|B_\bQ\rangle,  
\label{eq:gsedef}
\eea
\ese
in the variational state $|B_\bQ\rangle$ labeled by $B_\bQ$, where $H_f=H_K+H_F$
is the fermion part of the Hamiltonian with its kinetic and Feshbach resonance 
parts given by
\bse
\bea
&&\hspace{-1.5cm}
H_K = \sum_{\bk,\sigma} (\epsilon_k -\mu_\sigma) \ch_{\bk\sigma}^{\dagger} \ch_{\bk\sigma}^{\phdag},
\label{eq:onekin}
\\
&&\hspace{-1.5cm}H_F
 = \!\sum_{\bk}\!\big( \Delta^*_\bQ \ch_{\bk+\frac{\bQ}{2}\downarrow}^{\phdag}\ch_{-\bk+\frac{\bQ}{2}\uparrow}^{\phdag}
\!+\!\ch_{-\bk+\frac{\bQ}{2}\uparrow}^{\dagger}\ch_{\bk+\frac{\bQ}{2}\downarrow}^{\dagger}\Delta^\phstar_\bQ
\big),
\label{eq:onefbr}
\eea
\ese
with the pair field $\Delta_\bQ \equiv g B_\bQ$.

These averages can be  computed from the fermion Green function,
which can be easily obtained from the coherent-state action, $S_f$. The
latter is constructed from $H_f$ in a standard way~\cite{Negele}, using the 
fermion anticommutation rules 
\be
\label{eq:ac}
\{\ch_{\bk\sigma}^\phdag,\ch_{\bk'\sigma'}^\dagger\} = \delta_{\sigma,\sigma'} \delta_{\bk,\bk'} ,
\ee
by first
writing $H_f$ in the Bogoliubov-de Gennes form
\bea
&&\hspace{-.5cm}H_f= \sum_{\bk} 
\Psih^\dagger(\bk)
\begin{pmatrix}
\xi_{-\bk+\frac{\bQ}{2}\uparrow} & \Delta^\phstar_\bQ \cr \Delta^*_\bQ  &
-\xi_{\bk+\frac{\bQ}{2}\downarrow}
\end{pmatrix}\Psih(\bk) \nonumber \\
&&\hspace{-.5cm}\qquad \qquad 
+ \sum_\bk \xi_{\bk + \frac{\bQ}{2} \downarrow},
\label{eq:hkf}
\eea
where $\xi_{k\sigma} \equiv \epsilon_{k} - \mu_\sigma$ and 
 the Nambu spinor 
\be
\label{eq:nambu}
\Psih(\bk)\equiv\begin{pmatrix} \ch_{-\bk+\frac{\bQ}{2}\uparrow}^{\phdag} \cr \ch_{\bk+\frac{\bQ}{2}\downarrow}^{\dagger} 
\end{pmatrix}.
\ee
Note that, to get Eq.~(\ref{eq:hkf}) in this desirable matrix form, the 
components of the Nambu spinor are defined with momentum-shifted
arguments.
From Eq.~(\ref{eq:hkf}) we construct a coherent-state path integral for the
partition function $Z=\int D\Psi D \Psi^{\dagger} \exp[-S_f]$, 
with effective action [with $\omega_n = \pi T (2n+1)$ the fermionic Matsubara frequency;
here $\hbar = 1$]
\bea
&&\hspace{-1cm}S_f = -\sum_{\omega_n} \sum_{\bk} \Psi^\dagger_\alpha(\bk,\omega_n)
G^{-1}_{\alpha\beta}(\bk,\omega_n)
\Psi^\phdag_\beta(\bk,\omega_n),
\label{bdgaction}
\eea
where 
\bea
&&\Psi(\bk,\omega_n) \equiv \frac{1}{\sqrt{\beta}}\int_0^\beta d\tau {\rm e}^{i\omega_n\tau} \Psi(\bk,\tau),
\eea
is an anticommuting Grassman field and 
\be
G^{-1}_{\alpha\beta}(\bk,\omega_n) =
-\begin{pmatrix}
-i\omega_n + \xi_{-\bk+\frac{\bQ}{2}\uparrow} & \Delta_\bQ^\phstar \cr \Delta_\bQ^*  &
-i\omega_n - \xi_{\bk+\frac{\bQ}{2}\downarrow} 
\end{pmatrix}.
\label{eq:greenpre}
\ee
From  Eqs.~(\ref{bdgaction}) and ~(\ref{eq:greenpre}) the Green function $G_{\alpha\beta}(\bk, \omega_n)$
[obtained by inverting Eq.~(\ref{eq:greenpre})]
 \bea
&&G_{\alpha\beta}(\bk,\omega_n)
 = \frac{1}{(i \omega_n - \xi_{\bk-\frac{\bQ}{2}\uparrow})(i\omega_n + \xi_{\bk+\frac{\bQ}{2}\downarrow}) -  |\Delta_\bQ|^2 }
\nonumber \\
&&\qquad\times
 \begin{pmatrix}
i\omega_n +  \xi_{\bk+\frac{\bQ}{2}\downarrow} &  \Delta_\bQ^\phstar \cr \Delta_\bQ^*  &
i\omega_n - \xi_{\bk-\frac{\bQ}{2}\uparrow}
\end{pmatrix},
\label{eq:green2}
\eea
is easily related to averages of the fermion fields through a Gaussian integration, giving
\bea
\label{eq:green}
&&\hspace{-.5cm}-G_{\alpha\beta}(\bk,\omega_n) = \langle \Psi_\alpha(\bk,\omega_n) \Psi^\dagger_\beta(\bk,\omega_n)\rangle,
\\
&&\hspace{-.5cm}= 
\begin{pmatrix}
\langle \ch_{-\bk+\frac{\bQ}{2},-\omega_n\uparrow}^\phdag  \ch_{-\bk+\frac{\bQ}{2},-\omega_n\uparrow}^\dagger\rangle
&  
\langle \ch_{-\bk+\frac{\bQ}{2},-\omega_n\uparrow}^\phdag  \ch_{\bk+\frac{\bQ}{2},\omega_n\downarrow}^\phdag\rangle
\cr  
\langle \ch_{\bk+\frac{\bQ}{2},\omega_n\downarrow}^\dagger  \ch_{-\bk+\frac{\bQ}{2},-\omega_n\uparrow}^\dagger\rangle
&
\langle \ch_{\bk+\frac{\bQ}{2},\omega_n\downarrow}^\dagger  \ch_{\bk+\frac{\bQ}{2},\omega_n\downarrow}^\phdag\rangle
\end{pmatrix}.\nonumber 
\eea

Armed with expressions Eq.~(\ref{eq:green}) and ~(\ref{eq:green2}), the averages in 
$\langle B_\bQ| H_F |B_\bQ\rangle$ can
be easily computed. Specializing to the zero temperature limit (continuous
$\omega_n$, with Matsubara sums replaced by integrals), with details
relegated to Appendix~\ref{SEC:gsetwochannel}, we find:
\bse
\bea
&&\hspace{-1cm} \langle H_K \rangle 
= \sum_\bk \txi_k + \sum_\bk \frac{\txi_k^2}{\te_k} 
\big[\Theta(-E_{\bk\uparrow}) - \Theta(E_{\bk\downarrow}) \big]
\nonumber \\
&&\hspace{-1cm}\quad
+ \sum_\bk \Big(
\frac{\bk \cdot \bQ}{2m}+h
\Big)
\big[1-\Theta(-E_{\bk\uparrow}) - \Theta(E_{\bk\downarrow}) \big],
\label{eq:hkinavg}
\\
&&\hspace{-1cm}\langle H_F\rangle = \sum_k \frac{|\Delta_\bQ|^2}{\te_k} 
\big[\Theta(-E_{\bk\uparrow}) - \Theta(E_{\bk\downarrow}) \big],
\label{eq:hfravg}
\eea
\ese
where $\Theta(x)$ is the Heaviside step function and  $E_{\bk\sigma}$ is the excitation
energy for hyperfine state $\sigma$ with (taking $\Delta_\bQ$ real)
\bse
\label{eq:energydefs}
\bea
\varepsilon_k &\equiv& \frac{k^2}{2m} - \mu + \frac{Q^2}{8m},
\label{eq:tildexi}
\\
E_k &\equiv& (\varepsilon_k^2 +\Delta_\bQ^2)^{1/2},
\label{eq:excitationenergy}
\\
E_{\bk\uparrow} &\equiv& E_k - h - \frac{\bk \cdot \bQ}{2m},
\label{ekuparrow}
\\
E_{\bk\downarrow} &\equiv& E_k + h + \frac{\bk \cdot \bQ}{2m}.
\label{ekdownarrow}
\eea
\ese
Combining Eqs.~(\ref{eq:hkinavg}) and (\ref{eq:hfravg})  with  Eq.~(\ref{eq:gsedef}), we 
thus have the following expression for the mean-field ground-state energy $E_G(\Delta_\bQ,\bQ)$:
\bea 
&&\hspace{-.3cm}E_G(\Delta_\bQ,\bQ) = \big(\frac{\epsilon_Q}{2} +
\delta_0-2\mu \big)\frac{\Delta_{\bQ}^2}{g^2} - \sum_k (\te_k-\txi_k )
\nonumber \\
&&\hspace{-.3cm} \quad +\sum_\bk \te_k \big(1+ \Theta(-E_{\bk\uparrow}) -
\Theta(E_{\bk\downarrow})) \big)
\nonumber \\
&&\hspace{-.3cm} \quad + \sum_\bk \big(\frac{\bk \cdot \bQ}{2m} + h\big) 
\big(1-\Theta(-E_{\bk\uparrow}) - \Theta(E_{\bk\downarrow})\big),\;\;\;\;\;
\label{eq:havgpre}
\eea
which can be put in the simpler form
\bea
&& E_G(\Delta_\bQ,\bQ)  = \big(\frac{\epsilon_Q}{2} +
\delta_0-2\mu \big)\frac{\Delta_{\bQ}^2}{g^2} - \sum_k (\te_k-\txi_k )
\nonumber \\
&& \qquad + \sum_\bk \big[
E_{\bk \uparrow} \Theta(-E_{\bk \uparrow})+ E_{\bk \downarrow} \Theta(-E_{\bk \downarrow})
\big].
\label{eq:havg}
\eea 
% 

%%%%%%%%%%%%%%%%%%%%%%%%%%%%%%%%%%%%%%%%%%%%%%%%%%%%%%%%%%%%%%%%%%%%%%%%%%%%

Using the Green function Eq.~(\ref{eq:green2}), together with Eq.~(\ref{eq:green}),
the expectation values of the total atom number $\langle \hat{N}\rangle$ and 
the number difference (magnetic moment) $\langle \Delta\hat{N}\rangle$, can also 
be computed:
\bea
\label{eq:globalnumdefpre}
&&\hspace{-1cm}N = \frac{2\Delta_\bQ^2}{g^2} +
\sum_k \big( \langle
 \ch_{\bk\uparrow}^{\dagger} \ch_{\bk\uparrow}^{\phdag} \rangle + \langle \ch_{\bk\downarrow}^{\dagger} 
\ch_{\bk\downarrow}^{\phdag} \rangle \big),
\\
&&\hspace{-.4cm}
= \frac{2\Delta_\bQ^2}{g^2} + \sum_k \big( 1- \frac{\varepsilon_k}{E_k}[ \Theta(E_{\bk\uparrow})-\Theta(-E_{\bk\downarrow})]\big),
\label{eq:globalnumdef}
\\
&&\hspace{-1cm}\Delta N = \sum_\bk \big( \langle
 \ch_{\bk\uparrow}^{\dagger} \ch_{\bk\uparrow}^{\phdag} \rangle - \langle \ch_{\bk\downarrow}^{\dagger} 
\ch_{\bk\downarrow}^{\phdag} \rangle \big),
\\
&&\hspace{-.4cm}= \sum_\bk \Big(\Theta(-E_{\bk\uparrow})-
\Theta(-E_{\bk\downarrow})\Big). \label{eq:globalmagdef}
\eea
These expressions will be important for eliminating the chemical potentials appearing in $E_G$
in favor of the experimentally-controlled atom number $N$ and polarization $\Delta N/N$.

Equation~(\ref{eq:havg}) is quite general, encompassing (as noted above) both uniform and periodic FFLO-type
paired states as well as  an unpaired normal state.  In various limits, however, it simplifies considerably.
In Appendix~\ref{SEC:BECBCSreview} 
 we review the most well-studied such limit, namely the conventional equal-population BEC-BCS crossover
at $h=0$ and $\bQ=0$.

  With our primary result Eq.~(\ref{eq:havg}) in hand, the remainder of the paper is
conceptually straightforward, simply amounting to minimizing $E_G(B_\bQ)$ over $B_\bQ$ and $\bQ$ to find
their optimum values as a function of detuning $\delta$, polarization $\Delta N$ (or $h$) and 
total atom number $N$ (or $\mu$).  Although this can be done numerically, considerable insight is obtained by approximate analytic
analysis, possible for a narrow Feshbach resonance, $\gamma\ll 1$.
 Before turning to this, in the next subsection we present another derivation of $E_G(B_\bQ)$, 
Eq.~(\ref{eq:havg}).

\subsection{Canonical transformation approach}

In this subsection, we present an alternate derivation of Eq.~(\ref{eq:havg}) for $E_G$, using
only the canonical commutation relations of the fermion operators $\ch_{\bk\sigma}$.  As in the 
preceding subsection, we focus on the fermion portion $H_f$ of the total 
Hamiltonian,
with the starting point Eq.~(\ref{eq:hkf}): 
\bea
H_f &=& \sum_{\bk} 
\Psih^\dagger(\bk)  \hat{H}_f   \Psih(\bk) + \sum_\bk \xi_{\bk + \frac{\bQ}{2} \downarrow},
\label{eq:hkf2}
\\
 \hat{H}_f&\equiv&
\begin{pmatrix}
\xi_{-\bk+\frac{\bQ}{2}\uparrow} & \Delta^\phstar_\bQ \cr \Delta_\bQ  &
-\xi_{\bk+\frac{\bQ}{2}\downarrow}
\end{pmatrix},
\label{eq:hkfmatrix}
\eea
where without loss of generality we have taken $\Delta_\bQ$ real.
 The matrix $ \hat{H}_f$ can be diagonalized using the unitary matrix $\Uh$
\bea
\Uh = \begin{pmatrix}
u_k & v_k \cr v_k & -u_k
\end{pmatrix}.
\label{eq:unitary}
\eea
Here, the coherence factors 
\bse
\label{eq:coherencefactors}
\bea
u_k &=& \frac{1}{\sqrt{2}} \sqrt{1+ \frac{\varepsilon_k}{E_k}},
\\
v_k &=& \frac{1}{\sqrt{2}} \sqrt{1- \frac{\varepsilon_k}{E_k}},
\eea
\ese
are analogues of those appearing in BCS theory~\cite{Schrieffer,deGennes,Tinkham},
whose form is constrained to preserve the canonical commutation relations, Eq.~(\ref{eq:ac}), that 
require $u_k^2 + v_k^2 = 1$.
Since, by construction, the left and right  columns of $\Uh$ are the eigenvectors of $\hat{H}_f$ (with eigenvalues
$E_{\bk \uparrow}$ and $-E_{\bk\downarrow}$, respectively), this accomplishes a diagonalization of $H_f$ in
Eq.~(\ref{eq:hkf2}).  We find:
\bea
\nonumber 
&&H_f = \sum_\bk \begin{pmatrix}
\alphah_{\bk \uparrow}^\dagger & \alphah_{\bk \downarrow}^\phdag 
\end{pmatrix}
 \begin{pmatrix}
E_{\bk \uparrow} &0 \cr 0 & -E_{\bk \downarrow}
\end{pmatrix}
\begin{pmatrix}
\alphah_{\bk \uparrow}^\phdag \cr \alphah_{\bk \downarrow}^\dagger 
\end{pmatrix} 
\\
&&\qquad \qquad  
+ \sum_\bk \xi_{\bk + \frac{\bQ}{2} \downarrow},
\eea
with the Bogoliubov (normal mode) operators 
$\alphah_{\bk\sigma}$ following from the operation of $\Uh$  on the Nambu spinor Eq.~(\ref{eq:nambu}):
\bse\label{eq:alphas}
\bea
\alphah_{\bk\uparrow} = u_k  \ch_{-\bk+\frac{\bQ}{2}\uparrow}^{\phdag} + 
v_k  \ch_{\bk+\frac{\bQ}{2}\downarrow}^{\dagger},
\\
\alphah_{\bk\downarrow}^\dagger = v_k  \ch_{-\bk+\frac{\bQ}{2}\uparrow}^{\phdag} -
u_k  \ch_{\bk+\frac{\bQ}{2}\downarrow}^{\dagger}.
\eea
\ese
Thus, using $\{\alphah_{\bk \downarrow}^\phdag ,\alphah_{\bk \downarrow}^\dagger\} = 1$, we have
\bea
&&
H_f = \sum_\bk \Big(
E_{\bk \uparrow}  \alphah_{\bk \uparrow}^\dagger\alphah_{\bk \uparrow}^\phdag
+ E_{\bk \downarrow}  \alphah_{\bk \downarrow}^\dagger\alphah_{\bk \downarrow}^\phdag
\Big)
\nonumber \\
&& \qquad 
+ \sum_\bk \big(\xi_{\bk + \frac{\bQ}{2} \downarrow} - E_{\bk \downarrow}
\big).
\label{eq:hkf3}
\eea 
Recall that our aim is to compute the ground-state energy.
If it were true that $E_{\bk \uparrow}$ and $E_{\bk \downarrow}$ were 
both positive, the first two terms in Eq.~(\ref{eq:hkf3}) would simply count 
the excitation energy associated with the 
excitation quanta $\alphah_{\bk\sigma}$ (Bogoliubov quasiparticles) above
the ground state defined as the vacuum of $\alphah_{\bk\sigma}$ particles, i.e., 
$\alphah_{\bk \sigma} |B_\bQ\rangle = 0$.  This would also allow an 
immediate identification of the constant part of $H_f$ [the second line of Eq.~(\ref{eq:hkf3})]
with the fermion contribution $E_{Gf} = \langle H_K \rangle +  \langle H_F \rangle$ 
to the ground-state energy. 

However, because the energies 
$E_{\bk \sigma}$  are {\it not\/} positive definite, the second line 
of Eq.~(\ref{eq:hkf3}) is {\it not\/} the ground
state energy when either of $E_{\bk \sigma}<0$.
It is nonetheless possible to write $H_f$ as a sum of the excitation and ground-state energies. 
To this end, we use step functions to separate the sum over momenta into regions that have 
$E_{\bk\sigma}>0$ and $E_{\bk\sigma}<0$:
\bea
&&\hspace{-1cm}\sum_\bk
 E_{\bk \sigma} \alphah_{\bk \sigma}^\dagger \alphah_{\bk \sigma}^\phdag
 = \sum_\bk\big[
 E_{\bk \sigma}\Theta(E_{\bk \sigma}) 
 \alphah_{\bk \sigma}^\dagger \alphah_{\bk \sigma}^\phdag 
\nonumber \\
&& \qquad  \qquad 
+ E_{\bk \sigma}\Theta(-E_{\bk \sigma}) 
 (1- \alphah_{\bk \sigma}^\phdag  \alphah_{\bk \sigma}^\dagger)\big],
\label{eq:inthkf}
\eea 
where, for the momenta satisfying $E_{\bk \sigma}<0$, we used the anticommutation
relation $\{\alphah_{\bk \sigma}^\phdag,\alphah_{\bk \sigma}^\dagger \} = 1$.
Using Eq.~(\ref{eq:inthkf}), we can write $H_f$ as 
\bea
\nonumber
&&
\hspace{-.5cm}H_f = \sum_{\bk\sigma} \Big(  E_{\bk \sigma}\Theta(E_{\bk \sigma}) 
 \alphah_{\bk \sigma}^\dagger \alphah_{\bk \sigma}^\phdag 
- E_{\bk \sigma}\Theta(-E_{\bk \sigma}) 
 \alphah_{\bk \sigma}^\phdag  \alphah_{\bk \sigma}^\dagger
\Big)
\\
&&
\hspace{-.5cm}\qquad \qquad + E_{Gf},
\label{eq:firstlinehkf}
\eea
with
\bea
&& \hspace{-.5cm}E_{Gf} \equiv \sum_\bk \big(   E_{\bk\uparrow} \Theta(-E_{\bk\uparrow}) + 
E_{\bk\downarrow} \Theta( -E_{\bk\downarrow}) 
\nonumber 
\\
&&
\hspace{-.5cm}\qquad \qquad 
+ \xi_{\bk + \frac{\bQ}{2} \downarrow} - E_{\bk \downarrow}
\big).
\eea
With these manipulations, each pair of operators in the sum in the first line of 
Eq.~(\ref{eq:firstlinehkf}) by construction multiplies  a {\it positive} excitation energy that is $|E_{\bk \sigma}|$.
Thus, they represent excitations above the ground state with the fermion part of the ground state
energy given by 
$\langle H_f \rangle = E_{Gf}$.
Using the definitions of $E_{\bk\sigma}$ (and $\sum_\bk \bk \cdot \bQ = 0$), 
it is straightforward to see that $E_{Gf}$ agrees with the 
result from the preceding 
subsection, i.e., $\langle H_K \rangle + \langle H_F \rangle$:
\be E_{Gf} =   \sum_k (\varepsilon_k - E_k )
 +\sum_\bk \big(
E_{\bk\uparrow} \Theta(-E_{\bk\uparrow}) + E_{\bk\downarrow} \Theta(-E_{\bk\downarrow}) 
\big),
\label{eq:havgfermion}
\ee
which is our result for the fermion contribution to the ground-state energy $E_G(B_\bQ)$, Eq.~(\ref{eq:havg}).

This method for computing the ground-state energy has the additional benefits that (i) it provides
a physical interpretation to the step functions appearing in $E_G(B_\bQ)$ and (ii) it allows
a straightforward computation of the fermion ground-state wavefunction $|B_\bQ\rangle$ that we now derive. 
Clearly, $|B_\bQ\rangle$ must be annihilated by the excitation part $H_f^e$ of the Hamiltonian $H_f$:
\be
\hspace{-.5cm}
H_f^e \equiv \sum_{\bk\sigma} \big(  E_{\bk \sigma}\Theta(E_{\bk \sigma}) 
 \alphah_{\bk \sigma}^\dagger \alphah_{\bk \sigma}^\phdag 
- E_{\bk \sigma}\Theta(-E_{\bk\sigma}) 
 \alphah_{\bk \sigma}^\phdag  \alphah_{\bk \sigma}^\dagger \big).
\label{eq:hfnought}
\ee
The step functions divide the momentum sum into 
three regions: (1) $\bk$ such that $E_{\bk \uparrow}>0$ and  $E_{\bk \downarrow}>0$,
(2) $\bk$ such that  $E_{\bk \uparrow}<0$ and (3) $\bk$ such that  $E_{\bk \downarrow}<0$.  From
the explicit expressions Eqs.~(\ref{ekuparrow}) and ~(\ref{ekdownarrow}) it is easy to see 
that the $E_{\bk \sigma}$ cannot both be negative; thus, these are the only three possibilities.

Consider momenta such that condition (1) is satisfied, which we denote $\bk_1$.  For such
momenta, the summand in Eq.~(\ref{eq:hfnought}) is 
$E_{\bk_1\uparrow}\alphah_{\bk_1 \uparrow}^\dagger \alphah_{\bk_1 \uparrow}^\phdag 
+  E_{\bk_1 \downarrow}
 \alphah_{\bk_1 \downarrow}^\dagger \alphah_{\bk_1 \downarrow}^\phdag $, which annihilates
factors of the form $(u_{\bk_1} + v_{\bk_1} \ch_{\bk_1 + \frac{\bQ}{2} \downarrow}^\dagger 
 \ch_{-\bk_1  + \frac{\bQ}{2}\uparrow}^\dagger$).
Of course, if we set $\bQ = 0$, this is the usual factor in the 
BCS ground-state wavefunction~\cite{Schrieffer,deGennes,Tinkham} that encodes pairing correlations;
thus, the states $\bk_1$ are {\it paired\/} at nonzero center of mass momentum $\bQ$. This can also 
be seen from the form of the ground-state energy  Eq.~(\ref{eq:havgfermion}): Since the $\Theta$ functions
in the second term of this equation {\it vanish\/} for states satisfying condition (1) they have the 
standard BCS contribution to the ground-state energy from the first term of Eq.~(\ref{eq:havgfermion}).

Now consider states in region (2).  
For them,  
the excitation energy
is $- E_{\bk_2 \uparrow}
 \alphah_{\bk_2 \uparrow}^\phdag  \alphah_{\bk_2 \uparrow}^\dagger + 
E_{\bk_2 \downarrow} \alphah_{\bk_2 \downarrow}^\dagger \alphah_{\bk_2 \downarrow}^\phdag$.
By examining the form of Eq.~(\ref{eq:alphas}), it is clear that an operator annihilated by
this summand is simply $\ch^\dagger_{-\bk_2+\frac{\bQ}{2}\uparrow}$.  Thus, states $\bk_2$ 
are simply filled by unpaired spin-$\uparrow$ fermions, i.e., they are unpaired.  States in region 3 are 
similarly unpaired, but with spin-$\downarrow$. Expressing $|B_\bQ\rangle$ 
as a product over these three  regions of momenta yields our variational fermion ground state wavefunction: 
\bea
&&\hspace{-1cm}|B_\bQ\rangle =\prod_{\bk \in \bk_3}  
\ch^\dagger_{\bk+\frac{\bQ}{2}\downarrow} 
\prod_{\bk \in \bk_2}  
\ch^\dagger_{-\bk+\frac{\bQ}{2}\uparrow} 
\nonumber \\
&&\hspace{-1cm}\qquad \qquad \times \prod_{\bk \in \bk_1} 
(u_{\bk} + v_{\bk} \ch_{\bk + \frac{\bQ}{2} \downarrow}^\dagger  \ch_{-\bk+\frac{\bQ}{2}\uparrow}^\dagger)|0\rangle,
\label{eq:groundstate}
\eea
with $|0\rangle$ the fermionic atom vacuum.

We now describe  our general strategy.  The ground 
state energy Eq.~(\ref{eq:havg}) is a function of $\Delta_\bQ$ (or equivalently $B_\bQ$) 
and $\bQ$. By minimizing it 
over $\Delta_\bQ$ and $\bQ$, we find possible ground states 
of the two-channel model at particular values of $\mu$, $h$ and detuning. 
Ground states  at fixed imposed atom number (relevant for atomic physics experiments)
 are stationary with respect to $\Delta_\bQ$ and $\bQ$, 
and satisfy the total number and polarization constraints:
\bse
\label{eq:fullygeneral}
\bea
0 &=& \frac{\partial E_G}{\partial \Delta_\bQ},
\label{eq:gapequationgeneral}
\\
0 &=& \frac{\partial E_G}{\partial\bQ},
\label{eq:momequationgeneral}
\\
N &=&  -\frac{\partial E_G}{\partial \mu},
\label{eq:numequationgeneral}
\\
\Delta N &=&  -\frac{\partial E_G}{\partial h},
\label{eq:polarizationequationgeneral}
\eea
\ese
the latter two being equivalent to Eqs.~(\ref{eq:globalnumdef}) and (\ref{eq:globalmagdef}).
We shall refer to Eqs.~(\ref{eq:fullygeneral}) as the gap, momentum, number and polarization equations, respectively.
It is crucial to emphasize that $E_G$ has numerous stationary points and we must be sure to ascertain that the particular 
stationary point we are studying is indeed a  {\it minimum\/}~\cite{Paonote} of $E_G$.  
  Thus, while often in studies of the BEC-BCS crossover at $h=0$ authors simply simultaneously solve the number and
gap equation, at $h\neq 0$ this can (and has in a number of works in the literature~\cite{Pao,Iskin,Sedrakian,Paonote,Comment,CommentNote})
lead to incorrect results, especially at negative detunings
where $E_G$ is particularly complicated.  Moreover, frequently we shall find that the ground-state at a particular 
$\delta$,
$N$, and $\Delta N$ (or even at fixed $\delta$, $N$, and $h$) is a {\it phase-separated mixture\/} of two of these 
phases, with fractions $x$ and $1-x$ that must be determined.
With these caveats in mind, we now turn to the analysis of Eqs.~(\ref{eq:havg}) and  (\ref{eq:fullygeneral}).
%%%%%%%%%%%%%%%%%%%%%%%%%%%%%%%%%%%%%%%%%%%%%%%%%%%%%%%%%%%%%%%%%%%%%%%%%%%%%%%%%%%%%%%%%%%%%%%%%%%%%%%%%%%%%%%%

%%%%%%%%%%%%%%%%%%%%%%%%%%%%%%%%%%%%%%%%%%%%%%%%%%%%%%%%%%%%%%%%%%%%%%%%%%%%%%%%%%%%%%%%%%%%%%%%%%%%%%%%%%%%%%%%
\section{Positive-detuning regime of two-channel  model at finite population difference}
\label{SEC:BCS}
In this section, we consider our system at $\delta>0$ with a finite population 
imbalance (polarization)
between the species undergoing pairing, encompassing the BCS  $\delta \gg 2\ef $
and crossover $0 < \delta <2\ef$ regimes.  
   As discussed in Sec.~\ref{SEC:MFT}, it is convenient
to first study the equivalent problem of pairing with a finite {\it chemical potential\/}
difference $\delta \mu = 2h$ between the species.  
What phases do we expect in this regime?  
For equal chemical potentials ($\mu_\uparrow = \mu_\downarrow = \mu$), the favored ground state is the 
paired superfluid (SF) state that (since $\mu >0$ in this regime) is associated with pairing near
the common Fermi surface at momentum $k_{\rm F} = \sqrt{2m\mu}$.  This paired SF state is of course described by 
the BCS ground state.
For very large chemical potential difference
$h$, we expect pairing to be  destroyed ($\Delta_\bQ = 0$) and the
ground-state to be a normal (N) Pauli-paramagnetic Fermi gas with $N_\uparrow> N_\downarrow$.
The remaining logical possibility is that, for intermediate values of $h$, we may have 
a phase exhibiting 
nonzero pairing
and a nonzero population difference ($\Delta N \neq 0$); 
it remains to be seen whether such a  ground state is stable anywhere in the
$\delta$-$\Delta N$ phase diagram.

 One set of possible ground states that accomplish this is the so-called FFLO class of states,~\cite{ff,lo} already
well-studied for a BCS-type superconductor.  The fact that our Feshbach resonance model
is closely related to the BCS model of interacting fermions ensures that we will indeed
find that FFLO-like states are stable over portions of the phase diagram at large positive 
detuning.

As we will show, within our  restricted ground-state ansatz, these three states (FFLO, SF and N) are
the {\it only homogeneous ground states \/} that are stable at positive detuning, with the phase diagram 
dominated by phase separated mixtures of SF and N or SF and FFLO.  
Furthermore, the FFLO state 
is only stable for a very narrow window of $h$ (or $\Delta N$) values that 
vanishes below $\delta_* \simeq 2\ef$.
The reason that FFLO states can only exist for a restricted narrow window
of parameters  is simply that, in order to accommodate
both pairing {\it and\/} a nonzero polarization $\Delta N$,
 the system must pair at a {\it finite wavevector\/}
$Q\sim k_{{\rm F}\uparrow} -  k_{{\rm F}\downarrow}$.
This corresponds to a moving superfluid that, at sufficiently high superfluid velocity 
(corresponding to a critical $Q \sim k_{{\rm F}\uparrow}- k_{{\rm F}\downarrow}\propto h$)
is unstable (via the Landau criterion) to quasiparticle proliferation, 
as seen from the form of $E_{\bk\sigma}$ [Eqs.~(\ref{ekuparrow}) and ~(\ref{ekdownarrow})].

\begin{figure}[bth]
\vspace{1.4cm}
\centering
\setlength{\unitlength}{1mm}
\begin{picture}(40,40)(0,0)
\put(-50,0){
\begin{picture}(0,0)(0,0)
\includegraphics{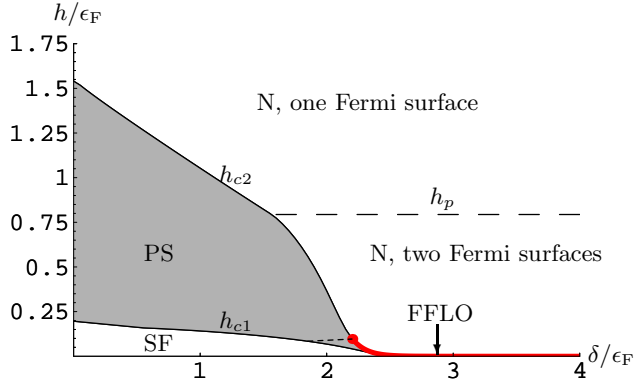}
\end{picture}}
\put(-12,50) {$h/\ef$}
\put(59,4.5) {$\delta/\ef$}
\put(0,6) {SF}
\put(10,9) {$h_{c1}$}
\put(10,28.5) {$h_{c2}$}
\put(38,25.5) {$h_{p}$}
\put(0,18) {PS}
\put(30,18) {N, two Fermi surfaces}
\put(15,38) {N, one Fermi surface}
\put(35,10) {FFLO}
\put(39,9.5) {\vector(0,-1){4}}
\end{picture}
\vspace{-.5cm}
\caption{(Color Online) 
Positive detuning ($\delta$)-chemical potential difference ($h$) 
phase diagram of the two-channel model for the case $\gamma = 0.1$
showing the superfluid phase (SF), normal phase (N),
FFLO phase (along red curve, too thin to see) and the regime of 
phase separation (PS).  Above the horizontal dashed line the N phase is
fully spin-polarized ($N_\uparrow = N$, $N_\downarrow = 0$), consisting of a single
Fermi surface, while below the dashed line the N phase has two Fermi surfaces.
 The dashed line in the shaded PS regime separates SF-N coexistence from
SF-FFLO coexistence and is derived in Sec.~\ref{SEC:regimeofphaseseparation}. 
}
\label{fig:bcshphasediagram}
%
%   phasediagram_twochannel.nb
%
\end{figure}

\begin{figure}[bth]
\vspace{1.4cm}
\centering
\setlength{\unitlength}{1mm}
\begin{picture}(40,40)(0,0)
\put(-50,0){
\begin{picture}(0,0)(0,0)
\includegraphics{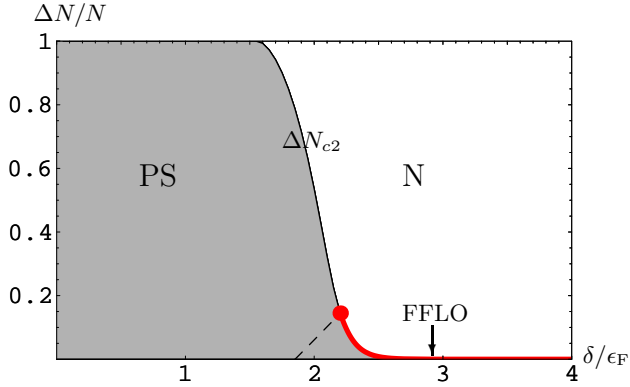}
\end{picture}}
\put(-14,50) {$\Delta N/N$}
\put(59,4.5) {$\delta/\ef$}
%\put(0,6) {SF}
\put(0,28) {\large{PS}}
\put(19,33) {$\Delta N_{c2}$}
\put(35,28) {\large{N}}
\put(35,10) {FFLO}
\put(39,9.5) {\vector(0,-1){4}}
\end{picture}
\vspace{-.5cm}
\caption{(Color Online) 
Positive detuning ($\delta$)-population difference ($\Delta N$) phase diagram of the two-channel model for the case $\gamma = 0.1$
 showing the regime of  phase separation (PS), the normal phase (N)
and the FFLO phase (along red curve). The SF phase is confined to the $\Delta N = 0$ 
axis.  As in Fig.~\ref{fig:bcshphasediagram}, the dashed line in the shaded PS regime separates SF-N coexistence from
SF-FFLO coexistence (the latter sharing a boundary with the FFLO phase).
}
\label{fig:bcsmphasediagram}
%
%   phasediagram_twochannel.nb
%
\end{figure}

From a pedagogical point
of view, it is easiest to proceed by first neglecting the FFLO state entirely,
setting $\bQ=0$ at the outset in our expression for $E_G$ (which simplifies it considerably),
before determining the phase diagram (Figs.~\ref{fig:bcshphasediagram} and \ref{fig:bcsmphasediagram}). 
 Afterwards, in Sec.~\ref{SEC:FFLO} we 
will return to the FFLO state, finding the regime of the phase diagram in which it is stable.

%-----------------------------
%
% fig%9
%
%-----------------------------
%-----------------------------
%
% fig%10
%
%-----------------------------

\subsection{Ground-state energy at $\bQ=0$}

In this  subsection we focus on the $\bQ=0$ case, which  greatly simplifies the ground-state energy
$E_G$.  Although this is in preparation for studying
possible uniform ground-states at $\delta>0$, the results of the present section will also apply to 
the negative-detuning BEC regime, to be studied in Sec.~\ref{SEC:BEC},
where, as we will show, the FFLO state is unstable and $\bQ = 0$ is the only possibility.
  With this in mind, in this 
subsection we shall make no approximations that rely on $\delta>0$.

%-----------------------------
%
% fig%11
%
\begin{figure}[bth]
\vspace{1.4cm}
\centering
\setlength{\unitlength}{1mm}
\begin{picture}(40,40)(0,0)
\put(-50,0){\begin{picture}(0,0)(0,0)
\includegraphics{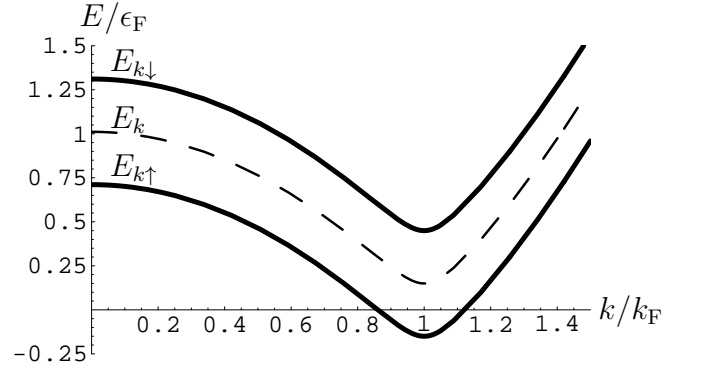}
\end{picture}}
\put(-10,48) {\large{$E/\ef$}}
\put(-6,42) {\large{$E_{k\downarrow}$}}
\put(-6,34.5) {\large{$E_{k}$}}
\put(-6,28) {\large{$E_{k\uparrow}$}}
\put(59,9.0) {\large{$k/\kf$}}
\end{picture}
\vspace{-.5cm}
\caption{Dashed line: Plot of quasiparticle excitation energy $E_k = \sqrt{\xi_k^2 + \Delta^2}$
for a BCS superconductor, normalized to $\ef$ for the case $\mu = \ef$ and $\Delta = 0.15 \ef$.  Solid
 lines: Spin-up and spin-down quasiparticle excitation energies for the case $h = 0.3 \ef$.}
\label{excitationenergyplot}
%
%  excitationenergy.nb 
%
\end{figure}
%------------------------------
%

  We start with the general expression Eq.~(\ref{eq:havgpre}) at $\bQ=0$,
for simplicity of notation dropping the zero subscript on $\Delta$ (i.e. $\Delta_0 \to \Delta$):
\bea
&&E_G = (\delta-2\mu) \frac{\Delta^2}{g^2} + 
\sum_k ( \xi_k - E_k +\frac{\Delta^2}{2\epsilon_k}) 
\nonumber \\ 
&& \qquad +\sum_k \te_k \big(1+ \Theta(-E_{k\uparrow}) - \Theta(E_{k\downarrow})) \big)
\nonumber \\
&& \qquad + h \sum_k 
\big(1- \Theta(-E_{k\uparrow}) - \Theta(E_{k\downarrow})\big), 
\label{eq:havgBCS}
\eea
where now $\xi_k = \epsilon_k - \mu$, the excitation energies are simply
\bse
\label{eq:bcsexcitationenergies}
\bea
E_{k\uparrow} &=& E_k -h, 
\\
E_{k\downarrow} &=& E_k +h,
\eea
\ese
plotted for $\mu>0$ in Fig.~\ref{excitationenergyplot}, and we have used Eq.~(\ref{eq:deltaintrop}) to exchange
the bare detuning $\delta_0$ for the physical detuning $\delta$. 

Without loss of generality we take $h>0$, so that $\Theta(E_{k\downarrow}) = 1$ and Eq.~(\ref{eq:havgBCS})
can be written in the simpler form 
\bea
&&E_G = (\delta-2\mu) \frac{\Delta^2}{g^2} + 
\sum_k ( \xi_k - E_k +\frac{\Delta^2}{2\epsilon_k}) 
\nonumber 
\\
&& \qquad  + \sum_k (E_k-h) \Theta(h-E_k).
\label{eq:havgBCSp}
\eea
The last $h$-dependent term is only nonzero when $h>E_k$ for some $k$.  For $\mu>0$, this means $E_G$ is
identically $h$-{\it independent\/} until the spin-up band crosses zero energy (depicted in Fig.~\ref{excitationenergyplot})
at $h>\Delta$.

Equation~(\ref{eq:havgBCSp}) may be further simplified by 
rewriting the $h$-dependent momentum sums in terms of the magnetization $\mg(h)$.  The simplest 
way to do this is to recall Eq.~(\ref{eq:polarizationequationgeneral}) for the population difference
\be
N_\uparrow - N_\downarrow =  -\frac{\partial E_G}{\partial h},
\label{eq:intdiffH2}
\ee
which can be written in terms of the magnetization $\mg(h)$ as 
[using Eqs.~(\ref{eq:magfullygeneraldef}); note we have taken the system volume
$\vol$ to be unity]:
\bse
\label{eq:magdef}
\bea
\label{eq:mbcspre}
&&\hspace{-1.5cm}\mg(h) 
= \int \frac{d^3 k}{(2\pi)^3} \Theta(h - E_k),
 \\
&&\hspace{-1.5cm} \;\;\;\;\;\;\quad= 
 \frac{2c}{3} \big[
(\mu + \sqrt{h^2 - \Delta^2})^{3/2} \Theta(\mu + \sqrt{h^2 - \Delta^2})
\nonumber \\
&&\hspace{-1.5cm}\qquad\qquad -
(\mu - \sqrt{h^2 - \Delta^2})^{3/2}\Theta(\mu - \sqrt{h^2 - \Delta^2})
\big], \label{eq:mbcs}
\eea
\ese
where in the last equality we evaluated the integral over momenta by first converting to 
an energy integral using
the usual prescription $\int \frac{d^3k}{(2\pi)^3} \cdots = \int N(\epsilon) d\epsilon\cdots$ 
with the three-dimensional density of states per spin
\bea
N(\epsilon) &=& c\sqrt{\epsilon}, 
\\
\label{eq:cdef}
c&\equiv& \frac{m^{3/2}}{\sqrt{2}{\pi^2}}.
\eea
Thus [using the condition that the magnetic contribution
 of Eq.~(\ref{eq:havgBCSp}) vanishes at $h=0$], we have the following general expression for the $\bQ=0$ ground-state energy
(also converting the first momentum sum to an integral):
\bea
&&\nonumber 
E_G = (\delta-2\mu) \frac{\Delta^2}{g^2} + 
\int \frac{d^3k}{(2\pi)^3} ( \xi_k - E_k +\frac{\Delta^2}{2\epsilon_k}) 
\\
&&\qquad \qquad 
- \int_0^h \mg(h') dh'.
\label{eq:havgBCS2}
\eea
  The steps connecting the last term of 
Eq.~(\ref{eq:havgBCSp}) to the last term of 
Eq.~(\ref{eq:havgBCS2}) can also be derived in a more straightforward approach by manipulating the
$h$-dependent momentum sums in Eq.~(\ref{eq:havgBCSp}), as shown in Appendix~\ref{SEC:appsums}.

Possible $\bQ=0$ ground-states correspond to minima of Eq.~(\ref{eq:havgBCS2}) [which thus satisfy the 
gap equation, Eq.~(\ref{eq:gapequationgeneral})] having magnetization $\mg$ given by Eq.~(\ref{eq:mbcs})
and density $n$ given by 
\be
n = -\frac{1}{\vol}\frac{\partial E_G}{\partial\mu} .
\ee

Quite frequently, we shall be interested in systems at fixed {\it imposed\/} total atom (as free atoms and molecules)
density $n$. This introduces 
another natural scale, $\ef$, that is the Fermi energy at asymptotically large detuning for $g\to0$:
\bea
\label{numdef}
n &=& \frac{4}{3}\,c\,\ef^{3/2}.
\eea
We use $\ef$ as a convenient energy scale to normalize quantities with dimensions of energy. Thus we define 
\bse
\label{eq:dimensionlessvariables}
\bea
\muh &\equiv& \frac{\mu}{\ef} ,
\\
\Deltah &\equiv& \frac{\Delta}{\ef},
\\
\deltah &\equiv& \frac{\delta}{\ef},\label{eq:deltahatdef}
\\
\hh &\equiv& \frac{h}{\ef},\label{eq:hhatdef}
\\
e_G &\equiv& \frac{E_G}{c \ef^{5/2}},
\label{eq:egdef}
\eea
\ese
with the last definition providing a convenient normalization for the ground-state energy.
We shall typically determine quantities (such as critical chemical potential differences
and polarizations) as a function of the normalized detuning $\deltah$.
To make contact with experiments that typically plot physical observables 
as a function of the inverse scattering length through the dimensionless parameter
$-(\kf \as)^{-1}$, we note that our normalized detuning $\deltah$ is proportional to
this parameter and given by [combining Eq.~(\ref{eq:as}) with 
Eqs.~(\ref{eq:gammadef})  and ~(\ref{eq:deltahatdef})]
\be
\label{eq:deltahas}
\deltah = -\frac{\pi}{2}   
\gamma
\frac{1}{\kf \as}.
\ee

Converting the momentum integral in Eq.~(\ref{eq:havgBCS2}) to an energy integral, 
 we have for $e_G$:
\bea
&&\hspace{-1cm}e_{G} =
\gamma^{-1} \Deltah^2(\deltah - 2\muh)  - \int_0^{\hh} \mgh(\hh') d\hh'
\nonumber \\
&&\hspace{-.7cm}
 +\int_0^\infty dx \sqrt{x} \big(x-\muh -\sqrt{(x-\muh)^2 + \Deltah^2} + \frac{\Deltah^2}{2x}\big),
  \label{eq:gsetotallygeneral}
\eea
where $\mgh(\hh)$ is the dimensionless form of Eq.~(\ref{eq:mbcs}):
\bea
 &&\hspace{-1cm}\mgh(\hh) \equiv \frac{2}{3} \big[
(\muh + \sqrt{\hh^2 - \Deltah^2})^{3/2} \Theta(\muh + \sqrt{\hh^2 - \Deltah^2})
\nonumber \\
&&\hspace{-1cm}\qquad -
(\muh - \sqrt{\hh^2 - \Deltah^2})^{3/2}\Theta(\muh - \sqrt{\hh^2 - \Deltah^2})
\big]. 
\label{eq:mbcsdimless}
\eea
For future reference, we note that the relations between $\mg(h)$, $\Delta N$ and $\mgh(\hh$)
can be summarized by
\be
\label{magrelation2}
\frac{\mg(h)}{n} = \frac{\Delta N}{N} = \frac{3}{4} \mgh(\hh).
\ee

With these definitions of dimensionless variables, the gap, number 
and polarization
equations [i.e., Eqs.~(\ref{eq:gapequationgeneral}), (\ref{eq:numequationgeneral}), and ~(\ref{eq:polarizationequationgeneral}),
respectively] become
\bse
\label{eq:gapnumequationgeneralnorm}
\bea
0 &=& \frac{\partial e_G}{\partial\Deltah},
\label{eq:gapequationgeneralnorm}
\\
\frac{4}{3} &=& -\frac{\partial e_G}{\partial\muh},
\label{eq:numequationgeneralnorm}
\\
\frac{4}{3}\frac{\Delta N}{N} &=& -\frac{\partial e_G}{\partial\hh}.
\label{eq:polequationgeneralnorm}
\eea
\ese
As we have already mentioned, the equations in the present section apply to the $\bQ=0$ ground-state
energy in the BCS regime (that we now proceed to study) and also in the BEC regime that we 
shall study in Sec.~\ref{SEC:BEC}.

\subsection{$\bQ=0$ phases of the ground state energy in the BCS regime}
\label{SEC:bcsandtransition}
In the present section, we study the phases of Eq.~(\ref{eq:gsetotallygeneral}) for $\hh \neq 0$ in the BCS limit of
large positive detuning. In this regime, $\muh\gg \Deltah$ and the integral in Eq.~(\ref{eq:gsetotallygeneral})
is well-approximated by (see Appendix~\ref{SEC:BECBCSreview}):
\bea
&&\hspace{-.5cm}\int_0^\infty dx \sqrt{x} \big(x-\muh -\sqrt{(x-\muh)^2 + \Deltah^2} + \frac{\Deltah^2}{2x}\big)
\nonumber \\
&&\hspace{-.5cm}\qquad\quad\simeq 
-\frac{8}{15} \muh^{5/2} 
-\sqrt{\muh}\Deltah^2\big(\frac{1}{2}-
\ln\frac{\Deltah} {8 {\rm e}^{-2} \muh}\big).
\label{approximategseintegral}
\eea
This yields:
\bea
&&e_G \simeq -\frac{\sqrt{\muh}}{2}\Deltah^2 +\Deltah^2 (\deltah - 2\muh) \gamma^{-1}
+\sqrt{\muh} \Deltah^2 \ln\frac{\Deltah} {8 {\rm e}^{-2} \muh}
\nonumber \\
&& \qquad -\int_0^{\hh} d\hh' \mgh(\hh')
 - \frac{8}{15} \muh^{5/2},
\label{eq:gsesc}
\eea
as the BCS-regime normalized ground-state energy that we shall analyze in the remainder of 
this section.

\subsubsection{Normal phase}
\label{subsectionnormal}
Zero-temperature phases are stationary points of $e_G$, satisfying the gap equation 
Eq.~(\ref{eq:gapequationgeneralnorm}).  Since $e_G$ is a function only of $\Deltah^2$, it is obvious 
that $\Deltah = 0$ is always such a stationary point.  This solution represents the normal (N) state
consisting of spin-up and spin-down Fermi surfaces characterized by chemical potentials
$\mu_\uparrow$ and $\mu_\downarrow$.  
The corresponding ground-state energy simply counts the energetic contributions from
these two Fermi seas [See Eq.~(\ref{eq:havgnormal4})]:
\bse
\bea
&&\hspace{-1.6cm}E_{G,N} \!=\! 
\sum_\bk \!\Big[\big(\xi_k\!  -\!  h\big)\Theta(h\!  -\!  \xi_k)
\!+\!\big(\xi_k\!+\!h\big)\Theta(-h\!-\!\xi_k)\Big],
\\
&&\hspace{-1.6cm}\qquad \!=  \!
 - \frac{4c}{15} \big[(\mu \! + \! h)^{5/2}\Theta(\mu \!+ \!h) \!
+ \! (\mu \! - \! h)^{5/2}\Theta(\mu \! -\!h)].
\label{519}
\eea
\ese
We note that in the normal state Eq.~(\ref{519}) is valid for an
arbitrary relation between $h$ and $\mu$ and therefore (in contrast to 
small $h\ll\mu$ approximations, valid in the superfluid regime, used below)
allows us to access the high-polarization regime of the normal state.

In terms of our dimensionless variables, the normalized ground-state energy 
$e_{G,N}$ is thus 
\bse
\bea
&&\hspace{-1.5cm}e_{G,N} \!=\!  - \frac{4}{15}  \Big[(\muh\! +\! \hh)^{\frac{5}{2}}\Theta(\muh\!+\!\hh)\! 
+\! (\muh\! -\! \hh)^{\frac{5}{2}}\Theta(\muh\!-\!\hh)],
\label{eq:gsenormalexact}
\\
&&\hspace{-1.5cm}\qquad \simeq -\frac{8}{15} \muh^{5/2} -\sqrt{\muh}\hh^2,\,\,\, \hh\ll\muh.
\label{eq:gsenormalapprox}
\eea
\ese
 Equation~(\ref{eq:gsenormalapprox}) 
applies in the linear regime $\hh \ll \muh$, with $\sqrt{\muh}\hh^2$ the harmonic
Pauli-paramagnetic contribution, and will  frequently be sufficient in
the low-polarization normal phase.  

The normal-state atom density $n_N$ and polarization $\mg_N$ 
at fixed $h$ and $\mu$ 
 [following from Eqs.~(\ref{eq:numequationgeneral}) and 
~(\ref{eq:polarizationequationgeneral})]  are also  particularly simple:
\bse
\label{mnhctwo}
\bea
\label{eq:nn}
&&\hspace{-1.6cm}n_N\! = \!\frac{2c}{3}\! \big[(\mu\!+\!h)^{3/2}\Theta(\mu+h)\! +\! (\mu\! -\!h)^{3/2}\Theta(\mu-h)\big],
\\
&&\hspace{-1.6cm}\mg_N\! =\! \frac{2c}{3} \big[(\mu\!+\!h)^{3/2}\Theta(\mu+h)\! -\! (\mu\! -\!h)^{3/2}\Theta(\mu-h)\big].
\label{eq:magnn}
\eea
\ese
If we impose a total atom density $n = \frac{4}{3}c\ef^{3/2}$ and population imbalance
$\Delta N$, Eqs.~(\ref{mnhctwo}) can be written in the dimensionless form
\bse
\label{mnhctwop}
\bea
&&\hspace{-1.5cm}1 = \frac{1}{2}\big[(\muh + \hh)^{\frac{3}{2}}\Theta(\muh+\hh) +
  (\muh- \hh)^{\frac{3}{2}}\Theta(\muh-\hh)\big],
\label{eq:dimnormnum}
\\
&&\hspace{-1.5cm}\frac{\Delta N}{N} \!=\! \frac{1}{2}\big[(\muh + \hh)^{\frac{3}{2}}\Theta(\muh+\hh)\! -\! 
 (\muh- \hh)^{\frac{3}{2}}\Theta(\muh-\hh)\big].
\label{eq:dimnormmag}
\eea
\ese
Note that the normal state undergoes a transition with increasing $h$ from a system with 
two Fermi surfaces with energies $\mu\pm h$ 
to a system that is fully polarized, with only one Fermi surface.  Clearly, this happens at 
$\hh = \muh$, namely $\mu_\downarrow = 0$ in terms of the dimensionful variables.
 According to Eq.~(\ref{eq:dimnormnum}), this implies the transition occurs at 
$\hh_p = 2^{-1/3}$, which, when inserted into Eq.~(\ref{eq:dimnormmag}) indeed yields $\Delta N = N$, 
i.e., a fully polarized
state.  This critical $\hh_p$ is displayed in Fig.~\ref{fig:bcshphasediagram}
as a horizontal dashed line inside the N phase.

%%%%%%%%%%%%%%%%%%%%%%%%%%%%%%%%

%-----------------------------
%
% fig%12
%
\begin{figure}[bth]
\vspace{1.4cm}
\centering
\setlength{\unitlength}{1mm}
\begin{picture}(40,40)(0,0)
\put(-50,0){\begin{picture}(0,0)(0,0)
\includegraphics{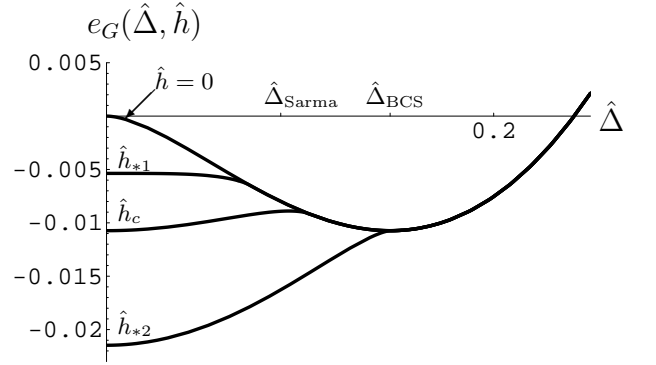}
\end{picture}}
\put(-9,48) {\large{$e_G(\Deltah,\hh)$}}
\put(0,40.5) {$\hh=0$}
\put(0,40.5) {\vector(-1,-1){4}}
%
%\put(-5,26) {$0.085$}
%\put(-5,21) {$0.104$}
%\put(-5,11.5) {$0.13$}
%
\put(-5,30) {$\hh_{*1}$}
\put(-5,23.0) {$\hc$}
\put(-5,8.0) {$\hh_{*2}$}
\put(59,35.0) {\large{$\Deltah$}}
\put(14.5,38.5) {$\Deltah_{\rm Sarma}$}
\put(28,38.5) {$\deltahbcs$}
\end{picture}
\vspace{-.5cm}
\caption{(Color online) 
 Plot of $e_G$ [Eq.~(\ref{eq:gsescapprox}), dropping the final $\Deltah$-independent
constant] at $\muh =1$ as a function of 
$\Deltah$ for four values of $\hh$ (labeled by $\hh$): For $\hh = 0$, no applied chemical potential
difference, $e_G(\Deltah)$ has a minimum at $\Deltah = \deltahbcs \approx 0.15$, representing the fully paired BCS state. With
increasing $\hh$, for $\hh = \hh_{*1}$ a local minimum develops at $\Deltah = 0$ that becomes degenerate with
the minimum at $\deltahbcs$ for $\hh = \hc$.  A local maximum is at $\Deltah = \Deltah_{\rm Sarma}$.
For $\hh>\hh_{*2}$, the minimum at $\deltahbcs$ disappears. 
}
\label{sarmaplots}
%
%  sarmaplots.nb 
%
\end{figure}
%------------------------------

\subsubsection{BCS superfluid phase and the superfluid to normal transition}
\label{eq:bcsandtrans}
At large positive detuning, fermionic attraction is weak and any pairing leads to 
$\Delta$ that is exponentially small in $(\deltah - 2\muh)/\gamma$.
We have already partially taken this into account, using the approximation Eq.~(\ref{approximategseintegral})
to obtain Eq.~(\ref{eq:gsesc}).
Next, we proceed by using the conditions $\Deltah \ll \muh$ and  $h \ll \muh$
to simplify $\mgh(\hh)$ in Eq.~(\ref{eq:mbcsdimless})  and 
 in the remaining integral in 
Eq.~(\ref{eq:gsesc}).  This yields
\be
 \mgh(\hh) \simeq 2\sqrt{\muh} \sqrt{\hh^2 - \Deltah^2}\Theta(\hh-\Deltah),
\label{eq:approxmagsarma}
\ee
and
\bea
&&e_G \simeq -\frac{\sqrt{\muh}}{2}\Deltah^2 +\Deltah^2 (\deltah - 2\muh) \gamma^{-1}
+\sqrt{\muh} \Deltah^2 \ln\frac{\Deltah} {8 {\rm e}^{-2} \muh}
\nonumber \\
&& \qquad -\sqrt{\muh} \big[
\hh\sqrt{\hh^2 -\Deltah^2} - \Deltah^2 \cosh^{-1} (\hh/\Deltah)\big]
\Theta(\hh - \Deltah)
\nonumber \\
&&\qquad - \frac{8}{15} \muh^{5/2},
\label{eq:gsescapprox}
\eea
that we have plotted as a function of $\Deltah$ for various $\hh$'s in Fig.~\ref{sarmaplots}.
To determine the stable ground state, we study minima of $e_G$, given
by the gap equation,
 $\partial e_G/\partial\Deltah = 0$ [Eq.~(\ref{eq:gapequationgeneralnorm})]:
\bea
&&0 \simeq \coupling^{-1} \Deltah (\deltah - 2\muh) + \Deltah\sqrt{\muh}\ln \frac{\Deltah}{8{\rm e}^{-2} \muh}
\nonumber \\
&&\qquad \qquad 
+ \Deltah \sqrt{\muh} \cosh^{-1} (\hh/\Deltah)\Theta(\hh - \Deltah).
\label{eq:sarmagapequation}
\eea
The nature of possible solutions depends on the value of $\hh$, as is clear
from the evolution of $e_G(\Deltah,\hh)$ with a set of increasing $\hh$ values at fixed
chemical potential $\muh$, illustrated in Fig.~\ref{sarmaplots}. For $\hh$ sufficiently
small, $0<\hh<\deltahbcs/2$, the BCS logarithm [see Eq.~(\ref{eq:gsescapprox})] dominates,
leading to a single BCS minimum at $\Deltah \simeq \deltahbcs$, with
\be\label{eq:baredeltanoughtmaintext}
\deltahbcs(\deltah,\muh) \equiv 8{\rm e}^{-2} \muh{\rm e}^{-\gamma^{-1}(\deltah - 2\muh)/\sqrt{\muh}},
\ee
with the normal state characterized by an unstable maximum at $\Deltah=0$,
discussed in Sec.~\ref{subsectionnormal}. The corresponding BCS ground state energy is given by
[inserting Eq.~(\ref{eq:baredeltanoughtmaintext}) into Eq.~(\ref{eq:gsescapprox})]
\be
e_{G,SF}\simeq -\frac{8}{15} \muh^{5/2} -\frac{\sqrt{\muh}}{2} \deltahbcs^2,
\label{eq:gsebcs}
\ee
with the terms multiplying $\Theta(\hh - \Deltah)$ dropping out and therefore not
influencing the location and depth of the BCS minimum for $\hh < \deltahbcs$. This behavior persists
until $\hh=\hh_{*1}=\deltahbcs/2$ at which point the normal state extremum at
$\Deltah=0$ develops into a metastable {\em local} minimum, separated from the
stable SF minimum at $\deltahbcs$ by a maximum at~\cite{Sarma} 
\be
\Deltah_{\rm Sarma} \simeq \deltahbcs \sqrt{\frac{2\hh}{\deltahbcs} - 1},\,\,\,
\hh_{*1} < \hh < \hh_{*2},
\label{eq:sarmastate}
\ee
where $\hh_{*2} = \deltahbcs$.
Equation~(\ref{eq:sarmastate}) follows from Eq.~(\ref{eq:sarmagapequation})
upon using  $\cosh^{-1}x = \ln \sqrt{x^2-1}+x$ and Eq.~(\ref{eq:baredeltanoughtmaintext}).
With a further increase in $\hh$ the maximum at $\Deltah_{\rm Sarma}$ moves out
towards the  $\hh$-independent BCS minimum, joining it at $\hh=\hh_{*2}$ as 
can be seen from Eq.~(\ref{eq:sarmastate}), so that for $\hh>\hh_{*2}$ the only stationary
point of $e_G$ is at $\Deltah=0$. 

In contrast to a number of erroneous conclusions in the literature~\cite{Pao,Iskin,Sedrakian,Paonote} (that
identified $\Delta_{\rm Sarma}$ with a magnetized superfluid ground state by
studying the gap equation without checking the corresponding energy),
this maximum at $\Delta_{\rm Sarma}$, as first shown by Sarma~\cite{Sarma},
clearly does not correspond to any physical (stable) phase of an
attractive Fermi gas.

The BCS ground state remains a global minimum for $0 < \hh <
\hc(\muh)$.
For $\hh > \hc(\muh)$, 
 the energy $e_{G,N}$ 
of the normal state $\Deltah=0$ minimum
 drops below $e_G(\deltahbcs)$.
The fermion gas then undergoes a first-order transition to the normal ground
state at the critical chemical potential difference $\hc(\muh)$ given by 
equating the normal ground-state energy $e_{G,N}$ [Eq.~(\ref{eq:gsenormalexact})]
with the exact~\cite{footnoteexact} superfluid-state energy $e_{G,SF}$,
\bea
\label{eq:egsffull}
&&\hspace{-.5cm}e_{G,SF} = 
\gamma^{-1} \Deltah^2(\deltah - 2\muh)
 \\ 
\nonumber 
&&\hspace{-.5cm}\qquad \qquad + 
\int_0^\infty dx \sqrt{x} \big(x-\muh -\sqrt{(x-\muh)^2 + \Deltah^2} + \frac{\Deltah^2}{2x}\big),
\eea
(obtained from Eq.~(\ref{eq:gsetotallygeneral}) by setting $\mgh =0$). This yields 
\begin{widetext}
\be
\label{eq:533exact}
 - \frac{4}{15}  \Big[(\muh\! +\! \hc)^{5/2}\Theta(\muh\!+\!\hc)\! 
+\! (\muh\! -\! \hc)^{5/2}\Theta(\muh\!-\!\hc)]
= 
\gamma^{-1} \Deltah_0^2(\deltah - 2\muh)+
 \int_0^\infty dx \sqrt{x} \big(x-\muh -\sqrt{(x-\muh)^2 + \Deltah_0^2} + \frac{\Deltah_0^2}{2x}\big),
\ee
\end{widetext}
with the right-hand side evaluated at its  minimum $\Deltah = \Deltah_0$ satisfying   
\bea
&&\hspace{-1.2cm}0 = \frac{\partial e_{G,SF}}{\partial\Deltah}\big|_{\Deltah = \Deltah_0} ,
\\
&&\hspace{-.9cm}\!\!=\frac{2}{\gamma}(\deltah - 2\muh)\! +\! \int_0^\infty dx \sqrt{x} \big(
\frac{1}{x} - \frac{1}{\sqrt{(x-\muh)^2 + \Deltah_0^2}}
\big),
\label{eq:533exact2}
\eea
where in the second line we canceled the  $\Deltah = 0$ solution.  Equations~(\ref{eq:533exact}) 
and ~(\ref{eq:533exact2}) then implicitly give $\hc(\muh)$ [or, equivalently, $\muh_c(\hh)$] that we will use shortly. 
In the limit of $\Deltah_0 \ll \muh$ the right-hand side of Eq.~(\ref{eq:533exact}) 
  reduces to Eq.~(\ref{eq:gsebcs}) with $\Deltah_0$ given by $\deltahbcs$, yielding
\bea
&&\hspace{-1.2cm}\frac{4}{15} [(\muh+ \hc)^{5/2} + (\muh - \hc)^{5/2}\Theta(\muh - \hc) - 2 \muh^{5/2}]
\label{generalhc}
 \\
&&\qquad \qquad 
\hspace{-1.2cm}\simeq \frac{1}{2}\sqrt{\muh}\deltahbcs^2
=
 32{\rm e}^{-4} \muh^{5/2} {\rm e}^{-2\gamma^{-1} (\deltah  - 2\muh)/\sqrt{\muh}}.\nonumber
\eea

Deep in the BCS regime ($\delta \gg 2 \ef$) the molecular condensate 
and the corresponding gap $\deltahbcs$, given by Eq.~(\ref{eq:baredeltanoughtmaintext})
 above, are
exponentially small.  We expect (and self-consistently find) that the
critical Zeeman-field boundary $\hc(\delta)$ tracks the gap
$\deltahbcs(\deltah,\muh)$ and therefore must also be exponentially small in
$(\deltah-2\muh)/\gamma$. 

As we will see below (when we consider the more experimentally-relevant ensemble of fixed atom number), 
the crossover regime
$\delta < 2\ef$ corresponds to $\mu$ locking to a value slightly below
$\delta/2$, with $\deltah/2 -\muh\approx \curO(\gamma\ln\gamma)\ll 1$. This is still
sufficient to ensure that, in the crossover regime, the BCS estimate of
$\deltahbcs(\muh,\deltah)$ and $e_{G,SF}$ remain valid [since  $\deltahbcs\ll \muh$
allowing the integral in Eq.~(\ref{eq:egsffull}) to be approximated by Eq.~(\ref{approximategseintegral})], 
but with $\deltahbcs(\mu,\delta)/\muh\approx \curO(\gamma^{1/2})$
(rather than exponentially small in $1/\gamma$).

Consequently, in the BCS ($\deltah \gg 2\muh$) and crossover
[$\deltah-2\muh\approx \curO(\gamma\ln\gamma)$] regimes the condition
$\hc\ll\muh$ is well-satisfied. This allows an accurate approximation of
the normal state energy $e_{G,N}$, Eq.~(\ref{eq:gsenormalexact}), and correspondingly the
left-hand side of Eq.~(\ref{generalhc}) by its lowest order Taylor expansion in
$\hc/\muh$. This {\em linear} Pauli paramagnetic approximation then gives
an accurate prediction for the critical Zeeman field
\bse
\label{eq:hcboth}
\bea
\label{eq:hcpre}
&&\hspace{-1cm}\hc(\muh) \simeq \deltahbcs/\sqrt{2}, \,\,\,\text{for $\hc\ll \muh$,}
\\
&&\hspace{-.8cm}\qquad \simeq 4\sqrt{2}{\rm e}^{-2} \muh{\rm e}^{-\gamma^{-1}(\deltah - 2\muh)/\sqrt{\muh}},
\label{eq:hc}
\eea
\ese
that in the BCS regime self-consistently satisfies the condition $\hc \ll
\muh$ used to obtain it.

For $\muh$ close to $\deltah/2$, Eq.~(\ref{eq:533exact}) is approximately [to $\curO(\gamma)$] satisfied
for {\it any\/} $\hh$ since we can neglect the left-hand side and the second term on the right-hand side.  This
behavior is also reflected in the approximate formula (taken beyond its strict regime 
of validity) Eq.~(\ref{eq:hc}) that exhibits rapid variation
for $\muh$ near $\deltah/2$.  Thus, in this regime at the transition the critical $\muh$ is 
approximately given by 
\be
\muh_c(\hh,\deltah)\approx \deltah/2.
\label{eq:star}
\ee
Hence, below we will use Eq.~(\ref{eq:star})  as an accurate (in
 $\gamma\ll 1$, narrow resonance limit) form for the high Zeeman field regime
$\hc(\muh,\deltah)$, where the BCS condition  $\Deltah \ll \muh$ is violated.
This will be essential to determine the upper boundary $\hctwo(\deltah)$ of the coexistence
region in the crossover regime.

%%%%%%%%%%%%%%%%%%%%%%%%%%%%%%%%%%%%%%%%%%%%%%%%%%%%%%%%%%

It is important to note that, because of the step functions in the ground-state energy
Eq.~(\ref{eq:havgBCSp}), the curve of $e_G(\Deltah,\hh)$ is identically $\hh$-independent for
$\Deltah > \hh$. Consequently the BCS minimum at $\deltahbcs$ (stable for $0 < \hh
< \hc$) and the corresponding energy $e_G(\deltahbcs)$ are
strictly $\hh$-independent, ensuring that the BCS state is indeed
characterized by a vanishing magnetization,
\be
\mg_{SF} =  -\frac{\partial E_G}{\partial h}=  0,
\ee
and density given by
\be
\label{eq:nsf}
n_{SF} \simeq \frac{2\Delta^2}{g^2}
+ \frac{4c}{3} \mu^{3/2} +
\frac{c\Delta^2}{\sqrt{\mu}} \big(
\frac{5}{4} - \frac{1}{2} \ln \frac{\Delta}{8{\rm e}^{-2} \mu} 
\big).
\ee

Since $\hh$ only enters through a term multiplied by a step function,
$\Theta(\hh-\Deltah)$, $\hh$-independence of the stable minimum at $\deltahbcs$
persists until $\hh=\deltahbcs$, beyond which the energy at this minimum
{\it would\/} have become $\hh$-dependent (resulting in a magnetized paired
superfluid) had the minimum survived.  However, as we saw above, the
BCS minimum becomes unstable to the normal state for $\hh\geq\hc
\simeq\deltahbcs/\sqrt{2}$, {\it before} the point of this
magnetized-superfluid condition of $\hh=\deltahbcs$ is reached.

As expected at a first-order transition with a tuned chemical potential
(rather than tuned density), our system exhibits jumps in the
density at $\hc$. 
This occurs because the densities of the superfluid and normal
states at the transition $\hc(\muh)$ [using Eq.~(\ref{eq:hcpre}) and converting to dimensionful 
quantities]
\bea
\label{eq:nsf2}
&&\hspace{-1.4cm}n_{SF}\simeq \frac{2\Delta_{BCS}^2}{g^2}\! +\! \frac{4c}{3}\mu^{3/2}\! + \! 
\frac{c\Delta_{BCS}^2}{\sqrt{\mu}} \big(\frac{5}{4}\! - \!\frac{1}{2}\ln \frac{\Delta_{BCS}}{8{\rm e}^{-2}\mu}
\big),
\\
&&\hspace{-1.4cm}n_N\simeq  \frac{4c}{3}\mu^{3/2} +\frac{c\Delta_{BCS}^2}{4\sqrt{\mu}},
\label{eq:nn2}
\eea
are different.
Thus, upon increasing $\hh$ past $\hc(\muh,\deltah)$ there is a density discontinuity equal to
\be
n_{SF} - n_N \simeq \frac{2\Delta_{BCS}^2}{g^2}+ \frac{c\Delta_{BCS}^2}{\sqrt{\mu}} \big(1-
\frac{1}{2}\ln \frac{\Delta_{BCS}}{8{\rm e}^{-2}\mu} 
\big).
\label{eq:jump}
\ee
Similarly, since the BCS paired superfluid is characterized by a vanishing
magnetization and the normal state by $m_N$ given by Eq.~(\ref{eq:magnn})
there is a jump discontinuity in $m$ equal to $m_N$ evaluated at
$h_c$.

\subsection{First-order SF-N transition at fixed density: phase separation and
coexistence}
\label{SEC:firstorderBCS}
Because we are primarily interested in applications of the theory to
degenerate trapped atomic-gas experiments, where it is the atom
number $N$ (rather than chemical potential $\mu$) that is fixed (also it is
$\Delta N =N_\uparrow -N_\downarrow$ rather than $h$ that is imposed, but for pedagogical purposes
we delay the analysis of this fixed $\Delta N$ ensemble until Sec.~\ref{SEC:fixedpopbcs}), in
this section we study the above transition in the  fixed {\em average} density 
$n=N/\vol$  and imposed $h$ ensemble\cite{CommentsONn}.

However, there is no guarantee that the system's true ground
state corresponds to one of our assumed spatially homogeneous ground
states. In fact, it is clear from the discussion at the end of the last
subsection that the existence of the density discontinuity Eq.~(\ref{eq:jump}) at
$h_c(\delta)$
 implies that, at $h=h_c(\delta)$, a
homogeneous atomic gas with density $n$ anywhere in the range between $n_N$
and $n_{SF}$ is in fact unstable to phase separation. As we will
see below, for densities in this range the ground state is an
inhomogeneous coexistence of  normal and superfluid phases at the
critical chemical potential $\mu_c(h,\delta)$, with corresponding densities
$n_N(\mu_c(h,\delta),\delta)$, $n_{SF}(\mu_c(h,\delta),\delta)$, appearing in
fractions $1-x(h,\delta,n)$ and $x(h,\delta,n)$, determined by the constraint
that the {\em average} density is the imposed one (or equivalently the
total number of atoms is $N$). 

%-----------------------------
%
% fig%13
%
\begin{figure}[bth]
\vspace{1.4cm}
\centering
\setlength{\unitlength}{1mm}
\begin{picture}(40,40)(0,0)
\put(-50,0){\begin{picture}(0,0)(0,0)
\includegraphics{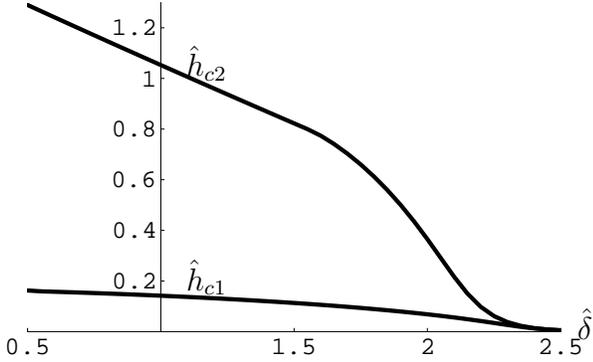}
\end{picture}}
\put(5,39) {\large{$\hh_{c2}$}}
\put(57,4) {\large{$\deltah$}}
\put(5,10.5) {\large{$\hcone$}}
\end{picture}
\vspace{-.5cm}
\caption{ Upper curve is $\hh_{c2}$ [Eq.~(\ref{eq:hctwoequationsbcs})]
and lower curve is $\hcone$ [Eq.~(\ref{eq:hconebcs})], solved numerically, 
that bound the coexistence region in the BCS and crossover regimes. }
\label{fig:hcbcs} 
%
%  phasediagram_twochannel.nb 
%
\end{figure}
%------------------------------

Thus, as illustrated in Fig.~\ref{fig:hcbcs}  (and the full phase diagram, Fig.~\ref{fig:hphasetwo}), at
fixed average density $n$, the critical Zeeman field $h_c(\delta)$ splits
into  lower- and upper-critical fields, $h_{c1}(\delta)$ and $h_{c2}(\delta)$,
bounding the coexistence region from below and above, respectively. To
understand how these emerge in detail, we imagine increasing the
Zeeman field $h$ (at fixed  $n$ and detuning $\delta$ within the BCS or
crossover regimes) from low values starting from the singlet
superfluid state as the global minimum.  As $h$ is increased, the
chemical potential $\mu_{SF}(n,\delta)$ (which, because the BCS superfluid is
a singlet, is in fact $h$ independent), determined by the superfluid equation
of state (written in terms of dimensionless variables), satisfies
\bse
\label{eq:hconebcs}
\bea
&&\hspace{-1.7cm}\frac{4}{3}\! =\! \frac{5\Deltah^2}{4\sqrt{\muh_{SF}}} \!
 +\!\frac{4}{3} \muh_{SF}^{3/2} \!+ \! \frac{2\Deltah^2}{\gamma}  
- \frac{\Deltah^2}{2\sqrt{\muh_{SF}}}\ln\frac{\Deltah}{8{\rm e}^{-2} \muh_{SF}},
\label{eq:musc}
\eea
to keep the density fixed at $n$. In this
process, as $\hh$ is increased, the ground-state energy
function $e_G(\hh,\muh,\deltah)$ changes according to Eq.~(\ref{eq:gsetotallygeneral}) with 
the relative
level of the normal and superfluid minima changing (see Fig.~\ref{sarmaplots}). At a
sufficiently large Zeeman field $\hcone(\deltah,n)$, given by
\be
\hcone =  \hc(\muh_{SF}),
\label{deltanoughtmixed}
\ee
\ese
the two minima become degenerate.
This would then naively imply that the
system should jump to the normal ground state at these values of
$\muh=\muh_{SF}(n,\deltah)$, $h$, and $\delta$. However, as seen from Eqs.~(\ref{eq:nsf2})
and (\ref{eq:nn2}), since
the normal ground state has a density that (at the same chemical
potential $\mu$) is distinct from that of the superfluid state, this discontinuous
transition to the normal state would not keep the density fixed at the
imposed value $n$.

%-----------------------------
%
% fig%14
%
\begin{figure}[bth]
\vspace{3.5cm}
\hspace{-.1cm}
\centering
\setlength{\unitlength}{1mm}
\begin{picture}(40,80)(0,0)
\put(-55,55){\begin{picture}(0,0)(0,0)
\includegraphics{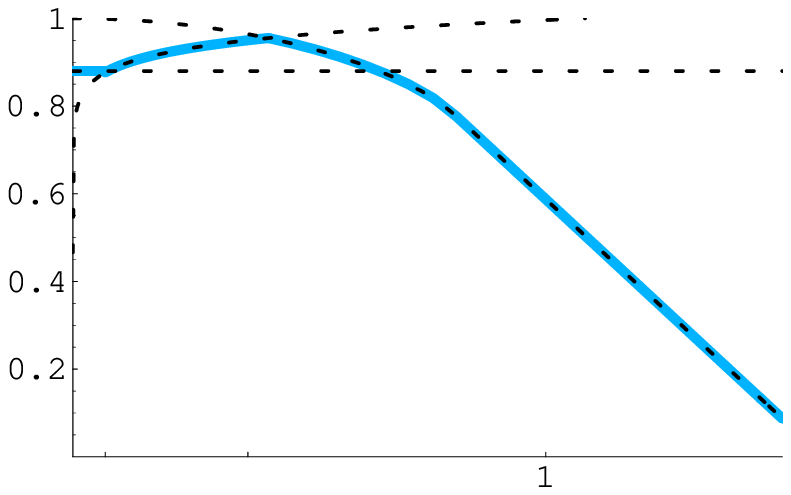}
\end{picture}}
\put(-15.5,107) {\large{$\muh$}}
\put(-13,101) {(a)}
\put(-6,105.5)  {$\muh_N$}
\put(20,105.5)  {$\muh_c$}
\put(35,100.5)  {$\muh_{SF}$}
\put(59.0,59) {\large{$\hh$}}
\put(-12,56.5) {$\hcone$}
\put(2.5,56.5) {$\hctwo$}
\put(-55,0){\begin{picture}(0,0)(0,0)
\includegraphics{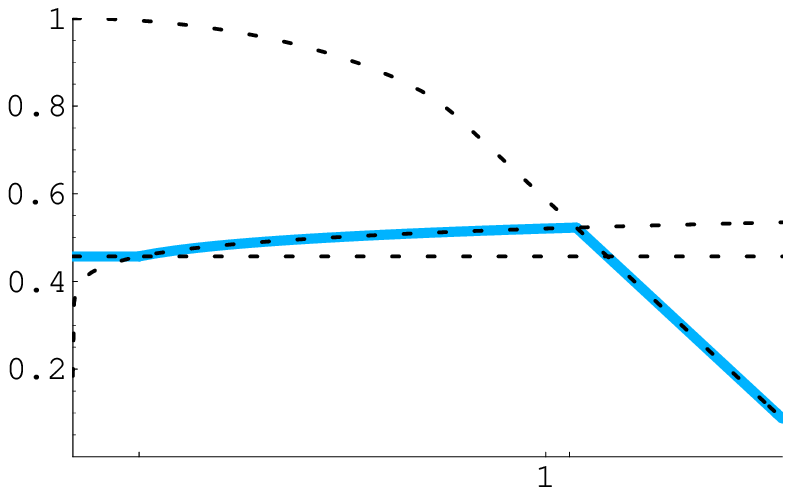}
\end{picture}}
\put(-13,45) {(b)}
\put(-15.5,52) {\large{$\muh$}}
\put(59.0,4) {\large{$\hh$}}
\put(45,30) {$\muh_c$}
\put(50,26.5) {$\muh_{SF}$}
\put(-8.5,1.5) {$\hcone$}
\put(35,1.5) {$\hctwo$}
\put(20,44) {$\muh_N$}
%
%--------------------------------
%
\end{picture}
\vspace{-.5cm}
\caption{(Color online) Plots of the normalized chemical potential $\muh(\hh,\deltah)$ (solid blue line) as a function of 
 the normalized chemical potential difference $\hh$ at (a) high detuning $\deltah = 2.0$ 
and (b) low detuning $\deltah = 1.0$.  For $\hh<\hcone$, the system is in the SF
phase with $\muh = \muh_{SF}$.  For $\hcone<\hh<\hctwo$, the system is in the mixed
phase, with $\muh$ constrained to lie on the first-order critical boundary $\muh= \muh_c(\hh)$.  
For $\hctwo<\hh$, the system is in 
the pure N phase, with $\muh  = \muh_N(\hh)$ [the solution to Eq.~(\ref{eq:dimnormnum})]. 
 }
\label{fig:muvshplots} 
%
%  morelongmanuplots.nb (both)
%
\end{figure}
%------------------------------

The only solution to this dilemma (which is generic to first-order transitions) 
is for the system to get pinned at the coexistence curve $\muh_c(\hh,\deltah)$,
Eq.~(\ref{eq:533exact}), defined by equality of the normal-state and superfluid-state
minima.
 A subsequent increase in $\hh > \hcone$ changes $\mu$ along the critical curve $\mu_c(h,\delta)$,
keeping the normal and superfluid ground states degenerate.
 This, however,
leads to a chemical potential $\muh_c(\hh,\deltah)$, illustrated in Fig.~\ref{fig:muvshplots},  that no
longer allows the density of either of the pure ground states, N and
SF to be equal to $n$. However, the total imposed atom number can still be
satisfied by a mixture of coexisting SF and N states~\cite{Bedaque} in
respective proportions $x(\hh,\deltah)$ and $1-x(\hh,\deltah)$, defined by
\be
\frac{N}{\vol} =  
x(\hh,\deltah)n_{SF}(\muh_c(\hh),\deltah))
+
[1-x(\hh,\deltah)] n_N(\hh,\muh_c(\hh)) .
\ee
This evolution of the chemical potential according to $\muh_c(\hh)$
continues until $\hh$ has increased sufficiently, so that the number
constraint equation can be satisfied by a pure normal state that minimizes $E_G$.  The
corresponding value of the Zeeman field is precisely the upper-boundary of
the coexistence region, with
\bse
\label{eq:hctwoequationsbcs}
\bea
&&\hspace{-1.4cm}\hctwo = \hc(\muh_N),
\label{deltanoughtmixed2}
\\
&&\hspace{-1.4cm}1 = \frac{1}{2}\big[(\muh_N+ \hctwo)^{\frac{3}{2}} \!+ \! 
(\muh_N- \hctwo)^{\frac{3}{2}}\Theta(\muh_N-\hctwo)\big],
\label{eq:mun}
\eea
\ese
and $\hc(\muh)$ given by Eq.~(\ref{eq:533exact}) in the preceding subsection.

Hence, as is clear from the above discussion,  the intersections of the
$\muh_c(\hh)$ curve with $\muh_{SF}(n,\deltah)$ and $\muh_N(\hh,n)$ determine $\hcone(n,\deltah)$
and $\hctwo(n,\deltah)$.  The full
evolution of the chemical potential with $\hh$, from the SF state through
the coexistence region and to the normal state for detuning in the BCS
and crossover regimes, is illustrated in Fig.~\ref{fig:muvshplots}.

The two pairs of equations Eq.~(\ref{eq:hconebcs}a,b) and (\ref{eq:hctwoequationsbcs}a,b) 
for $\hcone$ and $\hctwo$ can be straightforwardly
solved numerically. We do this for the case of $\gamma=0.1$, with
results illustrated in Fig.~\ref{fig:hcbcs}.  In the next subsections, we obtain accurate analytic approximations
for $\hcone$ and $\hctwo$ in the BCS and crossover regimes.

%%%%%%%%%%%%%%%%%%%%%%%%%%%%%%%%%%%%%%%%%%%%%%%%%%%%%%%%%%%%%%%%

\subsubsection{$h_{c1}$ and $h_{c2}$ for the BCS limit $\delta\gg 2\ef$}
\label{SEC:hconetwohighdetuning}

As discussed in Sec.~\ref{SEC:bcsandtransition},
at large $\deltah$ we will have $\hcone,\hctwo \ll \muh$, allowing the use of  Eq.~(\ref{eq:hcboth})
for  $\hc(\mu)$, valid in the linear-response (to $\hh$) limit.  Starting with $\hcone$, for $\deltah \gg 2$ 
the normalized gap
$\Deltah$ is exponentially small so that we may neglect the  first term on
the right side of Eq.~(\ref{eq:musc}).  The last
term in Eq.~(\ref{eq:musc}) may be simplified using the gap equation, yielding 
\be
\label{eq:musc2}
\frac{4}{3} \approx
 \frac{4}{3} \muh_{SF}^{3/2} +  \frac{2\Deltah^2}{\gamma} + \frac{\Deltah^2}{2\muh_{SF} \gamma}(\deltah - 2\muh_{SF}).
\ee
Solving Eq.~(\ref{eq:musc2}) to leading order in small $\gamma$ gives:
\be
\muh_{SF} \approx 1- \frac{1}{4} (\deltah+2 ) \Deltah^2 \gamma^{-1}.
\label{eq:musc3}
\ee
It is convenient to define the BCS gap Eq.~(\ref{eq:baredeltanoughtmaintext})
at $\muh_{SF} = 1$:
\be
\label{eq:deltaf}
\Deltah_F[\gamma,\deltah]  \equiv 8{\rm e}^{-2}{\rm e}^{-\gamma^{-1}(\deltah - 2)}.
\ee
With this definition, to leading order it is valid to replace $\Deltah^2$ on the right side of Eq.~(\ref{eq:musc3}) 
with $\Deltah_F^2$.  Taking advantage of $\muh_{SF} \approx 1$, we 
Taylor expand $\Deltah$ in small $\muh_{SF} - 1$:
\be
\label{eq:deltaderivative}
\Deltah[\gamma,\deltah,\muh_{SF}] \!\simeq \! \Deltah_F[\gamma,\deltah]\! + \!(\muh_{SF} - 1)\Deltah'[\gamma,\deltah,1],
\ee
where
\bse
\bea
\Deltah'[\gamma,\deltah,1]\! &=&\!\! 8{\rm e}^{-2} {\rm e}^{-\gamma^{-1}(\deltah -2)}\big(
\frac{\deltah}{2\gamma}\! +\! \frac{1}{\gamma} + 1
\big),\\
 &\simeq& 4{\rm e}^{-2} \gamma^{-1}(\deltah + 2){\rm e}^{-\gamma^{-1}(\deltah -2)},
\\
&\simeq& \frac{1}{2}\gamma^{-1}(\deltah + 2) \Deltah_F[\gamma,\deltah],
\label{eq:deltaprime}
\eea
\ese
with the prime denoting differentiation with respect to $\muh$.  Using Eq.~(\ref{eq:deltaderivative})
along with Eq.~(\ref{eq:deltaprime}) and Eq.~(\ref{eq:musc3}), Eq.~(\ref{deltanoughtmixed}) 
becomes 
\bse
\label{eq:hconebcstwo}
\bea
\hcone &\approx & \frac{\Deltah[\gamma,\deltah,\muh_{SF}]}{\sqrt{2}},
\\
 &\approx & \frac{1}{\sqrt{2}}\Big[\Deltah_F - \frac{\gamma^{-2}}{8}(\deltah+2)^2\Deltah_F^3\big],
\\
&\approx & \frac{1}{\sqrt{2}} \Deltah_F[\gamma,\deltah] \exp\big[- \frac{\deltah^2}{8\gamma^2} \Deltah_F^2
\big],\label{eq:hconebcstwop}
\eea
\ese
where in the final result we have taken $\deltah \gg 2$ and 
 re-exponentiated the second factor, valid since $\Deltah_F$ is exponentially small
for $\gamma\ll 1$.  In the asymptotic large $\deltah$ limit, $\hcone$
thus decays exponentially with $\deltah$, as seen in Fig.~\ref{fig:hcbcs}.  

Similarly, $\hctwo$ can be obtained by 
 solving Eq.~(\ref{eq:hctwoequationsbcs}) iteratively utilizing the fact that, 
in the large-$\deltah$ limit, $\hctwo$ is exponentially small while $\muh_N \approx 1$.  Thus, in 
Eq.~(\ref{eq:mun}) we can expand in $\hctwo/\muh_N\ll 1$, yielding 
\be
1 \approx \muh_N^{3/2} +  \frac{3\hctwo^2}{8\sqrt{\muh_N}},
\label{eq:mun2}
\ee
which has the zeroth order solution $\muh_N \approx 1$; at this order
Eq.~(\ref{deltanoughtmixed2})
yields $\hctwo \approx \Deltah_F/\sqrt{2}$.  Inserting the latter expression
into Eq.~(\ref{eq:mun2}) yields a leading-order correction to $\muh_N$:
\be
\muh_N \approx 1- \frac{1}{4}\hctwo^2 \approx 1- \frac{1}{8} \Deltah_F^2.
\label{munint}
\ee
Using this inside  Eq.~(\ref{deltanoughtmixed2})  together with the expansion 
Eq.~(\ref{eq:deltaderivative}) (but with $\muh_N$ instead of $\muh_{SF}$)
our final leading-order result for $\hctwo$ is:
\bse
\label{eq:hctwobcstwo}
\bea
\label{eq:hctwobcspre}
\hctwo &\approx & \frac{1}{\sqrt{2}}\Big[\Deltah_F  - \frac{1}{16\gamma}(\deltah+2)\Deltah_F^3\Big],
\\
&\approx & \frac{1}{\sqrt{2}} \Deltah_F[\gamma,\deltah] \exp\big[- \frac{\deltah}{16\gamma} \Deltah_F^2
\big],
\label{eq:hctwobcs}
\eea
\ese
that decays exponentially with increasing $\deltah$ as exhibited in Fig.~\ref{fig:hcbcs}.  Note that, since 
$\deltah^2/8\gamma^2 \gg \deltah/16 \gamma$ for large $\deltah$, $\hctwo> \hcone$ in the large
detuning limit, i.e., the curves in Fig.~\ref{fig:hcbcs} never cross. 

\subsubsection{$h_{c1}$ and $h_{c2}$ for the crossover limit $0<\delta< 2\ef$}

We now compute $\hcone(\deltah)$  and $\hctwo(\deltah)$ at small $\deltah$.  To obtain $\hcone$, 
we first note that at large detuning in the 
 SF state at $\hh=0$, the full gap and
number equations are obtained from Eqs.~(\ref{eq:gapequationgeneralnorm})  and (\ref{eq:numequationgeneralnorm}) with Eq.~(\ref{eq:gsesc})
for $e_G$ [see also Eqs.~(\ref{eq:gapzeroh2}) and ~(\ref{eq:numzeroh2})]:
\bse
\label{eq:lowdelta}
\bea
\label{eq:gaplowdelta}
&&\hspace{-1cm}0\simeq 2\Deltah (\deltah - 2\muh )\gamma^{-1} + 
\sqrt{\muh} \Deltah \ln \frac{\Deltah} {8 {\rm e}^{-2} \muh},
\\
&&\hspace{-1cm}
\frac{4}{3}\simeq
\frac{5}{4} \frac{\Deltah^2}{\sqrt{\muh}} + \frac{4}{3} \muh^{3/2} + \frac{2\Deltah^2}{\gamma}
- \frac{\Deltah^2}{2\sqrt{\muh}}\ln \frac{\Deltah}{8{\rm e}^{-2} \muh}.
\label{eq:numlowdelta}
\eea
\ese
As  $\deltah$ is reduced below $2$, the system undergoes a crossover from the BCS
regime (where $\muh$ is pinned near unity)
to the BEC regime where $\Deltah$ is no longer exponentially small (although we still have
 $\Deltah \ll 1$) and $\muh$ begins to track $\deltah/2$, as atoms pair up into 
Bose-condensed molecules (see Fig.~\ref{BECBCSplot}). 
  How is this reflected in Eqs.~(\ref{eq:lowdelta})? As $\deltah$ drops
below $2$, $\Deltah$ grows such that  $\Deltah/\muh$ becomes $\curO(\gamma^{1/2})$,
 so that we may neglect the final $\Deltah \ln \Deltah/\muh$ term on 
the right 
side of Eqs.~(\ref{eq:lowdelta}).  Taking $\gamma \ll 1$ in Eq.~(\ref{eq:numlowdelta}) (so that the term
 $5\Deltah^2/4\sqrt{\muh}$ may be neglected) thus yields the following 
approximate solutions to Eqs.~(\ref{eq:lowdelta}):
\bse
\label{eq:lowdelta2}
\bea
\label{eq:mulowdeltaapprox}
\muh &\approx& \deltah/2,
\\
\Deltah &\approx& \sqrt{\frac{2\gamma}{3}} \sqrt{1- (\deltah/2)^{3/2}}.
\label{eq:gaplowdeltaapprox}
\eea
\ese
Note that the role of the equations has been reversed, with the gap equation fixing $\muh$ to be 
close to $\deltah/2$ [to $\curO(\gamma\ln\gamma)]$, and the number equation fixing $\Deltah$; we shall also see such behavior
in the asymptotic BEC regime at negative detuning.  Using
Eq.~(\ref{eq:gaplowdeltaapprox}) inside 
 Eq.~(\ref{eq:hcpre}) [still using the $\hh_c\ll\muh$ expression since, as we shall verify a posteriori,
$\hcone$ remains  small in this regime], 
we find for $\hcone = \hc(\muh_{SF})$:
\be
\hcone \approx \sqrt{\frac{\gamma}{3}} \sqrt{1- (\deltah/2)^{3/2}}.
\ee

Next, we compute $\hctwo$ in the crossover regime.  As in the preceding subsection, $\hctwo$ is
determined by combining the solution to the gap equation at low detunings ($\delta\simeq 2\mu$, 
which approximately solves the equation for $\hc$ at low detunings, as discussed in 
Sec.~\ref{eq:bcsandtrans})
with the normal-state
chemical potential $\muh_N$, given by Eq.~(\ref{eq:mun}). 
 We shall denote the solution to this equation 
 as $\hh^{(N)}(\muh)$, defined by
\be
1 = \frac{1}{2}\big([\muh + \hh^{(N)}(\muh)]^{3/2} 
+  [\muh - \hh^{(N)}(\muh)]^{3/2}\Theta[\muh-h^{(N)}(\muh)]\big).
\label{eq:hndef}
\ee
Although Eq.~(\ref{eq:hndef}) cannot be solved analytically for arbitrary $\muh$, 
 we {\it can\/} find solutions 
 for $\muh< \hh$ 
[so that the second term on the right side of Eq.~(\ref{eq:hndef}) vanishes] and
in the limit $\muh \gg \hh$:
\bea
\label{eq:hnlow}
\hh^{(N)}(\muh) &=& 2^{2/3} -\muh  \hskip 0.5cm\text{for $\muh <2^{-1/3}$,}
\\
&\approx&  \sqrt{\frac{8}{3}} \sqrt{1-\muh^{3/2}}  \hskip 0.5cm \text{for $\muh \to 1$.}  
\eea
Having defined $\hh^{(N)}(\muh)$, it is straightforward to combine it with the approximate solution
 $\muh_c \approx \deltah/2$ [Eq.~(\ref{eq:star})]to obtain $\hctwo(\deltah)$:
\be
\hctwo \approx \hh^{(N)}(\deltah/2).
\ee

For low detunings ($\deltah<\deltah_p\approx 2^{2/3} \approx 1.59$), we can use Eq.~(\ref{eq:hnlow}) for 
$\hh^{(N)}$, giving our final low-detuning result for $\hctwo$:
\be
\hctwo \approx 2^{2/3} - \frac{\deltah}{2}, 
\label{eq:hctwofinallowdetuning}
\ee
a linear-in-$\deltah$  behavior that is clearly seen in the numerically-determined curve shown in 
Fig.~\ref{fig:hcbcs}.
%

%-----------------------------
%
% fig%15
%
\begin{figure}[bth]
\vspace{1.4cm}
\centering
\setlength{\unitlength}{1mm}
\begin{picture}(40,40)(0,0)
\put(-50,0){\begin{picture}(0,0)(0,0)
\includegraphics{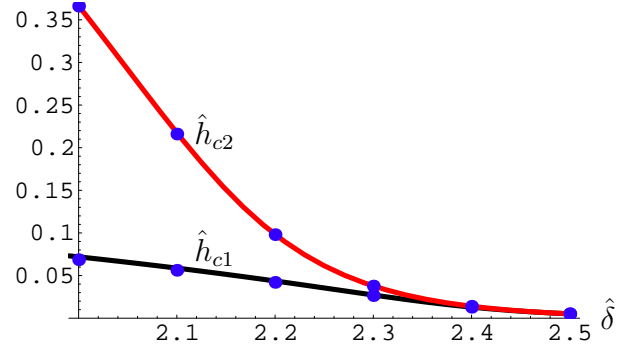}
\end{picture}}
\put(5,29) {\large{$\hh_{c2}$}}
\put(59,4.5) {\large{$\deltah$}}
\put(5,13.5) {\large{$\hcone$}}
\end{picture}
\vspace{-.5cm}
\caption{(Color online)  Upper (red) curve is $\hh_{c2}$ [Eq.~(\ref{eq:hctwoequationsbcs})]
and lower (black) curve is $\hcone$ [Eq.~(\ref{eq:hconebcs})]. For comparison, the solid points on
each curve are the same curves computed using the method described in Sec.~\ref{sec:mixedbcs}.}
\label{fig:hcbcsp} 
%
% 14Nov.nb  
%
\end{figure}
%------------------------------

\subsection{Mixed state in the BCS regime}
\subsubsection{Phase fractions} 
\label{sec:mixedbcs}

We have argued in the preceding section that, for  $\hcone(\deltah)<\hh<\hctwo(\deltah)$ 
(the region between the two curves in Fig.~\ref{fig:hcbcs}), and 
neglecting the possibility of the FFLO state, our system is a mixed state
in which the N and SF ground states coexist in $1-x(\deltah,\hh)$ and $x(\deltah,\hh)$
fractions, respectively.  
  To show this directly, here we  study 
the properties of this phase-separated coexistence state  by computing  its energy and determining the optimum
fractions of SF and N.~\cite{Bedaque}
We note that our analysis from the start ignores the interfacial energy~\cite{DeSilva06}
between the two coexisting (N and SF) phases.  For a macroscopic phase separation, this energy contribution is subdominant 
in the thermodynamic limit.  However, ignoring it precludes us from determining the spatial SF-N profile in
this regime. 
In the SF regions, $\Deltah$ is given by Eq.~(\ref{eq:baredeltanoughtmaintext}),
with the chemical potential given by $\muh_c(\hh,\deltah)$, Eq.~(\ref{eq:533exact}) [inverting $\hc(\muh,\deltah)$].
The energy of the mixed system in the canonical ensemble (appropriate for fixed density) is (normalized to $c\ef^{5/2}$)
\bea\label{eq:energymixed}
&&\hspace{-1cm}e_{G,N+SF} = -\big(\sqrt{\muh} \frac{\Deltah^2}{2} +
 \frac{8}{15} \muh^{5/2}\big) x  
  \\
&&\hspace{-1cm}\qquad\qquad-\big(\sqrt{\muh} \hh^2 + \frac{8}{15} \muh^{5/2}\big)(1-x)
+\frac{4}{3} \muh,
\nonumber
\eea
where in the second term we used Eq.~(\ref{eq:gsenormalapprox})
for the normal-state contribution to the energy, valid since $\Deltah\ll1$ on the BCS side.  The
final term comes from switching from the grand-canonical to the canonical ensemble 
(recall that in our notation the total {\it normalized\/} density is $4/3$).

The total density of atoms is similarly constructed from contributions from 
the normal and paired regions
\bea
&&\frac{4}{3} = x\big(\frac{5}{4} \frac{\Deltah^2}{\sqrt{\muh}} 
+ \frac{4}{3} \muh^{3/2} + 2\gamma^{-1} \Deltah^2\big)
\nonumber \\
&&\qquad
+ (1-x)\big(\frac{4}{3} \muh^{3/2} + \frac{\hh^2}{2\sqrt{\muh}}\big).\label{eq:numbermixed}
\eea
In the BCS regime, $\Deltah \ll \muh$, so that we may neglect the
term proportional to $\Deltah^2/\sqrt{\muh}$ on the right side of Eq.~(\ref{eq:numbermixed}).  
The remaining terms may be approximately solved for $\muh$,
yielding
\be
\muh \simeq 1- \gamma^{-1} \Deltah^2 x - \frac{\hh^2}{4}(1-x).
\label{eq:mumixed}
\ee 

To determine the properties of the mixed state we look 
for $x(\deltah,\hh)$ that minimizes $e_{G,N+SF}(x)$, Eq.~(\ref{eq:energymixed}), using the numerically-determined 
simultaneous solution to Eq.~(\ref{eq:mumixed}) and Eq.~(\ref{eq:baredeltanoughtmaintext}) for $\Deltah$ and $\muh$. 
 In the absence of a mixed state, the optimum $x$
would jump discontinuously from $x=1$ to $x=0$ as $\hh$ is increased.  The existence of a solution
where the optimum $0<x(\deltah,\hh)<1$ therefore indicates a stable phase-separated state.

Defining  them to be the endpoints of the region where $x$ is greater than zero or less than unity (indicating
a mixed state) yields an alternate procedure for finding $\hcone(\deltah)$ and $\hctwo(\deltah)$
as given in Eqs.~(\ref{eq:hconebcs}) and (\ref{eq:hctwoequationsbcs}).
The two methods of determining $\hcone(\deltah)$ and $\hctwo(\deltah)$ are displayed in 
 Fig.~\ref{fig:hcbcsp}, showing excellent agreement.

%%%%%%%%%%%%%%%%%%%%%%%%%%%%%%%%%%%%%%%%%%%%%%%%%%%%%%%%%%%%%%%%%%%%%%%%%%%%%%%%%%%%%%%%%%%%%%%%%%%%%%%%%%%%%%
\subsubsection{Atom density}
\label{SEC:regphasesep}
Another way to characterize the regime of phase separation 
is to study the total atom density $n$  as  a function of chemical potential
and verify that, in the mixed regime $\hcone<h<\hctwo$, it is impossible to adjust $\muh$ to 
attain the imposed density by either of the {\it pure\/} SF or N phases. 
The normalized SF-state density $\hat{n} = n/c\ef^{3/2}$
is given by  [c.f. Eq.~(\ref{eq:nsf})]
\bea
\nonumber 
&&\hat{n}_{SF}[\gamma,\deltah,\muh,\Deltah]
  = \frac{5}{4} \frac{\Deltah^2}{\sqrt{\muh}} + \frac{4}{3} \muh^{3/2} + 2\gamma^{-1} \Deltah^2
\\
&&\qquad \qquad \qquad 
- \frac{\Deltah^2}{2\sqrt{\muh}}\ln\frac{\Deltah}{8{\rm e}^{-2} \muh},
\label{eq:numbermixed1}
\eea
with $\Deltah[\gamma,\deltah,\muh]$ given by Eq.~(\ref{eq:baredeltanoughtmaintext}).
In the N state, the normalized density is [c.f. Eq.~(\ref{eq:nn})]
\bea
\label{eq:numbermixed2pre}
&&\hspace{-1cm}\hat{n}_{N}[\hh,\muh]  =\frac{2}{3}\big[(\muh  + \hh)^{3/2} \!+\! (\muh  - \hh)^{3/2}\Theta(\muh-\hh)\big],
\\
&&\hspace{-.4cm}\qquad \simeq\frac{4}{3} \muh^{3/2} + \frac{\hh^2}{2\sqrt{\muh}}.
\label{eq:numbermixed2}
\eea
Using these expressions we have 
\bea
\nonumber 
&&\hat{n}[\hh,\muh] = \hat{n}_N[\hh,\muh]
\Theta[\muh_c(\hh)-\muh]
\\   
&&\qquad \qquad
+ \hat{n}_{SF}[\gamma,\deltah,\muh,\Deltah]\Theta[\muh-\muh_c(\hh)],
\label{eq:hatn}
\eea
where  $\muh_c(\hh)$ is implicitly defined  as the solution of $\hc(\muh_c,\deltah) = \hh$, with $\hc(\muh)$ 
given by Eq.~(\ref{eq:533exact}). In the simplest BCS regime $\muh_c(\hh)$ satisfies
\be
\label{eq:mucritical}
\hh \simeq 4\sqrt{2}{\rm e}^{-2} \muh_c{\rm e}^{-\gamma^{-1}(\deltah - 2\muh_c)/\sqrt{\muh_c}}.
\ee
%

%-----------------------------
%
% fig%16
%
\begin{figure}[bth]
\vspace{1.4cm}
\centering
\setlength{\unitlength}{1mm}
\begin{picture}(40,40)(0,0)
\put(-50,0){
\begin{picture}(0,0)(0,0)
\includegraphics{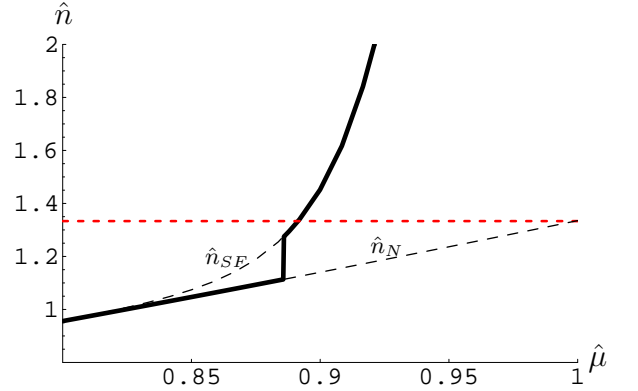}
\end{picture}}
\put(-12,50) {\large{$\hat{n}$}}
\put(59,4.5) {\large{$\muh$}}
\put(30,19.5) {$\hat{n}_N$}
\put(8,18) {$\hat{n}_{SF}$}
\end{picture}
\vspace{-.5cm}
\caption{(Color online) Plot of the normalized density $\hat{n}$ [solid line, Eq.~(\ref{eq:hatn})] vs. $\muh$ for $\hh =.06$ with 
parameters $\deltah = 2$  and $\gamma = 0.1$.  Thin dashed lines denote $\hat{n}_{SF}$ and $\hat{n}_N$.
Since the imposed physical atom density 
($\hat{n} = 4/3$, horizontal dashed red line) is intersected by the solid curve for $\muh$ above the critical 
$\muh$ (at which $\hat{n}$ is discontinuous), the system is in the SF state.}
\label{fig:nvsmu1}
%
%  BCSnvsmu.nb  
%
\end{figure}
%------------------------------

In Fig.~\ref{fig:nvsmu1}, we plot $\hat{n}$ vs. $\muh$ for $\hh=.06$ with
parameter values $\gamma = 0.1$ and
$\deltah = 2.0$.  At these values of the parameters, $\hcone\approx 0.068$ at the physical density, thus, we expect 
the SF state to be stable.  This is reflected in Fig.~\ref{fig:nvsmu1} 
in the fact that the $\hat{n}(\muh)$ curve intersects the physical density $\hat{n} = 4/3$ (horizontal dashed 
red line) 
for $\muh > \muh_c$.

%-----------------------------
%
% fig%17
%
\begin{figure}[bth]
\vspace{1.4cm}
\centering
\setlength{\unitlength}{1mm}
\begin{picture}(40,40)(0,0)
\put(-50,0){
\begin{picture}(0,0)(0,0)
\includegraphics{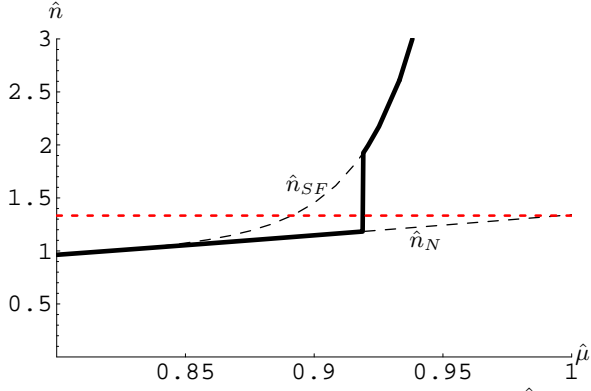}
\end{picture}}
\put(-12,50) {$\hat{n}$}
\put(58,4.5) {$\muh$}
\put(36,19.5) {$\hat{n}_N$}
\put(19.5,27) {$\hat{n}_{SF}$}
\end{picture}
\vspace{-.5cm}
\caption{(Color online) Same as Fig.~\ref{fig:nvsmu1}  but with $\hh = 0.13$
raised into the regime of phase separation. 
At this $\hh$, neither the SF nor the N yields the imposed density $\hat{n} = 4/3$, as illustrated by
$\hat{n} = 4/3$ falling into the discontinuous region of coexistence.}
\label{fig:nvsmu2}
%
%  BCSnvsmu.nb
%
\end{figure}
%---------------  

With increasing $\hh$, the discontinuity in $\hat{n}$  moves to higher values of $\muh$. 
 The regime of phase separation occurs when the physical density $\hat{n} = 4/3$ intersects this discontinuity,
as shown in Fig.~\ref{fig:nvsmu2} for $\hh = 0.13$, 
which satisfies $\hcone < \hh < \hctwo \approx 0.37$ (all other parameters being the same).
Thus, we see that at this $\hh$, no value of $\muh$ yields $\hat{n} = 4/3$.   At the critical $\muh$ (where there is
a discontinuity in $\hat{n}(\muh)$), $\muh_c \approx 0.919$, 
there are two possible homogeneous values of $\hat{n}$, that can be read off Fig.~\ref{fig:nvsmu2} 
or determined from Eqs.~(\ref{eq:numbermixed1}) and (\ref{eq:numbermixed2}) 
[using Eq.~(\ref{eq:baredeltanoughtmaintext}) for $\Deltah$].  These are  $\hat{n}_N \approx 1.18$ 
in the N state and $\hat{n}_{SF} \approx 1.92$ in the SF state.  Using these values allows us
to determine the fractions $x$ and $1-x$ of the system that are in the SF and N phases, respectively, 
using
\be
\label{eq:hatnmixed}
\frac{4}{3} = x \hat{n}_{SF} + (1-x) \hat{n}_N,
\ee
which yields $x\approx 0.20$, i.e., most of the system is in the N phase.  

%-----------------------------
%
% fig%18
%
\begin{figure}[bth]
\vspace{1.4cm}
\centering
\setlength{\unitlength}{1mm}
\begin{picture}(40,40)(0,0)
\put(-50,0){
\begin{picture}(0,0)(0,0)
\includegraphics{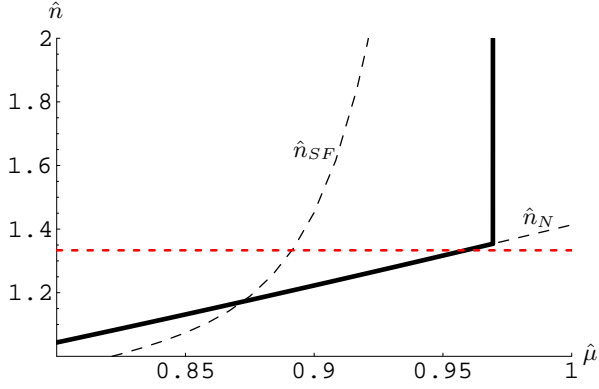}
\end{picture}}
\put(-12,50) {$\hat{n}$}
\put(59,4.5) {$\muh$}
\put(51,22.5) {$\hat{n}_N$}
\put(20.5,31.5) {$\hat{n}_{SF}$}
\end{picture}
\vspace{-.5cm}
\caption{(Color online) Same as Figs.~\ref{fig:nvsmu1} and ~\ref{fig:nvsmu2}, but with  
$\hh = 0.40>\hh_{c2}$, so that 
the normal state is stable.  }
\label{fig:nvsmu3}
%
%  BCSnvsmu.nb 
%
\end{figure}
%-----------------------------

With further increasing $\hh$, the SF fraction decreases continuously until, 
above $\hh_{c2}$ (when the vertical segment representing the mixed phase moves higher than $\hat{n} = 4/3$),
the system enters the pure N phase.  This is depicted in Fig.~\ref{fig:nvsmu3}, 
which, for $\hh = 0.40$, is slightly above $\hh_{c2}\approx 0.37$. 
In Fig.~\ref{fig:nvsmu3} we did not display the entire vertical range of the discontinuity in $\hat{n}$,
because at such large $\hh$ for the system to be in the SF state requires 
a very large $\Deltah$, that, by virtue of the term  $\Deltah^2/\gamma$ in Eq.~(\ref{eq:numbermixed1}),
 translates into a very large jump in $\hat{n}$
 (for this case, to $\hat{n} \approx 8.2$)
that is off the scale of Fig.~\ref{fig:nvsmu3}.

\subsection{Fixed population difference in the BCS and crossover regimes}
\label{SEC:fixedpopbcs}
Having characterized the regime of phase separation bounded by the critical curves $\hcone(\deltah)$
and $\hctwo(\deltah)$ at fixed density, we next convert these boundaries to critical
population differences  $\Delta N_{c1}(\deltah)$ and $\Delta N_{c2}(\deltah)$,
first focusing on the positive 
detuning (BCS and crossover) side of Fig.~\ref{fig:mphasetwo}.  In fact, since at $\hcone(\deltah)$ 
the system undergoes a first-order transition from the unmagnetized SF phase to the 
regime of phase separation, $\Delta N_{c1} = 0$ in the BCS regime.

%-----------------------------
%
% fig%19
%
\begin{figure}[bth]
\vspace{1.4cm}
\centering
\setlength{\unitlength}{1mm}
\begin{picture}(40,40)(0,0)
\put(-50,0){
\begin{picture}(0,0)(0,0)
\includegraphics{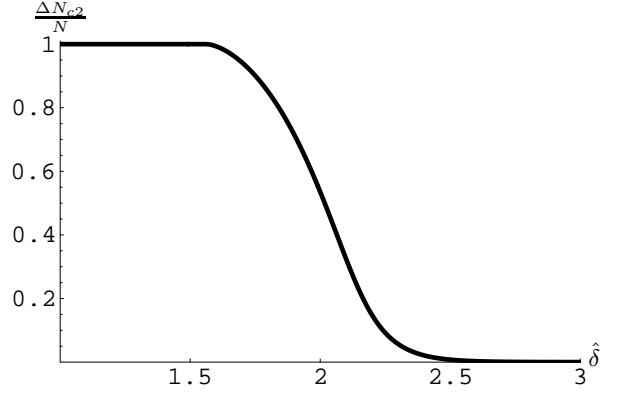}
\end{picture}}
\put(-15,50) {$\frac{\Delta N_{c2}}{N}$}
\put(59,4.5) {$\deltah$}
\end{picture}
\vspace{-.5cm}
\caption{Plot of the upper-critical polarization 
$\frac{\Delta N_{c2}}{N}$ as a function of detuning $\deltah$ for $\gamma = 0.1$, determined by Eq.~(\ref{mctwoovern})
along with the numerical solution to Eq.~(\ref{eq:hctwoequationsbcs}).  }
\label{fig:mctwobcs}
%
%   phasediagram_twochannel.nb
%
\end{figure}
%-----------------------------

Thus, the singlet BCS-BEC superfluid phase is confined to the $\Delta N/N = 0$ axis, and 
on the BCS side of the resonance an arbitrarily small 
population difference puts the system in the mixed state
(although
for a very small population imbalance the  fraction of the system in the SF state will
be close to unity).
 With increasing population difference, 
 eventually the SF fraction disappears and the system enters the N phase at $\Delta N_{c2}$.  
To compute $\Delta N_{c2}$ merely requires us to combine our 
result for $\hctwo(\deltah)$ [Eq.~(\ref{eq:hctwoequationsbcs})]
with the expression for $\Delta N(\hh)$ in the normal state, i.e., Eq.~(\ref{eq:dimnormmag}), that gives 
\be
\frac{\Delta N_{c2}}{N}=  \frac{1}{2}[(\muh_N + \hctwo)^{3/2}  - (\muh_N-\hctwo)^{3/2}\Theta(\muh_N - \hctwo)],
\label{mctwoovern}
\ee
with $\muh_N$ given by Eq.~(\ref{eq:dimnormnum}).
Along with Eq.~(\ref{eq:hctwoequationsbcs}), Eq.~(\ref{mctwoovern}) provides an accurate determination 
(see Fig.~\ref{fig:mctwobcs}) 
of 
$\Delta N_{c2}(\deltah)$ on the BCS side of the resonance [with accuracy only limited by the
approximations used in computing Eq.~(\ref{deltanoughtmixed2})].
We now determine explicit (but approximate) analytic expressions for $\Delta N_{c2}/N$ in the large and 
small $\deltah$ limits.

\subsubsection{$\frac{\Delta N_{c2}(\deltah)}{N}$ in the BCS regime of $\deltah\gg 2$}

For large $\deltah$, $\hctwo(\deltah)$ is exponentially small according to Eq.~(\ref{eq:hctwobcs}) while
$\muh_N \approx 1$ according to Eq.~(\ref{munint}).  Thus, in this limit we can expand 
Eq.~(\ref{mctwoovern}) in small $\hctwo/\muh_N$, finding:
\be
\frac{\Delta N_{c2}}{N}  \approx \frac{3}{2} \sqrt{\muh_N} \hctwo,
\label{mcovernlargedelta}
\ee
with the linear dependence on $\hctwo$ simply reflecting the Pauli paramagnetism~\cite{Negele}
of the N phase at small $h$.  Inserting Eqs.~(\ref{munint}) and (\ref{eq:hctwobcspre}) into Eq.~(\ref{mcovernlargedelta})
(and taking $\deltah \gg2$ in the latter), we find 
\bse
\label{mcovernlargedelta23}
\bea
\frac{\Delta N_{c2}}{N}
 &\approx& \frac{3\Deltah_F}{2\sqrt{2}} 
(1- \frac{1}{6} \Deltah_F^2)(1-\frac{\deltah}{16 \gamma}\Deltah_F^2),
\label{mcovernlargedelta2}
\\
&\approx&  
\frac{3\Deltah_F}{2\sqrt{2}} \exp\big[- \frac{\deltah}{16\gamma} \Deltah_F^2
\big],
\label{mcovernlargedelta3}
\eea
\ese
where in the final result we kept only leading-order terms in Eq.~(\ref{mcovernlargedelta2}) (i.e.,
we took $\deltah/\gamma \gg 1$) and re-exponentiated the last factor. 

\subsubsection{$\frac{\Delta N_{c2}(\deltah)}{N}$ in the crossover regime of $0<\deltah< 2$}

Next, we turn to intermediate detuning in the crossover regime.   We recall from Sec.~\ref{SEC:hconetwohighdetuning} that, at low $\deltah$,
$\hctwo$ is approximately given by solving the number equation [Eq.~(\ref{eq:mun})] along with the 
approximate result $\muh_N \approx \deltah/2$ valid in this regime.  With $\hctwo(\deltah)$ determined 
by the number equation
\bea
\label{mctwolowdelta1}
&&\hspace{-1.5cm}
1=  \frac{1}{2} \big[\big(\frac{\deltah}{2} + \hctwo\big)^{\frac{3}{2}}  + 
\big(\frac{\deltah}{2}   -\hctwo\big)^{\frac{3}{2}}\Theta\big( \frac{\deltah}{2}   - \hctwo\big)\big],
\eea
the upper-critical polarization is given by 
\bea
&&\hspace{-1.8cm}
\frac{\Delta N_{c2}}{N} 
 =  \frac{1}{2} \big[\big(\frac{\deltah}{2}    + \hctwo\big)^{\frac{3}{2}}  -
 \big(\frac{\deltah}{2}   -\hctwo\big)^{\frac{3}{2}}\Theta\big( \frac{\deltah}{2}    - \hctwo\big)\big].
\label{mctwolowdelta2}
\eea
Physically, Eq.~(\ref{mctwolowdelta1}) encodes the instability of the polarized Fermi gas to
a molecular superfluid as the point ($\hctwo$ and the corresponding $\Delta N_{c2}$) at which 
the N and SF chemical potentials are equal to $\muh_c \approx \deltah/2$, Eq.~(\ref{eq:star}).
This leads to Eq.~(\ref{mctwolowdelta1}), and Eq.~(\ref{mctwolowdelta2})  
translates $\hctwo$ to $\Delta N_{c2}/N$, determined by the polarizability of the 
N state. 

We proceed by solving Eq.~(\ref{mctwolowdelta1}) for $\hctwo$  and using the result inside Eq.~(\ref{mctwolowdelta2})
to obtain the critical polarization.  
In the case where $\hctwo>\deltah/2$, this is particularly simple, as the second step function in
Eq.~(\ref{mctwolowdelta1}) vanishes (as we found already in Sec.~\ref{SEC:hconetwohighdetuning}),
giving Eq.~(\ref{eq:hctwofinallowdetuning}) for $\hctwo$.  Using this inside 
Eq.~(\ref{mctwolowdelta2}) yields 
\be
\label{eq:lowdetuningmc2}
 \frac{\Delta N_{c2}}{N}= 1, \,\,\,\deltah < \deltah_p \simeq 2^{2/3}.
\ee
Thus, at such low detunings $\delta < \delta_p$, the normal state is only stable 
at full (100\%) polarization.  To understand this in more detail consider starting at large
$h$ in a fully-polarized (spin-$\uparrow$) normal state with $\Delta N = N$ ($m=n$) 
and $\mu_\downarrow = \mu-h <0$.  As $h$ and $\Delta N$ are reduced one of two scenarios
is possible depending on the value of the detuning $\delta$: (1) For large $\delta > \delta_p$,
upon lowering $h$, $\mu_\downarrow$ becomes positive first, converting spin-$\uparrow$ atoms to 
spin-$\downarrow$, partially depolarizing the Fermi sea. (2) For low $\delta< \delta_p$, 
case (1) is preempted by $\mu_N$ exceeding $\mu_{c}\simeq \delta/2$ causing the system 
to undergo a first-order transition to the SF state. 

We have denoted by $\delta_p$ the critical detuning below which $\Delta N_{c2}(\delta)=N$.
It is given by the solution of $\hctwo(\deltah_p) = \hh_p= 2^{-1/3}$ [the intersection of 
the horizontal dashed line in Fig.~\ref{fig:bcshphasediagram} with $h_{c2}(\delta)$],
giving
\be
\deltah_p \simeq 2^{2/3}.
\ee
As $\deltah$ is increased above $\deltah_p$, $\Delta N_{c2}/N$ drops continuously 
below unity, before starting to decrease exponentially according to our large $\deltah$ prediction
Eq.~(\ref{mcovernlargedelta3}).  To calculate $\Delta N_{c2}(\deltah)$ in the vicinity of $\deltah_p$, 
we take $\hctwo$ to be close to its value for $\deltah<2^{2/3}$:
\be
\hctwo = 2^{2/3} -  \frac{\deltah}{2} - \epsilon.
\label{eq:hctwoeps}
\ee
with $\epsilon$ small.  Inserting Eq.~(\ref{eq:hctwoeps}) into Eq.~(\ref{mctwolowdelta1}) 
and expanding to leading order in $\epsilon$ yields 
\be
1 \approx  1-\frac{3}{2} \frac{\epsilon}{2^{2/3}} + \frac{1}{2} (\deltah - 2^{2/3})^{3/2} \big(1+ 
\frac{3}{2} \frac{\epsilon}{\deltah - 2^{2/3}}\big),
\ee
which can be easily solved for $\epsilon$ and inserted into Eq.~(\ref{eq:hctwoeps})
to find $\hctwo(\deltah)$:
\be
\label{eq:hctwoeps2}
\hctwo \approx 2^{2/3} - \frac{\deltah}{2} - \frac{2^{2/3}}{3} (\deltah - 2^{2/3})^{3/2},\,\,\,\deltah \agt 2^{2/3}
\ee
Inserting Eq.~(\ref{eq:hctwoeps2}) into Eq.~(\ref{mctwolowdelta2}) and again expanding in small deviation 
$\deltah -2^{2/3}$, we find 
\bea
\frac{\Delta N_{c2}}{N} &\simeq&  1 - (\deltah - 2^{2/3})^{3/2}\Theta(\deltah- 2^{2/3}),
\\
&\simeq &
1 - (\deltah - \deltah_p)^{3/2}\Theta(\deltah- \deltah_p),
\eea
where we added a step function to emphasize that the second term is only nonzero for 
$\deltah > \deltah_p$.  
Our analysis in this section has neglected interactions, valid for a narrow resonance.  The leading-order
effects of finite $\gamma$  will add corrections to the approximate relation $\muh = \deltah/2$ and hence 
slightly adjust $\deltah_p$ (in the numerically generated curve  of Fig.~\ref{fig:mctwobcs}, $\deltah_p \approx 1.57<2^{2/3}$), 
but the qualitative picture remains the same.

\subsection{Coexisting fractions $x(\Delta N)$ in the mixed state}
\label{SEC:regimeofphasesep}

As we have seen, at positive detuning for $0< \Delta N <\Delta N_{c2}$,
resonantly interacting fermions phase separate into an s-wave paired
singlet SF and a spin-polarized N state with fractions $x(\delta,\Delta N)$
and  $1-x(\delta,\Delta N)$, respectively.

%-----------------------------
%
% fig%20
%
\begin{figure}[bth]
\vspace{1.4cm}
\centering
\setlength{\unitlength}{1mm}
\begin{picture}(40,40)(0,0)
\put(-50,0){
\begin{picture}(0,0)(0,0)
\includegraphics{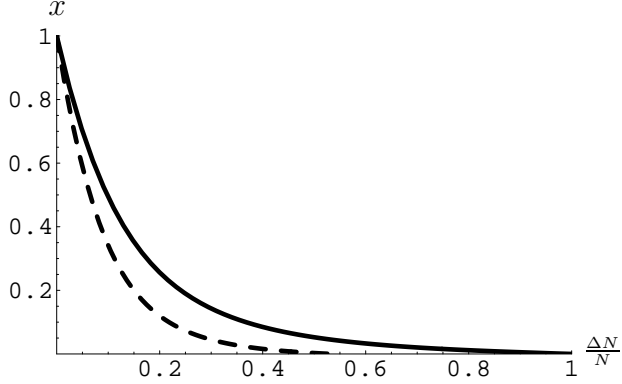}
\end{picture}}
\put(-12,50) {\large{$x$}}
\put(59,4.5) {$\frac{\Delta N}{N}$}
\end{picture}
\vspace{-.5cm}
\caption{(Color online) Plot of the fraction $x(\delta,\Delta N)$ of mixed state that is in the SF phase as  
a function of polarization for detuning $\deltah = 1.5$ (solid line) and
 $\deltah = 2$ (dashed line).
 }
\label{fig:fractionplot}
%
% fractionphasesep.nb
%
\end{figure}
%------------------------------

To calculate $x(\delta, \Delta N)$, we begin with Eq.~(\ref{eq:hatnmixed}) 
for the normalized density $\hat{n}$ expressed in terms
of the densities $\hat{n}_{SF}$ and $\hat{n}_{N}$ in the SF and N regions
[Equations~(\ref{eq:numbermixed1}) and (\ref{eq:numbermixed2pre}), respectively].  Using
the overall density constraint $\hat{n} = 4/3$,
we solve Eq.~(\ref{eq:hatnmixed}) for $x$ to obtain
\be
\label{eq:fracbcs}
x(\deltah,\hh) = \frac{\frac{4}{3}  - \hat{n}_N[\hh,\muh_c(\hh)]}{\hat{n}_{SF}[\deltah,\muh_c(\hh),\sqrt{2}\hh]
- \hat{n}_N[\hh,\muh_c(\hh)]},
\ee
with $\muh_c(\hh)$  given by Eq.~(\ref{eq:mucritical}).
Here, in the last argument of $\hat{n}_{SF}[\deltah,\muh_c(\hh),\Deltah(\muh_c)]$ we have substituted
$\Deltah(\muh_c) = \sqrt{2}\hh$, valid everywhere in the mixed phase. 

To compute $x(\deltah, \Delta N)$, we combine $x(\deltah,\hh)$ above with a computation of $\Delta N(\deltah, \hh)$ 
as a function of $\hh$ in the mixed phase. 
 Since any population difference can only occur in the N regions of fractional volume 
$1-x$, we have
\bea
&&\frac{\Delta N(\gamma,\deltah,\hh)}{N} = \frac{1}{2} \big[1-x(\gamma,\deltah,\hh)\big]\big[
(\muh_c(\hh) +\hh)^{3/2} 
\nonumber \\
&&\qquad\qquad
- (\muh_c(\hh) -\hh)^{3/2}\Theta(\muh_c(\hh) -\hh)
\big].\label{eq:fracbcs2}
\eea
In Fig.~\ref{fig:fractionplot}, we present a numerical solution $x(\deltah,\Delta N)$
of Eqs.~(\ref{eq:fracbcs}) and  (\ref{eq:fracbcs2}),
for $\deltah = 1.5$ (solid line) and $\deltah = 2.0$ (dashed line), with $\gamma = 0.1$.
The two curves are qualitatively
similar, each describing a continuous depletion of $x(\deltah,\Delta N)$ from unity to zero.
  The main notable difference between the
curves is that  for  $\deltah = 1.5$, $\frac{\Delta N_{c2}}{N} = 1$, while 
for   $\deltah = 2.0$  the dashed curve reaches $x=0$ at $\frac{\Delta N_{c2}}{N} \approx  0.535<1$, 
 characterizing a transition to a non-fully-polarized N state. 
The rapid initial drop of $x$ with $\frac{\Delta N}{N}$ reflects the fact that, when most of the system is paired at low polarization, 
the only way to polarize is to convert regions of the system from paired to unpaired (i.e.~to decrease $x$), while for 
large polarization when much of the system is already in the normal phase, to attain higher polarization the system can further 
polarize already normal sections.  

%%%%%%%%%%%%%%%%%%%%%%%%%%%%%%%%%%%%%%%%%%%%%%%%%%%%%%%%%%%%%%%%%%%%%%%%%%%%%%%%%%%%%%%%%%%%%%%%%%%%%%%%%%%%%%

\section{Negative-detuning regime of two-channel model at finite population difference}
\label{SEC:BEC}

In the present section we extend our analysis to the  $\delta<0$ BEC regime.  As for
$\delta>0$, here too we minimize the ground-state energy Eq.~(\ref{eq:havg}) with 
respect to $\Delta_\bQ$ and $\bQ$, subject to the total atom number and 
imposed spin-polarization constraints. 
 As we shall show in Sec.~\ref{SEC:FFLO}, 
the FFLO state is only stable above a critical detuning $\delta_* \approx 2\ef$ [given by Eq.~(\ref{eq:deltastar})
below], i.e., in the BCS regime.  Physically this reflects the fragility of the FFLO state
that is driven by atomic Fermi-surface mismatch, absent in the BEC regime where $\mu\approx \delta/2<0$.
 Therefore, in studying the BEC regime at the outset we shall focus on $\bQ=0$ pairing order as we did in 
Sec.~\ref{SEC:BCS}.
Thus, the ground state energy that we shall analyze is still given by Eq.~(\ref{eq:havgBCS2}), with its
dimensionless form given by Eq.~(\ref{eq:gsetotallygeneral}).   

%-----------------------------
%
% fig%21
%
\begin{figure}[bth]
\vspace{1.4cm}
\centering
\setlength{\unitlength}{1mm}
\begin{picture}(40,40)(0,0)
\put(-50,0){
\begin{picture}(0,0)(0,0)
\includegraphics{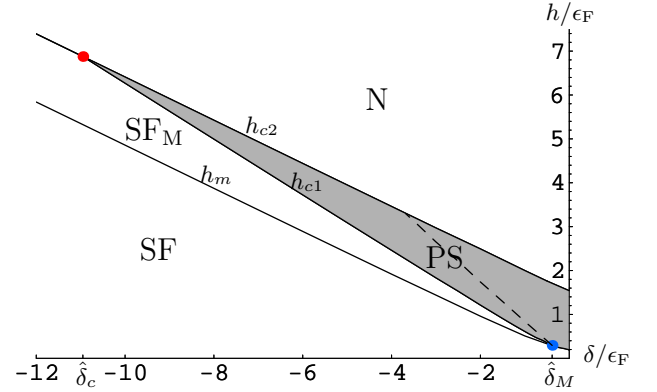}
\end{picture}}
\put(54,50) {$h/\ef$}
\put(59,4.5) {$\delta/\ef$}
\put(-2,34) {\large{SF$_{\rm M}$}}
\put(0,18) {\large{SF}}
\put(38,17) {\large{PS}}
\put(30,38) {\large{N}}
\put(8,28.5) {$h_{m}$}
\put(20,27.5) {$h_{c1}$}
\put(14,35) {$h_{c2}$}
\put(-8.5,1.5) {$\deltah_c$}
\put(53.75,1.5) {$\deltah_M$}
\end{picture}
\vspace{-.5cm}
\caption{(Color Online) 
Negative-detuning phase diagram of the two-channel model for the case $\gamma = 0.1$
at fixed chemical potential difference $h$ 
showing regions of singlet superfluid (SF), magnetic superfluid (\sfm), 
phase separation (shaded, PS) and  normal phase (N). 
To the right of  the dashed line in the PS regime,  SF and N states coexist,
while to the left of the dashed line in the PS regime \sfm and N states coexist. $\delta_c$
is a tricritical point.}
\label{fig:bechphasediagram}
%
%   phasediagram_twochannel.nb
%
\end{figure}
%-----------------------------
%-----------------------------
%
% fig%22
%
\begin{figure}[bth]
\vspace{1.4cm}
\centering
\setlength{\unitlength}{1mm}
\begin{picture}(40,40)(0,0)
\put(-50,0){
\begin{picture}(0,0)(0,0)
\includegraphics{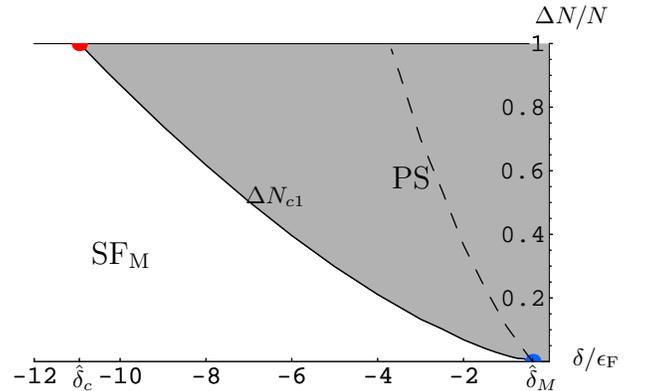}
\end{picture}}
\put(54,50) {$\Delta N/N$}
\put(59,4.5) {$\delta/\ef$}
%\put(0,6) {SF}
\put(-5,18) {\large{SF$_{\rm M}$}}
\put(35,28) {\large{PS}}
\put(15.5,26) {$\Delta N_{c1}$}
\put(-7.5,1.5) {$\deltah_c$}
\put(52.75,1.5) {$\deltah_M$}
\end{picture}
\vspace{-.5cm}
\caption{(Color Online) 
Negative-detuning phase diagram of the two-channel model for the case $\gamma = 0.1$
at fixed spin  population difference $\Delta N$, showing regions  of magnetic superfluid (\sfm) and  
phase separation (shaded, PS).  The fully-polarized normal phase is confined to the upper
boundary $\Delta N = N$, while the unpolarized SF state is confined to  the $\Delta N = 0$ axis.
To the right of  the dashed line, the PS regime consists of coexisting SF and N states; 
to the left of the dashed line the PS regime consists of coexisting \sfm and N states.}
\label{fig:becmphasediagram}
%
%   phasediagram_twochannel.nb
%
\end{figure}
%-----------------------------

Before proceeding to our detailed analysis, we 
  briefly summarize our main results, the negative-detuning phase diagrams 
at fixed chemical potential difference $h$ and fixed spin imbalance $\Delta N$,
Figs.~\ref{fig:bechphasediagram} and
~\ref{fig:becmphasediagram}, respectively.  The three critical
$h$'s ($h_m$, $h_{c1}$ and $h_{c2}$) in
Fig.~\ref{fig:bechphasediagram} separating the singlet superfluid (SF), magnetic
superfluid (\sfm), normal (N)  states and phase-separation (PS) regime were computed by
numerically solving the stationarity 
and number constraint conditions [i.e., Eqs.~(\ref{eq:gapequationgeneralnorm})
and 
Eqs.~(\ref{eq:numequationgeneralnorm})], always ensuring that the
solution  is a minimum of the normalized ground-state energy $e_G$,
Eq.~(\ref{eq:gsetotallygeneral}). 
In the PS region, it is possible to solve the stationarity and number constraint equations, but the 
solution is {\it not\/} a minimum of $e_G$.~\cite{Pao,Iskin,Sedrakian,Paonote}
 Then, to obtain Fig.~\ref{fig:becmphasediagram},
these critical $h$'s were converted to critical population 
differences using Eq.~(\ref{eq:polequationgeneralnorm}).  The dashed lines in 
these figures separate different types of PS regime and are derived approximately
in Sec.~\ref{SEC:firstordersfmn}.

The main aim of this section is to present details of analytic calculations that complement
this numerical analysis of the gap and number-constraint equations and lead to the phase diagrams
in Figs.~\ref{fig:bechphasediagram} and ~\ref{fig:becmphasediagram} in the narrow Feshbach resonance limit, 
$\gamma \ll 1$.  These
computations are aided by the fact that, for $\gamma \ll 1$, $\mu \simeq \delta/2$ and hence $\mu<0$
for $\delta<0$. 
Taking $\mu<0$ 
yields an  important simplification to the equation for the ground-state energy
and consequently for the  gap equation
and the number and polarization constraints, since only one of the terms in Eq.~(\ref{eq:mbcsdimless}) for the 
dimensionless magnetization contributes, yielding 
\be 
\label{eq:mequation}
\mgh(\hh) = \frac{2}{3} \Big(\sqrt{\hh^2 -\Deltah^2} - |\muh|\Big)^{3/2}\Theta\Big(
\sqrt{\hh^2-\Deltah^2} -|\muh|\Big).
\ee
As in the preceding section,  we proceed by inserting Eq.~(\ref{eq:mequation}) into Eq.~(\ref{eq:gsetotallygeneral}) 
for the normalized ground-state energy $e_G$.  Stationary points of $e_G$ satisfy the gap equation, Eq.~(\ref{eq:gapequationgeneralnorm}).
In finding solutions to the gap equation we  always verify that such stationary points are actually {\it minima} of $e_G$ 
rather than
 saddle points or local maxima.
Failure to do this in a number of recent theoretical works~\cite{Pao,Iskin,Paonote,Sedrakian} has led to erroneous results.
  To impose  constraints on the total atom number and spin population difference, we use 
Eqs.~(\ref{eq:numequationgeneralnorm}) and ~(\ref{eq:polequationgeneralnorm}).

We will show that an accurate 
quantitative description of the BEC regime, for the case of a narrow resonance, can be found by expanding 
the normalized ground-state energy $e_G$ to leading order in $\gamma$.   Although a full description requires keeping terms up
to order $\curO(\gamma^3)$ (as we show in Sec.~\ref{SEC:detailedanalysis} below), many essential features are correctly 
described in the  leading-order  $\gamma = 0$ limit.

\subsection{Zero-coupling approximation}
\label{SEC:zerocoupling}
In the present section, we analyze the BEC regime in the zero-coupling limit $\gamma \to 0$ ($g\to 0$).
As we shall show,    many of the essential 
features of the phase diagram are captured in this extreme narrow resonance limit. Although
the Feshbach resonance interconversion term that is proportional to $g$ is required for pairing 
and equilibration between atoms and molecules,
on 
the BEC side of the resonance the molecular density is determined by the number equation and therefore,
in equilibrium, is finite even in the $g\to 0$ limit.  

 To treat this limit, we change variables from $\Deltah$ (which vanishes at $g\to 0$) 
to the normalized molecular condensate order parameter $\Bh = \Deltah/\sqrt{\gamma}$.  The 
physical molecular boson density $n_m$ is related to $\Bh$ via 
\be
\label{eq:relationbhat}
n_m = |B|^2 = N(\ef) \ef |\Bh|^2 = c \ef^{3/2}  \Bh^2,
\ee
where for simplicity we have taken $\Bh$ to be real.
Inserting this variable change into Eq.~(\ref{eq:gsetotallygeneral}), we find, expanding to leading
order in small $\gamma$:
\be 
 e_G = 
  \Bh^2  (\deltah - 2\muh)  - \int_0^{\hh} \mgh(\hh') d\hh',
\label{gsezerobec}
\ee
where  to the same order the magnetization Eq.~(\ref{eq:mequation}) reduces to
\be
\label{eq:mzerocoupling}
\mgh(\hh) = \frac{2}{3}(\hh-|\muh|)^{3/2}\Theta(\hh-|\muh|).
\ee
Performing the integral over $\hh$ in Eq.~(\ref{gsezerobec}), we find
\be 
e_G=
  \Bh^2  (\deltah - 2\muh)  - \frac{4}{15} (\hh-|\muh|)^{5/2}\Theta(\hh-|\muh|).
\label{becweakcouplingenergy}
\ee
The number and gap equations are then given by:
\bse
\label{eq:weakcouplinggapnumber}
\bea
\frac{4}{3} &=& 2 \Bh^2 + \frac{2}{3}(\hh-|\muh|)^{3/2}\Theta(\hh-|\muh|),
\\
0&=& 2\Bh (\deltah - 2\muh).
\label{eq:weakcouplinggap}
\eea
\ese
In the normal (N) phase, $\Bh = 0$, 
Eq.~(\ref{eq:weakcouplinggap}) is automatically satisfied, and the number equation
reduces to
\be
\hh-|\muh| = 2^{2/3},
\label{eq:munormal}
\ee
for the normal-state chemical potential.
 Returning to
dimensionful quantities clarifies the meaning of  Eq.~(\ref{eq:munormal}):
\be
(h+\mu)^{3/2} = 2\ef^{3/2}.
\label{eq:munormalunits}
\ee
Since $h +\mu = \mu_\uparrow$, this simply states that the normal-phase spin-$\uparrow$
density $n_\uparrow = \frac{2}{3} c \mu_\uparrow^{3/2}$ is equal to the total 
fermion density, $n$. 

%-----------------------------
%
% fig%23
%
\begin{figure}[bth]
\vspace{3.2cm}
\centering
\setlength{\unitlength}{1mm}
\begin{picture}(40,65)(0,0)
%\put(-10,50){\begin{picture}(0,0)(0,0)
%\special{psfile=sfmpic1.eps vscale = 110 hscale=110}
%\end{picture}}
%
%
\put(-20,0){\begin{picture}(0,0)(0,0)
\includegraphics{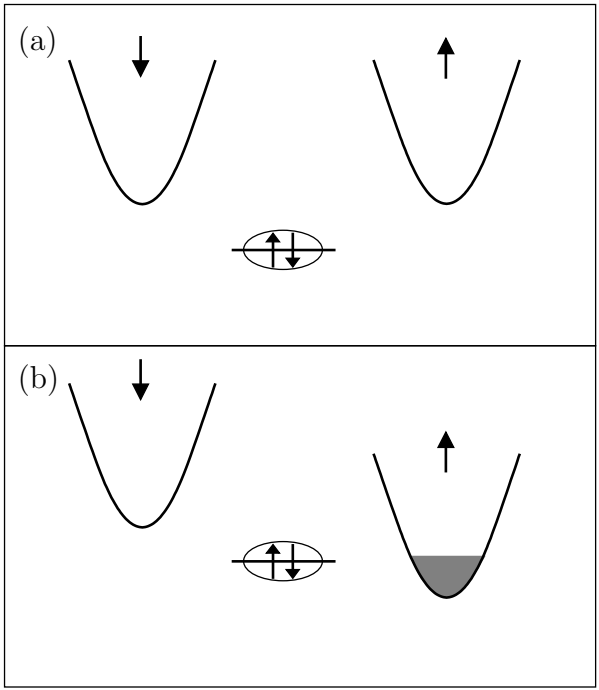}
\end{picture}}
\put(-18,85) {\large{(a)}}
\put(-18,40) {\large{(b)}}
\end{picture}
%\vspace{-.5cm}
\caption{ A schematic of the population of atomic states
(parabolas, labeled by spin) and the 
molecular level (line) in the BEC regime ($\delta<0$) at (a) $h=0$  
and (b) $h> h_m$.}
\label{fig:cartoonsfm}
%
%  sfmpic1.fig
% 
\end{figure}
%------------------------------

In the superfluid phase, $\Bh \neq 0$, and Eqs.~(\ref{eq:weakcouplinggapnumber}) give
\bse
\bea
  \Bh^2 &=& \frac{2}{3} \Big[1-\frac{1}{2} \big( \hh - |\muh| \big)^{3/2}\Theta(\hh - |\muh|)
\Big],
\label{eq:numbecpre} 
\\
\muh &=& \frac{\deltah}{2},
\label{eq:gapbec}
\eea
\ese
which combine to yield 
\bea
  \Bh^2 &=& \frac{2}{3} \Big[1-\frac{1}{2}
 \big( \hh - \frac{|\deltah|}{2} \big)^{3/2}\Theta(\hh -  \frac{|\deltah|}{2})
\Big],
\label{eq:numbec}
\eea
an expression that is only sensible when the right side is positive.  What do the preceding
expressions tell us about the BEC regime at $\hh \neq 0$?  For $\hh< |\deltah|/2$, 
corresponding to Zeeman field less than the molecular binding energy, 
 $\Bh^2 = 2/3$,
which, in dimensionful units, gives the molecular density 
$n_m = n/2$.  
This is the usual singlet molecular superfluid 
(SF) phase in the BEC limit, in which all the atoms are bound into diatomic molecules
[See Fig.~\ref{fig:cartoonsfm}(a)].  Although including 
nonzero interactions will deplete the molecular density somewhat, clearly this qualitative picture 
will still hold.

However, for $\hh> |\deltah|/2$ the Zeeman field exceeds the molecular binding 
energy,  and the molecular density
continuously depletes with increasing $\hh$ according to Eq.~(\ref{eq:numbec})
as molecules break up into a fully-polarized Fermi sea with $\mu_\uparrow = h+ \delta/2>0$
[See Fig.~\ref{fig:cartoonsfm}(b)].  Using Eq.~(\ref{eq:gapbec}) 
and  Eq.~(\ref{eq:mzerocoupling}), we find the corresponding magnetization
\be
\mgh(\hh) = \frac{2}{3}(\hh-\frac{|\deltah|}{2})^{3/2}\Theta(\hh-\frac{|\deltah|}{2}),
\ee
{\it continuously\/}  increasing from zero
beyond $\hh>|\deltah|/2$.  
 We refer to this uniform state, 
consisting of both condensed molecules and spin-up polarized fermions, as the magnetic 
superfluid  (\sfm) state.  

At low energies, we expect the \sfm to be very similar to 
a uniform state of bosons and one species of fermion which has been previously explored in a 
different context~\cite{Viverit}. 
 In the present context, 
the \sfm state exists for $\hh_m < \hh <\hctwo$, with (in the $\gamma \to 0$ limit)
\bse
\label{eq:hszerocoupling}
\bea
\hh_m & = & |\deltah|/2,
\label{eq:hmzerocoupling}
\\
\hctwo & = & 2^{2/3} + |\deltah|/2,
\label{eq:hc2zerocoupling}
\eea
\ese
the latter defined by where $\Bh^2 = 0$ according to Eq.~(\ref{eq:numbec}). 
The approximate linear dependences of $\hh_m$ and $\hctwo$ on $\deltah$ are also clearly seen 
(away from the unitarity point $\delta = 0$) in
the numerically-determined  finite $\gamma = 0.1$ phase diagram Fig.~\ref{fig:bechphasediagram}.

Expressing Eq.~(\ref{eq:numbec})
for $\Bh(\hh)$ in terms of $n_m$, $n$ and the physical magnetization, $\mg$:
\be
n_m = \frac{1}{2} (n - \mg), 
\label{eq:bosdensityzerocoupling}
\ee
further clarifies its meaning.  At $\hh_m$, $\mgh(\hh_m) = 0$ and all the atoms are confined 
into Bose-condensed molecules.  Applying a sufficiently large $\hh>\hh_m$ such that
$\mu_\uparrow= h+\delta/2>0$ populates the spin-up atomic band as illustrated in
Fig.~\ref{fig:cartoonsfm}, depleting the number of molecular bosons according to 
Eq.~(\ref{eq:bosdensityzerocoupling}).  
 At $\hh = \hctwo$, $\mg = n$ and the system is fully polarized.
  As we shall see in the next sections,
this simple picture of molecular bosons depairing into free polarized atoms
once the fermion band dips below the molecular level remains
qualitatively correct for  $\gamma>0$.

\subsection{Weak-coupling description of \sfm state}
\label{sec:sfm}

In the present section, we extend the above $\gamma \to 0$ analysis 
 to higher order in $\gamma$.  As we will show, the qualitative picture
of the preceding subsection remains the same: A depletion of the condensate (and concomitant
population of the spin-$\uparrow$ Fermi sea)  with increasing $\hh$ starting
at $\hh_m$, with a continuous depletion until $\hctwo$ beyond which
the molecular density vanishes and all the atoms occupy the 
spin-$\uparrow$ band. 
The properties of the homogeneous magnetic superfluid \sfm,
that consists of a molecular superfluid and a spin-$\uparrow$ Fermi sea,
remain the same. 
 The major {\it qualitative\/} modification from the preceding section 
is the continuation of the first-order phase transition curve $\hcone(\deltah)$, that we found in Sec.~\ref{SEC:BCS},
 into the BEC regime.  
Recall that, on the positive-detuning side of the resonance, $\hcone(\deltah)$ denotes the chemical 
potential difference above which
the SF phase is unstable,  via a first-order transition, to phase separation.  At fixed
polarization, this translates to  $\Delta N_{c1} = 0$: For any $\Delta N\neq 0$  in the BCS regime 
the system phase separates.

As we shall see, in  the BEC regime at moderate negative 
detunings, the \sfm state is similarly unstable to phase separation at $\hcone$.
However, since the \sfm state is {\it polarized\/}, the corresponding fixed-polarization
boundary $\Delta N_{c1} \neq 0$.  Remarkably,
our naive formula for $\hctwo$, Eq.~(\ref{eq:hc2zerocoupling}), remains quantitatively correct, but, close
to the resonance position, should be more generally
interpreted as the chemical potential difference below which the spin-polarized N state is unstable to phase separation.
At fixed atom density $n$, 
the sequence of phases and regimes with increasing $\hh$ is, then, SF$\to$\sfm$\to$PS$\to$N.
At large negative detuning, $\delta = \delta_c \simeq -10.6 \ef$, $\hcone$ {\it intersects\/} $\hctwo$, 
so that the first-order behavior ends and, for $\delta<\delta_c$, we find
the sequence of phases SF$\to$\sfm$\to$N separated by continuous transitions.

 Before computing the $\hcone(\deltah)$ curve, we first analyze the \sfm phase in 
more detail by studying the $\curO(\gamma^2)$ expression for $e_G$, obtained by expanding the arguments
of the integrals in Eq.~(\ref{eq:gsetotallygeneral}) in small $\gamma$ and evaluating the integrals term by term.
We find:
\be 
e_G =  -\frac{4}{15}(\hh-|\muh|)^{5/2}\Theta(\hh-|\muh|)  - \hat{V}_2\Bh^2 + \frac{1}{2}\hat{V}_4 \Bh^4, 
\label{becefweakcoupling}
\ee
with
\bse
\bea
&&\hspace{-.5cm}\hat{V}_2[\gamma,\deltah,\hh,\muh] \equiv 2\muh - \deltah - \gamma\sqrt{|\muh|}F_2(\hh/|\muh|),
\label{eq:v2}
\\
&&\hspace{-.5cm}\hat{V}_4[\gamma,\hh,\muh] \equiv \frac{\gamma^2 \pi}{32 |\muh|^{3/2}}F_4(\hh/|\muh|) ,
\label{eq:v4}
\eea
\label{vee}
\ese
where we defined the functions $F_2(x)$ and  $F_4(x)$:
\bea
&&\hspace{-1.2cm}F_2(x) \equiv \frac{\pi}{2} + \big[
\sqrt{x-1} -\tan^{-1}\sqrt{x-1}
\big]\Theta(x-1),
\label{eq:f2def}
\\
&&\hspace{-1.2cm}F_4(x)\equiv 1 - 
\frac{2}{\pi x^2} \big[
\sqrt{x-1} (x + 2) 
\nonumber 
\\
&&\qquad\qquad 
+ x^2 \tan^{-1} \sqrt{x-1}
\big]\Theta(x-1).
\label{eq:f}
\eea
To leading order in $\gamma$, the  number [Eq.~(\ref{eq:numequationgeneralnorm})] and gap 
[Eq.~(\ref{eq:gapequationgeneralnorm})]
equations are:
\bse
\bea
&&\hspace{-.5cm}\frac{4}{3} = 2\Bh^2+
\frac{\pi\gamma \Bh^2}{4\sqrt{|\muh|}} + \frac{2}{3} (\hh - |\muh|)^{3/2}\Theta(\hh - |\muh|)
\nonumber \\
&&\hspace{-.5cm}\qquad 
 - \frac{\gamma \Bh^2}{2\sqrt{|\muh|}}\tan^{-1}  \sqrt{\hh/|\muh|-1}\Theta(\hh - |\muh|),
\label{eq:elbec1approx}
\\
&&
\hspace{-.5cm}\hat{V}_2 = \Bh^2 \hat{V}_4.
\label{eq:elbec2approx}
\eea
\ese
On the BEC side of the resonance the saddle-point equations approximately 
switch roles, with the number equation [Eq.~(\ref{eq:elbec1approx})]  approximately determining the boson density
$\Bh^2$ and the gap equation [Eq.~(\ref{eq:elbec2approx})]
approximately determining the chemical potential $\muh$, as can be seen by solving these equations:
\bse
\bea 
&&\hspace{-.7cm}\Bh^2 = \frac{2}{3}\frac{1-\frac{1}{2} \big( \hh - |\muh| \big)^{3/2}}
{1+ \frac{\gamma}{8\sqrt{|\muh|}}\big(\pi-2\tan^{-1}\sqrt{\hh/|\muh|-1}\big)},
\label{eq:numbecapprox}
\\
 &&\hspace{-.7cm}\muh  = \frac{\deltah}{2}  + \frac{\gamma\sqrt{|\muh|}}{2}\Big[
\frac{\pi}{2}  + 
\sqrt{\hh/|\muh|-1}
 \nonumber \\
 &&\hspace{-.7cm}\qquad \qquad \qquad 
-\tan^{-1}\sqrt{\hh/|\muh|-1}
\Big],
\label{eq:gapbecapprox}
\eea
\ese
with the second equation determining $\muh(\deltah)$ through $\hat{V}_2 \approx 0$, 
correct to $\curO(\gamma)$ since $\Vh_4$ is $\curO(\gamma^2)$.
 At this level of approximation,
Eq.~(\ref{eq:gapbecapprox}) is independent of Eq.~(\ref{eq:numbecapprox})
allowing us to
solve  Eq.~(\ref{eq:gapbecapprox}) to determine $\muh(\deltah)$ which can then be 
used to compute $\Bh^2(\deltah)$ through Eq.~(\ref{eq:numbecapprox}). 
  To order $\gamma$, $\muh$ is determined by 
simply iterating Eq.~(\ref{eq:gapbecapprox}), 
which amounts to inserting the zeroth order result 
$\muh \approx \deltah/2$ into the second term.  This gives to $\curO(\gamma)$
\bea
&&\muh \simeq \muh_1(\deltah,\hh)  \equiv \frac{\deltah}{2}  + \frac{\gamma\sqrt{|\deltah|}}{2\sqrt{2}}\Big[
\frac{\pi}{2}  + 
\sqrt{2\hh/|\deltah|-1}
 \nonumber \\
 &&\qquad \qquad \qquad 
-\tan^{-1}\sqrt{2\hh/|\deltah|-1}
\Big],\label{eq:gapbecapprox2}
\eea
while to the same order  $\Bh^2$ is given by 
\be
\Bh^2 = \frac{2}{3}\frac{1-\frac{1}{2} \big( \hh - |\muh_1(\deltah,\hh)| \big)^{3/2}}
{1+ \frac{\gamma\sqrt{2}}{8\sqrt{|\deltah|}}(\pi-2\tan^{-1}\sqrt{2\hh/|\deltah|-1})},
\label{eq:numbecapprox2}
\ee
describing the condensate depletion as a function of $\hh$ and $\deltah$.

These expressions can be easily translated to the experimentally-relevant fixed-population 
ensemble.  
Expanding Eq.~(\ref{eq:mequation}) to $\curO(\gamma)$, we obtain 
\be
\mgh\simeq \frac{2}{3}(h-|\muh|)^{3/2} - \frac{\gamma \Bh^2}{2\hh} \sqrt{\hh/|\muh|-1},
\ee
the lowest-order finite-$\gamma$ correction to Eq.~(\ref{eq:mzerocoupling}).
This can be turned into an $\curO(\gamma)$ expression for the quantity $(\hh-|\muh|)^{3/2}$ 
appearing in Eq.~(\ref{eq:numbecapprox}):
\bea
\frac{2}{3}(\hh-|\muh|)^{3/2} &\simeq& \mgh + \frac{\gamma \Bh^2}{2\hh} \big(\frac{3\mgh}{2}
\big)^{1/3},
\\
&\simeq& \mgh + \frac{\gamma \Bh^2}{2|\muh|} \big(\frac{3\mgh}{2}
\big)^{1/3},
\label{intstepbecnum}
\eea
where in the second line we took $\hh \approx |\muh|$, valid to leading order in small $\mgh$.  Since,
as we shall see, the \sfm state is generally only stable for $\hh \agt |\muh|$, this is approximately
valid.
Inserting Eq.~(\ref{intstepbecnum}) into Eq.~(\ref{eq:numbecapprox}), and using $\muh \approx \deltah/2$ 
[correct in this expression to $\curO(\gamma)$] we find
\begin{widetext}
\bse
\bea
&&\hspace{-.75cm}\Bh^2 \simeq \frac{\frac{2}{3} - \frac{\mgh}{2}}
{1+ \frac{\gamma}{2|\deltah|}\big(\frac{3\mgh}{2}\big)^{\frac{1}{3}} + \frac{\sqrt{2}\gamma}{8\sqrt{|\deltah|}} 
\big(
\pi - 2\tan^{-1} \frac{(3\mgh)^{1/3} 2^{1/6}}
{\sqrt{|\deltah|}}\big)},
\\
&&\hspace{-.75cm}\frac{n_m}{n} =  \frac{\frac{1}{2} \big(1  - \frac{\Delta N}{N} \big)}
{1+ \frac{ \gamma}{2^{\frac{2}{3}}|\deltah|}\big(\frac{\Delta N}{N}\big)^{\frac{1}{3}} +
 \frac{\sqrt{2}\gamma}{8\sqrt{|\deltah|}} 
\Big[
\pi - 2\tan^{-1} 
\frac{2^{5/6}}{\sqrt{|\deltah|}}
\big(\frac{\Delta N}{N}\big)^{\frac{1}{3}} 
\Big]},
\label{eq:densitycanon}
\eea
\ese
\end{widetext}
where in Eq.~(\ref{eq:densitycanon}) we used Eqs.~(\ref{numdef}),  (\ref{magrelation2}), and (\ref{eq:relationbhat}) 
to express the final result for the physical molecular density $n_m(\Delta N)$.
This corrects, to  $\curO(\gamma)$, Eq.~(\ref{eq:bosdensityzerocoupling}) of the preceding section.

%-----------------------------
%
% fig%24
%
\begin{figure}[bth]
\vspace{1.4cm}
\centering
\setlength{\unitlength}{1mm}
\begin{picture}(40,40)(0,0)
\put(-50,0){\begin{picture}(0,0)(0,0)
\includegraphics{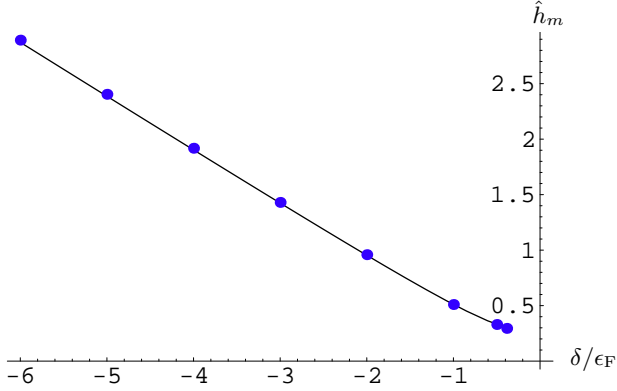}
\end{picture}}
\put(53,50) {$\hh_m$}
\put(58,4.5) {$\delta/\ef$}
\end{picture}
\vspace{-.5cm}
\caption{(Color online) Plot of $\hh_m$, the normalized chemical potential difference
 above which the system enters the \sfm phase (i.e., $\Delta N \neq 0$) as a function
of detuning $\delta$. 
The analytical low-$\gamma$ prediction of Eq.~(\ref{eq:hmagnetic}) (solid line) shows excellent
agreement with  a numerical computation (points), here done for $\gamma = 0.1$.}
\label{fig:h}
%
%  longmanuBECplots.nb 
%
\end{figure}
%------------------------------

Finally, we consider the $\curO(\gamma)$ corrections to Eqs.~(\ref{eq:hszerocoupling})
for $\hh_m$ and $\hh_{c2}$.  The SF-\sfm transition at $\hh_m$ occurs when
the magnetization $\mgh(\hh)$ becomes nonzero. By examining Eq.~(\ref{eq:mequation}) 
it is clear that $\hh_m$ is exactly given by
\bse
\bea\label{eq:hmexactpre}
\hh_m &=& \sqrt{|\muh|^2 + \Deltah^2},
\\
&=& \sqrt{|\muh|^2 + \gamma\Bh^2},
\label{eq:hmexact}
\eea\ese
with $\muh$ and $\Bh$ given by their values in the SF state (assuming $\deltah<\deltah_M$, so that
this continuous transition is not preempted by a first-order SF-N transition).
Expressions for $\muh$ and $\Bh$ in the SF phase are derived in Appendix~\ref{SEC:BECBCSreview},
 Eqs.~(\ref{eq:eomzeroh2p}) and ~(\ref{eq:deltazeroh}).  Inserting these expressions into Eq.~(\ref{eq:hmexact}), we find,
to leading order in $\gamma$ (plotted in Fig.~\ref{fig:h}), 
\bse
\bea
\label{eq:hmagneticpre}
\hh_{m}  &\approx & \sqrt{|\muh|^2 + 2\gamma/3},
\\
&\approx & 
\sqrt{\frac{\deltah^2}{4} + \gamma \big(
\frac{2}{3} - \frac{\pi |\deltah|^{3/2}}{4\sqrt{2}}
\big) },
\label{eq:hmagnetic}
\eea
\ese
where we displayed the intermediate result Eq.~(\ref{eq:hmagneticpre}) because it will
be useful below.

To compute the upper-critical chemical potential difference
$\hh_{c2}$, below which the normal  (N) state is unstable to pairing or phase separation, we
note that stability of the N phase requires $\hat{V}_2<0$, so that $e_G$ has positive 
upward curvature at $\Bh = 0$ (see Eq.~\ref{becefweakcoupling}). A second-order transition to the  $\Bh \neq 0$ SF state 
occurs when $\hat{V}_2$ changes sign.  We shall proceed to compute $\hctwo$ by
finding the location of this assumed  second-order  N-\sfm transition.  However,  as we shall see, 
for 
$\deltah> \deltah_c$ 
this second-order transition is preempted by a first-order transition to the regime of phase separation.
 We study this first-order behavior 
in more detail in the next section. For now, we proceed with the second-order
assumption that locates the boundary $\hctwo$ to a good accuracy,
as will be seen in the following sections. 
 The corresponding condition $\hat{V}_2 =0$ is actually equivalent to  our
$\curO(\gamma)$ expression Eq.~(\ref{eq:gapbecapprox}) for the gap equation. Combining this with 
Eq.~(\ref{eq:munormal}) for the normal state chemical potential yields:
\bse
\bea
&&\hspace{-1cm} \hh_{c2} = 2^{2/3} + \frac{|\deltah|}{2}
 - \frac{\gamma}{2} \sqrt{\hh_{c2} - 2^{2/3}} \big(
\frac{\pi}{2}\nonumber
\\
&& \hspace{-1cm} \qquad \quad
 + \frac{2^{1/3}}{\sqrt{\hh_{c2} - 2^{2/3}}} 
- \tan^{-1}  \frac{2^{1/3}}{\sqrt{\hh_{c2} - 2^{2/3}}} \big),\label{eq:hc2correctedpre}
 \\
&&\hspace{-.7cm} \quad   \approx 2^{2/3} + \frac{|\deltah|}{2}\nonumber
\\
&&\hspace{-1cm}  \qquad\quad  - \frac{\gamma|\deltah|^{1/2}}{\sqrt{8}} 
\big(
\frac{\pi}{2} + \frac{2^{5/6}}{|\deltah|^{1/2}}
- \tan^{-1} \frac{2^{5/6}}{|\deltah|^{1/2}} 
\big),
\label{eq:hc2corrected}
\eea
\ese
with the second expression, obtained by approximating $\hctwo \approx 2^{2/3} + |\deltah|/2$ on 
the right side of Eq.~(\ref{eq:hc2correctedpre}),
 correct to $\curO (\gamma^1)$.  As 
derived, Eq.~(\ref{eq:hc2corrected}) denotes the chemical
potential difference at which a putative second-order N to \sfm
phase transition occurs.  As noted above, 
this only occurs for sufficiently low detunings with a first-order 
transition occurring for higher detunings. Despite this, we shall see that the 
critical $\hh$ for the first-order instability (which we also denote $\hctwo$)
to good accuracy is  given by Eq.~(\ref{eq:hc2corrected}). 
We will demonstrate this assertion in Sec.~\ref{SEC:transitionphasesep}.

%%%%%%%%%%%%%%%%%%%%%%%%%%%%%%%%%%%%%%%%%%%%%%%%%%%%%%%%%%%%%%%%%%%%%%%%%%%%%%%%%%%%%%%%%%%%%%%
%%%%%%%%%%%%%%%%%%%%%%%%%%%%%%%%%%%%%%%%%%%%%%%%%%%%%%%%%%%%%%%%%%%%%%%%%%%%%%%%%%%%%%%%%%%%%%%%

\subsection{Transition to the regime of phase separation}
\label{SEC:transitionphasesep}
As  mentioned above, the regime of phase separation we found on the BCS side of the resonance
persists into the BEC regime, bounded by the $\hcone(\deltah)$ and $\hctwo(\deltah)$ curves.
A notable qualitative difference on the BEC side is the existence of a homogeneous magnetized superfluid
(\sfm) state. 

%-----------------------------
%
% fig%25
%
\begin{figure}[bth]
\vspace{1cm}
\centering
\setlength{\unitlength}{1mm}
\begin{picture}(40,30)(0,0)
%\put(-10,50){\begin{picture}(0,0)(0,0)
%\special{psfile=sfmpic1.eps vscale = 110 hscale=110}
%\end{picture}}
%
%
\put(-5,0){\begin{picture}(0,0)(0,0)
\includegraphics{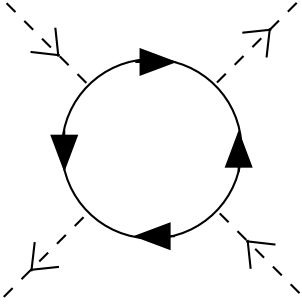}
\end{picture}}
\end{picture}
%\vspace{-.5cm}
\caption{Feynman diagram corresponding to the molecular scattering amplitude, with the solid lines
indicating fermionic atom propagators and the dashed lines indicating scattering molecules.}
\label{fig:bubble} 
%
%  bubble.fig
%
\end{figure}
%------------------------------

The condition for the instability of 
the \sfm to phase separation at $\hcone(\deltah)$ is of course precisely the same as we used on the BCS side: a first-order
transition to the N phase in the grand-canonical ensemble
at fixed $\hh$ and $\muh$.  Although finding the location of this first-order transition
is most accurately done by examining the precise structure of the ground-state energy as 
a function of applied chemical potential difference, to a good approximation it is signaled
by a {\it vanishing\/} of the molecular scattering length $a_m$.  To show this, we compute
the molecular scattering length by using its relation $a_m = \frac{m}{2\pi} T_m$
to the molecular $T$-matrix $T_m$~\cite{LandauQM}, the latter given to leading order (i.e.~the Born approximation) 
by the diagram in Fig.~\ref{fig:bubble} .  Direct calculation of this diagram (relegated to Appendix~\ref{tmatrix})
yields 
\bea
a_m &=& \frac{cg^4m}{64 |\mu|^{3/2}} F_4(h/|\mu|),
\\
&=& \frac{\sqrt{2}\pi^2 \ef \gamma^2}{64 |\mu|^{3/2}\sqrt{m}} F_4(h/|\mu|),
\eea
involving the function $F_4(x)$ [defined in Eq.~(\ref{eq:f})] that we  have already introduced in the 
definition of the quartic term $\hat{V}_4$ in the ground-state energy Eq.~(\ref{becefweakcoupling}).  
This relation to $\hat{V}_4$ is expected as $a_m$ measures the repulsion between molecular bosons.

Stability of a molecular Bose gas requires 
$a_m>0$.~\cite{Fetter}
 However, we find that while $a_m(h=0)>0$, it decreases 
monotonically with $h/|\mu|$ (as illustrated in Fig.~\ref{molscatlenplot}),
vanishing at $h/|\mu| \approx 1.30$.~\cite{commentLG}
Based on a detailed analysis of $e_G(\Bh,\hh)$ (see Sec.~\ref{SEC:detailedanalysis} below), we associate this vanishing
of $a_m(\hh)$ with a first-order \sfm-N transition driven by the increased density of unpaired 
single-spin species atoms.  Since (as discussed previously; see Sec.~\ref{SEC:firstorderBCS}) at fixed atom number 
the system phase separates at such a first-order transition we identify this critical $\hh$ 
with $\hcone$. 
 Combining the value $x \approx 1.30$ at which $F_4(x)$ 
vanishes with the relation $\muh \approx \deltah/2$ that is valid in the \sfm state, we find
\bse
\bea
\hcone &\approx& 1.30 |\muh(n,\deltah)|,
\label{eq:hmixedapproxpre}
\\
\label{eq:hmixedapprox}
&\approx& 0.65 |\deltah|,
\eea
\ese
At sufficiently negative $\deltah$, $\hcone(\deltah)$ {\it  intersects\/} 
$\hctwo(\deltah)$.  Using the $\curO(\gamma^0)$ result Eq.~(\ref{eq:hc2zerocoupling}) for $\hctwo$,
we find the corresponding detuning
\be
\label{eq:deltac}
\deltah_c \approx   -\frac{2^{2/3}}{0.15}   \approx -10.6. 
\ee
  The implication of this 
is that, for $\deltah< \deltah_c$, the molecular bosons in the \sfm phase  are repulsive 
for any $\hh$ 
(and thus are stable).   For $\deltah<\deltah_c$,
the \sfm undergoes a {\it continuous\/} transition at $\hctwo$ to a fully-polarized 
atomic gas.

In the more experimentally-relevant ensemble of fixed atom numbers,
we translate $\hcone$ to the lower critical polarization $\Delta N_{c1}$
at which the \sfm state is unstable to phase separation.  Using 
Eq.~(\ref{eq:hmixedapprox}) along with Eqs.~(\ref{eq:mzerocoupling})
and (\ref{magrelation2}) (as well as $\muh  \approx \deltah/2$) we find 
\bse
\bea
\label{eq:magmixedapproxpre}
\frac{\Delta N_{c1}}{N}& \simeq& \frac{1}{2} (0.15)^{3/2} |\deltah|^{3/2},
\\
& \simeq& 0.029 |\deltah|^{3/2}. \label{eq:magmixedapprox}
\eea
\ese
We next turn to a more detailed demonstration of the preceding discussion by a careful
analysis of the ground-state energy $e_G$.

%%%%%%%%%%%%%%%%%%%%%%%%%%%%%%%%%%%%%%%%%

\subsection{Detailed analysis of the SF-\sfm and \sfm-N transitions}
\label{SEC:detailedanalysis}
In the present section, we present a more precise determination of the 
first-order transition instability described 
in the preceding 
subsection.  Doing so will provide more accurate (in $\gamma$) predictions 
for $\hh_{c1}(\deltah)$, $\Delta N_{c1}(\deltah)$ and  $\hctwo(\deltah)$ 
(note that, as found for sufficiently low detuning on the BCS side, 
 $\Delta N_{c2} = 1$ on the BEC side), as well as a more 
complete description of the first-order \sfm-N transition.  We start by expanding $e_G(\Bh)$ to higher order in $\Bh$: 
\be
\label{eq:egbec}
e_G \approx  - \hat{V}_2 \Bh^2 + \frac{\hat{V}_4}{2} \Bh^4
+ \frac{\hat{V}_6}{3} \Bh^6
+ \curO (\Bh^8),
\ee
where for simplicity we have dropped a $\Bh$-independent term.  Here,
$\hat{V}_2$ and $\hat{V}_4$ are given in Eq.~(\ref{vee}) and  
\bse
\bea
\label{eq:v6}
&&\hspace{-1.5cm}\hat{V}_6[\hh,\muh] = \frac{3\gamma^3}{32 |\muh|^{7/2}} F_6(\hh/|\muh|),
\\
&&\hspace{-1.5cm}F_6(x) \equiv  - \frac{5\pi}{64} + \Big[\frac{1}{3x^3\sqrt{x-1}} 
+ \frac{5}{32}\tan^{-1}\sqrt{x-1}
\nonumber \\ 
&&\hspace{-1.5cm}\quad
+ \frac{\sqrt{x-1}}{96x^4} (-48+8x+10 x^2 + 15x^3)\Big] \Theta(x-1),
\label{eq:f6}
 \eea
\ese
is the the sixth-order coefficient.

\subsubsection{SF-\sfm transition at $\hh_m$}

We first recall that, for low $\hh<\hh_m\approx |\muh|$, it is sufficient to limit $e_G$ to quartic order in $\Bh$.   
Standard minimization gives a nontrivial SF solution $\Bh^2 = \Vh_2/\Vh_4$ and
energy $e_{G,SF} = -\Vh_2^2/\Vh_4$ that [because of the step functions $\Theta(\hh/|\muh|-1)$]
are explicitly $\hh$-independent.  Hence, $\mgh = -\frac{\partial e_{G,SF}}{\partial \hh} = 0$ 
as  expected in the singlet SF state.  

For a more accurate description (and for later study of the \sfm-N
transition; see below) we include the $\Vh_6\Bh^6$ term in the expansion of
$e_G(\Bh)$. Although $\Vh_6$ has a form  similar
to $\hat{V}_2$ and $\hat{V}_4$, namely a constant plus an $\hh$-dependent
correction that vanishes for $\hh < \hh_m$, examination of Eq.~(\ref{eq:f6}) reveals the function 
$F_6(x)$ to be divergent at $x\to 1^+$. This divergence arises from the approximate way 
we evaluate the integral of $\mgh(x)$ (the Zeeman energy) appearing in Eq.~(\ref{eq:gsetotallygeneral}):
\be
-\int_{\hh_m}^{\hh}  dh' \mgh(\hh')
= -\frac{2}{3} \int_{\hh_m}^{\hh} dh'
 \Big(\sqrt{\hh^2 -\gamma \Bh^2} - |\muh|\Big)^{3/2},
\label{eq:magintegral}
\ee
perturbatively in $\gamma \Bh^2/(\hh^2-\muh^2)$.  Recall that 
 $\hh_m \equiv \sqrt{\muh^2+\gamma \Bh^2}$.  Hence, strictly speaking we should
only use these expressions in the regime where  $\gamma \Bh^2/(\hh^2-\muh^2)\ll 1$.

To elucidate the nature of the transition exhibited by $e_G(\Bh,\hh,\deltah)$, Eq.~(\ref{eq:egbec}),
we determine the location of its minima, given by 
\be
\label{eq:gapdetailed}
0 = - \Vh_2\Bh + \Vh_4 \Bh^3 + \Vh_6 \Bh^5,
\ee
which (in addition to the trivial stationary point at $\Bh = 0$ corresponding to the N state) has two nontrivial 
solutions
\be
\Bh^2_\pm = \frac{\Vh_4}{2\Vh_6}\Big[-1 \pm \sqrt{1+ 4\Vh_2 \Vh_6/\Vh_4^2}
\Big],
\label{eq:becstationary}
\ee
that yield physical stationary points of $e_G$ only when the right side
of Eq.~(\ref{eq:becstationary}) is real and positive.

%-----------------------------
%
% fig%26
%
\begin{figure}[bth]
\vspace{1.4cm}
\centering
\setlength{\unitlength}{1mm}
\begin{picture}(40,40)(0,0)
\put(-55,0){\begin{picture}(0,0)(0,0)
\includegraphics{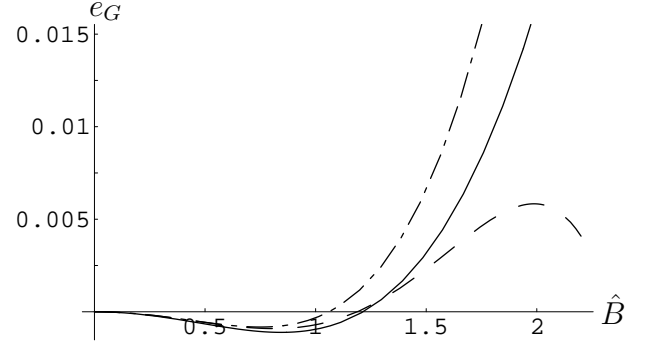}
\end{picture}}
\put(-9,47) {\large{$e_G$}}
\put(59,6) {\large{$\Bh$}}
\end{picture}
\vspace{-.5cm}
\caption{(Color online) Plot of $e_G$ as a function of $\Bh$ for $\hh < \hh_m$, comparing fourth-order 
(dot-dashed), sixth-order (dashed) and exact numerical (solid) expressions. The sixth-order
case approximates well the physical minimum at low $\Bh$ but has a maximum at larger $\Bh$ 
that is an artifact of the sixth-order truncation.}
\label{fig:sixeightcompare}
%
%  sixthorderp.nb 
% 
\end{figure}
%------------------------------

For sufficiently weak chemical potential difference, $\hh < \hh_m$, the
Landau expansion coefficients are $\Vh_2 > 0$, $\Vh_4 > 0$ and $\Vh_6 < 0$. Hence
as long as $4\Vh_2|\Vh_6| < \Vh_4^2$, the $+$ and $-$ solutions in
Eq.~(\ref{eq:becstationary}) correspond to a minimum and a maximum, respectively with the
complete shape illustrated in Fig.~\ref{fig:sixeightcompare} (dashed line). By plotting the numerical
solution of the non-Taylor expanded expression for $e_G$ in Eq.~(\ref{eq:gsetotallygeneral})
for this
range of parameters ($\deltah < 0$, $\hh<\hh_m$) and comparing with the quartic
and sixth order (in $\Bh$) approximation to $e_G$, it is clear that while the $+$
solution minimum represents the physical, stable singlet-SF ground
state, the $-$ solution maximum is simply an artifact of truncation of $e_G$ at sixth
order.  However, the $+$ solution provides an accurate approximation to the 
true SF minimum, as shown in Fig.~\ref{fig:sixeightcompare}.

The SF-\sfm transition occurs at $\hh_m\approx |\muh|$.
For $\hh > \hh_m$ two qualitative changes take place in the coefficients
$\Vh_n(\hh,\deltah)$. Firstly, because of the positive argument of the step
functions, the coefficients $\Vh_n(\hh,\deltah)$ and therefore the minimum
superfluid solution $\Bh_+$ and the corresponding energy $e_G(B_+)$ become
nontrivial functions of the chemical potential difference $\hh$. This then
immediately gives a finite magnetization (population imbalance) $\mgh = -
de_G/d\hh\neq 0$, characteristic of the \sfm ground state. Secondly, for
$\hh\agt\hh_m$ the coefficient $\Vh_6 > 0$. 
Hence the shape of the $e_G(\Bh)$ function and the corresponding
nature of the transition out of the \sfm state (\sfm - N transition)
are determined by the signs of the $\Vh_2$ and $\Vh_4$ coefficients.
%-----------------------------
%
% fig%27
%
\begin{figure}[bth]
\vspace{1.4cm}
\hspace{-1.4cm}
\centering
\setlength{\unitlength}{1mm}
\begin{picture}(40,40)(0,0)
\put(-50,0){\begin{picture}(0,0)(0,0)
\includegraphics{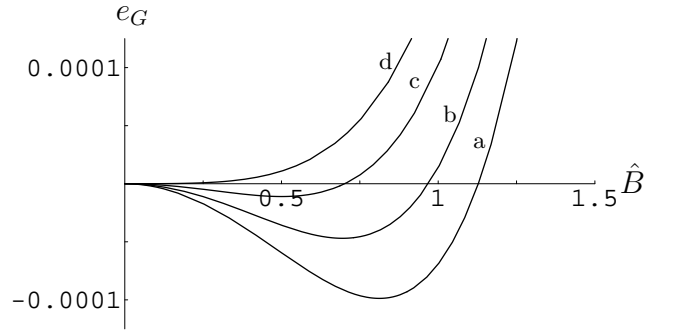}
\end{picture}}
\put(-3,48) {\large{$e_G$}}
\put(64,25) {\large{$\Bh$}}
\put(44.5,31) {a}
\put(40.5,34.5) {b}
\put(36,39) {c}
\put(32,41.5) {d}
\end{picture}
\vspace{-.5cm}
\caption{Evolution of the BEC-regime normalized ground-state energy $e_G(\Bh)$ [Eq.~(\ref{eq:egbec})],
for parameters such that a continuous \sfm$\to$N transition occurs.  Here, 
 $\gamma = 0.1$, $\deltah = -2$, $\muh = -0.924$.  For curves a, b, and c, $\hh<\hc$ 
with $\hh =1.0$,$1.007$, and $1.014$ respectively and the minimum at $\Bh \neq 0$ represents
the \sfm phase.  For curve d, $\hh = 1.020 >\hc$ and the minimum is at $\Bh = 0$.}
\label{fig:becenergysequence1}
\end{figure}
%------------------------------

\subsubsection{Second-order \sfm-N transition: $\deltah < \deltah_c$}
\label{SEC:secondordersfmn}
Generically a transition is continuous if there is only a single
minimum in the energy function that vanishes continuously as a parameter
is tuned. As can be seen from Eq.~(\ref{eq:becstationary}), this is possible for $e_G(\Bh)$ when
$\Vh_4 > 0$ and $\Vh_6 > 0$, in which case 
\be
 \Bh_{+}^2 =  \Bh_{{\rm SF}_{\rm M}}^2  = \frac{\Vh_4}{2\Vh_6}\big[ \sqrt{1+ 4\Vh_2 \Vh_6/\Vh_4^2} -1
\big],
\label{eq:becstationary3}
\ee
 (which vanishes continuously as $\Vh_2 \to 0$) characterizes the \sfm state
while  $\Bh_{-}^2 < 0$ and therefore no longer corresponds to an  extremum of $e_G$, 
as $\Bh_{-}$ is complex. 

The evolution of $e_G(\Bh)$ with $\Vh_2$ for such a continuous transition is illustrated in 
Fig.~\ref{fig:becenergysequence1}. The
continuous \sfm - N transition is then determined by the vanishing of
$\Vh_2(\hh,\deltah)$ when the \sfm minimum at $\Bh_+$ vanishes into the normal
state maximum at $\Bh=0$, with this taking place at $\hc$ satisfying
\be
F_2(\hc/|\muh|)=\frac{2\muh-\deltah}{\gamma\sqrt{|\muh|}}.
\label{hcsecondgeneral}
\ee

At fixed imposed density, the condition of  $\Vh_4(\hc,\deltah) > 0$ for such a continuous transition 
is satisfied for $\deltah < \deltah_c$ with $\deltah_c$ approximately given by Eq.~(\ref{eq:deltac}).
When combined with the normal-state chemical potential
Eq.~(\ref{eq:munormal}), this $\hc$ is equal to $\hctwo(\deltah)$  that we have 
previously calculated
[Eq.~(\ref{eq:hc2corrected})], plotted in the phase diagram (see Figs.~\ref{fig:hphasetwo}
and \ref{fig:bechphasediagram}).  As
discussed above, this transition corresponds to a point at which all
of the molecules have dissociated into a fully polarized (single
species) gas of atoms, with the molecular condensate $n_m$,
characteristic of the \sfm state, vanishing. As such, experimentally
(where it is $\Delta N$ rather than $h$ that is controlled), the fully
polarized normal state corresponds to a single point $\Delta N/N =1$.

\subsubsection{First-order \sfm-N transition: $\deltah_c < \deltah < \deltah_M$}
\label{SEC:firstordersfmn}
In contrast to the above behavior, for $\deltah > \deltah_c$ at fixed density, the continuous
transition at $\Vh_2=0$ [given by $\hc$, Eq.~(\ref{hcsecondgeneral})] is preempted by a first-order
transition.  This happens because $\Vh_4(\hh,\deltah)$ changes sign (with
$e_G(\Bh)$ still well-defined, stabilized at large $\Bh$ by the positive $\Vh_6$ 
term), becoming negative {\it before\/} $\Vh_2$ has a chance to vanish. We
note that as long as $\Vh_2 > 0$ (and $\Vh_6 > 0$, valid for $\hh\agt\hh_m$  that we are
considering here), independent of the sign of $\Vh_4$ there is a single $\Bh \neq 0$
minimum of  Eq.~(\ref{eq:becstationary}) that characterizes the \sfm state. However, for $\Vh_4
> 0$ the \sfm state is given by $\Bh_+$ solution, while, for $\Vh_4 < 0$, 
it is the $B_{-}$ solution that corresponds to the \sfm
phase. The change in this behavior takes place when $\Vh_4=0$,
corresponding to
\be
\hh_*\simeq  1.30\muh,
\ee
[which, incidentally, is the $\curO(\gamma^0)$ approximation to $\hcone$ given
in Eq.~(\ref{eq:hmixedapproxpre})] and $\Bh_{SF,*}=\Bh_{-}(\Vh_4=0)=(\Vh_6/\Vh_2)^{1/4}$.

%-----------------------------
%
% fig%28
%
\begin{figure}[bth]
\vspace{1.4cm}
\hspace{-1.4cm}
\centering
\setlength{\unitlength}{1mm}
\begin{picture}(40,40)(0,0)
\put(-50,0){\begin{picture}(0,0)(0,0)
\includegraphics{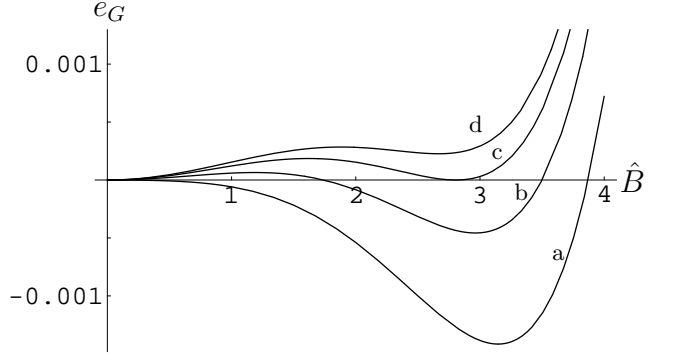}
\end{picture}}
\put(-6,49.5) {\large{$e_G$}}
\put(64,26) {\large{$\Bh$}}
\put(55,17) {a}
\put(50,25) {b}
\put(47,30.5) {c}
\put(44,34) {d}
\end{picture}
\vspace{-.5cm}
\caption{Evolution of the BEC-regime normalized ground-state energy $e_G(\Bh)$ [Eq.~(\ref{eq:egbec})],
for parameters such that a first-order \sfm$\to$N transition occurs.  Here, 
 $\gamma = 0.1$, $\deltah = -2$, $\muh = -0.921$ and $\hh$ takes four different values: 
a) $\hh = 1.317$, so that $\Vh_2>0$ and  $\Vh_4<0$, 
b) $\hh = 1.322$, so that $\Vh_2<0$ and  $\Vh_4<0$,
c) $\hh = 1.325 = \hc$, and d) $\hh =1.326> \hc$. }
\label{fig:becenergysequence}
%
%  10JuneBECtrans.nb 
%
\end{figure}
%------------------------------

Once $\Vh_4 < 0$, as illustrated in Fig.~\ref{fig:becenergysequence},
 the shape of $e_G(\Bh)$
changes qualitatively when $\Vh_2$ changes sign. With both $\Vh_2$ and $\Vh_4$
negative the normal-state extremum at $\Bh=0$ turns into a {\it local minimum\/}.
(Compare curves a and b in Fig.~\ref{fig:becenergysequence}.)
 Concomitantly, both extrema in Eq.~(\ref{eq:becstationary}) become important, 
with the $\Bh_+$ solution giving the barrier (maximum) separating the \sfm $\Bh_-$
minimum from the normal-state minimum at $\Bh=0$. At low $\hh<\hc$ (but larger
than $\hh_*$), the energy $e_G(\Bh_-)$
of the \sfm minimum is lower than that of the normal state (for which $e_G=0$) and the 
\sfm remains a stable ground state. However, with increasing $\hh$ the \sfm
minimum rises and reaches the normal-state energy at the first-order condition 
\bea
0 =  - \Vh_2\Bh_{-}^2 + \frac{1}{2}\Vh_4 \Bh_{-}^4 + \frac{1}{3}\Vh_6 \Bh_{-}^6,
\label{eq:energydetailed}
\eea
giving
\bse
\bea
\Vh_4 &=& -\frac{4}{3} \Vh_6 \Bh^2_-,
\label{eq:firstordercondition}
\eea
or, equivalently (including arguments for clarity),
\bea
\Vh_2[\deltah,\hc,\muh] \Vh_6[\hc,\muh] &=& -\frac{3}{16} \Vh_4^2[\hc,\muh],
\label{eq:firstordercondition2}
\eea
\ese
as the condition for the critical point $\hc(\deltah,\muh)$ of the first-order
\sfm-N transition.

As is usual for first-order transitions and as discussed in Sec.~\ref{SEC:BCS} for the
BCS regime, the state of the system for $\hh > \hc$ is determined by
whether the system is kept at fixed chemical potential or at fixed
average density (atom number). For the former case the system simply jumps
from the \sfm to the normal (fully spin-polarized) state, 
and
the density and magnetization are free to adjust discontinuously to this
sudden change in state at $\hc$. 

On the other hand, for atom number fixed at $N$, a discontinuous change
of state is not an option available to the system as this would
require a change in $N$ (and $\Delta N$) that are fixed. As explained in
Sec.~\ref{SEC:firstorderBCS}, upon increasing $\hh$, the system tunes its chemical potential to
remain on the critical curve $\hc(\deltah,\muh)$ and instead phase separates
into coexisting \sfm and normal states, in proportions so as to satisfy
the imposed total number constraint.

We conclude this subsection by noting that the regime of phase separation (PS) at negative detuning 
consists of \sfm-N coexistence only for sufficiently low $\deltah$. At larger $\deltah$, the \sfm phase
ceases to exist and the coexistence is between N and SF phases
 (as we find at low detunings
in the BCS regime).  To determine the boundary separating these possibilities, we note that, in the 
regime of phase separation, $\hh= \hc(\muh)$ with $\hc$ approximately given by Eq.~(\ref{eq:firstordercondition2}).
Implementing the constraint $4/3 = x \hat{n}_{SF} + (1-x) \hat{n}_N$ (where here SF can refer to the SF or
\sfm states) allows us to study, numerically, the negative-detuning PS regime.  The dashed 
line in Fig.~\ref{fig:bechphasediagram}
denotes where, in the PS regime, $\hc = \hh_m$ (indicating a continuous SF-\sfm transition inside the PS regime) and
the dashed line in the fixed-polarization phase diagram 
Fig.~\ref{fig:becmphasediagram} is thus obtained by converting this $\hh$ boundary to polarization $\Delta N/N$.

\subsection{Finite $\gamma$ corrections to the \sfm - N phase-separation boundaries}

Here we use results of the previous section for the \sfm - N first-order 
transition (occurring for $\deltah_c < \deltah < \deltah_M$) to compute the 
phase-separation boundaries, correcting our previous $\curO(\gamma^0)$ results. 
At fixed chemical potential difference
(rather than polarization), the boundaries are $\hcone(\deltah)$ and
$\hctwo(\deltah)$, to zeroth order in $\gamma$ given by Eqs.~(\ref{eq:hmixedapprox}) and 
(\ref{eq:hc2zerocoupling}), respectively.

First focusing on $\hcone(\deltah)$, corrections to its zeroth order
expression Eq.~(\ref{eq:hmixedapprox}) have two sources. One is the correction to the
approximation of the chemical potential by $\muh\approx \deltah/2$, 
only valid to $\curO(\gamma^0)$. The other source of approximation is the
location of the transition by $\Vh_4=0$, that is more accurately given by
Eq.~(\ref{eq:firstordercondition2}).

To determine $\muh(\deltah,\hh)$ more accurately we turn to its defining
equation, Eq.~(\ref{eq:gapbecapprox}), that can be written as
\bea
\label{eq:approxmusfm}
&&\deltah - 2\muh = - \frac{\gamma \sqrt{|\muh|}\pi}{2} 
 \\
&&\qquad 
- \gamma\sqrt{|\muh|} \Big[\sqrt{\hh/|\muh|-1}
-\tan^{-1}\sqrt{\hh/|\muh|-1}
\Big].\nonumber
\eea
To first order in $\gamma$ and near $\hcone$ (of interest to us) Eq.~(\ref{eq:approxmusfm}) can
be considerably simplified by replacing $\hh$ by $\hcone$, with the latter
approximated by its zeroth order value in $\gamma$, Eq.~(\ref{eq:hmixedapproxpre}). This gives
\be
\deltah - 2\muh \approx - \frac{\gamma \sqrt{|\muh|}\pi}{2} - 0.047 \gamma  \sqrt{|\muh|},
\label{eq:approxmusfm2}
\ee
which, when solved for $\muh$  gives
\be
\muh \approx \muh^{(1)}(\deltah)\equiv  - \frac{1}{64} \Big(
-\gamma'\pi + \sqrt{(\gamma' \pi)^2 + 32 |\deltah|}
\Big)^2,\label{eq:approxmusfm4}
\ee
where $\gamma'\equiv\gamma[1+ 2(0.047)/\pi]$.

%-----------------------------
%
% fig%29
%
\begin{figure}[bth]
\vspace{1.4cm}
\centering
\setlength{\unitlength}{1mm}
\begin{picture}(40,40)(0,0)
\put(-50,0){\begin{picture}(0,0)(0,0)
\includegraphics{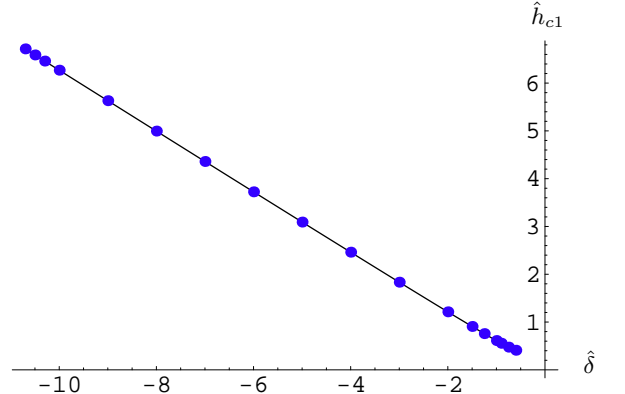}
\end{picture}}
\put(52,51) {$\hcone$}
\put(59,4.5) {$\deltah$}
\end{picture}
\vspace{-.5cm}
\caption{(Color online) Plot of $\hcone(\deltah,\gamma = 0.1)$, the normalized lower chemical potential 
difference above which the system enters
the regime of phase separation.  The points are a numerical computation and 
the solid line is Eq.~(\ref{eq:hconefinal}).
}
\label{fig:hconebec}
%
%  longmanuBECplots.nb 
%
\end{figure}
%------------------------------

%-----------------------------
%
% fig%30
%
\begin{figure}[bth]
\vspace{1.4cm}
\centering
\setlength{\unitlength}{1mm}
\begin{picture}(40,40)(0,0)
\put(-50,0){\begin{picture}(0,0)(0,0)
\includegraphics{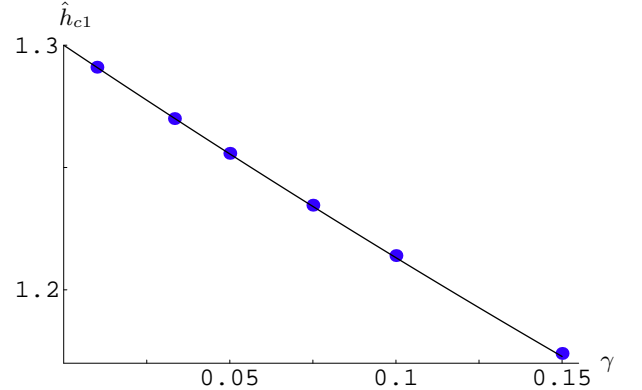}
\end{picture}}
\put(-13,50) {$ \hcone$}
\put(59,4.5) {$\gamma$}
\end{picture}
\vspace{-.5cm}
\caption{(Color online) Plot of $\hcone(\deltah,\gamma)$ at one particular detuning ($\deltah = -2$) as a function of 
resonance width parameter $\gamma$.
}
\label{fig:hconevscoupling}
%
%  longmanuBECplots.nb 
%
\end{figure}
%------------------------------

The second correction to $\hcone$ comes from using the proper equation,  
Eq.~(\ref{eq:firstordercondition}),
(rather than $\Vh_4=0$) as the location of the first-order transition and the
lower-field phase boundary $\hcone(\delta)$. We write  $\hcone = \hcone^{(0)} + \hcone^{(1)}$,
with $\hcone^{(0)}$ the $\curO(\gamma^0)$ expression given by Eq.~(\ref{eq:hmixedapproxpre})  and 
$\hcone^{(1)}$ the correction of $\curO(\gamma)$. Using 
Eqs.~(\ref{eq:v4}) and (\ref{eq:v6}), as well as the $\curO(\gamma^0)$ expression
for $\Bh^2$
[i.e.~Eq.~(\ref{eq:numbec})] along with Eq.~(\ref{eq:firstordercondition})  leads to
\bea
&&\hspace{-.5cm}F_4\Big[\frac{\hcone^{(0)}+ \hcone^{(1)}}{|\muh|}\Big] \nonumber
\\
&&\hspace{-.5cm}\qquad= -\frac{8\gamma}{3\pi|\muh|^2} 
[1-\frac{1}{2}(\hcone^{(0)} - |\muh|)^{\frac{3}{2}}]F_6\Big[\frac{\hcone^{(0)}}{|\muh|}\Big],
\eea
where on the right side we neglected $\hcone^{(1)}$, valid to order $\curO(\gamma)$.
 Expanding the left-hand side in small $\hcone^{(1)}$,
and using the fact that $F_4[\hcone^{(0)}/|\muh|]= F_4[1.30] =0$
[Eq.~(\ref{eq:hmixedapproxpre}) above], $F'_4[1.30] =-0.74$ and  $F_6[1.30] = 0.134$,
we find
\be
\hcone^{(1)} \approx \frac{0.154 \gamma}{|\muh|} [1-0.082 |\muh|^{3/2}],
\label{eq:hconefirst}
\ee
yielding 
\be
\hcone \approx 1.30 |\muh|  + \frac{0.154 \gamma}{|\muh|} [1-0.082 |\muh|^{3/2}]  . 
\label{eq:hconefinalpre}
\ee
Now using $\muh(\deltah)$ in  Eq.~(\ref{eq:hconefinalpre}) we finally obtain to $\curO(\gamma)$
\bea
\hspace{-.5cm}\hcone(\deltah) \approx 
1.30|\muh^{(1)}(\deltah)|+ \frac{\gamma 0.308}{|\deltah|} [1-0.029 |\deltah|^{3/2}],
\label{eq:hconefinal}
\eea
that we plot in Fig.~\ref{fig:hconebec}, as a function of detuning $\delta$ (on the BEC
side of $\delta < 0$) and compare it to a numerical determination of $\hcone$ 
directly from the full ground-state energy Eq.~(\ref{eq:gsetotallygeneral}).

Also, as illustrated in Fig.~\ref{fig:hconevscoupling}, for $\deltah=-2$, the linear decrease of
$\hcone$  with increasing $\gamma$ exhibited in Eq.~(\ref{eq:hconefinal}) also compares well for small
$\gamma$
with the $\gamma$ dependence of the numerical solution (points). We expect
the resulting increase in the regime of phase separation to remain
qualitatively correct beyond this narrow-resonance ($\gamma\ll 1$) limit.

The above result for $\hcone(\deltah)$ can be easily translated into a critical
polarization $\Delta N_{c1}(\delta)$ that is relevant for fixed spin-species number difference (fixed polarization)
 by inserting $\hcone$, $\muh(\deltah)$ and $\Bh^2$ inside $\mgh(\hh)$, Eq.~(\ref{eq:mequation}),
that gives for the lower-critical polarization 
\bea
&&\hspace{-1.5cm}\frac{\Delta N_{c1}}{N}= \frac{1}{2}  \Big(\sqrt{\hcone^2 -\frac{2}{3}\gamma(1-0.029|\deltah|^{3/2})}
 - |\muh^{(1)}|\Big)^{\frac{3}{2}},
\eea
above which the system phase separates into coexisting \sfm and N
states.

We can also use Eq.~(\ref{eq:hconefinal}) to compute the critical detuning $\deltah_M$
beyond which the \sfm phase ceases to exist and the PS is replaced by N-SF coexistence
 (as on the BCS side of the Feshbach resonance). It is
determined by the point where $\hh_m(\deltah)$ and $\hcone(\deltah)$ intersect,
i.e., $\hh_m(\deltah_M)=\hcone(\deltah_M)$, or equivalently $\Delta N_{c1}(\deltah_M)=0$.
Using the former condition, to $\curO(\gamma)$ we find
\bea
1.69 |\muh|^2 + 0.40 \gamma [1-0.082|\muh|^{3/2}] = |\muh|^2 + \frac{2}{3} \gamma,
\label{eq:quadraticmu}
\eea
that gives to leading order in $\gamma$
\be
\muh_M=-0.62\sqrt{\gamma},
\ee
for the chemical potential at this intersection.
Using the zeroth order relation $\muh=\deltah/2$  gives, to leading order in
$\gamma$,
\bea
\deltah_M \approx -1.24\sqrt{\gamma},
\label{eq:deltam}
\eea
a result that compares favorably (see Fig.~\ref{fig:bechphasediagram}) with the numerical
solution (solid point). 

%-----------------------------
%
% fig%31
%
\begin{figure}[bth]
\vspace{1.4cm}
\centering
\setlength{\unitlength}{1mm}
\begin{picture}(40,40)(0,0)
\put(-50,0){\begin{picture}(0,0)(0,0)
\includegraphics{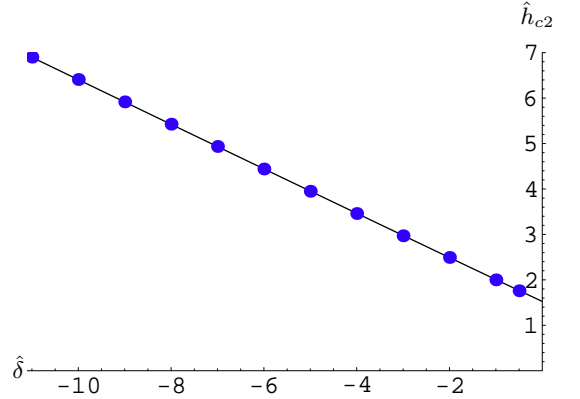}
\end{picture}}
\put(55.5,51) {$\hctwo$}
\put(-12,4.0) {$\deltah$}
\end{picture}
\vspace{-.5cm}
\caption{(Color online) Plot of $\hctwo(\deltah)$, the normalized upper-critical chemical potential difference
 above which the system enters
the normal (N) phase.  The points are a numerical computation and 
the solid line is Eq.~(\ref{eq:hc2corrected})).
}
\label{fig:hctwobec}
%
%  longmanuBECplots.nb 
%
\end{figure}
%------------------------------

We turn next to the calculation of the upper-critical boundary,
$\hctwo(\deltah)$, of the phase-separation region. This is defined by where the normal-state
chemical potential $\muh_N(\hh,n) = 2^{2/3} - \hh$
[Eq.~(\ref{eq:munormal})] intersects the first-order condition
Eq.~(\ref{eq:firstordercondition2}).  This yields the self-consistent set of equations
\bse
\bea
\label{eq:hctwoexact}
\Vh_2[\deltah,\hctwo,\muh_N]&=& -\frac{3}{16} \frac{\Vh_4[\hctwo,\muh_N]^2}{\Vh_6[\hctwo,\muh_N]},
\\
\muh_N &=& 2^{2/3} - \hctwo,
\label{eq:munormal2}
\eea
\ese
an expression that applies for $\delta\agt \delta_c$ (where the sixth-order expansion applies).  
Recall that, close to the first-order transition $\hc$, $\Vh_4$ vanishes allowing a secondary 
minimum in $e_G$ to form at $\Bh = 0$.  This vanishing implies that the right side of 
Eq.~(\ref{eq:hctwoexact}) is numerically small near $\hc$ and may therefore be neglected 
(despite  being formally of the same order in $\gamma$ as the left side).  This reduces 
Eq.~(\ref{eq:hctwoexact}) to our previous approximate result Eq.~(\ref{eq:hc2corrected}),
derived assuming a second-order \sfm-N transition at the point where $\Vh_2$ changes sign,
and shows why Eq.~(\ref{eq:hc2corrected}) accurately determines $\hctwo$ even in the
first-order regime, as shown in Fig.~\ref{fig:hctwobec}. 

 In fact, Eq.~(\ref{eq:hctwoexact})
shows that the true $\hctwo$ is slightly higher than that predicted by $\Vh_2=0$, since 
$\Vh_2<0$ in the N state and the right side is negative.  This is as expected, as the
first-order transition (that takes place for $\deltah > \deltah_c$) always
precedes the spinodal point $\Vh_2=0$ at which the metastability of the
normal state is lost.

\subsection{Fixed chemical potential}
\label{SEC:BECfcp}
%-----------------------------
%
% fig%32
%
\begin{figure}[bth]
\vspace{1.4cm}
\hspace{-1.4cm}
\centering
\setlength{\unitlength}{1mm}
\begin{picture}(40,40)(0,0)
\put(-50,0){\begin{picture}(0,0)(0,0)
\includegraphics{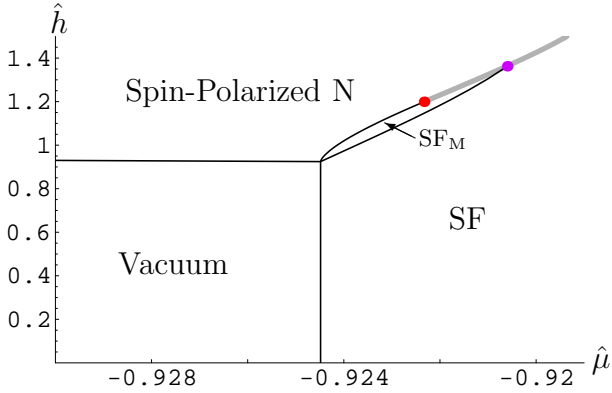}
\end{picture}}
\put(-8,49) {\large{$\hh$}}
\put(64,4.5) {\large{$\muh$}}
\put(1,17) {\large{Vacuum}}
\put(45,23) {\large{SF}}
\put(2,40) {\large{Spin-Polarized N}}
\put(41,34) {\sfm}
\put(40.5,35) {\vector(-2,1){4}}
\end{picture}
\vspace{-.5cm}
\caption{(Color online) Negative-detuning phase diagram in the 
grand-canonical ensemble at detuning $\deltah = -2 $
and width $\gamma=0.1$ showing superfluid (SF), magnetic superfluid (\sfm, thin region
indicated by arrow), vacuum and normal spin-$\uparrow$ phases. Black curves are 
continuous transitions and the gray curve is a first-order transition.
 The \sfm phase undergoes
either a continuous [to the left of the red point at $(-0.922,1.20)$] or first-order 
[to the right of the red point] transition to the N phase.  To the right of the purple point 
at $(-0.9206,1.36)$, the \sfm phase ceases to exist and there is a direct first-order
SF-to-N transition.}
\label{fig:phasediagramfixedmu}
%
%  2NovBECgrandcan.nb 
%
\end{figure}
%------------------------------

Having determined the BEC-regime phase diagram for the physically-relevant case of  
fixed total atom density and population difference, for completeness   
in the present section we compute the phase diagram in the grand-canonical ensemble
of fixed $\muh$ and $\hh$.    Of course, in the grand-canonical
ensemble there is no regime of phase separation. Thus, our main task
is to find the critical
curves for the first- and second-order phase transitions, 
focusing on the physically interesting regime of $\muh$ near
$\deltah/2$. This can be done by utilizing results derived in
Secs.~\ref{SEC:secondordersfmn} and \ref{SEC:firstordersfmn}.
Starting in the singlet SF state at $\hh=0$, characterized by an
effective bosonic chemical potential $\muh_m=2\muh-\deltah+\curO(\gamma)$, and
lowering $\muh$ from $> \deltah/2+\curO(\gamma)$ to $< \deltah/2+\curO(\gamma)$, the system
undergoes a continuous transition from a molecular SF to a vacuum of
molecules (and atoms, since $\muh < 0$). More generally it is defined by the 
vanishing of the quadratic coefficient $\Vh_2[\deltah,\hh,\muh]$, giving
\be
\muh_c(\deltah,\hh<\hh_m)=
 - \frac{1}{64} \Big(
-\gamma\pi + \sqrt{(\gamma \pi)^2 + 32 |\deltah|}
\Big)^2.
\ee
Since, for small $\hh < \hh_m$, the Landau coefficients $\Vh_\alpha$ are
 $h$-independent, the above result holds at $\hh < \hh_m$, leading to a
strictly vertical SF-vacuum phase boundary (see Fig.~\ref{fig:phasediagramfixedmu}). 

Starting in this vacuum state ($\muh < \muh_c$), with increasing $\hh$ the system
undergoes a transition to the spin-polarized N state as $\muh_\uparrow$
changes from negative to positive leading to a finite density of
spin-up fermions (but still $\Bh=0$, since $\muh < \muh_c$). We
indicate this phase boundary in Fig.~\ref{fig:phasediagramfixedmu} by a (nearly) horizontal line.

Starting instead in the SF state at $\muh > \muh_c$ and increasing $\hh$, a
continuous SF-\sfm transition takes place as $\mgh$ becomes nonzero at $\hh_m$.  Using
Eq.~(\ref{eq:hmexact}) for $\hh_m$ (approximately equal to $|\muh|$ near the SF-to-Vacuum
transition but strongly deviating from it at larger $\muh$) along with 
Eq.~(\ref{eq:becstationary}), $\Bh_+^2$, for the normalized molecular density, 
yields the SF-to-\sfm boundary shown in Fig.~\ref{fig:phasediagramfixedmu}.

As discussed in Sec.~\ref{SEC:detailedanalysis} there are two possibilities for exiting the
resulting \sfm state into the N state. (i) If $\Vh_2$ changes sign (from positive to
negative) {\it before\/} $\Vh_4$ becomes negative, the \sfm-N transition is
second-order, and given by Eq.~(\ref{hcsecondgeneral}).
 (ii) If $\Vh_4$ becomes negative while $\Vh_2$ is still positive,
then the \sfm-N transition is first-order. The tricritical
point~\cite{ChaikinLubensky}
(whose existence was also recently confirmed and extended to finite
temperature in Ref.~\onlinecite{Parish}) $\Vh_4(\hh_c,\muh)=0$ separating these two scenarios
is indicated by a red solid point, with $\Vh_4 < 0$ (and transition
first-order) and $\Vh_4 > 0$ (and transition second-order) to the right and
left of this point, respectively.  The gray curve
indicates the first-order \sfm-to-N transition [computed using Eq.~(\ref{eq:firstordercondition2})].

At larger $\muh$, to the right on the figure, $\hh_m$ intersects the first-order curve 
at the purple point at  $(-.9206,1.36)$.  To the right of this point, the \sfm phase 
ceases to exist and there is a direct SF-to-N first-order transition.  Note that, in
this regime, our sixth-order approximation begins to be quantitatively invalid. However,
this basic qualitative picture can be validated by directly (numerically) minimizing 
the full normalized ground-state energy Eq.~(\ref{eq:gsetotallygeneral}).

\subsection{Bogoliubov sound velocity in \sfm phase}
\label{SEC:sound}
In our earlier computation of $e_G(\Bh)$, Eq.~(\ref{becefweakcoupling}), among other quantities, we
have obtained an effective four-boson coupling $\Vh_4(\hh,\deltah)$ that at
zero molecular density is equivalent to the molecular T-matrix
proportional to the molecular scattering amplitude.
Thus, $\Vh_4(\hh,\deltah)$ is related to the molecular scattering length
$a_m$, which is a measure of the effective
molecular interaction induced by the Feshbach resonance coupling to atoms. We have found
that, inside the \sfm phase, $\Vh_4(\hh,\deltah)$ is positive but decreases with
increasing $\hh$ and in fact nearly (to $\curO(\gamma))$ vanishes at the boundary
$\hcone(\deltah)$ to phase separation; see Eq.~(\ref{eq:hmixedapprox}) and the discussion in
Sec.~\ref{SEC:transitionphasesep}. Physically this represents a {\em repulsive} molecular Bose
gas whose two-body repulsion nearly vanishes with increasing $\hh$ at the transition to phase
separation at $\hcone(\deltah)$.

 A striking observable 
consequence of this is a concomitant suppression of the Bogoliubov sound velocity $u(\delta,\hh)$ with increasing 
$\hh$ or population difference 
$\Delta N$. 
The simplest way to obtain $u$ is to use the standard result~\cite{Negele} 
\be
u^2= \frac{1}{2m} \frac{\partial P}{\partial |B|^2},
\ee
(recall the boson mass is $2m$), where $P$ is the pressure. 

 Since the grand-canonical
energy $E_G$ is equal to $-P\vol$ with $\vol$ the system volume, we have, plugging $\Bh^2\simeq \Vh_2/\Vh_4$
into Eq.~(\ref{eq:egbec}) (neglecting $\Vh_6$),
\bea
P  &=& \frac{1}{2N(\ef)}  B^4 \hat{V}_4,
\eea
where for clarity we have temporarily reverted to dimensionful quantities. 
Evaluating the derivative, and reverting back to the dimensionless boson density $\Bh^2 = B^2/c\ef^{3/2}$, 
we have 
\be
u^2 = \frac{\gamma^2 \pi \ef \Bh^2}{32 m |\muh|^{3/2}}F_4(\hh/|\muh|).
\ee
As we are primarily interested in fixed density, we insert the fixed-density expressions 
$\muh \approx \deltah/2$ [valid to $\curO(\gamma)$ on the BEC side of the
resonance] and Eq.~(\ref{eq:numbec}) for $\Bh^2$, yielding
\bea
&&\hspace{-1cm}u^2 \simeq u_0^2 \Big[1-\frac{1}{2} \Big( \hh - \frac{|\deltah|}{2} \Big)^{3/2}
\Theta\Big(h - \frac{|\deltah|}{2}\Big)
\Big]
F_4\Big(\frac{2\hh}{|\deltah|}\Big),
\eea
with $u_0$ the Bogoliubov sound velocity for $h=0$ (i.e., in the SF phase)
\be
u_0 = \frac{2^{3/4}\sqrt{\pi}\gamma}{8\sqrt{3}} \frac{\vf}{|\deltah|^{3/4}}.
\label{eq:unought}
\ee
Finally converting to fixed total density and population difference, we have 
\bea
\hspace{-1cm}u\simeq u_0 \sqrt{1-\frac{\Delta N}{N}}\sqrt{F_4\Big(1+{\textstyle\frac{2^{5/3}}{|\deltah|} 
\big(\frac{\Delta N}{N}\big)^{2/3}}\Big)},
\eea
for the sound velocity as a function of applied polarization.  Note that there are two ways $u$ can vanish with increasing
$\Delta N$, depending on whether the detuning is larger or smaller than $\delta_c$, and, correspondingly, whether
the transition is first or second order.
 For $\deltah>\deltah_c$, $u$ (nearly; see Ref.~\onlinecite{sounddiscussion}) vanishes
 at the first-order instability of the \sfm state to phase separation, when $F_4$ vanishes, 
as we have already discussed.  For  $\deltah<\deltah_c$, however, $F_4$ remains positive at the second-order
\sfm to N transition. In this case, $u$ vanishes at this second-order transition simply because the molecular density 
vanishes as $\Delta N\to N$.

%-----------------------------
%
% fig%33
%
\begin{figure}[bth]
\vspace{1.4cm}
\hspace{-1.0cm}
\centering
\setlength{\unitlength}{1mm}
\begin{picture}(40,40)(0,0)
\put(-50,0){\begin{picture}(0,0)(0,0)
\includegraphics{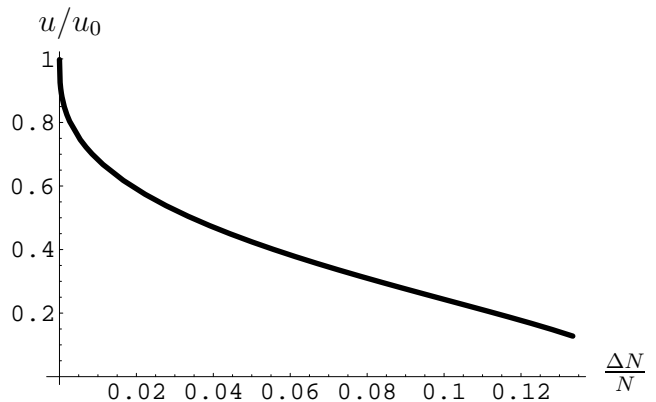}
\end{picture}}
\put(-15,51) {\large{$u/u_0$}}
\put(60,4.5) {\large{$\frac{\Delta N}{N}$}}
\end{picture}
\vspace{-.5cm}
\caption{ Bogoliubov sound velocity $u/u_0$ as a function of polarization for the case 
$\deltah = -3$. }
\label{fig:velocityplotlong}
%
%  velocity_longmanu.nb *) 
%
\end{figure}
%------------------------------

In Fig.~\ref{fig:velocityplotlong}, we plot $u/u_0$ as a function of $\Delta N/N$ for the case $\deltah = -3$
and $\gamma = 0.1$, showing its suppression near the first-order transition
at $\frac{\Delta N_{c1}}{N} \approx 0.133$, where it exhibits  a jump discontinuity 
to zero (before $\Vh_4(\Delta N)$ actually vanishes~\cite{sounddiscussion}) upon entering the PS regime.

\section{FFLO state of two-channel  model}
\label{SEC:FFLO}

Until this point, we have focused on  ground states of the two-channel Hamiltonian Eq.~(\ref{eq:meanfieldhamiltonianpre}) assuming
 $\bQ=0$ pairing.  However, it is well-known~\cite{ff,lo} that the BCS model 
 (which our model corresponds
to in the large positive detuning limit) under applied chemical potential difference 
possesses an alternative minimum, characterized by both 
periodic off-diagonal long-range  pairing order and
nonzero magnetization.  Such states, which we generally refer to as FFLO states, break translational
and rotational symmetries (and hence are examples of a supersolid~\cite{AndreevLifshitz,Chester,Leggett70,KimChan})
 and exhibit a compromise between the tendency to pair (due to attractive interactions) 
and the tendency to magnetize (due to the applied chemical potential difference $h$). 

As we will show, deep in the BCS regime $\delta\gg 2\ef$, where the resonant two-channel model
reduces to the BCS model, we reproduce well-known results in the literature~\cite{ff,lo}.  Our
main contribution is the extension of these results on FFLO ground states to a resonant
model at arbitrary detuning $\delta$.  We thereby predict the evolution of the FFLO state
[i.e., we predict $Q(\delta,h)$ and the FFLO phase boundary]
for $\delta$ outside the well-studied BCS regime.  In terms 
of the phase diagram, our main prediction is that the FFLO state is unstable, i.e., squeezed out
by the phase-separation regime and the normal state, for $\delta<\delta_* \simeq 2\ef$.
 
%-----------------------------
%
% fig%34
%
\begin{figure}[bth]
\vspace{1.4cm}
\centering
\setlength{\unitlength}{1mm}
\begin{picture}(40,40)(0,0)
\put(-50,0){
\begin{picture}(0,0)(0,0)
\includegraphics{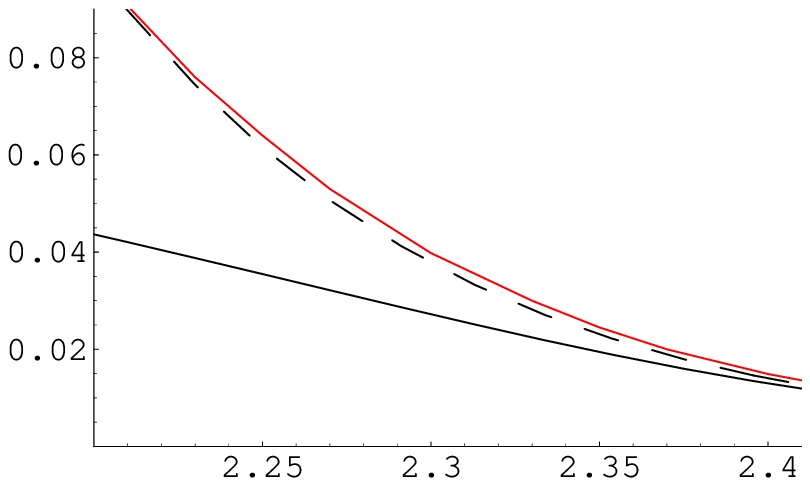}
\end{picture}}
\put(7,37) {\large{$\hh_{\rm FFLO}$}}
\put(5,30) {\large{$\hh_{c2}$}}
\put(7,18) {\large{$\hcone$}}
\put(60,4.5) {\large{$\deltah$}}
\end{picture}
\vspace{-.5cm}
\caption{(Color online) Plot of the $\hfflo$ phase boundary, along with  $\hctwo$ and $\hcone$, for $\gamma = 0.1$,
as a function of normalized detuning $\deltah$.  The FFLO phase
is stable for $\hctwo<\hh<\hfflo$, as shown in the phase diagram Fig.~\ref{fig:hphasetwo}.}
\label{fffig}
\end{figure}
%------------------------------

As we have already noted, we expect that FFLO-type ground states of Eq.~(\ref{eq:meanfieldhamiltonianpre})
are adequately modeled~\cite{SingleQ} by assuming single-harmonic pairing, namely
taking the pairing order to be  $\Delta(\br) = \Delta_\bQ {\rm e}^{i \bQ\cdot \br}$.
With this assumption, it is straightforward (as we already have done in Sec.~\ref{SEC:MFT}) to find the ground
state energy $E_G(\Delta_\bQ,\bQ)$ as a function of wavevector $\bQ$ and gap $\Delta_\bQ$, and then to minimize
$E_G(\Delta_\bQ,\bQ)$ over these parameters.  

Our analysis is very similar to that of FF~\cite{ff}, though generalized to the 
present two-channel resonant Hamiltonian Eq.~(\ref{eq:meanfieldhamiltonianpre}). Thus, our results differ from those of FF
only
away from the deep BCS limit. For $\delta \gg 2\ef$, however, our results map onto those of FF, 
finding that, instead of a phase transition from the SF to the N state  at $h_c \simeq \deltabcs/\sqrt{2}$
(see Fig.~\ref{fig:hphasetwo}), the FFLO state intervenes, so that there is a first-order SF to 
FFLO transition, followed by a second-order~\cite{MFTcomment} FFLO to N transition at $h_{\rm FFLO}$.

We now determine the region of the phase diagram in which the FFLO state has the lowest energy.  The 
governing (variational) ground-state energy is given by Eq.~(\ref{eq:havgpre})
that we redisplay here for convenience:
\bea &&\hspace{-.5cm}E_G(\Delta_\bQ,\bQ) 
 = \big(\frac{\epsilon_Q}{2} +
\delta_0-2\mu \big)\frac{\Delta_{\bQ}^2}{g^2} - \sum_k (\te_k-\txi_k )
\nonumber \\
&&\hspace{-.5cm} \qquad +\sum_\bk \te_k \big(1+ \Theta(-E_{k\uparrow}) -
\Theta(E_{k\downarrow})) \big)
\nonumber \\
&& \hspace{-.5cm}\qquad + \sum_\bk \big(\frac{\bk \cdot \bQ}{2m} + h\big) 
\big(1-\Theta(-E_{k\uparrow}) - \Theta(E_{k\downarrow})\big).\;\;\;\;\;
\label{eq:havgpre2}
\eea
For the $\bQ=0$ case, we have already evaluated the first momentum sum appearing in Eq.~(\ref{eq:havgpre2}). Examining
Eqs.~(\ref{eq:tildexi}) and ~(\ref{eq:excitationenergy}) reveals that, for this sum, having $\bQ\neq 0$
simply shifts the chemical potential  $\mu \to \tilde{\mu} \equiv \mu - Q^2/8m$. 
The remaining \lq\lq excluded\rq\rq\ sums in
Eq.~(\ref{eq:havgpre2}) (second and third lines) can also be evaluated [within 
the standard approximation of replacing $N(\epsilon) \to N(\epsilon_F)$, 
valid in the atomic degenerate limit of $\Delta/\mu \ll 1$] following FF; we do this in Appendix~\ref{app:ex}.
This yields
\bea
&&E_G  \approx -\frac{8}{15} N(\tilde{\mu})\tilde{\mu}^2
+ 
\big(\frac{\epsilon_Q}{2} + \delta - 2\mu \big)\frac{\Delta_\bQ^2}{g^2} 
\nonumber \\
&& \qquad
+2N(\tilde{\mu})\Big(-\frac{\Delta_\bQ^2}{4}+ \frac{\Delta_\bQ^2}{2} \ln \frac{\Delta_\bQ}{8 {\rm e}^{-2} \tilde{\mu}}\Big)
  \label{finenergy}\\
&& \qquad +  \frac{N(\tilde{\mu}) \Delta_\bQ^2}{2\Qb} 
\big[
I(\Qb+\dbmu) - I(\dbmu - \Qb) + I(\Qb-\dbmu)
\big],\nonumber
\eea
where $I(x) $ is given by 
\bea
&&I(x) \equiv  \big[-\frac{1}{3} (x^2-1)^{3/2}  - \sqrt{x^2-1}
\nonumber\\ 
&&\qquad \quad 
+ x\cosh^{-1} x\big]\Theta(x-1),
\label{eq:iff}
\eea
and, using notation similar to that of FF, we have defined the rescaled wavevector $\Qb$ and chemical potential 
difference $\bar{h}$:
\bse
\label{qbhb}
\bea
\label{eq:qbar}
\Qb &\equiv& \frac{\tkf Q}{2m \Delta_\bQ},
\\
\dbmu &\equiv& \frac{h}{\Delta_\bQ},
\eea
\ese
with $\tkf = \sqrt{2m \tilde{\mu}}$
the Fermi wavevector associated with the shifted chemical potential. 
We note that $E_G$, Eq.~(\ref{finenergy}), is consistent with FF, differing only in the $\epsilon_Q$ term
that accounts for the molecular kinetic energy absent in the BCS-like model of FF.

From this point on, the determination of the phase diagram and the corresponding phases is
conceptually simple, given by the minimization of $E_G(\Delta_\bQ,\bQ)$ with respect to $\bQ$ and $\Delta_\bQ$
(while satisfying the total number constraint).  However, because the general structure of Eq.~(\ref{finenergy})
is quite complicated, to do this generally is technically 
challenging and best done numerically.  Nevertheless, analytical solutions are possible in a number
of important limits. 

We now briefly overview the structure of $E_G(\Delta_\bQ,\bQ)$, Eq.~(\ref{finenergy}).
At small $Q$, only the  terms $\propto [I(\Qb+\dbmu) - I(\dbmu - \Qb)]$ are nonzero; 
in the $Q\to 0$ limit these conspire so that Eq.~(\ref{finenergy})
reduces to our previous result Eq.~(\ref{eq:gsescapprox}), from Sec.~\ref{SEC:bcsandtransition}.  
At small $h$, the SF state studied in that section is still a stable minimum of Eq.~(\ref{finenergy})
at $\Delta_\bQ = \deltabcs$, $\bQ=0$, and remains so until the critical chemical potential difference
$h_c$ (that is remarkably close to the critical $h$ studied in Sec.~\ref{SEC:bcsandtransition},
approximated by $h_c \simeq \deltabcs/\sqrt{2}$).   Already below $h_c$, 
$E_G(\Delta_\bQ,\bQ)$ exhibits a secondary {\it local\/} minimum at the FFLO-state
$(\Delta_\bQ,\bQ)$ and, for $h>h_c$, there is a 
first-order transition into this magnetized finite-$\bQ$ FFLO state that is nearly degenerate with
the $\Delta_\bQ = 0$ normal state.   The FFLO state is only stable for a narrow window of $h$
values (and sufficiently large detuning), undergoing a continuous transition to the N state
at $h_{\rm FFLO}$.

\subsection{Second-order N-FFLO transition}

Before studying the first-order SF-FFLO transition, it is convenient
to first look at the simpler transition from the N state, that takes
place upon lowering $h$ from a large value. As was first shown by
FF\cite{ff}, this N-FFLO transition is in fact continuous in
mean-field theory~\cite{MFTcomment} of the 
standard Landau type.

This is possible because, for sufficiently large $Q$ and small $\Delta_\bQ$, the
excluded-sum terms [i.e.~the last line of Eq.~(\ref{finenergy})] convert the double minimum
form of $E_G(\Delta_\bQ)$ (characteristic of the first-order transition that would
take place at $Q=0$) to a single, $h,\delta$-dependent minimum that
 leads to a continuous N-FFLO transition.

The continuous nature of the N-FFLO transition allows us to accurately
study its details by expanding $E_G(\Delta_\bQ,\bQ)$ in small $\Delta_\bQ$.  Since
we expect the transition to take place at finite $h$ (that we call
$h_{\rm FFLO}$) and finite $\bQ$, its Landau expansion in small $\Delta_\bQ$ is
characterized by the large $\Qb$, $\bar{h}$ limit as both are
proportional to $1/\Delta_\bQ$ according to Eq.~(\ref{qbhb}). As we will verify a posteriori,
the existence of the continuous N-FFLO transition furthermore requires, near $h_{\rm FFLO}$,  the
system to be in the \lq\lq doubly-depaired\rq\rq (or \lq\lq D\rq\rq) regime (in the notation of
FF), characterized by 
$\Qb+\dbmu >1$ and $\Qb-\dbmu >1$,
i.e., a regime in which excluded-sum contributions for both atom
species (spin up and down) are nonzero. In this regime, the term $\propto I(\dbmu - \Qb)$ vanishes
and, to leading order, the other two excluded-sum terms give
\bea
&&I(\Qb+\dbmu) + I(\Qb-\dbmu) \approx -\frac{2}{3} \Qb^3 - 2\Qb\dbmu^2-\Qb 
\nonumber \\
&& \qquad \qquad + \Qb\ln 4(\Qb+\dbmu)(\Qb-\dbmu)
\nonumber \\
&& \qquad \qquad + \dbmu \ln \frac{\Qb+\dbmu}{\Qb-\dbmu} + \frac{1}{4}\Big(
\frac{\Qb}{\Qb^2-\dbmu^2}
\Big),
\label{iexp}
\eea
where we utilized the expansion of $I(x)$ for large value of its argument:
\be
I(x) \approx \Theta(x-1)\Big[
-\frac{x^3}{3} + \big(
-\frac{1}{2} + \ln 2x\big) x + \frac{1}{8x} 
\Big].
\ee

As we will see, the above expansion Eq.~(\ref{iexp}) [to be used inside Eq.~(\ref{finenergy})] is
sufficient to get a Landau expansion of $E_G(\Delta_\bQ,\bQ)$ up to fourth order in
$\Delta_\bQ$, required to capture the second-order N-FFLO transition. We
furthermore expand the first term in Eq.~(\ref{finenergy}) in
small $Q$, finding
\be
\frac{8}{15} N(\tilde{\mu})\tilde{\mu}^2 \approx \frac{8}{15} N(\mu)\mu^2 + \frac{1}{3} N(\mu)\Qb^2 \Delta_\bQ^2.
\label{smallqexpansion}
\ee
It is easy to convince oneself that to lowest order necessary to
capture this transition, in all other terms $\tilde{\mu}$ can be
approximated by its unshifted $(Q=0)$ value $\mu$. Now, reverting to our
dimensionless variables $\Deltah_\bQ$, $\muh$, and $\hh$ [defined in Eqs.~(\ref{eq:dimensionlessvariables})],
defining a new dimensionless momentum
\be
\label{eq:pdef}
\Qh \equiv \sqrt{\muh}\frac{Q}{\kf} ,
\ee
and combining Eqs.~(\ref{smallqexpansion}) and (\ref{iexp}) inside $E_G$, Eq.~(\ref{finenergy}), we obtain the
sought-after quartic (in $\Deltah_\bQ$) Landau expansion for the normalized
ground-state energy $e_G$ [Eq.~(\ref{eq:egdef})]:
\bea
\label{approxgseff}
&&\hspace{-.5cm}e_G \approx  -\frac{8}{15}\muh^{5/2}
+\frac{\gamma^{-1} \Qh^2 \Deltah_\bQ^2}{2\muh}  +\sqrt{\muh}\Big[  -\Deltah_\bQ^2 - \hh^2
 \\
&& \hspace{-.5cm}
+
\frac{\Deltah_\bQ^2}{2} \ln \frac{4(\Qh+\hh)(\Qh-\hh)}{\deltahbcs^2} + \frac{\hh \Deltah_\bQ^2}{2\Qh} \ln \frac{\Qh+\hh}{\Qh-\hh} 
+ \frac{\Deltah_\bQ^4}{8} \frac{1}{\Qh^2-\hh^2}\nonumber
\Big],
\eea
where we used the zero-field ($\hh=0$) BCS gap 
\be
\label{deltanought}
\deltahbcs = 8{\rm e}^{-2} \muh {\rm e}^{-\gamma^{-1}(\deltah - 2\muh)/\sqrt{\muh}},
\ee
to simplify the final expression. We note that in $e_G$, Eq.~(\ref{approxgseff}), an
important cancellation of the nonanalytic $\Deltah^2_\bQ \ln \Deltah_\bQ$ terms
has taken place. Namely, the $\Deltah_\bQ^2 \ln \Deltah_\bQ$ term, characterizing the $\hh=0$
BCS ground state energy (guaranteeing that a continuum, $\hh=0$ degenerate
Fermi gas with arbitrarily weak attractive interactions is always
unstable to paired superfluidity) is exactly canceled by such a term
arising from the finite-$\hh$ excluded-sum contribution. 

Minimizing $E_G$ over $\Deltah_\bQ$ (the gap equation), $\Qh$ (which ensures
that the total momentum vanishes in equilibrium~\cite{Takada}) and
differentiating with respect to  
$\muh$ [to obtain the number equation, Eq.~(\ref{eq:numequationgeneralnorm})] we obtain
near the N-FFLO transition
\bse\label{fflomomnumgap}
\bea
&&\Deltah_\bQ^2 \simeq 2\hh^2(\lambda^2 - 1) \Big[2
-\frac{\gamma^{-1} \hh^2}{\muh^{3/2}} \lambda^2 
\nonumber \\
&&\quad 
- \ln \frac{4\hh^2(\lambda^2-1)}{\deltahbcs^2} - \frac{1}{\lambda} \ln\frac{\lambda+1}{\lambda-1} 
\Big],
\label{eq:gapff} 
\\
 &&0\simeq\frac{\gamma^{-1} \hh^2}{\muh^{3/2}} \lambda^2 + 1 - \frac{1}{2\lambda} \ln \frac{\lambda + 1}{\lambda-1} ,
\label{momsecond}
\\
&&\frac{4}{3} \simeq \frac{4}{3} \muh^{3/2} + 2\gamma^{-1} \Deltah_\bQ^2 + \frac{\hh^2}{2\sqrt{\muh}},  
\label{eq:numapproxff}
\eea
\ese
where we only kept terms to leading order in $\Deltah_\bQ$ and $\gamma$ and
defined $\lambda=\Qh/\hh$. The simultaneous numerical solution to these 
equations yields $\Deltah_\bQ$ and $\bQ$ in the FFLO phase, as well
as the critical chemical potential difference $\hfflo$ above which $\Deltah_\bQ \to 0$.

%-----------------------------
%
% fig%35
%
\begin{figure}[bth]
\vspace{1.4cm}
\centering
\setlength{\unitlength}{1mm}
\begin{picture}(40,40)(0,0)
\put(-50,0){
\begin{picture}(0,0)(0,0)
\includegraphics{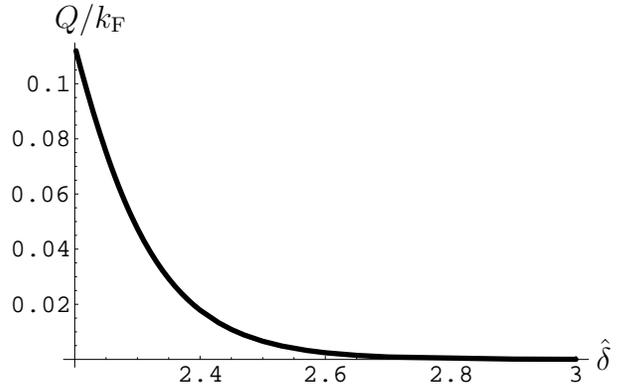}
\end{picture}}
\put(-12,50) {\large{$Q/\kf$}}
\put(60,4.5) {\large{$\deltah$}}
\end{picture}
\vspace{-.5cm}
\caption{(Color online) Plot of wavevector $Q(\deltah)$ of FF order, normalized to Fermi wavevector $\kf \equiv 
\pf/\hbar$ for 
$\gamma = 0.1$ as a function of Feshbach resonance detuning $\deltah$.   }
\label{qhatplot}
%
%  FFplots.nb
%
\end{figure}
%------------------------------
%-----------------------------
%
% fig%36
%
\begin{figure}[bth]
\vspace{1.4cm}
\centering
\setlength{\unitlength}{1mm}
\begin{picture}(40,40)(0,0)
\put(-50,0){
\begin{picture}(0,0)(0,0)
\includegraphics{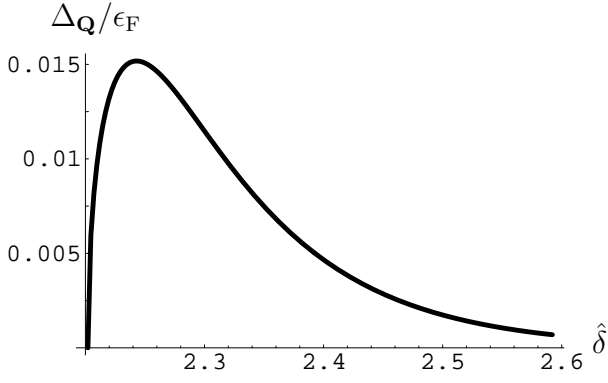}
\end{picture}}
\put(-12,50) {\large{$\Delta_\bQ/\ef$}}
\put(60,6.5) {\large{$\deltah$}}
\end{picture}
\vspace{-.5cm}
\caption{(Color online) Plot of maximum gap $\Delta_\bQ(\delta,h_{c2})$ in FFLO phase as a function of
Feshbach resonance detuning for the case $\gamma = 0.1$.}
\label{ffgapplot}
%
%  FFplots.nb
%
\end{figure}
%------------------------------
In Fig.~\ref{fffig} we plot, for $\gamma=0.1$, $\hfflo$ as a  
function of normalized detuning, along with $\hcone$ and $\hctwo$.  In Fig.~\ref{qhatplot},
we plot the FFLO wavevector $Q(\deltah,\hfflo)$ at the transition [$Q(\deltah,\hh)$ is only
weakly $\hh$-dependent near the transition].  Indeed, the fact that $Q$ is finite at the 
transition is consistent with our  large $\bar{Q}$, $\bar{h}$ expansion above.
To get an idea of the typical magnitude 
of pairing in the FFLO phase, in Fig.~\ref{ffgapplot} we plot $\Deltah_Q(\deltah,\hctwo)$,
i.e., the strength of pairing at the FFLO-to-phase-separation phase boundary.
To gain some intuition for these numerical results we examine
analytically the  solution to Eqs.~(\ref{fflomomnumgap}) in the large detuning 
($\deltah \gg 1$) 
%
%and narrow resonance ($\gamma \ll 1$) 
%
limit. In this regime, we can
safely neglect the first term of Eq.~(\ref{momsecond}), as near the transition
$\hh^2\approx \hfflo^2 \ll \gamma$, with the latter inequality arising from
simplest estimate of $\hfflo \sim \deltahbcs$. Our dropping of this first term in
Eq.~(\ref{momsecond}) can be traced back to a neglect of the molecular dispersion,
reducing our two-channel model to a single-channel one, equivalent to
that studied by FF. With this simplification the momentum equation
reduces to:
\be
\label{eq:lambda}
0 = 1- \frac{1}{2\lambda} \ln \frac{\lambda+1}{\lambda-1},
\ee
that is solved by $\lambda\approx 1.200$. Using this inside Eq.~(\ref{eq:gapff})  (neglecting the 
term  $\gamma^{-1} \hh^2\lambda^2/\muh^{3/2}$ in the latter) gives
\be
\Deltah_\bQ^2 \approx 2\hh^2(\lambda^2-1) \ln \frac{\deltahbcs^2}{4\hh^2 (\lambda^2-1)}.  
\ee
The vanishing of $\Deltah_\bQ(\hh,\deltah)$ then determines the critical
point $\hfflo$ for the N-FFLO continuous transition
\bse
\bea
\label{eq:ffresultpre}
\hfflo &\simeq& \frac{\deltahbcs}{2\sqrt{\lambda^2-1}} \simeq \eta \deltahbcs,
\\
 &\simeq& \eta  \Deltah_F[\gamma,\deltah] \exp\big[- \frac{\eta^2\deltah}{8\gamma} \Deltah_F^2\big]
\label{eq:ffresult},
\eea
\ese
with the constant of proportionality $\eta = 0.754$ in (expected) agreement
with FF~\cite{ff}.  In the final line 
Eq.~(\ref{eq:ffresult}), we used the number equation for $\muh$ [Eq.~(\ref{eq:numapproxff}), which, 
as expected, precisely reduces
to the N-state number equation at the transition where $\Deltah_\bQ \to 0$; c.f. Eq.~(\ref{eq:mun2})]
to find $\hfflo$ at fixed imposed density.  Here, we 
remind the reader that $\Deltah_F$ is defined in Eq.~(\ref{eq:deltaf}).

To this order of  approximation (i.e., that of FF~\cite{ff}) 
the $\hfflo(\delta)$ (for the
N-FFLO transition) and $\hc(\delta)\approx \deltabcs(\deltah)/\sqrt{2}$
(for the FFLO-SF transition; see below) phase boundaries
maintain a constant ratio as a function of $\deltah$. As we saw in Sec.~\ref{SEC:bcsandtransition}, 
at fixed number [determined by in addition imposing Eq.~(\ref{eq:numapproxff})] the critical curve
$\hc(\deltah,\muh)$ splits into  upper-critical and lower critical boundaries 
$\hctwo(\deltah,N)=\hc(\deltah,\muh_{\rm FFLO}(\deltah,N))$ and 
$\hcone(\deltah,N)=\hc(\deltah,\muh_{SF}(\deltah,N))$, respectively,
bounding the SF-FFLO coexistence region.  For later
reference we note that, because $\Deltah_{\bQ,FFLO} \ll \Deltah_{BCS}\ll \gamma$, the
chemical potential $\muh_{\rm FFLO}$ in the FFLO state is accurately given by
its normal state value, $\muh_N$, the latter given by Eq.~(\ref{eq:numapproxff}), with
$\Deltah_\bQ\approx 0$, giving, at small $\hh$, $\muh_{\rm FFLO}\approx 1$.  Just below
the N-FFLO transition, the order parameter $\Deltah_\bQ$ grows in the
expected generic mean-field way~\cite{ff}
\be
\frac{\Deltah_{\bQ,FFLO}}{\deltahbcs}\simeq \sqrt{\frac{\hfflo - \hh}{\hfflo}}
\,\,\, {\rm for\/}\,\,\, \hh<\hfflo,
\label{deltanearfflotrans}
\ee
with the characteristic $\Qh$ reducing linearly with $\hh$ according to $\Qh \approx
\lambda \hh$. 

%-----------------------------
%
% fig%37
%
\begin{figure}[bth]
\vspace{1.4cm}
\hspace{-1.0cm}
\centering
\setlength{\unitlength}{1mm}
\begin{picture}(40,40)(0,0)
\put(-50,0){\begin{picture}(0,0)(0,0)
\includegraphics{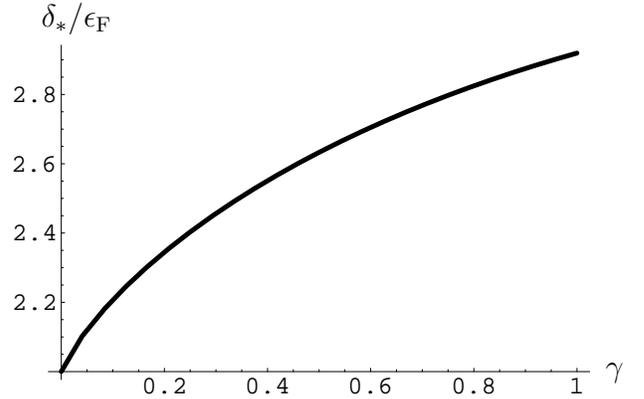}
\end{picture}}
\put(-15,51) {\large{$\delta_*/\ef$}}
\put(60,4.5) {\large{$\gamma$}}
\end{picture}
\vspace{-.5cm}
\caption{Critical detuning $\delta_*$ beyond which the FFLO phase is
stable at nonzero polarization, valid for narrow resonance (small $\gamma$).
}
\label{fig:deltastarplot}
%
%  FFplots.nb 
%
\end{figure}
%------------------------------

We now go beyond the above FF approximation for $\hfflo(\deltah,\muh)$ by including
the (so-far neglected) molecular dispersion in Eqs.~(\ref{eq:gapff}) and ~(\ref{momsecond}).
Combining these equations gives the order parameter
\be
\Deltah_\bQ^2 \simeq 2\hh^2(\lambda^2 - 1) \Big[3
-\frac{3}{2\lambda}\ln\frac{\lambda+1}{\lambda-1}
 - \ln \frac{4\hh^2(\lambda^2-1)}{\deltahbcs^2} \Big],
\label{eq:gapff2}
\ee
that, when solved simultaneously with Eq.~(\ref{momsecond}), gives the more accurate,
higher-order $\hfflo(\deltah,\muh)$ boundary. The latter is best determined
numerically, as illustrated in Fig.~\ref{fffig}.  Its new important feature
[beyond the lower order result in Eq.~(\ref{eq:ffresultpre})] is that $\hfflo(\deltah,\muh)$ is
no longer proportional to $\hc(\deltah,\muh)$, crossing it [and therefore
$\hctwo(\deltah,N)$] at a detuning $\deltah_*$ (see Fig~\ref{fig:hphasetwo}), that is
determined by the condition
$\hfflo(\deltah_*,\muh)=\hc(\deltah_*,\muh)\simeq\deltahbcs(\deltah_*,\muh)/\sqrt{2}$.
Inserting this condition into Eq.~(\ref{eq:gapff2}) gives $\lambda$ at this crossing,
which we call $\lambda_*$:
\be
0 = 3(1- \frac{1}{2 \lambda_*} \ln \frac{\lambda_*+1}{\lambda_* -1}) - \ln \big[2 (\lambda_*^2-1)\big] .
\label{eq:lambdastar}
\ee
Solving Eq.~(\ref{eq:lambdastar}) numerically gives $\lambda_*\approx 1.159$, only slightly
lower than its FF value of $\lambda\approx 1.200$ in the asymptotic BCS
regime. Inserting $\lambda_*$ into Eq.~(\ref{momsecond}) 
yields the following implicit expression for $\delta_*$ at which $\hfflo = \hc$:
\bse
\bea
\label{eq:deltanoughtfirst}
&&\hspace{-1cm}\frac{1}{2}\deltahbcs[\delta_*,\muh]^2 = \frac{\muh^{3/2}\gamma}{\lambda_*^2}
\Big[\frac{1}{2 \lambda_*} \ln \frac{\lambda_*+1}{\lambda_*-1}-1
\Big],
\\
&&\hspace{-.5cm}\qquad\qquad\simeq 0.094 \muh^{3/2} \gamma,
\label{eq:deltanoughtsecond}
\eea
\ese
with $\delta_*$ entering via $\deltahbcs$, the BCS gap at $\hh = 0$.
Using the explicit expression Eq.~(\ref{eq:baredeltanoughtmaintext}) in Eq.~(\ref{eq:deltanoughtsecond}) and
approximating $\muh_{\rm FFLO}\approx \muh_N\approx 1$, then gives
\be
\deltah_* \approx 2 - \frac{\gamma}{2} \ln 0.159 \gamma,
\label{eq:deltastar}
\ee
where we emphasize that, for $\gamma < 1$ (where the preceding
expressions are valid), $\deltah_*$ is bounded by $2\ef$ from  below and is an
increasing function of Feshbach resonance width $\gamma$, illustrated in Fig.~\ref{fig:deltastarplot}.

Using the solution of Eqs.~(\ref{eq:gapff}) and (\ref{momsecond})   for $\lambda$ together with the
definition of $\lambda=\Qh/\hh$ and Eq.~(\ref{eq:pdef}), we can obtain the wavevector
$Q(\delta,h)$ characterizing the FFLO state. Approximating $\hh$ by its
critical value $\hfflo(\delta)\approx \eta \deltahbcs$ and
reintroducing Planck's constant, we find
\be
\label{eq:qresult}
Q\approx \frac{2\eta \lambda \deltabcs}{\hbar \vf}.
\ee
As we have mentioned, along the phase boundary $\hfflo(\deltah)$ a more precise numerical
solution for $\Qh(\deltah)$ is given in Fig.~\ref{qhatplot}.

We close this subsection by noting that the N-FFLO phase boundary
$\hfflo(\deltah)$ can be easily converted into a critical polarization
boundary $\Delta N_{\rm FFLO}(\deltah)$ by using $\hfflo(\deltah)$
 inside the
expression for the normal-state spin imbalance $\Delta N(\hh)$, Eq.~(\ref{eq:dimnormmag}).
 Doing this numerically gives $\Delta N_{\rm FFLO}(\deltah)$, plotted along
with $\Delta N_{c2}$ in the phase diagram Fig.~\ref{fig:mphasetwo},
 that for large detuning is given by
\be
\frac{\Delta N_{\rm FFLO}}{N} \approx \frac{3\eta}{2}  \Deltah_F  \exp\big[- \frac{\eta^2\deltah}{8\gamma} \Deltah_F^2
\big],
\label{eq:deltanfflo}
\ee
a result that we will estimate [along with Eq.~(\ref{eq:qresult})] in the context of recent experiments in
Sec.~\ref{SEC:concludingremarks}.  

%-----------------------------
%
% fig%38
%
\begin{figure}[bth]
\vspace{5.3cm}
\hspace{-.1cm}
\centering
\setlength{\unitlength}{1mm}
\begin{picture}(40,80)(0,0)
\put(-55,55){\begin{picture}(0,0)(0,0)
\includegraphics{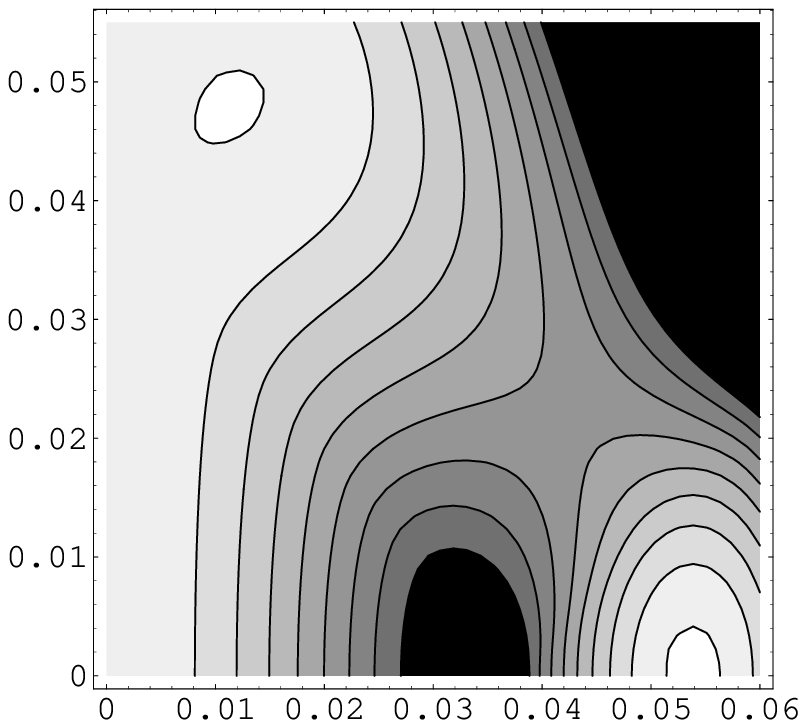}
\end{picture}}
\put(-12,122) {(a)}
\put(52.5,62) {\large{$\Deltah_\bQ$}}
\put(-14,127) {\large{$\Qh$}}
\put(-3.0,116) {FFLO}
\put(40,65) {SF}
\put(-55,-20){\begin{picture}(0,0)(0,0)
\includegraphics{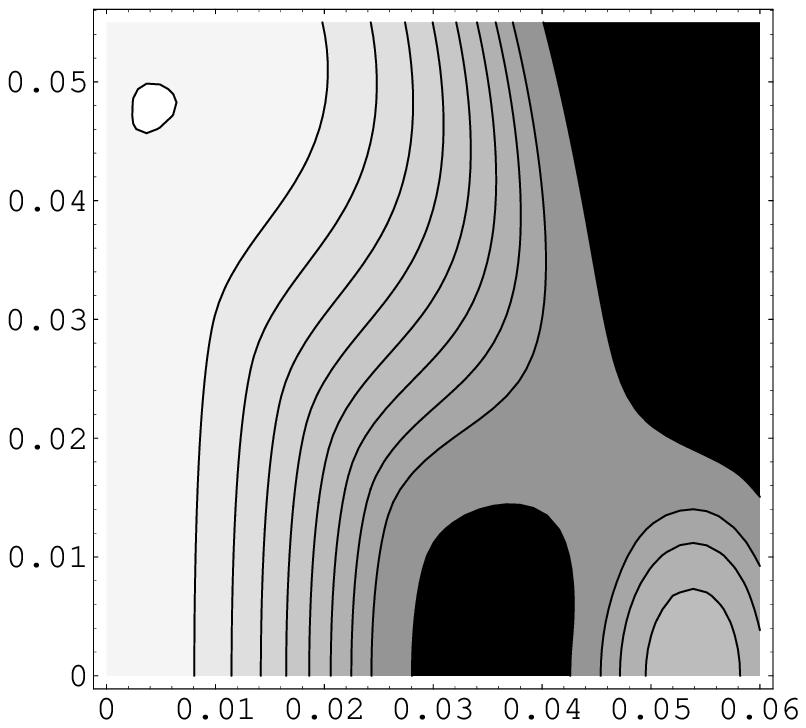}
\end{picture}}
\put(-12,46) {(b)}
\put(-14,52) {\large{$\Qh$}}
\put(40,-9) {SF}
\put(52.5,-12) {\large{$\Deltah_\bQ$}}
\put(-2,43.5)  {${\rm FFLO}$}
\put(-2,44.5) {\vector(-2,-1){4}}
%--------------------------------
%
\end{picture}
\vspace{1.5cm}
\caption{(Color online) Contour plots of the ground-state energy $E_G$ (darker
regions denote higher $E_G$) as a function
of the normalized gap $\Deltah_\bQ$ and the normalized wavevector $\Qh$.  For a), $\hh \agt \hh_c$
so that a first-order transition from SF to FFLO has just occurred.  For b), 
$\hh \alt \hfflo$, so the FFLO minimum is approaching the $\Deltah_\bQ = 0$ (N
state) axis continuously and the local SF-state minimum has moved
to higher energy. Note that the N phase occurs for {\it any\/} $\Qh$ at $\Deltah_\bQ=0$. }
\label{fig:fflocontour} 
%
%  1June.nb (both)
%
\end{figure}
%------------------------------

\subsection{First-order FFLO-SF transition and associated phase separation}
\label{SEC:regimeofphaseseparation}
The FFLO state arises as a result of a delicate balance between the
normal state (selected by atom species imbalance) and the
singlet-paired superfluid state (preferred by the attractive pairing
interaction). It is characterized
by an order parameter that is significantly smaller than that of a
singlet BCS superfluid and by a finite magnetization
$\mg=\Delta N/V$
\be
\mg(h) = \int \frac{d^3 k}{(2\pi)^3} \big[\Theta(-E_{\bk\uparrow})- \Theta(-E_{\bk\downarrow})\big], 
\ee
that is quite close to that of the normal state
\be
\mg_N \simeq 2c\sqrt{\mu}h.
\label{eq:mgnormalfflo}
\ee
Hence, in many of its properties the FFLO state is quantitatively quite
close to the normal state. This is illustrated in Fig.~\ref{fig:fflocontour}  by the
proximity of the FFLO energy minimum to the normal state, $\Deltah_\bQ=0$. 

As can be seen from the evolution of $e_G$ with decreasing $\hh$ below
$\hfflo(\deltah,\muh)$ (see Fig.~\ref{fig:fflocontour}b) , a secondary local BCS minimum arises at finite
$\Deltah_{\bQ=0}=\deltahbcs$. Upon further decreasing $\hh$, this BCS singlet SF
minimum becomes degenerate  (Fig.~\ref{fig:fflocontour}a) with the FFLO one, and the resonant Fermi
gas undergoes a first-order FFLO-SF transition by \lq\lq jumping\rq\rq\ from the
FFLO minimum to this BCS one.

From the quantitative similarity of the normal and FFLO states'
energetics, it is not surprising that we find this FFLO-SF transition
at 
\be
\hc(\deltah,\muh) \simeq \frac{2\eta}{2\eta^2+1}\deltahbcs\simeq 0.706\deltahbcs,
\ee
obtained using the approximate formula  Eq.~(\ref{deltanearfflotrans}) 
for the order parameter near $\hfflo$.  This is extremely close to the 
N-SF first-order transition at $\hc(\deltah,\muh) \approx \deltahbcs/\sqrt{2}\approx .707\deltahbcs$,
studied in Sec.~\ref{SEC:BCS}, where we ignored the existence of the FFLO state, validating
our approximation of $\hc(\deltah,\muh)$ by  $\deltahbcs/\sqrt{2}$ elsewhere in this section.
 Consistently, $\hc(\deltah,\muh)$ is also only slightly lower than
$\hfflo(\deltah,\muh)$ ensuring that the FFLO state is quantitatively  indeed quite close to the normal state, occupying only a
narrow sliver between $\hfflo(\deltah,\muh)$ and $\hc(\deltah,\muh)$ and limited to 
$\deltah> \deltah_*\approx 2$ on the phase diagram.

The analysis of the FFLO-SF first-order transition parallels our
analysis in Sec.~\ref{SEC:bcsandtransition} for the N-SF transition (where the FFLO state was
neglected). Because of the abovementioned similarity of the N and FFLO
states all the results derived in Sec.~\ref{SEC:bcsandtransition}
remain quantitatively accurate for the true FFLO-SF transition. To recap, for fixed
chemical potentials and decreasing $\hh$, the  system simply jumps at $\hc(\deltah)$ from the FFLO minimum
to the BCS SF minimum,  exhibiting density and
magnetization discontinuities given by Eqs.~(\ref{eq:jump}) and 
~(\ref{eq:mgnormalfflo}), respectively. The two superconducting order parameters
$\Deltah_\bQ$ and $\Deltah_0$ characterizing the FFLO and BCS SF states,
respectively, also experience jump discontinuities with decreasing $\hh$ at $\hc(\deltah,\muh)$,
with $\Deltah_\bQ$ jumping to $0$ and $\Deltah_0$ jumping to the finite BCS value
$\deltahbcs$.   Correspondingly, $Q$ jumps
to zero as $\hh$ reduced below $\hc(\deltah)$.

In contrast, (as also discussed in detail in Sec.~\ref{SEC:firstorderBCS}) at fixed atom
density $n$ (or number $N$), neither the SF nor the FFLO (nor N) states can
satisfy the number equation while remaining a global minimum of
$e_G$. Consequently, below $h_{c2}(\delta,N)=h_c(\delta,\mu_{\rm FFLO}(N)\approx
\mu_N)$ the gas phase separates into coexisting singlet-BCS SF and (at
large detuning)  FFLO state, in proportions $x(\delta,h)$ and
$1-x(\delta,h)$ that are well-approximated using Eq.~(\ref{eq:fracbcs}). This
coexistence region is bounded from below by a lower-critical Zeeman
field $h_{c1}(\delta,N)=h_c(\delta,\mu_{SF}(N))$, below which the system
transitions into a single-component  singlet SF.

%-----------------------------
%
% fig%39
%
\begin{figure}[bth]
\vspace{1.4cm}
\centering
\setlength{\unitlength}{1mm}
\begin{picture}(40,40)(0,0)
\put(-50,0){
\begin{picture}(0,0)(0,0)
\includegraphics{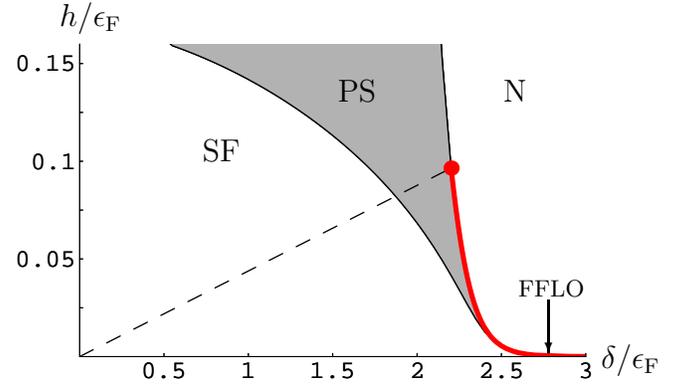}
\end{picture}}
\put(-12,50) {\large{$h/\ef$}}
\put(60,4.5) {\large{$\delta/\ef$}}
\put(7,32) {\large{SF}}
\put(25,40) {\large{PS}}
\put(47,40) {\large{N}}
\put(53,13.5) {\vector(0,-1){7}}
\put(49.0,14) {FFLO}
\end{picture}
\vspace{-.5cm}
\caption{(Color online) Portion of the positive-detuning phase diagram for $\gamma = 0.1$ at 
fixed detuning $\delta$ and chemical potential difference $h$ showing regions of normal phase
(N), FFLO (along red curve), singlet superfluid (SF) and phase-separation (shaded, PS).  The dashed line
connects the critical point $(\delta_*, \hfflo(\delta_*)$ with the origin.
Below the dashed line in the shaded PS regime, the phase separation
consists of coexisting SF and FFLO states, while above the dashed line  it consists of coexisting
SF and N states.}
\label{hfflophasesep}
%
%  FFLOphasesep.nb
%
\end{figure}
%------------------------------
%-----------------------------
%
% fig%40
%
\begin{figure}[bth]
\vspace{1.4cm}
\centering
\setlength{\unitlength}{1mm}
\begin{picture}(40,40)(0,0)
\put(-50,0){
\begin{picture}(0,0)(0,0)
\includegraphics{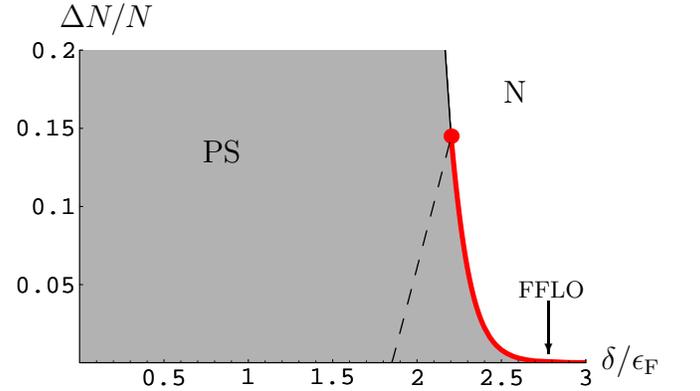}
\end{picture}}
\put(-12,50) {\large{$\Delta N/N$}}
\put(60,4.5) {\large{$\delta/\ef$}}
\put(7,32) {\large{PS}}
\put(47,40) {\large{N}}
\put(53,13.5) {\vector(0,-1){7}}
\put(49.0,14) {FFLO}
\end{picture}
\vspace{-.5cm}
\caption{(Color online)  Portion of the positive-detuning phase diagram for $\gamma = 0.1$ at 
fixed detuning $\delta$ and polarization $\Delta N$ showing regions of normal phase
(N), FFLO (along red curve) and phase-separation (shaded, PS).
Below the dashed line, the phase separation
regime consists of coexisting SF and FFLO states, while above the dashed line  it consists of coexisting
SF and N states.}
\label{hfflophasesep2}
%
%  FFLOphasesep.nb
%
\end{figure}
%------------------------------

We conclude this subsection by noting that, because of the existence of
the $\deltah_*$ detuning point on the $\hfflo(\deltah)$ boundary, the
phase separation (PS) regime
 between $\hctwo(\deltah)$ and $\hcone(\deltah)$ is somewhat
nontrivial. This arises from the fact that, for $\deltah \gg \deltah_*$ (deep
in the BCS regime), it is the FFLO state to which the BCS SF is
unstable above $\hc(\deltah,\muh)$.  Thus, at large detunings, we expect the PS 
to consist of SF-FFLO 
coexistence.  For $\deltah \ll \deltah_*$, on the other hand, we expect 
the regime of phase separation to consist of SF-N coexistence.

Our aim is to compute the boundary separating these two types of phase-separated mixture. 
We start by computing this boundary at fixed density and chemical potential difference,
shown in Fig.~\ref{hfflophasesep}, which can be done using a simple
(approximate) argument. Consider a point  $(\deltah_0, \hh_0) = (\delta_0/\ef,h_0/\ef)$
inside the region of phase separation.  The uniform phases comprising the mixture at 
this point must have the {\it same\/} detuning $\delta_0$ and chemical potential difference $h_0$ but
different densities.  Since both axes are normalized to  $\ef \propto n^{2/3}$, clearly such
uniform phases lie on a ray emanating from the origin and intersecting $(\deltah_0, \hh_0)$.  
This implies that the homogeneous states on either side of the regime of phase separation that 
intersect this ray will comprise the mixed state. 

This procedure can be used to determine the composition  of the mixed phase for any point in the 
regime of phase separation.  In Fig.~\ref{hfflophasesep} we have drawn a dashed line connecting
$(\deltah_*, \hh_{\rm FFLO}(\delta_*)) \approx (2.202,.0965)$ with the origin.  Clearly,
 for points inside the PS region that are above this dashed line
the coexistence is between SF and N while for points below the dashed line it is between SF and FFLO.

Having found a simple approximate scheme [it is approximate because the location of $(\deltah_*,\hh_{\rm FFLO})$
is not invariant with respect to changes of the density since $\gamma$ is a function of $\ef$] to 
determine the boundary between
SF-N coexistence and SF-FFLO coexistence, we now obtain it in a different way, defining the boundary as the
 place where 
the N portion of the SF-N coexistence undergoes a second-order transition (with decreasing chemical 
potential difference) 
to the FFLO phase.
The critical chemical potential difference   $\hfflo(\muh)$ at which this occurs 
 satisfies 
 Eq.~(\ref{momsecond}) 
and  Eq.~(\ref{eq:gapff}) with  $\Deltah = 0$ for the latter: 
\bse
\label{hffloofmu}
\bea
&&0 = \frac{\gamma^{-1} \hfflo(\muh)^2}{\muh^{3/2}} \lambda^2 
+ 1 - \frac{1}{2\lambda} \ln \frac{\lambda + 1}{\lambda-1},
\\
&&0 = 2
-\frac{\gamma^{-1} \hfflo(\muh)^2}{\muh^{3/2}} \lambda^2 
- \ln \frac{4\hfflo(\muh)^2(\lambda^2-1)}{\deltahbcs^2} 
\nonumber \\
&&\qquad \qquad 
- \frac{1}{\lambda} \ln\frac{\lambda+1}{\lambda-1},
\eea
\ese
where we include the chemical-potential argument in  $\hfflo(\muh)$ to emphasize that is distinct from
$\hfflo$ which we have already defined and which also satisfies the number equation; instead, here we must
combine it with conditions appropriate to the mixed phase.  

  In fact, we have already studied the gap and number
equations in the PS regime, assuming SF-N coexistence,  in Sec.~\ref{SEC:regimeofphasesep}, where we expressed
the total number constraint as an equation for the SF fraction $x$ in Eq.~(\ref{eq:fracbcs}); 
along with Eq.~(\ref{eq:fracbcs2}) for the normalized polarization and   Eq.~(\ref{eq:mucritical})
(which provides the first-order condition relating $\mu$ and $h$ in the mixed phase).  For a particular
$\Delta N/N$, such a procedure yields $\hh(\muh)$ in the PS regime assuming SF-N coexistence; where this 
crosses $\hfflo(\muh)$ given by Eq.~(\ref{hffloofmu}) leads to a phase boundary (plotted in 
Fig.~\ref{hfflophasesep2}) that marks the FFLO-N phase transition inside the PS regime.

%%%%%%%%%%%%%%%%%%%%%%%%%%%%%%%%%%%%%%%%%%%%%%%%%%%%%%%%%%%%%%%%%%%%%%%%%%%%%%%%%%%%%%%%%%%%%%%%%%%%%%%%%%%%%%
%%%%%%%%%%%%%%%%%%%%%%%%%%%%%%%%%%%%%%%%%%%%%%%%%%%%%%%%%%%%%%%%%%%%%%%%%%%%%%%%%%%%%%%%%%%%%%%%%%%%%%%%%%%%%%
%%%%%%%%%%%%%%%%%%%%%%%%%%%%%%%%%%%%%%%%%%%%%%%%%%%%%%%%%%%%%%%%%%%%%%%%%%%%%%%%%%%%%%%%%%%%%%%%%%%%%%%%%%%%%%

\section{One-channel model}
\label{SEC:OneChannelModel}
In the preceding sections, we have studied a resonantly-interacting Fermi gas across a 
Feshbach resonance as a function of detuning and chemical potential imbalance 
(or, equivalently, species asymmetry), summarized by the phase diagrams Figs.~\ref{fig:hphasetwo} 
and \ref{fig:mphasetwo}.  
We have limited our analysis to the two-channel model, Eq.~(\ref{eq:bareham}), for reasons 
discussed in Sec.~\ref{SEC:Hamiltonian}, namely, because it is characterized by a small parameter
$\gamma \propto \sqrt{\width/\ef}$ (with $\width$
the Feshbach resonance width), that justifies, in the narrow resonance $\gamma \ll 1$ 
limit, a perturbative mean-field analysis across the resonance.

%-----------------------------
%
% fig%41
%
\begin{figure}[bth]
\vspace{3.2cm}
%\hspace{-1cm}
\hspace{-1cm}
\centering
\setlength{\unitlength}{1mm}
\begin{picture}(40,80)(0,0)
\put(-55,55){\begin{picture}(0,0)(0,0)
\includegraphics{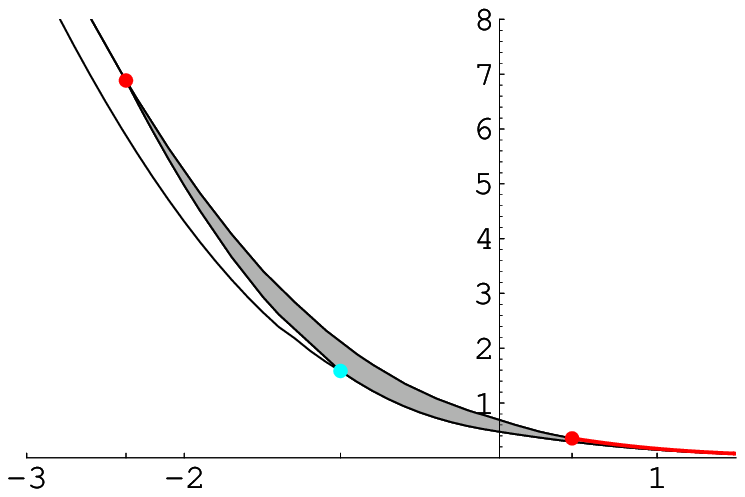}
\end{picture}}
\put(56,107) {(a)}
\put(26,107) {\large{$h/\ef$}}
\put(9,96) {Spin-Polarized Normal}
\put(58,60) {\large{$-\frac{1}{\kf \as}$}}
%
%\put(-7.75,57.5) {$c$}
%\put(13,57.25) {$M$}
%\put(37.75,57.25) {$*$}
%
\put(-7.75,56.5) {$\tilde{\delta}_{c}$}
\put(14.25,56.5) {$\tilde{\delta}_{M}$}
\put(37.5,56.5) {$\tilde{\delta}_{*}$}
%
%\put(-8.75,57.0) {$\hat{a}_{sc}^{-1}$}
%\put(12.25,57.0) {$\hat{a}_{sM}^{-1}$}
%\put(36.5,57.0) {$|\hat{a}_{s*}|^{-1}$}
%
\put(-8.5,89.5) {\vector(2,1){4}}
\put(-12.5,87) {SF$_{\rm M}$}
\put(20.5,74.5) {\vector(-2,-1){6}}
\put(21,74) {PS}
\put(-10,71.5) {\large{SF}}
\put(48,65.5) {\vector(0,-1){4}}
\put(44,66) {FFLO}
\put(-60,0){\begin{picture}(0,0)(0,0)
\includegraphics{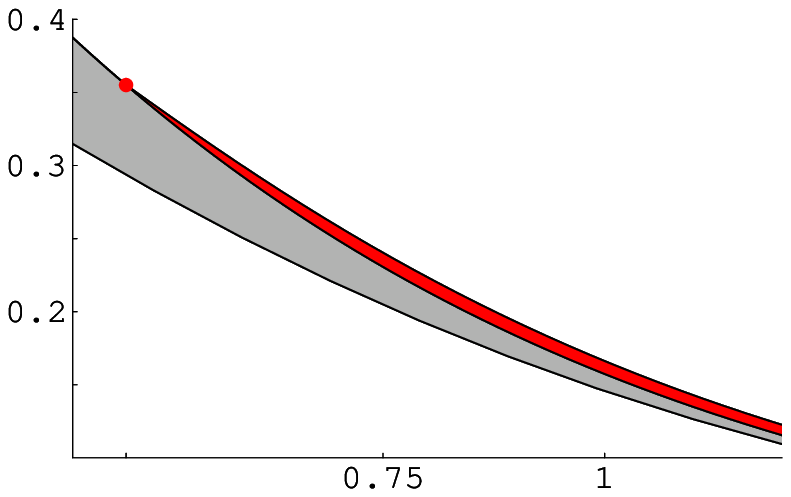}
\end{picture}}
\put(56,45) {(b)}
\put(-16,52) {\large{$h/\ef$}}
\put(60,4) {\large{$-\frac{1}{\kf \as}$}}
%
%\put(-9.5,2) {\large{$*$}}
%\put(-9.,1.5) {\large{$\hat{a}_{s*}^{-1}$}}
\put(-9.5,0.75) {\large{$\tilde{\delta}_*$}}
\put(27.75,24.5) {\vector(-2,-1){4}}
\put(28.5,23.5) {\large{FFLO}}
\put(-9,36) {\large{PS}}
\put(10,12) {\large{SF}}
\end{picture}
\vspace{-.25cm}
\caption{Chemical potential difference, $h$, vs. coupling $\frac{1}{\kf \as}$ 
phase diagram of the one-channel model showing regimes of FFLO, superfluid (SF), magnetized superfluid (\sfm),
and  phase separation (PS). Panel (a) is the global phase diagram and panel (b) is a zoom-in emphasizing the FFLO
regime.
The thick red dot at $(-2.37,6.89)$ 
in the BEC regime is a tricritical point separating first and
second order transitions.  On the horizontal axes three critical detunings are labeled: 
$\tilde{\delta}_{c}\equiv-(\kf a_{sc})^{-1} = -2.37$, 
$\tilde{\delta}_{M}\equiv-(\kf a_{sM})^{-1} = -1.01$ and $\tilde{\delta}_{*}\equiv-(\kf a_{s*})^{-1} = 0.46$.  
}
\label{fig:hsingle} 
%
%  hplotssingle.nb
%  fflosingle.nb 
%
\end{figure}
%------------------------------

In contrast,  the one-channel 
model (to which the two-channel model reduces in the broad-resonance $\gamma \gg 1$ limit)
is characterized by a gas parameter $n^{1/3} |\as|$ which diverges in 
the vicinity of the resonance where $\kf |\as| \to \infty$.  Hence, standard mean-field analysis 
(uncontrolled embellishments notwithstanding) 
is expected to become quantitatively unreliable across the resonance. 

%-----------------------------
%
% fig%42
%
\begin{figure}[bth]
\vspace{3.2cm}\hspace{-1cm}
\centering
\setlength{\unitlength}{1mm}
\begin{picture}(40,80)(0,0)
\put(-55,55){\begin{picture}(0,0)(0,0)
\includegraphics{globalmagphasesingle.eps}
\end{picture}}
\put(52,100) {(a)}
\put(-16,107) {$\frac{\Delta N}{N}$}
\put(60,59) {\large{$-\frac{1}{\kf \as}$}}
\put(41.75,78.5) {\vector(-2,-1){4}}
\put(42.5,77.5) {FFLO}
\put(36.5,95.5) {\large{N}}
\put(12,87.5) {\large{PS}}
\put(-11,67) {\large{SF$_{\rm M}$}}
%
%\put(-7.75,57.5) {$c$}
%\put(-14,57.5) {$c$}
%\put(7,57.25) {$M$}
%\put(31.75,57.25) {$*$}
%
%\put(-15,57.25) {$\hat{a}_{sc}^{-1}$}
%\put(6,57.0) {$\hat{a}_{sM}^{-1}$}
%\put(31.0,57.0) {$\hat{a}_{s*}^{-1}$}
%
\put(-13.75,56.75) {$\tilde{\delta}_{c}$}
\put(8,56.75) {$\tilde{\delta}_{M}$}
\put(31.50,56.75) {$\tilde{\delta}_{*}$}
\put(-55,0){\begin{picture}(0,0)(0,0)
\includegraphics{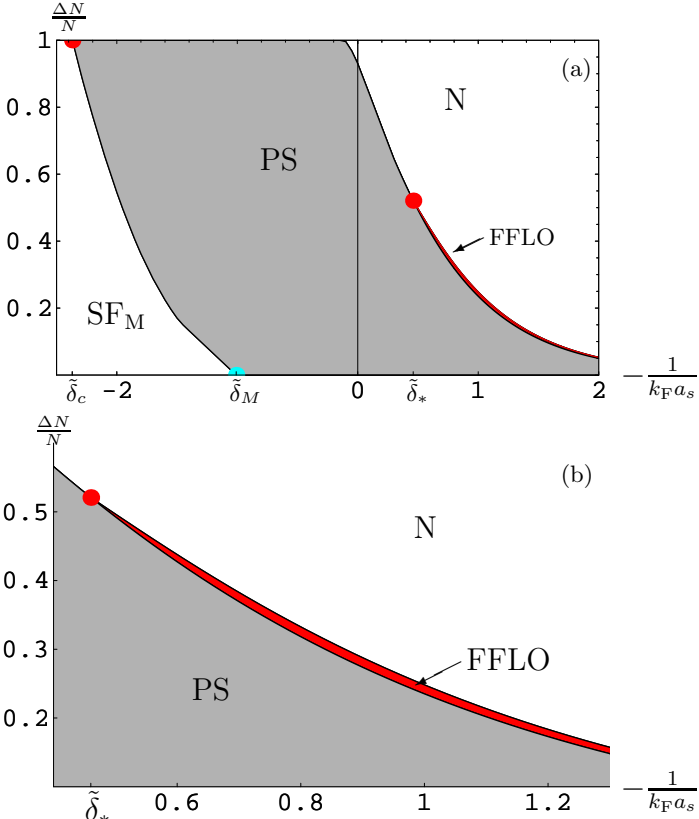}
\end{picture}}
\put(52,46) {(b)}
\put(-18,52) {$\frac{\Delta N}{N}$}
\put(60,4) {\large{$-\frac{1}{\kf \as}$}}
\put(38.75,22.0) {\vector(-2,-1){6}}
\put(39.5,21.0) {\large{FFLO}}
\put(32.5,38.5) {\large{N}}
\put(3,17) {\large{PS}}
\put(-11.25,1.0) {\large{$\tilde{\delta}_{*}$}}
\end{picture}
\vspace{-.5cm}
\caption{ Polarization $\Delta N/N$ vs. coupling $\frac{1}{\kf \as}$ 
phase diagram of the one-channel model showing regimes of FFLO, superfluid (SF), magnetized superfluid (\sfm),
and phase separation (PS).  As in the preceding figure, panel (a) is the global phase diagram, panel (b) is a 
zoom-in emphasizing the FFLO
regime, and 
on the horizontal axes the three critical detunings $\tilde{\delta}_{c}= -2.37$, $\tilde{\delta}_{M}= -1.01$, 
and $\tilde{\delta}_{*}=0.46$ are labeled.  Note that here (in contrast to a narrow Feshbach 
resonance $\gamma \ll 1$, 
Fig.~\ref{fig:mphasetwo}) at unitarity, $(\kf \as)^{-1} = 0$, the boundary between N and PS is at 
$\Delta N/N \simeq0.93<1$,
consistent with experiments~\cite{Zwierlein05}. }
\label{fig:globalmagphasesingle} 
%
%  magplotssingle.nb (both)
%  
\end{figure}
%------------------------------

However, most present-day experiments on $^{40}$K and  $^{6}$Li 
(e.g., the Feshbach resonance at 202 G in $^{40}$K or at 830 G in $^{6}$Li) are in the broad resonance limit
(see Appendix~\ref{SEC:scatteringamplitude})
and one is forced to analyze this analytically inaccessible limit.  This is best done directly
in a one-channel model.    Although mean-field theory
on the one-channel model is not expected to be accurate across a Feshbach resonance tuned to zero energy, it still may
be a qualitatively correct interpolation between the deep BCS and BEC regimes
where the gas parameter $n^{1/3} |\as|\ll 1$.  
 With this caveat in
mind, in the present section we study, within mean-field theory,
 the single-channel model as a function of  polarization and detuning (or, equivalently, s-wave scattering length
$\as$), with our results summarized by the phase diagrams 
Figs.~\ref{fig:hsingle} and 
\ref{fig:globalmagphasesingle} (which are in agreement with recent results~\cite{Gu,Chien,Parish}). 
Since the one- and two-channel models are so closely 
related (see Sec.~\ref{SEC:BCS}), 
many of the calculations will be very similar
to ones we have already presented in the context of the two-channel model.

We begin with the one-channel model Hamiltonian
Eq.~(\ref{eq:singlechannelintro}) in the grand-canonical ensemble
(with system volume $\vol = 1$)
\be
H = \sum_{\bk,\sigma}(\epsilon_k - \mu_\sigma) \ch_{\bk\sigma}^\dagger  \ch_{\bk\sigma}^\phdag 
+ \fermiint\sum_{\bk\bq\bp} \ch_{\bk\uparrow}^\dagger \ch_{\bp\downarrow}^\dagger \ch_{\bk+\bq\downarrow}^\phdag
\ch_{\bp-\bq\uparrow}^\phdag,
\label{eq:singlechannel}
\ee
applying a standard mean-field analysis, by first assuming an expectation
value 
\be
\label{eq:singlemeanfield}
\fermiint\langle \ch_\downarrow(\br) \ch_\uparrow(\br) \rangle = \Delta_\bQ {\rm e}^{i\bQ\cdot \br},
\ee
corresponding to pair condensation at a single wavevector $\bQ$, with $\Delta_\bQ$ and $\bQ$ to be 
self-consistently determined. 
With this mean-field assumption, $H$, Eq.~(\ref{eq:singlechannel}), reduces to the standard BCS mean-field
form:
\bea
&&\hspace{-.5cm}H = -\frac{|\Delta_\bQ|^2}{\fermiint}+ 
\sum_k (\epsilon_k - \mu_\sigma) \ch_{\bk \sigma}^\dagger  \ch_{\bk\sigma}^\phdag 
\nonumber \\
&&
\hspace{-.5cm}+ \sum_{\bk}\Big( \Delta_\bQ^\dagger \ch_{\bk+\frac{\bQ}{2}\downarrow}^{\phdag}\ch_{-\bk+\frac{\bQ}{2}\uparrow}^{\phdag}
+ \ch_{-\bk+\frac{\bQ}{2}\uparrow}^{\dagger}\ch_{\bk+\frac{\bQ}{2}\downarrow}^{\dagger}\Delta_\bQ^\phdag
\Big).
\eea
With the identification of $\Delta_\bQ$ with $gB_\bQ$, 
its main distinctions from the two-channel model Eq.~(\ref{eq:meanfieldhamiltonian})
are the lack of any dispersion in the $\Delta_\bQ$ field
[with $-\lambda^{-1}$ replacing $g^{-2}(\frac{1}{2}\epsilon_Q + \delta_0-2\mu)$] and the fact that the 
total atom-number
constraint here involves only $N = \sum_{\bk,\sigma} \langle c_{\bk \sigma}^\dagger c_{\bk \sigma}^\phdag \rangle$
rather than the analogous Eq.~(\ref{eq:globalnumdefpre}).

As for the two-channel model, after this mean-field approximation (treating $\Delta_\bQ$ as a c-number), 
everything else can be in principle computed exactly because the model is quadratic 
in fermion operators and can easily be
diagonalized. Equivalently, the theory can be formulated via a coherent-state path integral, where $\Delta_\bQ$
appears as a Hubbard-Stratonovich field used to decouple the quartic interaction~\cite{Kleinert} in $H$, 
Eq.~(\ref{eq:singlechannel}).  In this
approach, the mean-field approximation corresponds to a saddle-point treatment of the field $\Delta_\bQ$.

 All $T=0$ properties (on which we focus) follow directly from the corresponding ground state
energy $E_G(\Delta_\bQ)$, with $\Delta_\bQ$ appearing as a variational parameter.  
As in the two-channel case, we shall  find that for much of the phase diagram the ground-state 
is characterized by $\bQ = 0$.   
Anticipating this, we first focus on this $\bQ=0$ subspace, returning to the more general $\bQ\neq 0$ 
case in Sec.~\ref{SEC:FFLOsingle} [where we display the full $\bQ$-dependent ground-state energy
as  Eq.~(\ref{eq:havgsingle})].  With this simplification, we find the ground-state energy 
 (taking $\Delta_{\bQ=0}\equiv \Delta$ real):
\be
\label{eq:gsesinglechannel}
E_G = -\frac{\Delta^2}{\fermiint} + \sum_k (\xi_k - E_k) - \int_0^h \mg(h')dh' , 
\ee
with $\xi_k \equiv \epsilon_k -\mu$ and where the magnetization  
\bea
&&\mg(h) = \frac{2c}{3} \big[
(\mu + \sqrt{h^2 - \Delta^2})^{3/2} \Theta(\mu + \sqrt{h^2 - \Delta^2})
\nonumber \\
&&\qquad -
(\mu - \sqrt{h - \Delta^2})^{3/2}\Theta(\mu - \sqrt{h^2 - \Delta^2})
\big]. \label{eq:mag}
\eea
The short-scale (ultraviolet cutoff $\Lambda$) dependence in the first momentum sum of Eq.~(\ref{eq:gsesinglechannel}), as usual,
can be eliminated by re-expressing $E_G$ in terms of the s-wave scattering length using the relation 
  Eq.~(\ref{eq:singlechannelintroas}) between $\as$ and $\fermiint$.
This gives 
 (converting sums to integrals)
\be
\label{eq:gsesinglechannel2}
E_G = -\frac{m}{4\pi \as} \Delta^2 + \int \frac{d^3 k}{(2\pi)^3} (\xi_k - E_k + \frac{\Delta^2}{2\epsilon_k}) 
- \int_0^h \mg(h')dh', 
\ee
which is almost identical to our two-channel ground-state energy [c.f. Eq.~(\ref{eq:havgBCS2})].  We now proceed to 
study $E_G$, Eq.~(\ref{eq:gsesinglechannel2}), in a variety of  relevant regimes, finding the phase diagram at nonzero
chemical potential difference $h$ or population difference $\Delta N$.  We follow the standard 
practice of expressing all physical 
 quantities as a function of the dimensionless coupling $-\frac{1}{\kf \as} \propto \delta$ that \lq\lq measures\rq\rq\
the system's distance from the zero-energy resonance. In terms of this coupling, in the one-channel model
 the BCS regime occurs
for $-\frac{1}{\kf \as}\agt1$ (i.e., $\as$ small and negative), the BEC regime occurs for  $-\frac{1}{\kf \as}\alt-1$ 
(i.e., $\as$ small and positive), and the crossover between these regimes takes place for $\kf|\as|\gg 1$ 
across the unitary point where $\kf |\as| \to \infty$.

Armed with $E_G(\Delta,\mu,h)$, the determination of the phases and transitions is conceptually straightforward. 
We simply minimize $E_G$ over $\Delta$ and supplement the resulting gap equation, which ensures 
that $\Delta = \fermiint\langle \ch_\downarrow(\br) \ch_\uparrow(\br)\rangle$ is satisfied in the ground-state,
with the number equation ensuring that the total atom density equals the imposed density $n = \frac{4}{3} c\ef^{3/2}$.
The resulting equations
\bse
\label{singleeom}
\bea
0 &=& \frac{\partial E_G}{\partial\Delta},
\\
N &=& -\frac{\partial E_G}{\partial\mu},
\eea
\ese
are well-known in the standard BEC-BCS crossover at $h=0$ and equal spin population.  
At finite imposed population imbalance $\Delta N$ (equivalent to magnetization $m= \Delta N/\vol$) 
Eqs.~(\ref{singleeom}), must be solved simultaneously with
\be
\Delta N =  -\frac{\partial E_G}{\partial h},
\ee
ensuring the imposed $\Delta N$.  As we shall see,
however, caution must be exercised  in using these equations to map out the phase diagram, 
 since, at $h\neq 0$, they exhibit solutions that are local maxima of $E_G(\Delta)$ and
therefore do not correspond to a ground state of the system~\cite{Paonote}.  Failing to ensure that 
solutions to Eq.~(\ref{singleeom}) are indeed minima of Eq.~(\ref{eq:gsesinglechannel}) 
has led to a number of incorrect results in the literature (see, e.g., Ref.~\onlinecite{Pao,Iskin,CommentNote}).

\subsection{Single-channel model at $h=0$}

For completeness and point of reference we review the mean-field theory of the single-channel model at equal populations
by studying $E_G$ at $h=0$.
Using dimensionless variables $\Deltah \equiv \Delta/\ef$, $\muh \equiv \mu/\ef$ and $e_G = E_G/c \ef^{5/2}$
as in the two-channel model, 
\bea \label{eq:gsesingle3}
&&e_G = - \frac{\pi}{2\kf \as} \Deltah^2 
 \\
&& \qquad +
\int_0^\infty d\epsilon \sqrt{\epsilon}\big(\epsilon - \muh  - \sqrt{(\epsilon- \muh)^2 + \Deltah^2} 
 + \frac{\Deltah^2}{2\epsilon}\big). \nonumber
\eea
%

%-----------------------------
%
% fig%43
%
\begin{figure}[bth]
\vspace{1.4cm}
\hspace{-2cm}
\centering
\setlength{\unitlength}{1mm}
\begin{picture}(40,80)(0,0)
\put(-42.25,0){\begin{picture}(0,0)(0,0)
\includegraphics{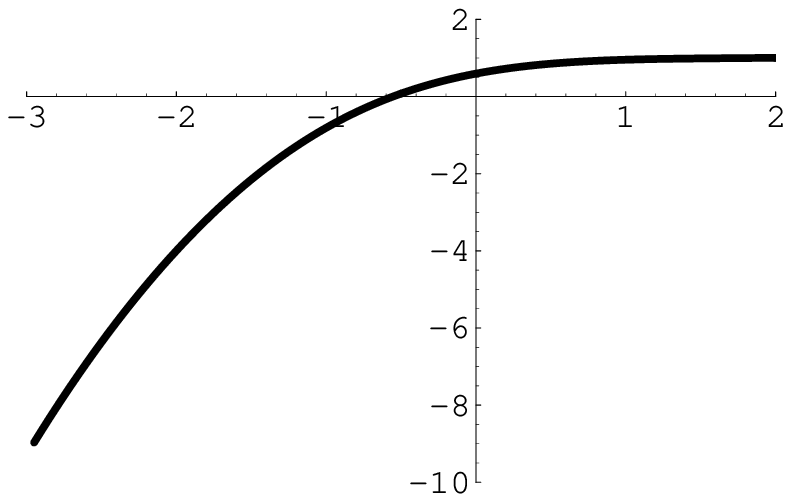}
\end{picture}}
\put(-42.25,47){\begin{picture}(0,0)(0,0)
\includegraphics{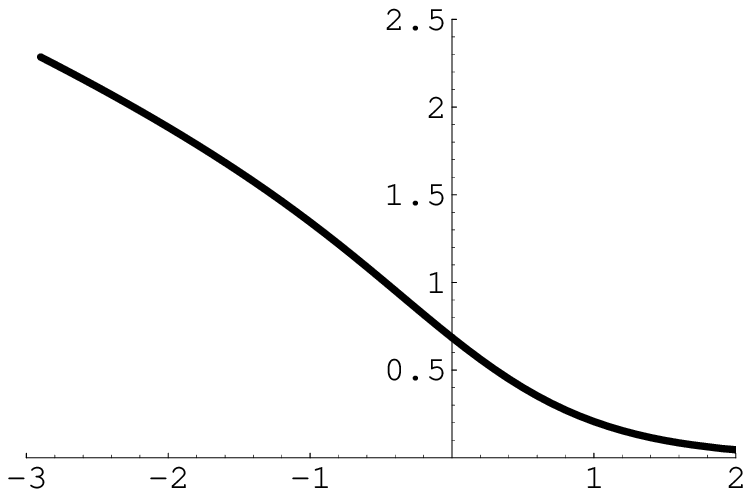}
\end{picture}}
\put(31,45) {{$\mu/\ef$}}
\put(63,34.5) {\large{$-\frac{1}{\kf \as}$}}
\put(63,50.5) {\large{$-\frac{1}{\kf \as}$}}
\put(31,92) {{$\Delta/\ef$}}
\put(-8,79) {(a)}
\put(-6,40) {(b)}
\end{picture}
\vspace{-.5cm}
\caption{(Color online) 
Plots of (a) $\Delta$ and (b) $\mu$ as a function of coupling $-(\kf \as)^{-1}$, within mean-field
theory for the single-channel model, each normalized to $\ef$.}
\label{fig:gapmusingle}
%
%  SingleNumerical.nb (both) 
%
\end{figure}
%------------------------------

The well-known BEC-BCS crossover behavior~\cite{Eagles,Leggett,Nozieres,sademelo}
 follows from minimizing $e_G$ at fixed total density. This corresponds to simultaneously solving
the gap and number equations (in dimensionless form) 
\bea
\label{gapsinglegen}
&&0 = \frac{\partial e_G}{\partial\Deltah},
\\
&&\frac{4}{3} = -\frac{\partial e_G}{\partial\muh},
\label{numsinglegen}
\eea
with numerical solutions given in Fig.~\ref{fig:gapmusingle}.
We can find simple analytic approximations to these solutions by 
evaluating the integral in Eq.~(\ref{eq:gsesingle3}) in the asymptotic BCS and BEC limits.
In the BCS regime  ($1 \alt - \frac{1}{\kf \as}$),  $\muh>0$ and 
$\Deltah\ll\muh$, yielding
for $e_G$ [using results from Appendix~\ref{SEC:BECBCSreview}]
\bea
&&\hspace{-1cm}e_G \simeq - \frac{\pi\Deltah^2}{2\kf \as}  - \frac{8}{15} \muh^{5/2}
-\sqrt{\muh} \Deltah^2 \big(\frac{1}{2} + \ln \frac{8{\rm e}^{-2} \muh}{\Deltah}
\big).
\eea
Using this approximation for $e_G$ inside Eqs.~(\ref{gapsinglegen})
and (\ref{numsinglegen}) in turn gives the gap and number equations 
\bse
\label{eq:gapmusinglebcs}
\bea\label{eq:gapsinglebcs}
\Deltah &\simeq& 8 {\rm e}^{-2} \muh \exp\big[
\frac{\pi}{2\kf \as \sqrt{\muh}}
\big],
\\
\frac{4}{3} &\simeq& \frac{4}{3} \muh^{3/2} + \frac{5}{4} \frac{\Deltah^2}{\sqrt{\muh}} - \frac{\Deltah^2}{2\sqrt{\muh}} 
\ln \frac{\Deltah}{8{\rm e}^{-2} \muh},
\eea
\ese
which, in the asymptotic BCS regime, are accurately solved by 
\bea
\Deltah &\approx&
\Deltah_F \equiv 
 8 {\rm e}^{-2}  \exp\big[
\frac{\pi}{2\kf \as }
\big],
\label{eq:deltahsingle}
\\
\muh &\approx& 1.
\eea

In the negative-detuning BEC regime of  $-\frac{1}{\kf \as}\alt-1$, 
in which  $\muh<0$ and   $|\muh|\gg\Deltah$, 
we find $e_G$ is approximated by
\be
e_G \simeq - \frac{\pi \Deltah^2 }{2\kf \as}
+\sqrt{|\muh|} \frac{\Deltah^2}{2}\big( \pi+ \frac{\pi}{32}\frac{\Deltah^2}{\muh^2}
\big),
\ee
that leads to the following approximate gap and number equations:
\bse
\label{eq:gapmusinglebec}
\bea
&&\frac{\pi}{2\kf \as} - \frac{1}{2} \pi \sqrt{|\muh|} \simeq \frac{\pi \Deltah^2}{32|\muh|^{3/2}},
\\
&& \frac{4}{3} \simeq \frac{\pi}{4\sqrt{|\muh|}}\Deltah^2 - \frac{3\pi}{128 |\muh|^{5/2}} \Deltah^4.
\eea
\ese
Accurate analytic approximations to these equations can be obtained by neglecting
terms of higher order in $\Deltah$. Thus, we find

\bea
\muh \approx  -\frac{1}{(\kf \as)^2},
\\
\Deltah \approx \sqrt{\frac{16}{3\pi\kf \as}},
\eea
where we see that the role of the gap and number equations is effectively reversed in the BEC regime,
with the gap equation approximately determining $\muh$ and the number equation 
approximately determining $\Deltah$.  In the next few subsections we will extend the above by-now
standard $h=0$ analysis to finite $h$ and finite imposed species imbalance $\Delta N$.

\subsection{BCS regime of one-channel model at $h\neq 0$}

In the present section, we analyze the positive-detuning side of the resonance ($\as<0$) within the 
one-channel model at finite $h$.  Our analysis will closely follow that of Sec.~\ref{SEC:BCS}
for the two-channel model; with the results
of that section in hand, we will be brief in deriving similar results for the one-channel model.
In addition, as in Sec.~\ref{SEC:BCS}, 
we will first neglect the possibility of the FFLO state in this section, returning to finite-$\bQ$ 
ground states in 
Sec.~\ref{SEC:FFLOsingle}.
Evaluating the momentum and $h$ integrals in Eq.~(\ref{eq:gsesinglechannel2}),
with approximations $\Delta \ll \mu$ and $h \ll \mu$ appropriate to the BCS regime, we obtain
\bea
&&\hspace{-.8cm}E_G \approx  -\frac{m}{4\pi \as} \Delta^2 -\frac{8c}{15}\mu^{5/2}
- c\sqrt{\mu} \Delta^2\big(
\frac{1}{2} - \ln \frac{\Delta}{8{\rm e}^{-2} \mu}
\big)
\nonumber \\
&&\hspace{-.8cm}- c\sqrt{\mu} \big[h\sqrt{h^2-\Delta^2} - \Delta^2\cosh^{-1}(h/\Delta)\big]\Theta(h-\Delta).
\label{eq:gsesinglechannelbcs}
\eea
We start in the grand-canonical ensemble at fixed $\mu$ and $h$.
As we have already seen, the first line of  Eq.~(\ref{eq:gsesinglechannelbcs}) ($E_G$ at $h=0$) has a minimum
at $\deltabcs$ given by 
\be
\deltabcs = 8 {\rm e}^{-2} \mu \exp\big[
\frac{\pi\sqrt{\ef}}{2\kf \as \sqrt{\mu}}
\big].
\label{eq:gapsinglebcs2}
\ee
We recall that $\deltabcs$ and the corresponding ground state energy $E_G(\deltabcs)$ remain
$h$-independent at low $h$, consistent with a singlet magnetization-free BCS ground state.  Physically,
this is due to the fact that the $h=0$ BCS state is gapped and $h$ couples to a conserved
quantity $\Delta N$.  Mathematically, because of the step function in Eq.~(\ref{eq:gsesinglechannelbcs}),
$E_G(\deltabcs)$ can only become $h$-dependent for $h>\deltabcs$.  However, before this can 
occur a first-order transition to the normal state $\Delta=0$ takes place at

\be
\label{eq:hcbcssingle}
h_c(\mu) = \frac{\deltabcs}{\sqrt{2}} =  4\sqrt{2} {\rm e}^{-2} \mu \exp\big[
\frac{\pi\sqrt{\ef}}{2\kf \as \sqrt{\mu}}
\big],
\ee
determined by 
equating $E_{G,N}=E_G(\Delta=0,h_c)$ and $E_{G,SF}=E_G(\deltabcs,h_c)$,
that is clearly smaller than $\deltabcs$. 
Hence, the BCS minimum remains stable and independent of $h$.

At fixed {\it density\/} $n$, the chemical potentials $\mu_{SF}(n)$ and $\mu_N(n)$ of the 
SF and N phases are bounded by the distinct atom-number constraint equations
\bea
&&\hspace{-.7cm} n \approx  \frac{4c}{3} \mu_{SF}^{3/2} + \frac{5}{4}\frac{c \deltabcs^2}{\sqrt{\mu_{SF}}}
- \frac{c\deltabcs^2}{2\sqrt{\mu_{SF}}} \ln \frac{\deltabcs}{8{\rm e}^{-2} \mu_{SF}},
\label{eq:musinglebcs3}
\\
\label{eq:musinglebcs2}
&&\hspace{-.7cm} n \approx \frac{4c}{3} \mu_N^{3/2} +\frac{c h^2}{2\sqrt{\mu_N}},
\eea
and therefore $\mu_{SF}\neq \mu_N$. As a result, the jump discontinuity at the first-order
transition at $h_c$ (for fixed $\mu$) opens up into a coexistence region for fixed $n$,
bounded by 
$h_{c1} = h_c(\mu_{SF})$ and $h_{c2} = h_c(\mu_{N})$, respectively determined by
the solutions of Eqs.~(\ref{eq:musinglebcs3}) and Eqs.~(\ref{eq:musinglebcs2}) 
inside Eq.~(\ref{eq:hcbcssingle}).
  As discussed at length in 
in Sec.~\ref{SEC:firstorderBCS}, since neither the SF nor the N can satisfy the atom number
constraint while minimizing $E_G$,  for $h_{c1}<h<h_{c2}$ the system is forced to phase separate into a 
coexisting mixture of SF and N states.

It is straightforward to find $h_{c1}$ and $h_{c2}$ by  numerically solving the sets of equations 
[i.e., Eqs.~(\ref{eq:musinglebcs3}) and ~(\ref{eq:hcbcssingle}), and  Eqs.~(\ref{eq:musinglebcs2}) 
and ~(\ref{eq:hcbcssingle})] that
define them. 
At large $(\kf |\as|)^{-1}$ in the BCS regime (large positive detuning), 
however, we can find accurate analytic approximations to these equations. To
do this, we first write the corresponding sets of equations in dimensionless form ($\hh \equiv h/\ef$,
$\Deltah \equiv \Delta/\ef$, $\muh \equiv \mu/\ef$) for $\hcone$:
\bse
\bea
&&\hspace{-1.5cm}\frac{4}{3} \approx  \frac{4}{3} \muh_{SF}^{3/2} + \frac{5\hcone^2}{2\sqrt{\muh_{SF}}}
- \frac{\hcone^2}{\sqrt{\muh_{SF}}} \ln \frac{\hcone}{4\sqrt{2}{\rm e}^{-2} \muh_{SF}},
\\
&&\hspace{-1.5cm}\hcone =  4\sqrt{2} {\rm e}^{-2} \muh_{SF} \exp\big[
\frac{\pi}{2\kf \as \sqrt{\muh_{SF}}}
\big],
\eea
\ese
and for $\hctwo$:
\bse
\label{eq:hctwoequations}
\bea
\frac{4}{3} &\approx&  \frac{4}{3} \muh_{N}^{3/2}  + \frac{\hctwo^2}{2\sqrt{\muh_N}},
\\
\hctwo &=&  4\sqrt{2} {\rm e}^{-2} \muh_{N} \exp\big[
\frac{\pi}{2\kf \as \sqrt{\muh_{N}}}
\big].
\eea
\ese
As in  Sec.~\ref{SEC:bcsandtransition} where we studied the 
 two-channel model, we solve these equations perturbatively about $\muh_{SF} \simeq 1$ and
 $\muh_{N} \simeq 1$, obviously a good approximation since $\hh_{c1}$ and $\hh_{c2}$ 
are exponentially small. 
 Expressing $\hcone$ and $\hctwo$ in terms of the gap Eq.~(\ref{eq:deltahsingle}) in that limit,
we have:
\bse
\bea
\hcone &\approx& \frac{\Deltah_F}{\sqrt{2}} \exp
\Big[ - \frac{\pi^2 \Deltah_F^2}{32 (\kf |\as|)^2}\Big],
\\
\hctwo &\approx& \frac{\Deltah_F}{\sqrt{2}} \exp
\Big[ - \frac{\pi^2 \Deltah_F^2}{32 \kf |\as|}\Big],\label{eq:hctwobcssingle}
\eea
\ese
with each expression only accurate to leading order  in the argument of 
the exponential.  Thus, we see that, in the asymptotic
large $(\kf |\as|)^{-1}$ limit, $0<\hcone < \hctwo$, although 
in the deep BCS regime 
they become exponentially close to zero and to each other.
We note that these expressions also agree exactly with
the corresponding two-channel results Eqs.~(\ref{eq:hconebcstwop})
and  Eqs.~(\ref{eq:hctwobcs}), using the relation Eq.~(\ref{eq:deltahas}) between the scattering length $\as$
and the detuning $\delta$.
   However, as we shall see, such a simple
correspondence only holds in the BCS regime and does not apply in the BEC regime.  Physically,
this correspondence occurs because the molecular dynamics that are accounted for in the two-channel model (but neglected in
the one-channel model) are unimportant in the BCS regime.

To connect to experiments in which $\Delta N$ rather than $h$ is imposed, $\hcone$ and $\hctwo$
can be easily converted into critical polarizations using 
 the general formula for the population difference:
\bea \label{eq:mbcssingle}
&&\frac{\Delta N}{N} = 
 \frac{1}{2} \big[
(\muh + \sqrt{\hh^2 - \Deltah^2})^{3/2} \Theta(\muh + \sqrt{\hh^2 - \Deltah^2})
\\
&&\qquad -
(\muh - \sqrt{\hh^2 - \Deltah^2})^{3/2}\Theta(\muh - \sqrt{\hh^2 - \Deltah^2})
\big]\Theta(\hh - \Deltah).\nonumber 
\eea
Since $\hcone < \deltahbcs$, clearly  Eq.~(\ref{eq:mbcssingle}) 
gives $\Delta N_{c1} = 0$.  This is consistent with our earlier finding that the BCS 
SF ground state cannot  tolerate any spin population difference in the BCS regime.  
Using $\Deltah = 0$ in the normal state we find 
\bea
&&\frac{\Delta N_{c2}}{N} = 
 \frac{1}{2} \big[
(\muh_N + \hctwo)^{3/2} \Theta(\muh_N + \hctwo)
\nonumber\\
&&\qquad \qquad 
-
(\muh_N - \hctwo)^{3/2} \Theta(\muh_N - \hctwo )
\big], \label{eq:mctwobcssinglepre}
\\
 &&\qquad \quad \approx
\frac{3}{2} \hctwo \sqrt{\muh_N},
\label{eq:mctwobcssingle}
\eea
with the second expression applying for $\hctwo \ll \muh_N$ (which is always valid in the BCS regime of interest).
Inserting Eq.~(\ref{eq:hctwobcssingle}) into Eq.~(\ref{eq:mctwobcssingle}),
and using $\muh_N \approx 1-\frac{1}{8}\Deltah_F^2$ [which follows from Eq.~(\ref{eq:hctwoequations}) to leading order],
we have 
\bea
\frac{\Delta N_{c2}}{N} &\approx& \frac{3 \Deltah_F}{2\sqrt{2}}\big( 1 - \frac{\pi \Deltah_F^2}{32 \kf |\as|}\big)
\big(1-  \frac{\Deltah_F^2}{8}\big) ,
\\
&\approx& \frac{3 \Deltah_F}{2\sqrt{2}}\exp\Big[ - \frac{\pi^2 \Deltah_F^2}{32 \kf |\as|}\Big],\label{eq:hctwobcssinglefinal}
\eea
where in the last line we used $(\kf |\as|)^{-1}\gg 1$ and re-exponentiated, valid to leading order.  
Again, Eq.~(\ref{eq:hctwobcssinglefinal}) agrees with the corresponding two-channel result Eq.~(\ref{mcovernlargedelta3})
using Eq.~(\ref{eq:deltahas}).
And, as in the two-channel model, for sufficiently large $(\kf |\as|)^{-1}$, there is a narrow regime of 
FFLO phase (neglected in this section) that 
intervenes between the regime
 of phase separation and the  polarized N state.  To remedy this, 
we must recompute the ground-state energy and number and gap equations including the possibility of a spatially-varying 
$\Delta(\br)$.  Before doing this (in Sec.~\ref{SEC:FFLOsingle}), we first study the strong-coupling BEC regime
at $\as>0$.

\subsection{BEC regime of single-channel model}
\label{SEC:BECsingle}
Here we analyze $E_G$, Eq.~(\ref{eq:gsesinglechannel2}), in the BEC regime of $\as>0$, still focusing on the
case of $\bQ=0$, which, as we saw in Sec.~\ref{SEC:FFLO}, is actually not a restriction as the ground state 
in the BEC regime is always at $\bQ=0$.  We also anticipate that $\mu<0$ in the BEC regime, remaining valid
even for finite $h$.  Physically, this well-known property of the BEC regime is due to the finite molecular
binding energy $E_b$ that fixes $\mu\approx -E_b/2$.  For $\mu<0$, the
 $\bQ=0$ ground-state energy for the single-channel model, Eq.~(\ref{eq:gsesinglechannel2}),
can be simplified considerably by expanding it to leading order in $\Delta/|\mu|$:
\be
E_G  = -\frac{4}{15} c (h -|\mu|)^{5/2} \Theta(h - |\mu|)
- V_2 \Delta^2  + \frac{1}{2} V_4 \Delta^4,
\label{eq:egbecsingle}
\ee
where the \lq\lq Landau\rq\rq coefficients are given by
\bse
\bea
 &&\hspace{-1cm}V_2 \equiv
\frac{m}{4\pi \as} -c \sqrt{|\mu|}F_2(h/|\mu|), 
\label{eq:v2single}
\\
&&\hspace{-1cm}V_4 \equiv \frac{c\pi}{32 |\mu|^{3/2}}F_4(h/|\mu|),
\label{eq:v4single}
\eea
\ese
with $F_2(x) $ and $F_4(x)$ given by Eqs.~(\ref{eq:f2def}) and ~(\ref{eq:f}),
respectively.
The gap and number equations then reduce to 
\bse
\label{eq:gapnumbecsingle}
\bea
V_2 &=& \Delta^2 V_4, 
\\
n &=& -\partial E_G/\partial\mu,
\label{eq:numbecsingle}
\eea
\ese
with $n$ the total fermion density.  We emphasize that, like in the two-channel model, the naive solution of 
Eq.~(\ref{eq:gapnumbecsingle}) is only a {\it minimum\/} of the ground-state energy for sufficiently
small $h$; at larger $h$ it is a local {\it maximum\/} and therefore does not correspond to a physical ground state.  

 To leading order in the small $\Delta/|\mu|$  limit,  
Eqs.~(\ref{eq:gapnumbecsingle}) reduce to 
\bea
\label{eq:gapbecsingle2}
V_2 &=& 0,
\\
n &=& \frac{c\pi}{4\sqrt{|\mu|}} \Delta^2 + \frac{2}{3} c (h-|\mu|)^{3/2}\Theta(h-|\mu|).
\label{eq:numbecsinglep}
\eea
First focusing on the gap equation, Eq.~(\ref{eq:gapbecsingle2}), and ignoring the weak $h/|\mu|$
dependence in the second term of $V_2$ we find (using $F_2(0) = \pi/2$)
\be
0= \frac{m}{4\pi \as} - \frac{1}{2}c \pi \sqrt{|\mu|}, 
\ee
which gives for the chemical potential~\cite{sademelo}
\be
\label{eq:musinglebec}
\mu = -\frac{1}{2m \as^2} = - \frac{E_b}{2}= - \frac{\ef}{(\kf \as)^2},
\ee
minus one-half the molecular binding energy $E_b = 1/m \as^2$, a  well-known result for $h=0$, 
that we see extends approximately to $h>0$.  With this result, Eq.~(\ref{eq:numbecsinglep}) becomes
\be
n = \frac{\sqrt{2}c\pi}{4\sqrt{E_b}}\Delta^2 + \frac{2}{3} c \big(h-\frac{E_b}{2}\big)^{3/2}\Theta\big(h-\frac{E_b}{2}\big).
\label{eq:numbecsingle2}
\ee
Clearly, the first and second terms on the right side of this number equation represent fermions bound
into molecules and free spin-up fermions, respectively, characteristic of the magnetic superfluid (\sfm) 
phase that we have discussed for the two-channel model in Sec.~\ref{SEC:BEC}.
This allows us to relate the molecular condensate wave function (order parameter) to $\Delta$  via~\cite{sademelo}
\be
\psi_m^2 \equiv \frac{c\pi}{4\sqrt{2E_b}} \Delta^2,
\label{eq:psim}
\ee
giving for the number equation 
\be
n = 2\psi_m^2 + \frac{2}{3} c \big(h-\frac{E_b}{2}\big)^{3/2}\Theta\big(h-\frac{E_b}{2}\big).
\label{eq:numbecsingle3}
\ee

It is also useful to define an effective molecular chemical potential 
\be
\label{eq:mubsingle}
\mu_m \equiv E_b + 2\mu,
\ee
 in terms of which 
our approximate solution to the gap equation above [i.e.~Eq.~(\ref{eq:gapbecsingle2})] 
simply corresponds to $\mu_m \approx 0$, as expected for a $T=0$ Bose condensate.
Re-expressing $E_G$ in terms of $\mu$ and $\psi_m$, we find 
\be
E_G \approx  -\frac{4}{15} c \big(h - \frac{E_b}{2}\big)^{5/2}\Theta\big(h-\frac{E_b}{2}\big)
- \Vt_2\psi_m^2 +\frac{1}{2} \Vt_4\psi_m^4,
\label{eq:gsemolsingle}
\ee
with the quadratic and quartic coefficients
\bea
&&\hspace{-1cm}\Vt_2 \equiv 
2E_b\big[1-\frac{2}{\pi}\big(1-\frac{\mu_m}{2E_b}
\big)F_2\big(\frac{2h}{E_b}\big)\big],
\\
\label{eq:v4barsingle}
&&\hspace{-1cm}\Vt_4\equiv
\frac{4\pi}{\sqrt{E_b} m^{3/2}} F_4\big(\frac{2h}{E_b}\big) .
\eea

Using Eqs.~(\ref{eq:psim}) and (\ref{eq:numbecsingle3}), we can obtain an approximate 
expression for the pair field $\Delta$ as a function of $h$:
\bse
\bea
&&\hspace{-1cm}\Delta^2 = \frac{\sqrt{8 E_b}}{c\pi} \Big[
n\!-\!\frac{2c}{3} \big(h-\frac{E_b}{2}\big)^{3/2}\Theta(h\!-\!\frac{E_b}{2})
\Big],
\\
&&\hspace{-1cm}\quad\,\,\,\,= \Delta_0^2 \Big[
1-\frac{1}{2} \big(\frac{h}{\ef}-\frac{E_b}{2\ef}\big)^{3/2}\Theta(h-\frac{E_b}{2})
\Big],
\label{eq:deltabecsingle}
\eea
\ese
with $\Delta_0$ the pair field at $h=0$:
\be
\label{eq:deltanoughtbecsingle}
\Delta_0 = \sqrt{\frac{2 n\sqrt{2 E_b}}{c\pi}}= \sqrt{\frac{4n\pi}{m^2 \as}}=\ef\sqrt{\frac{16}{3\pi \kf \as}},
\ee
where we used Eq.~(\ref{numdef}) for $n$.
The preceding equations determine the phase diagram  of the single-channel model 
Eq.~(\ref{eq:singlechannel}) on the BEC side of
the resonance.
We  determine the phase diagram at fixed 
$N$ and $h$, fixed $N$ and $\Delta N$, and fixed $\mu$ and $h$ in Secs.~\ref{SEC:BECnhsingle}, 
\ref{SEC:BECnmsingle} and \ref{SEC:fixedmusingle}, respectively, with the latter a 
prerequisite to studying {\it inhomogeneous\/} polarized paired Fermi condensates in a harmonic trap
within the local density approximation.

%-----------------------------
%
% fig%44
%
\begin{figure}[bth]
\vspace{1.4cm}
\hspace{-1.4cm}
\centering
\setlength{\unitlength}{1mm}
\begin{picture}(40,40)(0,0)
\put(-50,0){\begin{picture}(0,0)(0,0)
\includegraphics{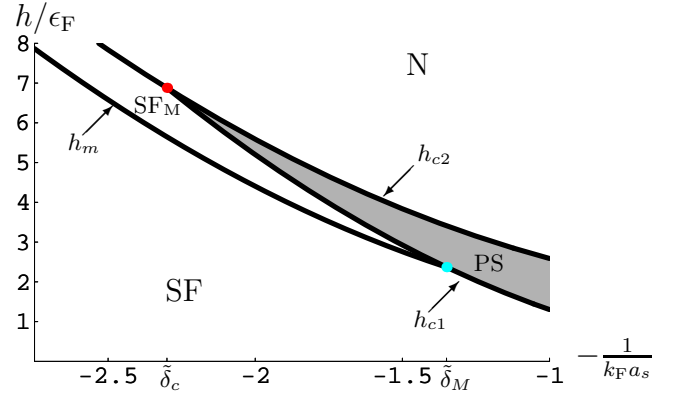}
\end{picture}}
\put(-12,49.5) {\large{$h/\ef$}}
\put(62.5,4.5) {\large{$-\frac{1}{\kf \as}$}}
\put(7.25,1.5) {$\tilde{\delta}_{c}$}
\put(44.5,1.5) {$\tilde{\delta}_{M}$}
\put(43,11.5) {\vector(1,1){4}}
\put(41,9.5) {$h_{c1}$}
\put(-3,35) {\vector(1,1){4}}
\put(-5,33) {$h_m$}
\put(41,31) {\vector(-1,-1){4}}
\put(41.5,31.5) {$h_{c2}$}
\put(3.75,38) {SF$_{\rm M}$}
\put(8,13) {\large{SF}}
\put(49,17) {PS}
\put(40,43) {\large{N}}
\end{picture}
\vspace{-.5cm}
\caption{(Color online)  Plot of {\it approximate\/} critical curves $h_m$ [Eq.~(\ref{eq:hmsingle2})],
$h_{c1}$ [Eq.~(\ref{eq:hc1single})] and $h_{c2}$ [Eq.~(\ref{eq:hc2single})] which separate
regions of  magnetized molecular superfluid (\sfm), a fully-polarized normal  Fermi gas (N), and a singlet
molecular superfluid (SF). The critical detunings $\tilde{\delta}_{c}\equiv-(\kf a_{sc})^{-1}$
and $\tilde{\delta}_{M}\equiv-(\kf a_{sM})^{-1}$ denote a tricritical point and the beginning
of the \sfm phase, respectively. This approximate phase digram agrees well with the numerically-determined 
mean-field phase diagram
for the single-channel model Fig.~\ref{fig:hsingle}.}
\label{fig:fixedhsingle}
%
%  hplotsbecsingle.nb 
%
\end{figure}
%------------------------------

\subsubsection{Phase diagram at fixed $\hh$ and density $n$}
\label{SEC:BECnhsingle}
The three critical chemical-potential differences
describing the phase diagram at fixed density and chemical potential difference are:
$h_{m}$, the SF-\sfm transition point, $h_{c1}$, defined by when the \sfm phase becomes
unstable to phase separation and 
$h_{c2}$, defined as the $h$ above which the purely fermionic polarized N state is stable.
By examining Eq.~(\ref{eq:mag}), we see that 
\be
h_m  = \sqrt{\mu^2+ \Delta^2},
\label{eq:hmsinglepre}
\ee
with $\mu$ and $\Delta$ given by their values in the SF phase, i.e.,  
Eq.~(\ref{eq:musinglebec})
 and Eq.~(\ref{eq:deltanoughtbecsingle}), respectively.
As we found for the two-channel model in  Sec.~\ref{SEC:transitionphasesep}, 
$h_{c1}$ is most easily (approximately) determined by finding the point where
$\Vt_4$ vanishes (so that molecules
are no longer repulsive), signaling the first-order instability. Examining Eq.~(\ref{eq:v4barsingle})
and recalling $F_4(1.30) = 0$, we see that this occurs at
$2h_{c1}/E_b = 1.30$.  Finally, $h_{c2}$ is defined as that $h$ at which the molecular 
density vanishes according to  Eq.~(\ref{eq:numbecsingle3}) and the system consists
of a fully-polarized Fermi gas.  Taken together, these
three critical $h$'s are then given by:
\bse
\bea
&&\hspace{-.75cm}h_m \simeq \sqrt{\frac{E_b^2}{4} + \frac{2n\sqrt{2E_b}}{c\pi}},
\label{eq:hmsingle}
\\
&&\hspace{-.75cm}\quad\,\,\,\,\simeq \ef\sqrt{\frac{1}{(\kf \as)^4} + \frac{16}{3\pi \kf \as}}, \label{eq:hmsingle2}
\\
&&\hspace{-.75cm}h_{c1} \simeq 0.65 E_b = 1.30 \frac{\ef}{(\kf \as)^2} ,
\label{eq:hc1single}
\\
&&\hspace{-.75cm}h_{c2} \simeq  \frac{1}{2} E_b  + \big(\frac{3n}{2c}\big)^{2/3} 
= \ef \big(2^{2/3} + \frac{1}{(\kf \as)^2}\big),
\label{eq:hc2single}
\eea
\ese
where we have expressed them in terms of the molecular binding energy  $E_b = 2\ef/(\kf \as)^2$.  
Note that, as can be seen from the way it is displayed  in
the phase diagram Fig.~\ref{fig:fixedhsingle}, $h_m$ is only defined for sufficiently large $(\kf \as)^{-1}$ 
(to the left in the figure), 
and ceases to be defined when it crosses the lower critical 
 $h_{c1}$ boundary at the critical 
scattering length $a_{sM}$ satisfying $(\kf a_{sM})^{-1} \approx 1.35$.
  Similarly,  $h_{c1}$
is only defined for small $(\kf \as)^{-1}$, until it crosses $h_{c2}$ at $a_{sc}$ given by 
$(\kf a_{sc})^{-1}\approx 2.30$; this is a tricritical point.
These two critical points are analogous to the critical detunings $\delta_M$ and $\delta_c$ in the two-channel model.

%

%-----------------------------
%
% fig%45
%
\begin{figure}[bth]
\vspace{1.4cm}
\hspace{-2cm}
\centering
\setlength{\unitlength}{1mm}
\begin{picture}(40,40)(0,0)
\put(-50,0){\begin{picture}(0,0)(0,0)
\includegraphics{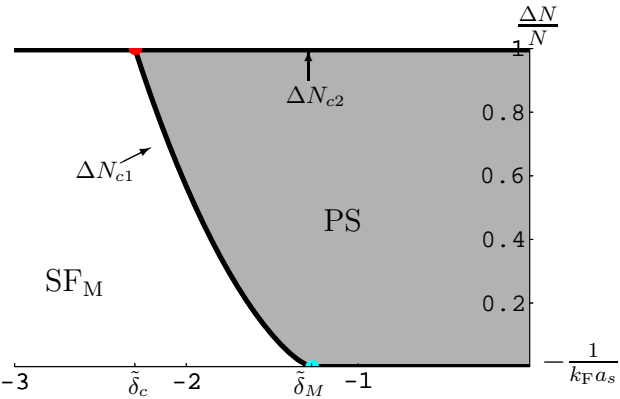}
\end{picture}}
\put(57.5,49.25) {\large{$\frac{\Delta N}{N}$}}
\put(61,4.5) {\large{$-\frac{1}{\kf \as}$}}
\put(6,1.35) {$\tilde{\delta}_{c}$}
\put(28,1.35) {$\tilde{\delta}_{M}$}
\put(5,32) {\vector(2,1){4}}
\put(-1.0,30) {$\Delta N_{c1}$}
\put(30,43) {\vector(0,1){4}}
\put(27,40) {$\Delta N_{c2}$}
\put(-5,15) {\large{SF$_{\rm M}$}}
\put(32,23) {\large{PS}}
\end{picture}
\vspace{-.5cm}
\caption{Approximate (analytic) polarization-scattering length phase diagram on BEC side, for 
the single-channel model, illustrating the critical polarization boundaries $\Delta N_{c1}$ 
and $\Delta N_{c2} = N$, with critical detuning values $\tilde{\delta}_c$ and $\tilde{\delta}_M$
defined as in the Fig.~\ref{fig:fixedhsingle}.
  This approximate phase digram agrees well with the numerically-determined 
mean-field phase diagram
for the single-channel model Fig.~\ref{fig:globalmagphasesingle}.}
\label{fig:mconebec}
%
%  hplotsbecsingle.nb 
%
\end{figure}
%------------------------------

\subsubsection{Phase diagram at fixed population difference}
\label{SEC:BECnmsingle}
In the present subsection, we convert these critical $h$'s to corresponding critical population differences 
(or magnetizations).  To do this, we simply use $\Delta N(h)$, Eq.~(\ref{eq:mequation}).  Of course, since
$h_m$ is defined by the $h$ at which the population difference $\Delta N$ increases from zero in a continuous fashion, 
$\Delta N_m = 0$.
Moreover, everywhere on the BEC side $\Delta N_{c2}/N = 1$, as can be seen by plugging Eq.~(\ref{eq:hc2single})
into Eq.~(\ref{eq:mequation}); thus, a normal Fermi gas with anything less than
complete polarization is unstable to pairing or phase separation in the deep BEC regime.  We are therefore left 
with the computation of $\Delta N_{c1}$, the population difference 
 at which  the  polarized one-channel model (in the \sfm state) is unstable
to phase separation.  Equation~(\ref{eq:mequation}) yields for $\Delta N_{c1}$ [using Eq.~(\ref{magrelation2});
recall Eq.~(\ref{eq:dimensionlessvariables})]:
\be 
\frac{\Delta N_{c1}}{N} = \frac{1}{2} \Big(\sqrt{\hcone^2 -\Deltah^2} - |\muh|\Big)^{3/2}\Theta\Big(
\sqrt{\hcone^2-\Deltah^2} -|\muh|\Big),
\label{eq:mconesingle}
\ee
where $\muh$ is given by Eq.~(\ref{eq:musinglebec}) and $\Deltah$ is given by 
its value at the transition, obtained by plugging $h_{c1}$ into 
Eq.~(\ref{eq:deltabecsingle}):
\be
\Deltah^2 \simeq \frac{16}{3\pi \kf \as} \Big[
1-\frac{1}{2}\big(\frac{1.30}{(\kf \as)^2} - \frac{1}{(\kf \as)^2}
\big)^{3/2}
\Big],
\label{eq:deltaoverefsingle}
\ee
which, along with Eqs.~(\ref{eq:mconesingle}) and (\ref{eq:hc1single}) yields
\be
\frac{\Delta N_{c1}}{N} \simeq \frac{1}{2}\Big(
\sqrt{
\frac{(1.30)^2}{(\kf \as)^4} - 
\Deltah^2}
-\frac{1}{(\kf \as)^2}
\Big)^{3/2},
\label{eq:mconesinglefin}
\ee
plotted in Fig.~\ref{fig:mconebec}.  This analytic approximation agrees well with the exact 
numerically-determined mean-field  phase
diagram for the single-channel model Fig.~\ref{fig:globalmagphasesingle}.

\subsubsection{Phase diagram at fixed chemical potential}

\label{SEC:fixedmusingle}
We conclude this subsection by computing the phase diagram in the grand-canonical ensemble at fixed
$\mu$ and $h$ that will be important for the extension of this bulk analysis to that of a trap
in Sec.~\ref{SEC:LDA}.  Our analysis here will mirror that of the BEC regime of the two-channel model in the 
grand-canonical ensemble presented in Sec.~\ref{SEC:BECfcp}.

%-----------------------------
%
% fig%46
%
\begin{figure}[bth]
\vspace{1.4cm}
\hspace{-2cm}
\centering
\setlength{\unitlength}{1mm}
\begin{picture}(40,40)(0,0)
\put(-50,0){\begin{picture}(0,0)(0,0)
\includegraphics{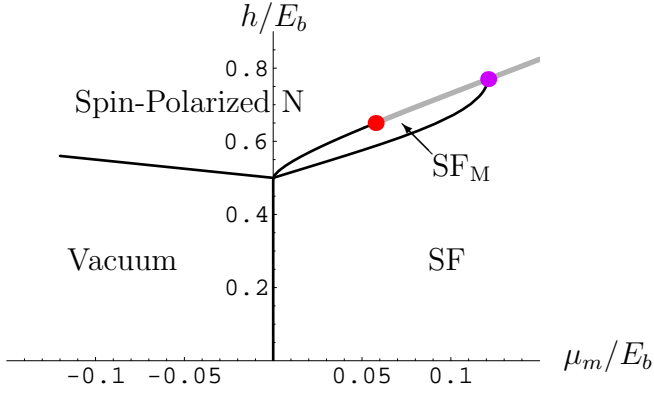}
\end{picture}}
\put(20,49.5) {\large{$h/E_b$}}
\put(63,4.5) {\large{$\mu_m/E_b$}}
\put(-3,17) {\large{Vacuum}}
\put(45,17) {\large{SF}}
\put(-2,38) {\large{Spin-Polarized N}}
\put(45.5,29.5) {\large{\sfm}}
\put(45.5,32.5) {\vector(-1,1){4}}
\end{picture}
\vspace{-.5cm}
\caption{(Color online) Phase diagram in the grand-canonical ensemble for the single-channel 
model Eq.~(\ref{eq:singlechannel}) as a function of molecular chemical potential $\mu_m$ and
atomic  chemical potential difference $h$,
 showing superfluid (SF), magnetic superfluid (\sfm), vacuum and
fully-polarized N phases. 
Black lines denote continuous $T=0$ transitions, while the gray curve denotes a 
first-order \sfm-N transition.  The red dot at (0.058,0.65) is a tricritical point separating 
first-order and second-order \sfm-N  transitions.  Beyond the purple dot at
 (0.121,0.77), the \sfm phase ceases to exist. 
}
\label{fig:phasefixedmusingle}
%
%  3NovBECLDA.nb 
%
\end{figure}
%------------------------------

To determine the phase diagram at fixed $\mu$ and $h$, it is most convenient to use 
the ground-state energy function Eq.~(\ref{eq:gsemolsingle})
as a starting point.
 We first consider the case of  $h=0$ in the SF phase.  Since
$E_G$ has the form of a conventional \lq\lq $\phi^4$\rq\rq\ theory~\cite{ChaikinLubensky}, with quadratic and quartic terms 
in the molecular field $\psi_m$, the vanishing of the quadratic coefficient signals
a second-order SF-to-Vacuum transition in a well-studied universality class~\cite{Uzunov,Sachdev}.  At $h=0$ the point 
where this occurs is given by $\Vt_{2,c} = \mu_{m,c} = 0$.  For small $h<h_m$, the same qualitative behavior,
of a SF-to-Vacuum transition at $\mu_{m,c}=0$, persists and leads to the vertical phase boundary in 
Fig.~\ref{fig:phasefixedmusingle}.  With increasing $h$, the vacuum phase undergoes a continuous transition 
to the spin-polarized N phase when $\mu_\uparrow$ becomes positive, yielding the phase boundary $h= (E_b-\mu_m)/2$.

Similarly, with increasing $h$ the SF phase acquires a nonzero magnetization, entering 
the \sfm state, when $\mg(h)$ becomes nonzero
at $h_m$ satisfying Eq.~(\ref{eq:hmsinglepre}).  Using $\mu = \frac{1}{2}(\mu_m - E_b)$ and the sixth-order
approximation for $\Delta^2$ [i.e., the analogue within the one-channel model 
of our result Eq.~(\ref{eq:becstationary}) for the two-channel model] with Eq.~(\ref{eq:hmsinglepre}) yields
the SF-\sfm phase boundary depicted in Fig.~\ref{fig:phasefixedmusingle}.

As we saw for the two-channel
model, for $h$ sufficiently close to $|\muh|$ (so that $\Vt_4>0$)  there is a continuous \sfm-N transition at 
$\Vt_2=0$, yielding the following extension of $\mu_{m,c}$ at higher $h$:
\be
\mu_{m,c} =E_b\Big( 2 - \frac{\pi}{F_2(2h/E_b)}\Big),
\label{eq:mucbecsingle}
\ee
a formula that applies to the left of the tricritical point at $(0.058,0.65)$ (where $\Vt_4$ vanishes).
Beyond this tricritical point, the fourth order expansion of the ground-state energy breaks down and 
we must incorporate the sixth-order term as we did in Sec.~\ref{SEC:detailedanalysis} for the 
two-channel model.  Of course, the close analogy between the one and two-channel models means
the sixth-order term is essentially the same.  Inclusion of this term (which
we shall derive in Sec.~\ref{SEC:ldabec} below) yields a first-order
\sfm-N transition at $h_c$ (satisfying the analogue of Eq.~(\ref{eq:firstordercondition2}) for
the one-channel model) at 
\be
\label{eq:mucritfirstsingle}
\mu_{m,c} \simeq E_b\Big(2-\frac{\pi}{F_2(2h/E_b)}\big[1- \frac{\pi^3}{256}\frac{(F_4(2h/E_b))^2}{F_6(2h/E_b)}
\big]
\Big), 
\ee
valid beyond the tricritical point 
\be
\mu_{m,c} > \mu_{m,tric.}  \simeq 0.058 E_b,
\ee
 and plotted
as a gray curve in Fig.~\ref{fig:phasefixedmusingle}.

As seen in the figure, at higher $\mu_m$, the first-order curve Eq.~(\ref{eq:mucritfirstsingle})
intersects the SF-\sfm transition curve $h_m$.  Although our small-$\Delta$ expansion of $E_G$
becomes quantitatively invalid in this regime, the existence of such an intersection at 
$\mu_m \approx 0.121 E_b$ (the purple point in Fig.~\ref{fig:phasefixedmusingle}) is correct,
as can be verified by an exact numerical minimization of the full ground-state energy.  
Beyond this point, the \sfm state ceases
to exist, and there is a direct first-order SF-N transition.

\subsection{Molecular scattering length and zeroth sound velocity}
\label{SEC:Soundvelsingle}

%-----------------------------
%
% fig%47
%
\begin{figure}[bth]
%\vspace{1cm}
\centering
\setlength{\unitlength}{1mm}
\begin{picture}(40,30)(0,0)
\put(-5,0){\begin{picture}(0,0)(0,0)
\includegraphics{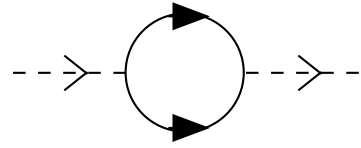}
\end{picture}}
\end{picture}
%\vspace{-.5cm}
\caption{Feynman diagram corresponding to the molecular self energy, in which the 
solid lines are  atomic  (fermion) propagators and the dashed lines are  
molecular propagators. 
}
\label{fig:selfenergy} 
%
% ** selfenergy.fig
%
\end{figure}
%------------------------------

In the present subsection, we compute the molecular scattering length $\as$ and the
corresponding  zeroth sound velocity within
the single-channel model. To do this, we need to first calculate 
the molecular dispersion induced by atom-molecule resonant interactions.  This 
requires a computation of  the leading-order contribution to the energy  due to spatial variations in $\Delta$ or 
$\psi_m$, which comes from computing the momentum-dependent part of the 
self energy (diagram Fig.~\ref{fig:selfenergy}):
\bea
\label{singleselfenergy}
&&\hspace{-1.2cm}\Sigma(q,\Omega) \!=\!  \int\frac{d^3 p}{(2\pi)^3} T\sum_\omega \frac{1}{i\omega - \xi_{p\uparrow}} 
\frac{1}{ i\omega- i\Omega + \xi_{\bq-\bp\downarrow}},
\eea
with the $\omega$ being  fermionic Matsubara frequencies  and $\xi_{p\sigma} \equiv \epsilon_p - \mu_\sigma$.  
As usual, we shall take $T \to 0$; furthermore, we only require $\Sigma(q,0)$ to leading (quadratic) order
in $q$. As in the gap equation, the short-scale (UV cutoff) dependence of 
$\Sigma(q,0)$ can be absorbed into the experimentally-measured scattering length $\as$.
Direct evaluation of Eq.~(\ref{singleselfenergy}) then yields (to quadratic order in $q$)
\bea
&&\hspace{-1.2cm}-\frac{1}{\lambda} + \Sigma(q,0) = -\frac{m}{4\pi \as} + c\sqrt{\mu} F_2(h/|\mu|)
\nonumber\\
&&\hspace{-1.2cm}\qquad \qquad + q^2\frac{c}{32m |\mu|^{1/2}}F_q(h/|\mu|),
\label{singleselfenergy2}
\eea
where
\bea
&&\hspace{-1cm}F_q(x) \equiv 1- \big[\frac{2}{\pi} \tan^{-1}\sqrt{x-1} 
\nonumber \\
&&\hspace{-1cm}\qquad \qquad 
- \frac{4}{3\pi} \sqrt{x-1} 
\big(\frac{1}{x^2} + \frac{1}{2x}
\big)
\big]\Theta(x-1),
\eea
and we recall that  $F_2(x)$ is given by Eq.~(\ref{eq:f2def}).  Clearly, 
the first two terms of Eq.~(\ref{singleselfenergy2}) simply represent the previously-computed 
coefficient $-V_2$ in 
Eq.~(\ref{eq:egbecsingle}).  The $q$-dependent part is new, and represents the energetic
cost of spatial variations of $\Delta$, or, more physically, the molecular kinetic energy.
 Thus, the generalization of $E_G$, Eq.~(\ref{eq:egbecsingle}), (restoring the system
volume $\vol$) to the case 
of a spatially-varying $\Delta$ is simply given by (neglecting an overall $\Delta$-independent constant term):
\be
\label{eq:egbecsinglespatial}
E_G \approx 
\int d^3 r \Big[
  \frac{c F_q(h/|\mu|)}{32m|\mu|^{1/2}} |\grad\Delta|^2  
- V_2 |\Delta|^2  + \frac{1}{2} V_4 |\Delta|^4\Big],
\ee
that in terms of $\psi_m$ reduces to the standard Ginzburg-Landau form [using the relation of $\Delta$ to 
$\psi_m$, Eq.~(\ref{eq:psim})]
\bea
\label{eq:egbecsinglespatial3}
E_G \approx 
\int d^3 r \Big[
 \frac{|\grad \psi_m|^2}{2m_b}
- \Vt_2 |\psi_m|^2 + \frac{1}{2} \Vt_4  |\psi_m|^2\Big].
\eea
The molecular mass $m_b$ is defined by 
\be
\frac{1}{m_b} = \frac{F_q(2h/E_b)}{2m},
\ee
and therefore is not simply $2m$ but is enhanced through the function $F_q(2h/E_b)$.
 Note that, unlike
the function $F_4(x)$ that vanishes indicating the instability of the \sfm phase,
$F_q(x)$ is positive in the regime of interest $x>1$.

This expression for $m_b$, together with $\Vt_4$, allows us to deduce the molecular 
scattering length $a_m$
\bea
\Vt_4 &=& \frac{8\pi \as F_4(2h/E_b)}{F_q(2h/E_b)m_b} ,
\\
&\equiv&  \frac{4\pi a_m}{m_b},
\eea
where in the second line we used the well-known relation~\cite{Fetter} between the quartic
coefficient and the molecular scattering length $a_m$ to identify
\be
\label{eq:amsingle}
a_m(h) =  2\as \frac{F_4(2h/E_b)}{F_q(2h/E_b)},
\ee
that decreases with increasing $h$ as we found in the two-channel model. 
We note here that, because Eq.~(\ref{eq:amsingle}) is computed only within 
the Born approximation (equivalent to our one-loop mean-field theory),
it is certainly not quantitatively accurate. This is consistent with
our caveat, made at the start of this section,  about the uncontrolled accuracy of the mean-field approximation
on the one-channel model.  The level of this quantitative inaccuracy can be assessed
by noting that our result for $a_m(h)$ reduces to~\cite{sademelo}  $a_m(h=0) = 2\as$
that is  known to be off the exact value in the dilute BEC limit
of $a_m(h=0) = 0.6 \as$ found by Petrov et al.~\cite{Petrov} 
(see also Refs.~\onlinecite{Brodsky,Levinsen}),
with the difference arising from strong molecular fluctuations about the mean-field theory
 solution~\cite{Levinsen}.
 Finding the generalization of the formula $a_m = 0.6 \as$ to finite chemical-potential 
difference [which would give the exact form of Eq.~(\ref{eq:amsingle})]
remains a challenging open problem.
%%%%

Using Eq.~(\ref{eq:amsingle}), it is straightforward to find the sound velocity $u$ inside  the 
SF and \sfm phases using the standard expression~\cite{Fetter} 
\be
u^2 = \frac{4\pi n_m a_m}{m_b^2}.
\ee
Following the analysis  of  Sec.~\ref{SEC:sound} and  using Eq.~(\ref{eq:amsingle}), we find 
\be
u^2 \simeq \frac{\pi n \as}{m^2} \big[1-\frac{1}{2} \big(\frac{h}{\ef} - \frac{E_b}{2\ef}\big)^{3/2}
\Theta(2h-E_b)\big]
F_q(\frac{2h}{E_b})F_4(\frac{2h}{E_b}),
\ee
at fixed chemical potential difference $h$, or, at fixed polarization, 
\bea
&&u^2 \simeq \frac{\pi n \as}{m^2}(1-\frac{\Delta N}{N})
F_q\big[(\kf \as)^2
\big(\frac{2\Delta N }{N}\big)^{2/3}
+1\big]\nonumber 
\\
&&\qquad\qquad\times
F_4\big[(\kf \as)^2
\big(\frac{2\Delta N }{N}\big)^{2/3}
+1\big].
\label{finalsinglesfvel}
\eea
With increasing $\Delta N$, there are two ways $u(\Delta N)$ can vanish according to 
Eq.~(\ref{finalsinglesfvel}): For moderate detunings, $(\kf \as)^{-1}$ close to unity, it
vanishes when the effective molecular interaction proportional to the 
 function $F_4$ vanishes at $\Delta N_{c1}$ approximately given by Eq.~(\ref{eq:mconesinglefin}), 
signaling the first-order transition to the regime of phase separation.
In contrast, for large negative detunings, deep in the BEC regime $\frac{1}{\kf \as} >\frac{1}{\kf a_{sc}}$,
 the velocity
$u(\Delta N)$ vanishes due to the vanishing molecular density $n_m \propto (1-\Delta N/N)$,
before $F_4$ does. In this regime the vanishing of $u$ 
simply signals the second-order transition to the normal state.  

This concludes our description of the single-channel model Eq.~(\ref{eq:singlechannel})
at finite chemical potential difference $h$ in the BEC regime.  Next, we study the FFLO regime
of the single-channel model, confined to the positive-detuning BCS regime.

\subsection{FFLO state in the single-channel model}
\label{SEC:FFLOsingle}
In the present subsection, we analyze the FFLO ground state  for the single-channel model 
Eq.~(\ref{eq:singlechannel}).  As in the preceding subsections, we shall be brief in 
our analysis as details closely resemble those for the two-channel model discussed in 
Sec.~\ref{SEC:FFLO}. For simplicity~\cite{SingleQ}, we take the single-$\bQ$ 
FF pairing ansatz for the ground state of the form $\Delta(\br) = \Delta_\bQ {\rm e}^{i \bQ \cdot \br}$.
 Following
our previous analysis, the ground state energy for this ansatz
is  [as before,  exchanging the coupling $\fermiint$ for the s-wave scattering length $\as$ using 
Eq.~(\ref{eq:singlechannelintroas})]:
\bea
&&E_G = -\frac{m}{4\pi \as}\Delta_\bQ^2 + 
\sum_k ( \txi_k - \te_k  + \frac{\Delta_\bQ^2}{2\epsilon_k}) \label{eq:havgsingle}
 \\ 
&& 
 \qquad + \sum_\bk \big[
E_{\bk \uparrow} \Theta(-E_{\bk \uparrow})+ E_{\bk \downarrow} \Theta(-E_{\bk \downarrow})
\big],\nonumber
\eea
where we recall that $\te_k$ is
given by Eq.~(\ref{eq:excitationenergy}),  $\txi_k$ is given by Eq.~(\ref{eq:tildexi}),
and $E_{\bk\sigma}$ is given by Eqs.~(\ref{ekuparrow}) and (\ref{ekdownarrow}).

As in the two-channel case, the FFLO state in the single-channel model is described (in
the grand-canonical ensemble) by the gap ($\partial E_G/\partial\Delta_\bQ =0$)
and momentum equations ($\partial E_G/\partial Q = 0$), that, using Eq.~(\ref{eq:havgsingle}), 
give
\bse
\label{eq:gapmomfflosingle}
\bea
&&0= -\frac{m}{2\pi \as} - \sum_k\big( \frac{1}{\te_k} - \frac{1}{\epsilon_k}\big)
\nonumber \\
 &&\qquad+ \sum_k  \frac{1}{\te_k} (1+\Theta(-E_{\bk\uparrow}) - \Theta(E_{\bk\downarrow})),
\label{eq:gapfflosingle}
\\
&&0= \frac{Q}{2}\sum_k \big[1+\frac{\txi_k}{\te_k} \big(\Theta(-E_{\bk\uparrow}) - \Theta(E_{\bk\downarrow})\big)\big]
\nonumber \\
&&\qquad -\frac{1}{Q}\sum_k \bk \cdot \bQ \big(\Theta(-E_{\bk\uparrow}) + \Theta(E_{\bk\downarrow})\big).
\label{eq:momfflosingle}
\eea
\ese
We remark that the right side of Eq.~(\ref{eq:momfflosingle}) can also be obtained by computing the expectation value
of the momentum operator with respect to our variational ground-state wavefunction~\cite{Takada}.

We now proceed to evaluate the momentum sums in Eq.~(\ref{eq:gapmomfflosingle}), focusing on the
small-$\Delta_\bQ$ limit.
 Starting with the gap equation, the first sum in Eq.~(\ref{eq:gapfflosingle}) is 
\be
\label{eq:sumfflogapsingleone}
\sum_k \big( \frac{1}{\te_k} - \frac{1}{\epsilon_k}\big) \simeq 2N(\tilde{\mu})\ln \frac{8{\rm e}^{-2} \tilde{\mu}}{\Delta_\bQ},
\ee
while the second, excluded, sum can be evaluated following the procedure of Appendix~\ref{app:ex}. We find,
assuming from the outset that we are in the doubly-depaired FFLO state occuring near the second-order 
FFLO-N transition at $h_{\rm FFLO}$~\cite{ff}:
\bea
\label{eq:sumfflogapsingletwo}
&& 
\sum_k \frac{1}{\te_k}  (1+\Theta(-E_{\bk\uparrow}) - \Theta(E_{\bk\downarrow}))
\nonumber \\
&& \qquad \qquad 
 \simeq \frac{N(\tilde{\mu})}{\Qb} \big[
G(\Qb+\bar{h}) + G(\Qb-\bar{h}) 
\big],
\eea
where $\bar{Q}\propto 1/\Delta_\bQ$ is given by Eq.~(\ref{eq:qbar}) and 
\be
G(x) \equiv (x\cosh^{-1} x - \sqrt{x^2-1})\Theta(x-1).
\ee
With this result, the gap equation Eq.~(\ref{eq:gapfflosingle}) becomes 
\bea
&& 0 = -\frac{m}{2\pi \as} - 2N(\tilde{\mu}) \ln \frac{8{\rm e}^{-2} \tilde{\mu}}{\Delta_\bQ} 
\nonumber \\ 
&& +\frac{N(\tilde{\mu})}{\Qb}\big[
G(\Qb+\bar{h}) + G(\Qb-\bar{h})
\big].
\label{eq:gapfflosingle2pre}
\eea
Near the second-order transition at $\hfflo$, we can expand the right side of Eq.~(\ref{eq:gapfflosingle2pre}) in
small $\Delta_\bQ$, using  Eqs.~(\ref{eq:sumfflogapsingleone}) and (\ref{eq:sumfflogapsingletwo}).  
To leading order, we find 
\bea
&& 0 = -\frac{m}{2\pi \as} - 2N(\tilde{\mu}) \ln \frac{8{\rm e}^{-2} \tilde{\mu}}{\Delta_\bQ} 
 + N(\tilde{\mu}) \Big(
\ln 4(\Qb^2 - \bar{h}^2) 
\nonumber \\
&& \qquad
+ \frac{\bar{h}}{\Qb} \ln \frac{\Qb+\bar{h}}{\Qb-\bar{h}} -2
+ \frac{1}{2}\frac{1}{\Qb^2-\bar{h}^2}
\Big).\label{eq:gapfflosingle2}
\eea
Expressing Eq.~(\ref{eq:gapfflosingle2}) in terms of the $h=0$ gap $\deltabcs$ in the 
single-channel model using
\be
\label{eq:gapequationsingle}
\frac{-m}{4\pi \as} = N(\mu) \ln \frac{8{\rm e}^{-2}\mu}{\deltabcs},
\ee
using $\tilde{\mu} \equiv \mu - Q^2/8m$, and switching to our
 dimensionless variables [Eqs.~(\ref{eq:dimensionlessvariables}) and (\ref{eq:pdef})]  
we have, to leading order in $\Deltah_\bQ$ and the 
dimensionless momentum $\Qh$, 
\bea
&&\Deltah_\bQ^2 \simeq 2(\Qh^2 - \hh^2)\big[
2 -  \ln \frac{4(\Qh^2 - \hh^2)}{\deltahbcs^2} - \frac{\hh}{\Qh}\ln \frac{\Qh+\hh}{\Qh-\hh} 
\nonumber \\
&&\qquad  \qquad  \qquad  
- \frac{\Qh^2}{8\muh^2}\ln \frac{8\muh}{\deltahbcs}
\big],
\label{gapffsingle}
\eea
which describes the magnitude of the pairing order parameter $\Delta_\bQ$ in the FFLO regime near $\hfflo$. 

Turning to the momentum equation, we split the first line of Eq.~(\ref{eq:momfflosingle}) into two sums:
\bea
&&\hspace{-.5cm}\sum_k \big[1+\frac{\txi_k}{\te_k} \big(\Theta(-E_{\bk\uparrow}) - \Theta(E_{\bk\downarrow})\big)\big]
\label{eq:intsumssinglefflo}
 \\
&&\hspace{-.5cm}
 = \sum_k\Big(1- \frac{\txi_k}{|\txi_k|} \Big) + 
\sum_k \Big( \frac{\txi_k}{|\txi_k|} +  \frac{\txi_k}{\te_k}\big(\Theta(-E_{\bk\uparrow}) - \Theta(E_{\bk\downarrow})\big)
\Big). 
\nonumber
\eea
The first sum can be easily evaluated exactly:
\be
\sum_k\Big(1- \frac{\txi_k}{|\txi_k|} \Big) = \frac{4}{3}N(\tilde{\mu})\tilde{\mu}.
\ee
 For the second sum in Eq.~(\ref{eq:intsumssinglefflo}), we make a standard approximation, replacing the 
density of states by its value at the Fermi surface, valid for a degenerate Fermi gas 
(appropriate in the BCS limit).  This gives:
\bea
&&\sum_k \big[1+\frac{\txi_k}{\te_k} \big(\Theta(-E_{\bk\uparrow}) - \Theta(E_{\bk\downarrow})\big)\big]
\nonumber \\
&& \qquad \simeq
\frac{4}{3}N(\tilde{\mu})\tilde{\mu} + N(\tilde{\mu})( \sqrt{\tilde{\mu}^2+\Delta_\bQ^2}-\tilde{\mu}),
\nonumber \\
&&
 \qquad \simeq
\frac{4}{3}c\tilde{\mu}^{3/2} + \frac{c\Delta_\bQ^2}{2\sqrt{\tilde{\mu}}}, 
\eea
with the last line applying for $\Delta_\bQ\ll\tilde{\mu}$.
  The second momentum sum  of 
Eq.~(\ref{eq:momfflosingle}) can also be evaluated within this degenerate  Fermi gas approximation, 
yielding
\bea
&&\hspace{-.8cm}\frac{1}{Q}\sum_k \bk \cdot \bQ \big(\Theta(-E_{\bk\uparrow}) + \Theta(E_{\bk\downarrow})\big)
\nonumber \\
&&\hspace{-.5cm} \simeq \frac{\Delta_\bQ \tilde{k}_{\rm F} N(\tilde{\mu})}{6\Qb^2} \Big[2[(\Qb+\bar{h})^2-1]^{3/2}
%\Theta(\Qb+\bar{h}-1)
-3\bar{h}\gamma(\Qb+\bar{h})
\nonumber \\
&&\hspace{-.5cm}\qquad 
+2[(\Qb-\bar{h})^2-1]^{3/2}
%\Theta(\Qb-\bar{h}-1)
+3\bar{h}\gamma(\Qb-\bar{h})\big],
\eea
where $\gamma(x) \equiv (x\sqrt{x^2-1}-\cosh^{-1} x)\Theta(x-1)$.  
Combining these terms yields for the number equation Eq.~(\ref{eq:momfflosingle}) (approximating $\tilde{k}_{\rm F}  \approx k_{\rm F}$)
\bea
&&0\simeq \frac{4}{3}\frac{m\Delta_\bQ \Qb\mu}{\kf} + \frac{m\Delta_\bQ^3\Qb}{2\kf \mu} 
 \\
&&  - \frac{\Delta_\bQ \kf}{6\Qb^2} \Big[2[(\Qb+\bar{h})^2-1]^{3/2}
%\Theta(\Qb+\bar{h}-1)
-3\bar{h}\gamma(\Qb+\bar{h})
\nonumber \\
&& \qquad 
+2[(\Qb-\bar{h})^2-1]^{3/2}
%\Theta(\Qb-\bar{h}-1)
+3\bar{h}\gamma(\Qb-\bar{h})\big],\nonumber
\eea
which, near $\hfflo$ (and switching to our standard dimensionless variables), yields
\be
\label{momffsingle} 
0\simeq 1- \frac{\hh}{2\Qh}\ln \frac{\Qh+\hh}{\Qh-\hh}+\frac{\Qh^2}{4\muh^2}.
\ee
%

%%%%%%%%%%%%%%%%%%%%%%%%%%%%%%%%%%%%%%%%%%%%%%%%%%%%%%%%

Our subsequent analysis of Eqs.~(\ref{gapffsingle}) and ~(\ref{momffsingle})  mirrors that of 
the two-channel gap and momentum equations [Eq.~(\ref{eq:gapff}) and Eq.~(\ref{momsecond}), 
respectively] from Sec.~\ref{SEC:FFLO}. We are primarily interested in 
the width of the FFLO regime in the phase diagram as a function of $(\kf |\as|)^{-1}$, 
which at fixed $\hh$ is bounded above by the second-order FFLO-N transition at $\hfflo$
and below by the first-order SF-FFLO transition.   
We express the gap and momentum equations
 in terms of the parameter $\lambda$, defined by $\Qh = \lambda \hh$
\bea
&&\hspace{-.5cm}\Deltah_\bQ^2 \simeq 2\hh^2(\lambda^2 - 1)\Big[
2 -  \ln \frac{4\hh^2(\lambda^2 - 1)}{\deltahbcs^2} 
\nonumber \\
&&\hspace{-.5cm}\qquad \qquad  \qquad 
- \frac{1}{\lambda}\ln \frac{\lambda+1}{\lambda-1} 
- \frac{\lambda^2\hh^2}{8\muh^2}\ln \frac{8\muh}{\deltahbcs}
\Big],
\label{gapffsingle2} 
\\
\label{momffsingle2} 
&&0\simeq 1- \frac{1}{2\lambda}\ln \frac{\lambda+1}{\lambda-1}+\frac{\lambda^2\hh^2}{4\muh^2}.
\eea

As we saw in Sec.~\ref{SEC:FFLO}, in the asymptotic BCS regime $(\kf |\as|)^{-1}\gg 1$, upon dropping
the exponentially small last term $\propto \hh^2$, Eq.~(\ref{momffsingle2}) is solved by
 $\lambda \approx 1.200$.~\cite{ff}  Inserting this $\lambda$ value into Eq.~(\ref{gapffsingle2})
and neglecting the subdominant last term of Eq.~(\ref{gapffsingle2}), and setting $\Deltah_\bQ = 0$,
we find~\cite{ff}
\be
\label{eq:hfflosingle}
\hfflo(\muh) \simeq \eta \deltahbcs,
\ee
with $\eta = 0.754$, as for the deep-BCS two-channel model.
This result 
applies for small $\kf |\as|$ in the BCS regime at fixed chemical potential.  Combining this with the number equation (which 
fixes  $\muh$ at $\muh \approx 1- \frac{\hfflo^2}{4} \approx 1- \frac{\eta^2\Deltah_F^2}{4})$
yields for $\hfflo$ and $\Delta N_{\rm FFLO}$:
\bse
\bea
\hfflo &\simeq & \eta \Deltah_F \exp\big[-
\frac{\pi\eta^2\Deltah_F^2}{16 \kf |\as|}
\big],
\\
\frac{\Delta N_{\rm FFLO}}{N} &\simeq & \frac{3}{2}\sqrt{\muh} \hfflo ,
\nonumber\\
&\simeq & \frac{3\eta}{2} \Deltah_F  \exp\big[-
\frac{\pi\eta^2\Deltah_F^2}{16 \kf |\as|}
\big]. 
\eea
\ese

However, 
as we showed in the two-channel case, close to the resonance $\hfflo$ approaches  $\hc  \approx 
\deltahbcs/\sqrt{2}$, with reduced $(\kf |\as|)^{-1}$, crossing it at a critical coupling strength $\kf a_{s*}$
that we now determine.  
To do this, we first re-express the last term in square brackets in Eq.~(\ref{gapffsingle2})
in terms of $\as$ using Eq.~(\ref{eq:gapequationsingle}), after setting $\Deltah_\bQ = 0$:
\bea
&&0 \simeq
2 -  \ln \frac{4(\lambda^2 - 1)}{\deltahbcs^2} - \frac{1}{\lambda}\ln \frac{\lambda+1}{\lambda-1} 
\nonumber \\
&&\qquad \qquad 
- \frac{\hfflo^2\lambda^2}{4\muh^2} 
+ \frac{\pi \lambda^2 \hfflo^2}{16\muh^2 \kf \as \sqrt{\muh}}.
\label{gapffsingle3} 
\eea
Numerically solving Eqs.~(\ref{gapffsingle3})  and ~(\ref{momffsingle2})
for $\hfflo(\kf \as)$, we find that $\hfflo$ crosses $\hc$ at   
\be
\frac{1}{\kf |a_{s*}|} \approx 0.46,
\ee
defining the abovementioned critical coupling strength beyond which the FFLO state ceases to be stable within our
mean field theory (as plotted in the phase diagram Fig.~\ref{fig:hsingle}).  Although the location of
this crossing is not guaranteed to be accurate (due to the absence of a small parameter to justify 
mean field theory near unitarity), we do expect the existence of the crossing to survive beyond mean-field
theory.

\section{Polarized superfluidity in a trap: Local Density Approximation}
\label{SEC:LDA}

The primary experimental application of our results on polarized paired superfluidity is 
that of trapped degenerate atomic gases.  It is thus crucial to extend our
results to take into account the effect of the potential $V_T(\br)$,
that in a typical experiment is well-approximated by a harmonic-oscillator potential.
While a full analysis of the effect of the trap is beyond the scope of this manuscript, in the
present section we study this problem within the well-known local density approximation (LDA).
We note that several recent studies 
(e.g., Refs.~\onlinecite{Pieri,Torma,Yi,Chevy,DeSilva,Haque,Imambekov})
 have also addressed polarized superfluidity in a trap.

For simplicity, and because of its more direct current experimental relevance, 
 in this section we focus on the single-channel model Eq.~(\ref{eq:singlechannel}).  
The generalization of this model to a trap is straightforward:
\bea
&&H = \sum_{\sigma=\uparrow,\downarrow}\int d^3r  \Big(\frac{|\grad \ch_\sigma(\br)|^2}{2m} +
(V_T(\br) - \mu_\sigma)
|\ch_\sigma(\br)|^2 
\Big)
\nonumber \\
&&\qquad \qquad
 + \fermiint\int d^3 r \ch_\uparrow^\dagger \ch_\downarrow^\dagger \ch_\downarrow^\phdag \ch_\uparrow^\phdag,
\label{eq:singlechannellda}
\eea
where $\ch_\sigma(\br)$ is a fermionic field operator with Fourier transform
$\ch_{\bk\sigma}$.
Henceforth, to be concrete, we shall focus on an isotropic harmonic trap 
$V_T(\br) = V_T(r)= \frac{1}{2} m \Omega_T^2 r^2$, although this simplification
can easily be relaxed.
Within LDA (valid for a sufficiently smooth trap potential $V_T(r)$, see our discussion in 
the introduction, Sec.~\ref{SEC:tf}), 
locally
the system is taken to be  well approximated as {\it uniform\/}, but with a local chemical
potential given by
\be
\label{eq:mulda}
\mu(r) \equiv \mu - \frac{1}{2} m \Omega_T^2 r^2, 
\ee
where the constant $\mu$ is the true chemical potential (a Lagrange multiplier) still enforcing the total 
atom number $N$. 
The spatially-varying spin-up and spin-down local chemical potentials are then [c.f. Eq.~(\ref{eq:mudefs})]:
\bea
\mu_\uparrow(r) &=& \mu(r) +h ,
\\
\mu_\downarrow(r) &=& \mu(r) - h,
\eea
with the chemical potential difference $h$ {\it uniform\/}.
Thus, within LDA we approximate the system's energy density by that of a uniform system with 
spatial dependence (via the trap) entering only through $\mu(r)$.  The ground state energy is then
simply a volume integral of this energy density.  Below we compute the resulting ground-state energy
and analyze spatial profiles that emerge from it throughout the phase diagram.  Within LDA, the phase
behavior as a function of chemical potential, $\mu$, translates into a spatial cloud profile through
$\mu(r)$, with critical phase boundaries $\mu_c$ corresponding to critical radii defined by
$\mu_c = \mu(r_c,h)$.~\cite{LDAvalidity}  As we first predicted~\cite{shortpaper}, this leads to a shell-like cloud
structure that has subsequently been observed experimentally~\cite{Zwierlein05,Partridge05,Shin} and reproduced theoretically
by a number of works~\cite{Torma,Chevy,Haque,DeSilva}. 

Below we study these shell structures in much more detail using LDA.  We note, however, that throughout
our discussion, sharp (discontinuous) features (like the shell structure) that are arise are an 
artifact of LDA (precisely where it is invalid) and are expected to be smoothed on microscopic (Fermi wave-)
length scales by the kinetic energy (or, surface tension~\cite{DeSilva06}).

\subsection{BCS regime}

As we have seen, for a bulk system in the BCS regime ($\as<0$) there are three possible homogeneous phases:
1) the singlet superfluid phase (SF), which is paired and has zero local magnetization 2) the normal phase (N),
and 3) the FFLO phase, which exhibits both
pairing and local magnetization, but is only stable  for a narrow window of chemical potential 
difference $\deltabcs/\sqrt{2} \alt h \alt0.754 \deltabcs$.  The narrowness of the window of FFLO
phase translates, within LDA, to a thin shell  $r_c<r<r_c+\delta r$ of FFLO phase in a trap.  Although the formalism
that we shall now present can be easily generalized to find this shell, 
we believe that 
the LDA approximation (which relies on slow variations of physical quantities) 
is not quantitatively trustworthy for such a thin region, especially considering
that the FFLO state itself varies
over a large length scale $Q^{-1} \approx (k_{{\rm F}\uparrow} - k_{{\rm F}\downarrow})^{-1}$.  
This is not to say that the FFLO state is not observable in a trap, but merely that theoretical
study of the FFLO state in a trap will require a more sophisticated technique than LDA~\cite{Torma}.
Thus, 
in the following analysis we shall generally neglect the FFLO phase, briefly returning to it at the end of
this section to estimate the expected width $\delta r$ of the FFLO phase in  a trap within LDA.
 
Taking advantage of our bulk results at fixed $\mu$ and $h$,
in the SF state, the mean-field LDA ground-state energy is given by
\bea
&&\hspace{-.5cm}E_{G,SF} = \int d^3r \Big[-\frac{8c}{15} \mu(r)^{5/2} -\frac{m}{4\pi \as} \Delta(r)^2 
\label{eq:gsesflda}
\\
&&\hspace{-.5cm}\qquad 
+ N(\mu(r))\big(
-\frac{1}{2}\Delta(r)^2 + \Delta(r)^2\ln \frac{\Delta(r)}{8{\rm e}^{-2}\mu(r)}
\big)\Big],\nonumber 
\eea
while in the N state it is given by
\bea
\hspace{-.6cm}E_{G,N} &=& -\frac{4c}{15}\int d^3r \Big[ (\mu(r)+h)^{\frac{5}{2}}+
   (\mu(r)-h)^{\frac{5}{2}} \Big],
\label{eq:gsenlda}
\\
&\approx& 
\int d^3r \Big[-\frac{8c}{15}\mu(r)^{5/2} -c\sqrt{\mu(r)}h^2 \Big],\label{eq:gsenlda2}
\eea
with the second line applying for $h\ll\mu(r)$ and we recall the dimensionful parameter $c$ is defined
in Eq.~(\ref{eq:cdef}).
The pairing field $\Delta(r)$ that locally minimizes Eq.~(\ref{eq:gsesflda}) within LDA is then simply given by 
\be
\label{eq:bcsgaplda}
\Delta(r)
\equiv 8{\rm e}^{-2} \mu(r) \exp\big[\frac{m}{4\pi \as N(\mu(r))} \big] .
\ee
Plugging this into Eq.~(\ref{eq:gsesflda}) yields 
\be
E_{G,SF} = \int d^3 r \Big[-\frac{8c}{15} \mu(r)^{5/2} -\frac{1}{2} c\sqrt{\mu(r)} \Delta(r)^2\Big],
\label{eq:gsesflda2}
\ee
so that, by comparing $E_{G,SF}$, Eq.~(\ref{eq:gsesflda2}), with $E_{G,N}$, Eq.~(\ref{eq:gsenlda2}), we find
 the critical chemical potential
difference 
\bea
h_c(r) &=&\frac{\Delta(r)}{\sqrt{2}},
\nonumber
\\
&=&4\sqrt{2}{\rm e}^{-2} \mu(r) \exp\big[\frac{m}{4\pi \as N(\mu(r))} \big],
\eea
at which the SF and N states {\it locally\/} have the same energy.  Thus, within LDA,
at fixed $h$ any regions of the system that satisfy $h<h_c(r)$ are in the SF state while those that 
satisfy $h>h_c(r)$ will be in the N state. Since $\mu(r)$ in Eq.~(\ref{eq:mulda}) decreases with increasing $r$ 
it is clear that, within LDA, the higher-density superfluid regions
 will be confined to the center of the trap (where $\mu$ and 
thus $h_c$ is largest), with the lower-density polarized N state expelled to the outside,
as illustrated in Fig.~\ref{fig:magplotlda}.  The resulting shell structure with radius 
$r_c(h)$ of the SF-N interface is implicitly given by 
\be
h  = 4\sqrt{2}{\rm e}^{-2} \mu(r_c) \exp\big[\frac{m}{4\pi \as N(\mu(r_c))} \big],
\label{eq:rclda}
\ee
a striking signature of the regime of phase separation in a trap. 

To describe the shell structure of the regime of phase separation for positive detuning at fixed atom number 
$N_\sigma$,
we  exchange  $\mu$ and $h$ for the total atom number $N = N_\uparrow + N_\downarrow$ and normalized 
population difference $\frac{\Delta N}{N} = \frac{N_\uparrow - N_\downarrow}{N_\uparrow + N_\downarrow}$.  
To do this we first note that
the local {\it total\/} atom density  in the SF and N phases is given, respectively, 
by [c.f.~Eqs.~(\ref{eq:numbermixed1}) and (\ref{eq:numbermixed2})]
\bse
\label{eq:numslda}
\bea
\label{eq:nsflda}
&&\hspace{-1.4cm} n_{SF}
  \simeq \frac{4c}{3} \mu(r)^{\frac{3}{2}} \! + \! \frac{5c\Delta(r)^2}{4\sqrt{\mu(r)}}  \!
-  \!\frac{c\Delta(r)^2}{2\sqrt{\mu(r)}}\ln\frac{\Delta(r)}{8{\rm e}^{-2} \mu(r)},
\\
&&\hspace{-1.4cm}
n_{N}  = n_{N\uparrow}(r) + n_{N\downarrow}(r),
\\
&&\hspace{-1.15cm}
\quad \simeq
\frac{4c}{3} \mu(r)^{\frac{3}{2}} + \frac{ch^2}{2\sqrt{\mu(r)}},
\label{eq:nnlda}
\eea
\ese
where the local density of spin-$\sigma$ atoms  in the N state is $n_{N\sigma}(r) \equiv \frac{2c}{3}\mu_\sigma(r)^{3/2}$.
Equations~(\ref{eq:numslda}) can also be 
easily obtained by functionally differentiating Eq.~(\ref{eq:gsesflda}) and Eq.~(\ref{eq:gsenlda2}) with respect 
to $\mu(r)$.  To determine the chemical potential
$\mu$ in the BCS regime, we note that the above 
expressions are each well-approximated by their first terms (which are identical).
This underscores the weakness of pairing $\Delta$ and of the corresponding depairing field 
$h_c$ in the BCS regime, with only a small fraction of states near the Fermi surface
paired, slightly modifying $\mu$.
We first consider the singlet BCS SF at $h=0$.  Since it is unmagnetized, $N_\uparrow = N_\downarrow$
and  the total atom number is given by
\be
\label{eq:nldapre}
N = \frac{4c}{3} \int d^3 r (\mu - \frac{1}{2} m \Omega_T^2 r^2)^{3/2},
\ee
where the integration is restricted to  $r<R_0$, with the Thomas-Fermi (TF) radius 
\be
\label{eq:rlda}
R_0 \equiv \sqrt{\frac{2\mu}{m\Omega_T^2}},
\ee
defined by where $\mu(r)$ 
vanishes and approximately delineating the edge of the system~\cite{LDAvalidity}.  
We shall see below that, at $h\neq 0$, the 
spin-up and spin-down clouds actually have different Thomas-Fermi radii $R_\uparrow \neq R_\downarrow$.  Evaluating the integral, 
we find
\be
\label{eq:nlda}
N = \frac{\pi^2 c}{6} \mu^{3/2} R_0^3 = \frac{\pi^2 c}{6} \mu^3 
\big(\frac{2}{m\Omega_T^2}\big)^{3/2},
\ee
that gives $\mu = (3N)^{1/3} \hbar \Omega_T$ (a result that is valid 
beyond LDA~\cite{Butts}), with Eqs.~(\ref{eq:rlda}) and
Eq.~(\ref{eq:nlda}) valid approximations for a cloud at small polarization.

\begin{figure}[bth]
\vspace{1.4cm}\hspace{-1.7cm}
\centering
\setlength{\unitlength}{1mm}
\begin{picture}(40,40)(0,0)
\put(-50,0){
\begin{picture}(0,0)(0,0)
\includegraphics{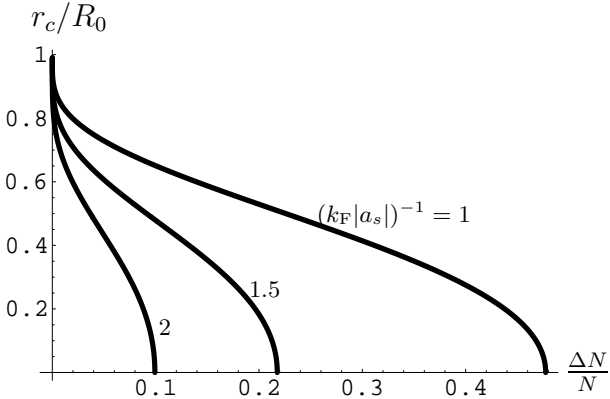}
\end{picture}}
\put(61,4.0) {\large{$ \frac{\Delta N}{N}$}}
\put(-10,51) {\large{$r_c/R_0$}}
\put(28,25) {{$ (\kf |\as|)^{-1} = 1  $}}
\put(19,15) {{$1.5$}}
\put(7,10) {{$2$}}
\end{picture}
\vspace{-.5cm}
\caption{Radius $r_c$ (normalized to the TF radius $R_0$) of the SF cloud core
as a function of imposed normalized population difference on the positive detuning 
BCS side of the resonance for coupling
strengths (by which they are labeled) $(\kf |\as|)^{-1} =2$, $(\kf |\as|)^{-1} =1.5$, and 
 $(\kf |\as|)^{-1} =1$. }
\label{fig:radiusldaplots}
%
%   25OctLDAnumbers2.nb 
%
\end{figure}

Next, we turn to the mixed state where $\mu>\mu_c(r,h)>0$, and, as a result, the 
abovementioned SF-N shell structure develops. Thus, for $r<r_c$ the system consists
of a SF core and for $r_c<r<R$ the system consists of a polarized N Fermi gas.  
Thus, since $n_\uparrow = n_\downarrow$ in the SF sphere,  the total
$N_\uparrow - N_\downarrow$  is determined by the difference of the total number of 
spin-up and spin-down atoms  in the normal shell outside $r_c$, which we 
label by $N_{N\uparrow}$ and $N_{N\downarrow}$.  The corresponding atom numbers in these
normal shells are given by:
\be
\label{eq:nsigma}
N_{N\sigma} = \frac{2c}{3} \int_{r_c<r<R_\sigma} d^3 r (\mu_\sigma - \frac{1}{2} m \Omega_T^2 r^2)^{3/2},
\ee
that lead to distinct Thomas-Fermi radii for spin $\uparrow$ and $\downarrow$:
\be
\label{eq:rsigma}
R_\sigma = \sqrt{\frac{2\mu_\sigma}{m\Omega_T^2}},
\ee
with the majority (taken as spin-$\uparrow$, for $h>0$) population occupying the larger volume.
Evaluating the integral in Eq.~(\ref{eq:nsigma}), we find 
\bse
\bea
\label{eq:nsigmalda}
&&\hspace{-1.6cm}N_{N\sigma} =  \frac{\pi^2 c}{12} \mu_\sigma^{3/2} R_\sigma^3\big[1-f(r_c/R_\sigma)],
\eea
with
\bea
&&\hspace{-1.6cm}f(x) \equiv \frac{2}{3\pi}\!\big[3\sin^{-1}x \!-\! x\sqrt{1-x^2}(8x^4\!-\!14x^2\!+\!3)
\big].
\eea
\ese
Equation~(\ref{eq:nsigmalda}) counts the number of spin-$\sigma$ atoms in a spherical shell
of inner radius $r_c$ and outer radius $R_\sigma$. 
For $0<x<1$, $f(x)$ is a monotonically increasing function of $x$ with $f(0) = 0$ and $f(1) = 1$ (so
that a shell of vanishing width naturally contains no atoms).  Subtracting $N_{N\downarrow}$ from
 $N_{N\uparrow}$, we have for the polarization 
\bea
&&\hspace{-.9cm} \Delta N =  N_{N\uparrow} - N_{N\downarrow},
\\
&&
\hspace{-.9cm}\qquad    = \frac{\pi^2 c}{12} \Big(\frac{2}{m\Omega_T^2}\Big)^{3/2} \Big(
\mu_\uparrow^3 \big[1-f\Big(\frac{r_c}{R_\uparrow}\Big)
\big]
\nonumber \\
&&
\qquad \qquad \qquad \qquad
-\mu_\downarrow^3 \big[1-f\Big(\frac{r_c}{R_\downarrow}\Big)
\big]
\Big).
\label{eq:popdifflda}
\eea
Expanding Eq.~(\ref{eq:popdifflda}) to leading order in $h/\mu$ (valid since $h_c(r) \ll \mu$
in the BCS regime), we have [using $R_{\uparrow,\downarrow} \simeq R_0(1\pm h/2\mu)$, from Eqs.~(\ref{eq:rlda}) and
(\ref{eq:rsigma})]:  
\be
\Delta N \simeq
 \frac{\pi^2 c}{12} \Big(\frac{2}{m\Omega_T^2}\Big)^{3/2} h\mu^2\Big(
6[1-f(x_c)] + x_c f'(x_c)\Big),
\ee
with $f'(x)$ the derivative of $f(x)$ and 
where we have defined $x_c \equiv r_c/R_0$.
Using Eq.~(\ref{eq:nlda}) for $N$, 
(valid for $h\ll \mu$)
we obtain
\be
\frac{\Delta N}{N} 
\simeq
\frac{h}{2\mu}
\big(
6[1-f(x_c)] + x_c f'(x_c)\big).
\ee
Using Eq.~(\ref{eq:rclda}) for $x_c$ to eliminate $h/\mu$, 
and defining $\kf$ via $\mu = \kf^2/2m$ (an approximation
corresponding to $\mu\approx \ef$, valid in the BCS regime), we thus have our final expression for $x_c$ in
the BCS regime:
\bea
&&\hspace{-.5cm}
\frac{\Delta N}{N} 
 \simeq
2\sqrt{2} {\rm e}^{-2} (1-x_c^2)\exp\big[
\frac{\pi}{2\kf \as \sqrt{1-x_c^2}}
\big]
\nonumber \\
&&\hspace{2cm}\times
\big(
6[1-f(x_c)] + x_c f'(x_c)\big).
\label{eq:xclda}
\eea
For $\Delta N = 0$, Eq.~(\ref{eq:xclda}) is solved by $x_c = 1$, i.e., the entire
system is in the SF phase. 
 With increasing difference  $\Delta N$ in the number  
of spin-up and spin-down atoms, however, 
 a thin shell of spin-polarized normal Fermi liquid forms 
on the outside of the cloud, corresponding to $x_c=r_c/R_0$  decreasing from unity. 
Although Eq.~(\ref{eq:xclda}) cannot be solved analytically for $x_c$, the radius of the inner SF sphere,
it is straightforward to determine it numerically, as shown in 
Fig.~\ref{fig:radiusldaplots} for coupling strengths $\kf |\as| = 2$, $\kf |\as| = 1.5$
and $\kf |\as| = 1$ (the latter falling outside the range of quantitative validity of the BCS approximation, but
expected to be qualitatively correct).  As illustrated in Fig.~\ref{fig:radiusldaplots}, $r_c(\Delta N)$ vanishes
at a critical population difference $\Delta N_c$ beyond which the cloud is completely in 
the N phase, exhibiting Pauli paramagnetism and no pairing even at zero
temperature.  This critical population difference increases with increasing coupling 
strength, indicating the increased strength of Cooper-pairing  as the Feshbach resonance is approached
from positive detuning.

%-----------------------------
%
% fig%48
%
%-----------------------------

When the system is in the mixed phase, $0<x_c<1$, the local magnetization 
$\mg(r) = n_\uparrow(r) -n_\downarrow(r)$ will exhibit an interesting
radius dependence that we now compute. We find for $\mg(r)$
\bea
&&\mg(r) = \frac{2c}{3} \mu^{3/2} \big[
(1- \frac{r^2}{R_0^2} + \frac{h}{\mu})^{3/2} 
-
(1- \frac{r^2}{R_0^2} - \frac{h}{\mu})^{3/2} 
\big]
\nonumber \\
&&\hspace{1cm}
\times \Theta(r-r_c),
\label{eq:magldapre}
\eea
with the heaviside step function enforcing that $\mg(r)$ vanishes in the SF state, where
$n_\uparrow(r) =n_\downarrow(r)$.  For $r>r_c$, in the normal shell, 
$\mg(r) = n_{N\uparrow}(r) - n_{N\downarrow}(r)$ is nonzero and given in terms of 
$x = r/R_0$  and  a population-difference 
scale (restoring $\hbar$ for clarity here)
\be
\mg_0 \equiv \frac{2c}{3} \mu^{3/2}
=\frac{2}{3} \frac{m^{3/2}\mu^{3/2}}{\sqrt{2} \pi^2 \hbar^3},
\ee
by
\bea
&&\mg(r) = \mg_0 \big[
(1- x^2 + \frac{h}{\mu})^{3/2} 
-
(1- x - \frac{h}{\mu})^{3/2} 
\big]
\nonumber \\
&&\hspace{1cm}
\times \Theta(x-x_c).
\label{eq:maglda}
\eea
To plot $\mg(r)$ for a particular coupling and population difference, we combine Eq.~(\ref{eq:maglda}) 
with Eq.~(\ref{eq:xclda}) for $r_c(\Delta N)$ and Eq.~(\ref{eq:rclda}) for $h$ at that particular population
difference and coupling.  In Fig.~\ref{fig:magplotlda}, we do this for coupling 
$(\kf |\as|)^{-1} = 1.5$ and two different values of the relative population difference: 
$\frac{\Delta N}{N}
= 0.15$ (dashed) and
$\frac{\Delta N}{N}= 0.20$ (solid).  We note that, since the 
spin-$\uparrow$ TF radius $R_\uparrow$ is slightly larger than $R_0$, $\mg(r)$ is nonzero even slightly beyond unity 
in the figure.

%-----------------------------
%
% fig%49
%
\begin{figure}[bth]
\vspace{1.4cm}\hspace{-1.7cm}
\centering
\setlength{\unitlength}{1mm}
\begin{picture}(40,40)(0,0)
\put(-50,0){
\begin{picture}(0,0)(0,0)
\includegraphics{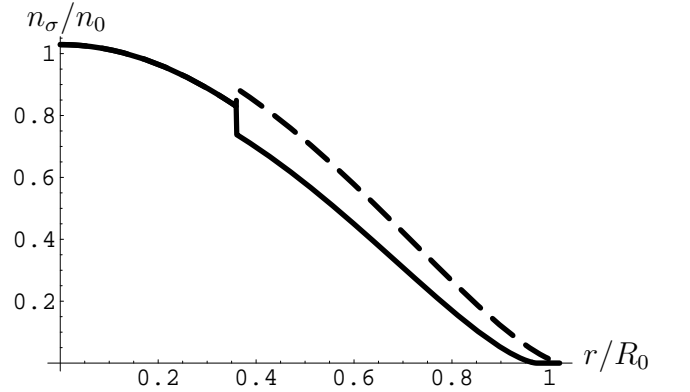}
\end{picture}}
\put(62,5.0) {\large{$ r/R_0$}}
\put(-12,50) {\large{$n_\sigma/n_0$}}
\end{picture}
\vspace{-.5cm}
\caption{Local fermion densities $n_\uparrow(r)$ and $n_\downarrow(r)$ (dashed and solid, respectively, normalized
to $n_0 \equiv \frac{4}{3} c\mu^{3/2}$)
as a function of radius (normalized to the $\Delta N = 0$ TF radius $R_0$) 
in the regime of phase separation
in a harmonic trap for coupling $(\kf |\as|)^{-1} = 1.5$ and 
$\frac{N_\uparrow - N_\downarrow}{N_\uparrow + N_\downarrow} = 0.15$.}
\label{fig:nupndownlda}
%
% ** 25OctLDAnumbers.nb 
%
\end{figure}
%-----------------------------

In Fig.~\ref{fig:nupndownlda}, we also plot the experimentally-accessible~\cite{Zwierlein05,Partridge05}
 individual spin-up and spin-down
densities $n_\uparrow(r)$ and  $n_\downarrow(r)$.   As expected, for $r<r_c$,
in the SF phase, 
$n_\uparrow(r) = n_\downarrow(r) = n_{SF}(r)/2$, with $n_{SF}$ given by Eq.~(\ref{eq:nsflda}).  To obtain
this plot, we need  $n_\uparrow(r)$  and $n_\downarrow(r)$ at a particular $\mu$.
Although in the above  analytical calculation of $\mu$ we have neglected the subdominant $\Delta$-dependent terms, 
in plotting $n_\uparrow(r)$
and  $n_\downarrow(r)$ in Fig.~\ref{fig:nupndownlda} we have included them. 

In the normal state, $n_\uparrow(r)$ and $n_\downarrow(r)$ are simply given by the two terms in 
Eq.~(\ref{eq:magldapre}):
\bea
&&\hspace{-1.3cm}
n_{N\uparrow}(r) = \frac{2c}{3} \mu^{3/2} (1+ \frac{h}{\mu} - x^2)^{3/2}\Theta(1+ \frac{h}{\mu} - x^2),
\\
&&\hspace{-1.3cm}
n_{N\downarrow}(r) = \frac{2c}{3} \mu^{3/2} (1- \frac{h}{\mu} - x^2)^{3/2}\Theta(1- \frac{h}{\mu} - x^2).
\eea
Then, to obtain $n_\sigma(r)$ 
for a particular $(\kf \as)^{-1}$ and population difference $\Delta N$, 
we first determine $x_c$ through Eq.~(\ref{eq:xclda})  and then plot 
$n_\sigma(r) = \frac{1}{2}n_{SF}(r)\Theta(r_c-r)+ n_{N\sigma}(r)\Theta(r-r_c)$.  
In Fig.~\ref{fig:nupndownlda} we plot the resulting $n_\uparrow(r)$ and $n_\downarrow(r)$ as a 
function of $r$ for $(\kf \as)^{-1} = 1.5$
and
 $\frac{\Delta N}{N} = 0.15$
(i.e.~the same parameters as the dashed curve of Fig.~\ref{fig:magplotlda}),
for which the SF-N boundary is at $x_c = 0.36$, i.e., at $r_c  = 0.36R_0$,

Before proceeding to the LDA in the negative-detuning BEC regime, we compute the width $\delta r$ of
the FFLO phase in a trap.  As we have noted above, because the FFLO phase intervenes between the SF and N phases,
for a homogeneous system at fixed $\mu$ and $h$, strictly speaking LDA predicts a thin spherical shell of 
FFLO between the SF and N.  To estimate the width of this shell, first imagine imposing a particular $h$ 
so that the system is polarized.  Now, $r_c$ defined by $h \approx \Delta(r_c)/\sqrt{2}$ [Eq.~(\ref{eq:rclda})] 
denotes the critical radius at which the system jumps from the SF phase to the FFLO phase with increasing radius.
Similarly, using Eq.~(\ref{eq:hfflosingle}), we see that at radius $r_c + \delta r$  defined by 
$h = h_{\rm FFLO} = \eta \Delta(r_c+ \delta r)$ 
the FFLO phase disappears continuously into the N phase.  To find $\delta r$, we simply expand each of these equations
to leading order in small $\delta r$ and equate them, which [using Eq.~(\ref{eq:rlda})] yields (at $\kf |\as|\gg 1$)
\be
\frac{\delta r}{R_0} \simeq \Big(1- \frac{1}{\sqrt{2}\eta}\Big)\frac{2\kf |\as| R_0}{\pi r_c} \big( 1 - \frac{r_c}{R_0})^{3/2}.
\label{eq:deltarfflo}
\ee
The first factor $(1-\frac{1}{\sqrt{2}\eta}) \simeq 0.062$ in Eq.~(\ref{eq:deltarfflo}) 
is numerically small by virtue of the thinness of the FFLO region of
the phase diagram, i.e., $\eta$ being close to $1/\sqrt{2}$.  The apparent divergence at $r_c\to 0$ is an
artifact of approximating $(r_c +\delta r)^2 \simeq r_c^2 + 2\delta r r_c$, valid for $\delta r \ll r_c$, that can be easily 
fixed.   Clearly, $\delta r\ll R_0$ simply because $\kf |\as| \ll 1$ in the BCS regime.  However, even for $\kf |a_s| = 1$, 
reading a typical value of $r_c/R_0\simeq 0.5$ for $\Delta N/N = 0.2$
from  Fig.~\ref{fig:radiusldaplots} yields $\delta r/R_0 \approx 0.05$.

\subsection{BEC regime}
\label{SEC:ldabec}

We now turn to the BEC regime in which $\as>0$.   As we have already discussed
in the preceding subsection, within LDA the phase structure in a trap follows from
 the phase diagram at fixed $\mu$ and $h$, with the local phase at position
$r$ determined by the local chemical potential $\mu(r)$ satisfying
Eq.~(\ref{eq:mulda}).  However, an important distinction is that, as we found
for the homogeneous case in Sec~\ref{SEC:BECsingle}, the chemical potential $\mu$
at the center of the trap is already negative in the BEC regime 
and therefore $|\mu(r)|$ does not vanish as in the BCS case.

\begin{figure}[bth]
\vspace{1.4cm}
%\hspace{-.3cm}
\centering
\setlength{\unitlength}{1mm}
\begin{picture}(40,40)(0,0)
\put(-55,0){\begin{picture}(0,0)(0,0)
\includegraphics{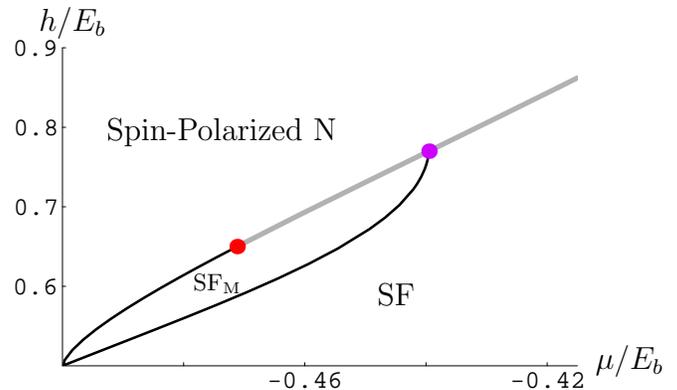}
\end{picture}}
\put(-15,49.5) {\large{$h/E_b$}}
\put(59,4.5) {\large{$\mu/E_b$}}
\put(30,13) {\large{SF}}
\put(-6,35) {\large{Spin-Polarized N}}
\put(5.5,15) {\sfm}
\end{picture}
\vspace{-.5cm}
\caption{(Color online) Mean-field phase diagram (equivalent to Fig.~\ref{fig:phasefixedmusingle})
for the single-channel 
model Eq.~(\ref{eq:singlechannel}) as a function of atomic chemical potential $\mu$ and
atomic  chemical potential difference $h$, each normalized to the molecular
binding energy $E_b$,  showing superfluid (SF), magnetic superfluid (\sfm), and
fully-polarized normal (N) phases. 
Thin black lines denote continuous $T=0$ transitions, while the gray thick curve denotes a 
first-order \sfm-N or SF-N transition.  The red dot at $(-0.47,0.65)$ is a tricritical point separating 
first-order and second-order \sfm-N  transitions while the purple dot at $(-0.44,0.77)$
shows the chemical potential above which the \sfm phase ceases to exist.}
\label{fig:muhphasenum}
%
%  22May.nb 
%
\end{figure}

  We start by recalling the BEC-regime phase diagram in the
grand-canonical ensemble Fig.~\ref{fig:phasefixedmusingle},
which we re-plot here (Fig.~\ref{fig:muhphasenum}) as a function of $h$ and the fermion
chemical potential  $\mu$ (zoomed-in to emphasize the \sfm phase).   
The black solid lines represent continuous phase transitions between
the \sfm and N phases (upper line), and between the SF and \sfm phases (lower line).  
The gray solid line denotes a first-order \sfm to N (or, to the
right, a first-order SF to N) transition.  Within LDA, a trapped
fermion gas with particle number $N$ and population difference $\Delta N$
is characterized by a certain chemical potential and chemical potential difference $(\mu,h)$ at the 
center of the trap that can be interpreted as a coordinate in Fig.~\ref{fig:muhphasenum}.  With
increasing radius, $\mu(r)$ changes according to Eq.~(\ref{eq:mulda}), tracing out a 
left-ward moving horizontal
line segment $(\mu(r),h)$ on Fig.~\ref{fig:muhphasenum}.  Thus, if a polarized Fermi gas is
in the SF phase at the center of the trap, with increasing radius it will generally go 
through the sequence of phases SF$\to$\sfm$\to$N, with the \sfm$\to$N transition 
continuous for $h<0.65E_b$ and  first order for $h>0.65E_b$ (the latter indicating jumps
in the local density as we have seen in the preceding subsection).

%-----------------------------
%
% fig%50
%
%------------------------------

The LDA ground-state energy, which determines  $\mu$ and $h$ as a function of $N$, $\Delta N$
and the scattering length $a_s$, is obtained by spatially integrating the local energy
density $E_G[\Delta(r),\mu(r)]$ over the cloud's volume.  We approximate
the full uniform-case energy density Eq.~(\ref{eq:gsesinglechannel2}) by expanding 
to sixth order in small $\Delta/E_b$ (with $E_b$ the binding energy $E_b = \hbar^2/m\as^2$).
Defining dimensionless quantities  $\Deltab(r) = \Delta(r)/E_b$, $\mub(r) = \mu(r)/E_b$ and $\hb = h/E_b$
(the latter not to be confused with the same symbol used in Sec.~\ref{SEC:FFLO} and Sec.~\ref{SEC:FFLOsingle})
we find 
\bea
&&\hspace{-.5cm}\frac{E_G}{cE_b^{5/2}}  = \int d^3r\Big(
-\frac{4}{15} c (h -|\mu(r)|)^{5/2} \Theta(h - |\mu(r)|)
\nonumber \\
&&\hspace{-.5cm} \qquad - \Vb_2(\hb,\mub(r))\Deltab(r)^2 + \frac{1}{2} \Vb_4(\hb,\mub(r))\Deltab(r)^4 
\nonumber \\
&&
 \hspace{-.5cm}\qquad +\frac{1}{3} \Vb_6(\hb,\mub(r))\Deltab(r)^6\Big),
\label{eq:egbecsinglelda}
\eea
with the coefficients 
\bea
\Vb_2(\hb,\mub)&=&  \frac{\pi}{2\sqrt{2}} - \sqrt{|\mub|}F_2(\hb/|\mub|),
\\
\Vb_4(\hb,\mub)&=& \frac{\pi}{32 |\mub|^{3/2}} F_4(\hb/|\mub|),
\\
\Vb_6(\hb,\mub)&=& \frac{3}{32 |\mub|^{7/2}} F_6(\hb/|\mub|),
\eea
where  $F_6(x)$ is given by Eq.~(\ref{eq:f6}).

The local pairing field $\Deltab(r)$ is given by the gap equation
[cf. Eq.~(\ref{eq:gapdetailed})]
\bea
&&0 =  - \Vb_2(\hb,\mub(r))\Deltab(r) + \Vb_4(\hb,\mub(r)) \Deltab^3(r) 
\nonumber \\
&&
\qquad \qquad 
+ \Vb_6(\hb,\mub(r)) \Deltab^5(r),
\eea
which has the trivial (normal-state) solution $\Deltab = 0$, as well as the 
nontrivial solution (suppressing the arguments of the $\Vb_\alpha$ for simplicity)
\be
\Deltab^2_\pm(r)  = \frac{\Vb_4}{2\Vb_6}\Big[-1 \pm \sqrt{1+ 4\Vb_2 \Vb_6/\Vh_4^2}
\Big].
\label{eq:ldabecstationary}
\ee
As discussed in Sec.~\ref{SEC:detailedanalysis}, the correct physical solution (locally 
corresponding to the SF or \sfm state) is given by the $+$ of Eq.~(\ref{eq:ldabecstationary})
for $\Vb_4>0$ and the $-$  of Eq.~(\ref{eq:ldabecstationary}) for $\Vb_4<0$.

  Next, we determine equations for the local number density and magnetization.  
For the former, we find (keeping only leading-order terms) 
$n(r) = cE_b^{3/2} \nb[\hb,\mub(r)]$,
with
\bea
&&\hspace{-1.5cm}
\nb[\hb,\mub(r)] \simeq  
\frac{2}{3} (\hb - |\mub(r)|)^{3/2} + \frac{\Deltab(r)^2}{2\sqrt{|\mub(r)|}}
\Big(\frac{\pi}{2} 
\nonumber \\
&&\hspace{-1.5cm}
\qquad \qquad \quad
- \tan^{-1} \sqrt{\hb/|\mub(r)|-1}\Theta(\hb - |\mub(r)|)
\Big),
\eea 
the dimensionless density.  Crucially, the spatial dependence of $\nb[\mub(r)]$ 
arises only via $\mub(r)$.  Similarly, the local magnetization $\mg(r) = cE_b^{3/2} \mb[\hb,\mub(r)]$
with 
\be
\mb[\hb,\mub(r)] = \frac{2}{3} \big(\sqrt{\hb^2 - \Deltab(r)^2}- |\mub(r)|
\big)^{3/2}.
\ee
In terms of $n(r)$ and $\mg(r)$, the total particle number $N$ and population 
difference $\Delta N$ are given by 
\bea
\label{eq:numbeclda}
N = \int d^3 r \,n(r), 
\\
\Delta N = \int d^3 r \,\mg (r), 
\label{eq:magbeclda}
\eea
constraints that determine $\mub$ and $\hb$ at a particular $N$ and $\Delta N$.

%-----------------------------
%
% fig%51
%
\begin{figure}[bth]
\vspace{1.4cm}
\hspace{-1.0cm}
\centering
\setlength{\unitlength}{1mm}
\begin{picture}(40,40)(0,0)
\put(-50,0){\begin{picture}(0,0)(0,0)
\includegraphics{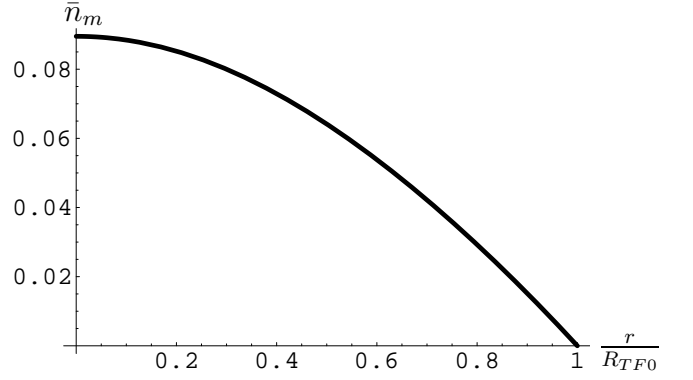}
\end{picture}}
\put(-12,49) {\large{$\nb_m$}}
\put(59,4.5) {\large{$\frac{r}{R_{TF0}}$}}
\end{picture}
\vspace{-.5cm}
\caption{Normalized molecular density $\nb_m= n_m/cE_b^{3/2}$ as a function of radius for an unpolarized
trapped fermion cloud, with parameters given in the text. 
 }
\label{fig:becldaunpol}
%
%  15June06.nb 
%
\end{figure}
%------------------------------

\subsubsection{$h=0$ case} 

We start by restricting attention to $\hb=0$, appropriate for $\Delta N = 0$.   For
this case, $\Deltab^2$, Eq.~(\ref{eq:ldabecstationary}), vanishes continuously as $\Vb_2 \to 0$, 
which in a homogeneous system corresponds to a second-order SF-to-Vacuum transition, 
as discussed in Sec.~\ref{SEC:fixedmusingle}.  In the present LDA context, 
it corresponds to the vanishing of the molecular density at the boundary of the system.
Thus, we can define the Thomas-Fermi radius $R_{TF0}$, where $\Vh_2$ vanishes, via
this condition:
\be
\Vb_2[0,\mub(R_{TF0})] = 0,
\ee
or, 
\be
\mub(R_{TF0}) = \mub_0 - \frac{m\Omega_T^2}{2E_b}R_{TF0}^2= -\frac{1}{2}.
\label{eq:mubequationlda}
\ee
Solving Eq.~(\ref{eq:mubequationlda}) for $R_{TF0}$ yields, [using Eq.~(\ref{eq:mulda})]
\be
R_{TF0} = \sqrt{\frac{E_b(2\mub_0 + 1)}{m\Omega_T^2}},
\ee
with the zero subscript on $\mub_0$ and $R_{TF0}$ indicating that they are for $h = 0$. 
We proceed to normalize the cloud radius $r$ to  $R_{TF0}$, defining $x = r/R_{TF0}$,
in terms of which the normalized chemical potential is 
\be
\mub_0(x) = \mub_0 - x^2 (\mub_0 + \frac{1}{2}).
\label{mubnoughtofx}
\ee
With this definition, Eq.~(\ref{eq:numbeclda}) becomes 
\be
\label{eq:numbeclda2}
N = N_0 (2\mub_0 + 1)^{3/2}\int  dx\,x^2 \bar{n}[\hb,\mub_0(x)],
\ee
with  
\be
N_0 = \frac{4\pi cE_b^3}{m^{3/2} \Omega_T^3} ,
\ee
a characteristic particle number scale.  One can easily estimate the parameter $N_0$ from typical experiments.  
For the case
of $^{40}$K, given typical values from the Jin group~\cite{Regal03,Regal}, we take scattering length 
$\as = 750 a_0$, ($a_0$ the Bohr radius) 
trap frequency $\Omega_T = 2\pi\times400 s^{-1}$.  With these parameters, $N_0 \simeq 2\times10^8$ and 
we must adjust $\mub_0$ to attain a realistic particle number.  
Numerically solving Eq.~(\ref{eq:numbeclda2}) to find the normalized chemical potential 
yields $\mub_0 = -0.465$ for $N = 10^5$.  Since the effective molecular chemical potential $\mu_m  = E_b + 2\mu$ 
 [Eq.~(\ref{eq:mubsingle})], which in dimensionless units is $\mub_m = 2\mub+1$, we see that 
the deviation of $\mub_0$ from $-0.5$ directly measures the effective molecular chemical potential.  
Consistently, we see from Eq.~(\ref{mubnoughtofx}) that, 
 at the boundary of the system ($x=1$), $\mub_0(x) =-\frac{1}{2}$ indicating a vanishing of 
the effective molecular chemical potential.  In Fig.~\ref{fig:becldaunpol},  we plot the effective normalized  molecular 
density $\nb_m = \pi \Deltab^2/8\sqrt{|\mub|}$ as a function of radius for this case, showing
the standard Thomas-Fermi profile for a molecular Bose condensate.

%-----------------------------
%
% fig%52
%
\begin{figure}[bth]
\vspace{3.2cm}
\centering
\setlength{\unitlength}{1mm}
\begin{picture}(40,65)(0,0)
\put(-52,42){\begin{picture}(0,0)(0,0)
\includegraphics{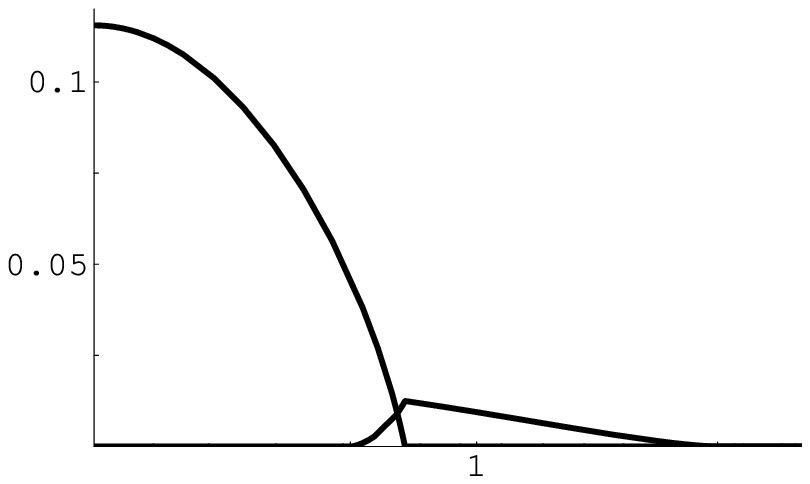}
\end{picture}}
\put(-52,-12){\begin{picture}(0,0)(0,0)
\includegraphics{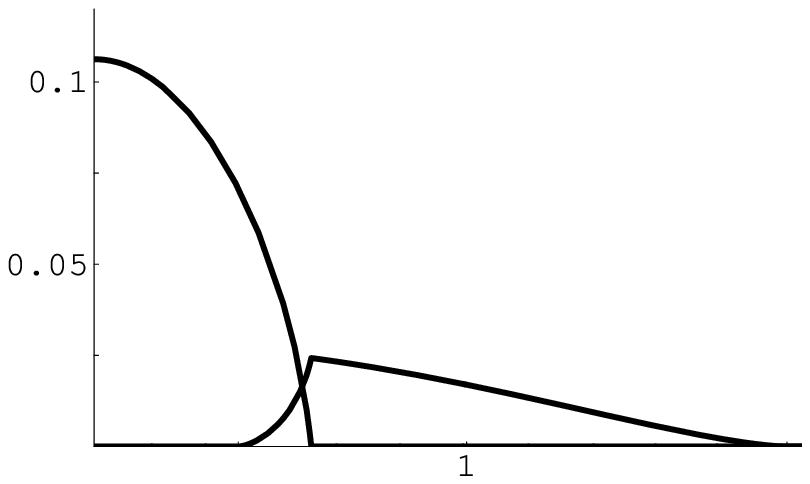}
\end{picture}}
\put(-18,92) {\large{(a)}}
\put(-18,37) {\large{(b)}}
\put(58,47) {\large{$\frac{r}{R_{TF0}}$}}
\put(-1,85) {\large{$\nb_m(r)$}}
\put(25,54) {\large{$\mb(r)$}}
\put(17,45) {$R_{TF}$}
\put(11,45) {$R_{f1}$}
\put(45,45) {$R_{f2}$}
\put(58,-7) {\large{$\frac{r}{R_{TF0}}$}}
\put(-3,27) {\large{$\nb_m(r)$}}
\put(25,0) {\large{$\mb(r)$}}
\put(7,-9) {$R_{TF}$}
\put(-1,-9) {$R_{f1}$}
\put(52,-9) {$R_{f2}$}
\end{picture}
\vspace{1cm}
\caption{Normalized molecular density $\nb_m= n_m/cE_b^{3/2}$  and normalized
magnetization $\mb = m/cE_b^{3/2}$ as a function of radius for a polarized
trapped fermion cloud, with  (a) $\Delta N/N =  0.26$ and (b)  $\Delta N/N =  0.73$ 
and other parameters given in the text. }
\label{fig:ldabeconetwo}
%
% 
% 15June06.nb 
% 
% 
\end{figure}
%------------------------------

\subsubsection{$h\neq0$ case} 

At $h\neq 0$, the system becomes locally magnetized, with the total population
difference given by Eq.~(\ref{eq:magbeclda}).  We now study our system at nonzero
population difference by simultaneously solving this along with the number equation.
In dimensionless form these are 
\bea
\label{eq:numbec2}
N  = N_0 (2\mub_0 + 1)^{3/2}\int  dx \,x^2 \bar{n}[h,\mub(x)],
\\
\label{eq:magbec}
\Delta N  = N_0 (2\mub_0 + 1)^{3/2}\int dx \, x^2 \bar{m}[h,\mub(x)],
\eea
where we note that the dependence of Eq.~(\ref{eq:magbec}) on $\mub_0$ arises because we 
are still measuring the radial coordinate in units of the unpolarized system, i.e., 
we continue to use $x = r/R_{TF0}$.  To maintain constant $N$, the chemical potential
$\mub$ deviates slightly from $\mub_0$ and the general formula for $\mub(x)$ as 
a function of the normalized radius is 
\be
\label{eq:mubofx}
\mub(x) = \mub - x^2(\mub_0 + \frac{1}{2} ).
\ee
Using Eq.~(\ref{eq:mubofx}) along with Eqs.~(\ref{eq:numbec2}) and (\ref{eq:magbec}),
it is straightforward to numerically study the cloud shape with increasing $\Delta N$,
using the same parameters as before.  

As shown in Fig~\ref{fig:ldabecintro} in Sec.~\ref{SEC:SummaryLDA}, the typical sequence
of phases with increasing radius that we find within LDA is SF$\to$\sfm$\to$N. As in
the positive-detuning BCS regime, this is due to the spin-polarized normal fermions
having been expelled 
to the outer shell of the system.  In this context, the \sfm phase
that is unique to negative detuning
 represents a  thin shell in which the singlet molecular bosons and outer  normal-phase
fermions  \lq\lq bleed\rq\rq\ into each other.  In Fig.~\ref{fig:ldabeconetwo} we plot 
the normalized molecular density $\nb_m(r)$ and normalized magnetization $\mb(r)$ 
as a function of radius for $\frac{\Delta N}{N} = 0.26$  (Fig.~\ref{fig:ldabeconetwo}a) 
and  $\frac{\Delta N}{N} = 0.73$ (Fig.~\ref{fig:ldabeconetwo}b). Figure~\ref{fig:ldabecintro}a
is for the same parameters but $\frac{\Delta N}{N} = 0.39$.

The three radii labeled on the horizontal axes of Figs.~\ref{fig:ldabeconetwo}a and ~\ref{fig:ldabeconetwo}b,
indicated in the cartoon picture Fig.~\ref{fig:ldabecintro}b of the superfluid shell structure, 
  are: $R_{f1}$, the radius
below  which $\mb(r)= 0$, $R_{TF}$, the radius below which $\nb_m(r) \neq 0$ and
the system is superfluid and $R_{f2}$, the radius above which $\mb(r) = 0$.  
Thus, for $r<R_{TF}$ and $r<R_{f1}$, the system is in the SF phase, consisting
of singlet molecular pairs while for 
 $R_{f1}<r<R_{TF}$ the system is in the \sfm phase with coexisting molecular pairs and 
single-species fermions.  For $R_{TF}<r<R_{f2}$ the system consists purely of 
single-species fermions. 

%-----------------------------
%
% fig%53
%
\begin{figure}[bth]
\vspace{1.4cm}
\hspace{-1.0cm}
\centering
\setlength{\unitlength}{1mm}
\begin{picture}(40,40)(0,0)
\put(-50,0){\begin{picture}(0,0)(0,0)
\includegraphics{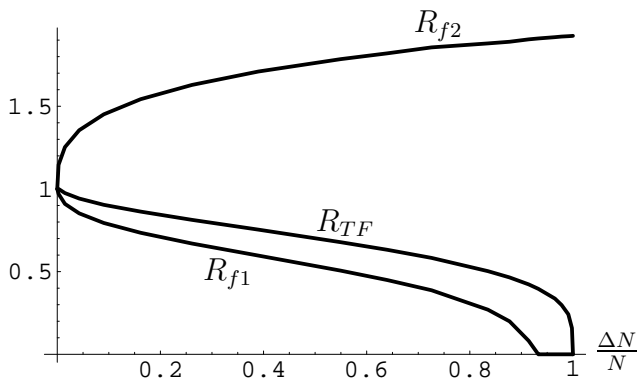}
\end{picture}}
\put(59,4.5) {\large{$\frac{\Delta N}{N}$}}
\put(35,48) {\large{$R_{f2}$}}
\put(7,15) {\large{$R_{f1}$}}
\put(22,21.5) {\large{$R_{TF}$}}
\end{picture}
\vspace{-.5cm}
\caption{Plots of the three radii $R_{f2}$ (outer boundary of N phase), 
$R_{TF}$ (outer boundary of \sfm phase)
and  $R_{f1}$ (outer boundary of SF phase) characterizing the polarized
cloud in the BEC regime within LDA.  }
\label{fig:iotfradii}
%
%  15June06.nb 
%
\end{figure}
%------------------------------

With increasing $\frac{\Delta N}{N}$, $R_{TF}$ and $R_{f1}$ decrease and  $R_{f2}$
rapidly increases as the system is converted from a molecular superfluid to
a single-species fermion gas.  This behavior, seen in comparing Figs.~\ref{fig:ldabeconetwo}a 
and ~\ref{fig:ldabeconetwo}b, is shown in detail in Fig.~\ref{fig:iotfradii}, in which we plot 
all three radii as a function of $\frac{\Delta N}{N}$.  We note in particular that, for 
very large $\frac{\Delta N}{N}$,  $R_{f1}\to 0$ with  $R_{TF}\neq 0$.  Thus, in this
regime, LDA predicts the sequence of phases with increasing radius to be 
\sfm$\to$N.

%%%%%%%%%%%%%%%%%%%%%%%%%%%%%%%%%%%%%%%%%%%%%%%%%%%%%%%%%%%%%%%%%%%%%%%%%%%%%%%%%%%%%%%%%%%%%%%%%%%%%%%%%%%%%%
%%%%%%%%%%%%%%%%%%%%%%%%%%%%%%%%%%%%%%%%%%%%%%%%%%%%%%%%%%%%%%%%%%%%%%%%%%%%%%%%%%%%%%%%%%%%%%%%%%%%%%%%%%%%%%
%%%%%%%%%%%%%%%%%%%%%%%%%%%%%%%%%%%%%%%%%%%%%%%%%%%%%%%%%%%%%%%%%%%%%%%%%%%%%%%%%%%%%%%%%%%%%%%%%%%%%%%%%%%%%%
%%%%%%%%%%%%%%%%%%%%%%%%%%%%%%%%%%%%%%%%%%%%%%%%%%%%%%%%%%%%%%%%%%%%%%%%%%%%%%%%%%%%%%%%%%%%%%%%%%%%%%%%%%%%%%

\section{Discussion and Conclusions}
\label{SEC:concludingremarks}

\subsection{Summary}

In this lengthy manuscript we have studied the rich zero-temperature phase
behavior of a two-species (pseudo-spin up and down) Fermi gas
interacting via a tunable Feshbach resonance and constructed its
phase diagram as a function of Feshbach resonance detuning $\delta$
and pseudo-spin
population imbalance  $\Delta N$ (and chemical potential difference $h$).
Our main results relied on a well-controlled mean-field analysis of the two-channel model 
of fermions interacting via a Feshbach resonance that
is quantitatively accurate in the narrow resonance limit. In addition, we have
complemented this with a study of the one-channel model, appropriate for
the wide resonance limit, for which
the mean-field approximation that we use is not quantitatively
justified near unitarity, but is expected to be qualitatively correct, finding
an expected qualitative agreement between the two models.

As described in the main text, for $h$ below a critical value that is
exponentially small in the BCS regime of $\delta \gg 2\ef$, the
fully-gapped singlet superfluid (SF) is stable and undergoes the (by now) standard
BCS-BEC crossover with reduced $\delta$. On the positive-detuning BCS side
of the resonance, for $h$ larger than $h_{c1}(\delta)$ the fixed atom-number
constraint forces the system to enter a regime of phase separation (PS) consisting of 
coexisting singlet SF and partially magnetized FFLO states for $h_{c1} < h < h_{c2}$.
At $h_{c2}$, for $\delta \gg 2\ef$, the system enters the periodically-paired FFLO phase,
before undergoing a continuous~\cite{MFTcomment} transtion to the normal Fermi gas (N) phase 
at  $h_{\rm FFLO}$.  
At lower detuning values $\delta\approx 2\ef$, 
the FFLO phase becomes unstable and the thin window of FFLO phase between $h_{c2}$ and $h_{\rm FFLO}$
is \lq\lq squeezed out\rq\rq.  In this crossover regime, upon increasing $h$, 
the SF undergoes a direct first-order transition to the N phase, with SF-N 
coexistence for $h_{c1}<h<h_{c2}$ for fixed atom number.

On the negative-detuning BEC side of the resonance, for $h > h_m(\delta)$ the fully-gapped molecular SF
undergoes a continuous transition to a homogeneous magnetized superfluid ground state (\sfm) 
composed of molecules and a single-species Fermi gas, with the latter responsible for 
the gapless atomic excitations and finite polarization characterizing the \sfm
state. Upon further increase of the chemical potential difference, $h$,
for $\delta>\delta_c$ the \sfm undergoes a first-order transition to
a fully-polarized N phase, with, for fixed particle number, a phase-separated
regime consisting of \sfm and N polarized states.
 In contrast, for $\delta < \delta_c$ there is a continuous  \sfm-N transition.  We
give a detailed description of these $T=0$ phases and phase boundaries
characterizing the phase diagram, with many of our predictions already
verified in recent experiments~\cite{Zwierlein05,Partridge05,Shin}.

\subsection{Relation to other work}

Our study of polarized resonantly-interacting Fermi gases builds on a
large body of work dating back to the seminal contributions of Clogston~\cite{Clogston}, 
Sarma~\cite{Sarma}, Fulde
and Ferrell~\cite{ff}, and Larkin and Ovchinnikov\cite{lo}.  These were followed
by many studies of FFLO and related exotic paired
superfluid states that, in additional to  off-diagonal long-range
order, also break spatial symmetries. These studies range from solid
state electronic systems~\cite{CeCoIn5} to nuclear matter (quark-gluon
plasma)~\cite{Alford,Bowers,RajagopalComment}, but until very recently
were generally confined to the weakly-interacting BCS regime.

This theoretical effort has recently seen a resurgence of activity,
stimulated by the discovery of tunable resonantly-paired superfluidity in
degenerate atomic Fermi gases~\cite{Regal,Zwierlein,Kinast,Bartenstein,Bourdel04,Chin}.
However, perhaps the first theoretical work to study atomic gases with unequal
spin populations in the BCS regime, by Combescot~\cite{Combescot01}, preceded these experiments.
Also preceding such experiments was the proposal by Liu and Wilczek~\cite{Liu} of the breached pair state,
that is closely related to the FFLO and \sfm phases.
Other notable theoretical work on spin-polarized Fermi gases 
in the BCS regime includes
that by Mizushima et al.~\cite{Mizushima}, who studied signatures
of FFLO phases in cold fermion experiments but did not determine the conditions
necessary (i.e., the phase boundaries) for the observation of such states
and that by Bedaque, et al.~\cite{Bedaque}, who
emphasized the first-order nature of the SF-N transition and the concomitant regime of 
phase separation (missed in an important work by
Gubankova et al.~\cite{Gubankova}, see Ref.~\onlinecite{Forbes} 
for a discussion).  

 Later, in a predominantly numerical work Carlson and Reddy~\cite{Carlson}
extended the study of polarized paired superfluids
 across the resonance to include the BEC regime, still not
including a trap. One important prediction in Ref.~\onlinecite{Carlson} is
numerical evidence for a uniform magnetized paired superfluid around
the unitary point, corresponding to a finite critical value of
polarization (species imbalance) to produce phase separation.
The unitary point under applied polarization was also studied by Cohen~\cite{Cohen}, 
who derived general conditions for phase separation in this regime.

These early microscopic studies and ideas that followed were compiled
into a general phenomenological phase diagram by Son and Stephanov~\cite{Son}.
The first microscopic {\em analytical} study of the problem across the
BCS-BEC crossover was done by Pao, et al.~\cite{Pao} within the single-channel
model via mean-field theory. On the BCS side their study of a uniform Fermi gas (that
ignored the interesting FFLO state), was consistent with earlier BCS-regime
studies~\cite{Sarma,Bedaque}. However, their extension to the
BEC regime was done incorrectly, leading to a qualitatively wrong
phase boundary below which the magnetized superfluid \sfm is
unstable.~\cite{Paonote}  The correct zero-temperature mean-field phase diagram 
for the single-channel  model (consistent with our earlier work on the
two-channel model~\cite{shortpaper}) was published by Gu et al.~\cite{Gu},
by Chien et al.~\cite{Chien},
and  by Parish et al~\cite{Parish}, although these authors also neglected
the possibility of an FFLO state.

Our work\cite{shortpaper} presented the first
analytical prediction of a complete (full range of detuning) phase
diagram for a resonantly-interacting Fermi gas within the more general
two-channel model.  Our work also included a detailed study of the
FFLO state (extending the original work of Fulde and Ferrell and Larkin and 
Ovchinnikov to a resonant tunable interaction) as well as a prediction regarding 
the experimental consequences of the trap. With regard to the former our
main contribution was a prediction of the phase boundaries ($h_{\rm FFLO}(\delta)$
and $h_{c2}(\delta)$) for the FFLO state, showing their crossing at  $\delta_*\approx
2\ef$ that leads to the elimination of the FFLO state for $\delta<\delta_*$. 
With regard to the trap that we studied within LDA, as summarized
in the inset of Fig. 1 and last section of our earlier
publication\cite{shortpaper}, with many additional details presented here, our main
prediction is that, in a trap, phase separation leads to the recently
observed~\cite{Zwierlein05,Partridge05,Shin}
 shell-like cloud profile, with interfacial
boundaries whose detailed dependence on detuning and population
imbalance we predict.

Our original study also clearly identified the source of the error of Pao et al.
in computing phase boundaries in the BEC regime. We unambiguously
showed that much of the BEC regime that they claimed to be a uniform
stable magnetized superfluid (\sfm in our notation),
was in fact a phase
separated regime, consisting of coexisting \sfm and fully polarized
normal Fermi gas. We have explicitly identified the error 
in Ref.~\onlinecite{Pao} by reproducing their (in our view erroneous) 
results.   The error stems from the fact that some solutions
to the gap equation correspond to saddle points or maxima of the variational ground state energy
$E_{G}(\Delta_\bQ)$.  Pao et al. only used the positivity of {\em local} magnetic
susceptibilities and the superfluid stiffness to check stability of the
identified solutions.  Since this  stability criterion is not stringent
enough, being necessary but not sufficient,
certain solutions of the gap equation identified as stable in Ref.~\onlinecite{Pao}
are in fact maxima of $E_G(\Delta_\bQ)$ and therefore in fact unstable~\cite{Comment}.
 In the presence of a first-order transition a careful
study of $E_G(\Delta_\bQ)$ (best done in the grand-canonical ensemble~\cite{Forbes},
with number constraints only imposed a posteriori) is necessary in
order to ensure that extrema solutions to the gap equation indeed
correspond to stable phases that {\it minimize\/} $E_G$. 

Following the subsequent experimental observations by Zwierlein et
al.~\cite{Zwierlein05} and by Partridge et al.~\cite{Partridge05}, qualitatively consistent
with our original predictions (e.g., the shell-like phase structure of the
atomic cloud as a signature of phase separation in a trap, the expulsion of
magnetization to the outer shell, the general location of phase boundaries
and their qualitative detuning and population imbalance dependence),
there has been an explosion of theoretical activity.  Many of
our $T=0$ predictions have been carefully verified and extended in
important ways by many works that
followed (see, however, Refs.~\cite{Pao,Iskin,Paonote}). These include nontrivial extensions to a finite
temperature\cite{Chien,Parish}, uncontrolled but elaborate
extensions to a broad resonance treated within the one-channel model
(beyond our original mean-field approximation),
as well as more detailed studies of the effects of the
trap\cite{Pieri,Torma,Yi,Chevy,DeSilva,Haque,DeSilva06,Imambekov}, and extensions of the FFLO\cite{Dukelsky} and
other related exotic states~\cite{Yang2,YangSachdev,SachdevYang}. A notable work by
De Silva and Mueller~\cite{DeSilva} demonstrated that consistency with
experiments~\cite{Partridge05} requires a breakdown of LDA in an
anisotropic trap; clearly LDA must also obviously break down at an
interface between coexisting phases. This was followed by a detailed
treatment of the trap beyond LDA\cite{DeSilva06}, an issue that 
was also addressed in Refs.~\onlinecite{Imambekov,Zwierlein06}.

The experimental observation\cite{Partridge05} (that, however, has not
been seen in the MIT experiment~\cite{Zwierlein05,Shin}, see also 
Refs.~\onlinecite{Zwierlein06,PartridgeComment}) of the existence of a
uniform magnetized superfluid at the unitary point (where $1/\kf a\rightarrow 0$) 
and a corresponding critical
polarization ($\Delta N_c/N\approx 0.09$) necessary to drive it to phase
separate has generated considerable theoretical interest. With the
exception of the theory by Ho and Zhai~\cite{HoZhai}, that attempts to account for
this feature via a phenomenological model of Bogoliubov quasi-particle
pairing, to our knowledge no model has been able to capture this
putative experimental feature; some support for it however exists in
the original Monte-Carlo work by Carlson and Reddy~\cite{Carlson}. From our
perspective, this seemingly qualitative feature reduces to a {\em
quantitative} question of the location of the critical detuning $\delta_M$ point
in Fig.~\ref{fig:mphasetwo}
 (or,
equivalently, the $a_{sM}$ point in Fig.~\ref{fig:globalmagphasesingle})
 Our work shows unambiguously that for a {\em
narrow} resonance ($\gamma \ll 1$) at $T=0$, $\delta_M$ most definitely falls
in the BEC regime (negative detuning), thereby excluding the uniform
magnetized \sfm state from the unitary point. 
Our mean-field theory predicts the critical scattering length for the broad-resonance 
one-channel model
to be close to $\kf a_{sM}\simeq 1$, again excluding the \sfm state from unitarity.
For reasons discussed in
the Introduction, Sec.~\ref{SEC:intro}, however, mean-field theory 
for a broad resonance (relevant to present-day experiments) 
is not quantitatively valid for a broad resonance.
Thus, it is quite possible that for finite
$T$~\cite{comment:AtfiniteT} and a broad resonance $\delta_M$ indeed shifts to a positive
detuning, with the experimental findings of $\Delta N_{c}$~\cite{Partridge05} then naturally
interpreted as the observation of the \sfm phase at the unitary point.

\subsection{Experimental predictions}

Most of our predictions were made both for a narrow resonance within a
two-channel model and for a broad resonance within a one-channel
model. As discussed in the Introduction, the former has an important
advantage that it is quantitatively accurate with the width of the resonance
as the small expansion parameter. However, unfortunately, experiments
are deep in the broad-resonance regime, where the two-channel model {\it exactly\/}~\cite{Levinsen}
reduces to a one-channel model, that in the interesting crossover
regime cannot be treated analytically in a quantitatively trustworthy
way. We do, however, expect it to give qualitatively correct
predictions. And, as mentioned earlier, indeed we find that recent
experimental findings are qualitatively consistent with our wide-resonance
predictions done within the one channel model.

However, as we discussed above, our perturbative mean-field analysis, 
in either the narrow or broad resonance limit, does not find a uniform
magnetized superfluid near the unitary point and a corresponding finite
critical population imbalance $\Delta N_{c}$. 

Another feature that is seen in experiments~\cite{Zwierlein05}
 is that at unitarity the
upper-critical boundary where coexistence ends and the normal state takes
over is given by $\Delta N_{c2}/N\approx 0.7 < 1$. That is, the transition from
phase separation is to a normal state that is only 70\% polarized. This seemingly
qualitative feature is actually a quantitative question. In contrast
to experiments, as we show in the main text (see Fig.~\ref{fig:mphasetwo})  our
quantitatively accurate narrow resonance two-channel model analysis
(that can answer such a question) at the resonance position unambiguously predicts
the transition directly to a fully-polarized normal state, i.e.,
$\Delta N_{c2}/N\approx 1$ at unitarity. However, the less quantitatively trustworthy
broad-resonance limit gives a prediction of $\Delta N_{c2}/N \simeq 0.93 < 1$,
more consistent with experiments. Pick your poison. 

We expect that many of our general predictions will continue being
fundamental to the understanding of specific experiments. Our predictions
that can be already directly tested (some have already been
qualitatively tested~\cite{Zwierlein05,Partridge05}) are for the atomic cloud phase
composition and spatial density and magnetization profiles.  Most of
these have already been discussed in Sec.~\ref{SEC:LDA}, summarized by
Figs.~\ref{fig:magplotlda},\ref{fig:ldabecintro}, and \ref{fig:nupndownlda}-\ref{fig:iotfradii},
where we give details of the phase separation-driven cloud shell structure, dependences of radii on detuning and
population imbalance, and detailed spatial gas profiles that can be
imaged by cloud expansion.

In principle, the FFLO state (that was the original impetus for our
study) should exhibit particularly striking experimental signatures,
associated with its simultaneous ODLRO and spontaneous breaking of
translational and orientational symmetry, encoded in $\Delta_\bQ$.
For a homogeneous cloud, 
the typical population imbalance required to enter the FFLO state
is given by Eq.~(\ref{eq:deltanfflo}), which in dimensionful units is
approximately (recall $\eta \simeq 0.754$)
\be
\label{eq:deltanff}
\frac{\Delta N_{\rm FFLO}}{N} \approx \frac{3\eta\deltabcs}{2 \ef},
\ee
which we can easily estimate using typical values of $\deltabcs$ and $\ef$ from experiments.
For example, the last data point of Fig.~2 of Ref.~\onlinecite{Chin} has Fermi temperature $T_{\rm F} = 1.2 \mu$K
and gap $\deltabcs/h \approx 1$ KHz (here $h$ is Planck's constant).  Converting the former to 
frequency units yields the Fermi energy $\ef \approx 25 $KHz and 
$\deltabcs/\ef \approx .04$, which, when inserted into Eq.~(\ref{eq:deltanff}), yields
$\Delta N_{FFLO}/N \approx .05$, a rather small polarization that will grow closer to the resonance.

Upon expansion (after projection onto a molecular condensate~\cite{Regal}), 
a {\it trapped\/} cloud in the FFLO (supersolid~\cite{AndreevLifshitz,Chester,Leggett70,KimChan})
 phase should exhibit peaks
\be
\label{eq:expansionsimplest}
n(\br,t)\propto {\cal F}(\br - \frac{\hbar t \bQ}{m}),
\ee
in the density profile $n(\br,t)$ at time $t$
 set by the FFLO wavevector
$\bQ$, reminiscent of a Bose superfluid state trapped in a periodic
optical potential~\cite{GreinerOL}, but contrasting from it by the {\em
spontaneous} (since translational symmetry is broken spontaneously)
nature of the peaks.  The width of the peaks (given by the function ${\cal F}(\br)$
that is a Gaussian for a Gaussian trapped cloud, see Appendix~\ref{app:free})
is set by the inverse spatial extent of the FFLO state.
Using the same typical numbers as above from the experiments of 
Ref.~\onlinecite{Chin}, the typical wavevector $Q$ [given by Eq.~(\ref{eq:qresult})] can be estimated 
to be $Q^{-1} \approx 5 \mu$m.

The formula Eq.~(\ref{eq:expansionsimplest}) assumes 
the simplest FFLO-type superfluid $B(\br) \propto {\rm e}^{i\bQ\cdot \br}$.
However, as we have 
discussed, in reality the true ground state in the FFLO regime of the phase diagram 
will likely be a more complicated (but nearly degenerate) state containing more 
Fourier modes~\cite{lo,Bowers,Combescot,Matsuo} $\bQ_n$, yielding a more complicated
density profile of the expanded gas given by the more general formula Eq.~(\ref{eq:fedd})
in Appendix~\ref{app:free}.

Also, the spontaneous anisotropy of the FFLO state
should manifest itself in the atom shot noise
distribution\cite{Altman,Greiner05,Lamacraft}, with peaked $\bk,-\bk$
correlations that are anisotropic around the Fermi surface with the axis
of symmetry spontaneously selected by the \bQ's characterizing the
FFLO ground state.

However, there might be serious impediments for such a direct
detection of the FFLO state in trapped atomic gases. As we showed this
phase is confined to a narrow sliver of the phase diagram on the BCS
side of the resonance. Within LDA, this narrow range of $\delta h$ 
in chemical potential difference translates into a thin FFLO shell at $r_c$ of
width [using Eq.~(\ref{eq:deltarfflo})]
\be
\delta r\sim  0.04 \kf |\as|\frac{R_0^2}{r_c}(1-\frac{r_c^2}{R_0^2}),
\ee
(applying for $r_c$ not too small) with the cloud radius $R_0$.
Now the (above-mentioned) width of the spontaneous Bragg peaks will be limited
from below by this finite shell width as $\delta Q=2\pi/\delta r$. This
places a requirement that $Q \gg 2\pi/\delta r$ in order to be able to
resolve the peaks. This is consistent with the condition of
applicability of LDA.  On the other hand in Sec.~\ref{SEC:FFLO} we found the
optimum $Q$ characterizing FFLO state is given by $Q\propto \deltabcs/\hbar\vf$ [Eq.~(\ref{eq:qresult})]
Hence it is clear that, generally,  $Q\delta r \ll 1$.

The identification of phases and the corresponding quantum and thermal phase
transitions should also be possible through thermodynamics by
measuring, for example, the heat capacity. Although we have not done
detailed quantitative studies of this, general
arguments~\cite{agd} predict that the \sfm and FFLO states at low $T$
should be characterized by a heat capacity that is linear in $T$ due to
gapless atomic excitations around the majority particle Fermi surface of the \sfm
state. This dependence should distinguish the \sfm and FFLO states from a
fully gapped BCS-BEC singlet SF state. On the other hand the finite
molecular BEC peak and ODLRO, along with their dependence on detuning
and polarization, should distinguish the \sfm and FFLO states from the
normal Fermi gas (N) state. Direct observation of gapless atomic
excitations via Bragg spectroscopy~\cite{bragg} should also be
possible. Phase transitions should also be readily identifiable via
standard thermodynamic anomalies, such as divergences of
susceptibilities across a transition, in addition to effects
observable in the density and magnetization profiles.

We hope that our work on this exciting subject of polarized resonant
superfluids will stimulate further careful and detailed experimental
studies to test our predictions.

%%%%%%%%%%%%%%%%%%%%%%%%%%%%%%%%%%%%%%%%%%%%%%%%%%%%%%%%%%%%%%%%%%%%%%%%%%%%%%%%%%%%%%%%%%%%%%%%%%%%%%%%%%%%%%

%------------------------------
\smallskip
\noindent
{\it Acknowledgments\/} ---  
%---------
We gratefully acknowledge useful discussions with S. Choi, J. Gaebler, V. Gurarie, D. Jin, A. Lamacraft,
J. Levinsen, E. Mueller, K. Rajagopal, C. Regal, J. Stewart, M. Veillette 
as well as  support from NSF DMR-0321848 and the Packard Foundation. DES also acknowledges the 
Aspen Center for Physics where part of this work was carried out. 
%--------------------------
\appendix

%%%%%%%%%%%%%%%%%%%%%%%%%%%%%%%%%%%%%%%%%%%%%%%%%%%%%%%%%%%%%%%%%%%%%%%%%%%%%%%%%%%%%%%%%%%%%%%%%%%%%%%%%%%%%%
\section{Scattering amplitude}
\label{SEC:scatteringamplitude}
For completeness, in this Appendix we review the properties of the s-wave scattering amplitude following 
Refs.~\onlinecite{LandauQM,agr}. 
The s-wave scattering amplitude is related to the phase shift $\phaseshift(k)$  
(not to be confused with the bare and renormalized Feshbach resonance detunings $\delta_0$ and $\delta$, respectively)
by
\be
\label{eq:scattamp}
f_0(k) = \frac{1}{2ik} \big({\rm e}^{2i\phaseshift} - 1
\big).
\ee
Clearly, the unitarity  requirement $| 2i k f_0 + 1|=1$ is automatically satisfied,
which implies that $f_0$ may be written as
\be
f_0 = \frac{1}{g_0(k) -i k},
\label{eq:unitarity}
\ee
with $g_0(k)$ a real function of $k^2$.  At low energies, expanding the denominator to leading order in $k^2$ yields
\be
f_0(k) =\frac{1}{-\as^{-1} + r_0 k^2/2 - ik},
\label{eq:unitarity2}
\ee
with $\as$ the s-wave scattering length and $r_0$ the effective range, that for our resonant two-channel 
model is actually negative.

The simplest way to compute the scattering amplitude (in vacuum, $\mu = 0$)  for the 
two-channel model Eq.~(\ref{eq:bareham}) is to note that it is proportional
to the molecular propagator $G_b(E) = \langle \bh(E)  \bh^\dagger(E)\rangle$,
\be
\label{eq:gb}
G_b(E) = \frac{1}{E - \delta_0 - \Sigma(0,E)},
\ee
 with  $\Sigma(q,E)$ the molecular self energy 
(diagram in Fig.~\ref{fig:selfenergy}; see Ref.~\onlinecite{agr}):
\be
\Sigma(q,E) =  g^2 \int \frac{d^3 p}{(2\pi)^3} T \sum_\omega \frac{1}{i\omega - \epsilon_p} 
 \frac{1}{i\omega -i\Omega +\epsilon_{\bq-\bp}}\big|_{i\Omega\to E+i0^{+}}. 
\ee
Taking the $q=0$ and low-$E$ limit, we obtain
\be
\label{eq:sigma}
\Sigma(0,E) \approx -i\frac{g^2 m^{3/2}}{4\pi} \sqrt{E} - g^2\int \frac{d^3 p}{(2\pi)^3} \frac{m}{p^2},
\ee
which, when used in Eq.~(\ref{eq:gb}), gives $G_b(E)$ in the low-energy limit.
Defining the physical (renormalized) detuning $\delta = \delta_0 - g^2 \int \frac{d^3 p}{(2\pi)^3} \frac{m}{p^2}$ to 
absorb the short (molecular) scale dependence in the  scattering amplitude, we have 
\bea
f_0(E) &=& -\frac{\sqrt{\Gamma_0/m}}{E - \delta + i \sqrt{\width} \sqrt{E}},
\\
f_0(k)&=& -\frac{\sqrt{\Gamma_0/m}}{k^2/m  - \delta + i \sqrt{\width} k/\sqrt{m}},
\label{fnoughtkint}
\eea
with 
\be
\width \equiv \frac{g^4 m^{3}}{16\pi^2},
\ee
a measure of the width of the resonance
and where an overall factor [that is the constant of proportionality between $G_b(E)$ and $f_0(E)$]
is fixed using the unitarity condition Eq.~(\ref{eq:unitarity}).
 In obtaining Eq.~(\ref{fnoughtkint}), we replaced $E \to k^2/{2m_r}$, 
with $m_r = m/2$ the reduced mass.
Comparing Eq.~(\ref{fnoughtkint}) to Eq.~(\ref{eq:unitarity2}), we identify
\bea
\label{eq:adefappendix}
\as^{-1} &=& - \frac{\delta \sqrt{m}}{\sqrt{\width}},
\\
\label{eq:rdefappendix}
r_0 &=& - \frac{2}{\sqrt{m}\sqrt{\width}},
\eea
in terms of the detuning $\delta$ and the width $\Gamma_0$.

\subsection{Experimental determination of parameters}
\label{SEC:expdetpar}
The detuning parameter $\delta$ is related to the difference in rest energy between
the closed channel and the open channel~\cite{Timmermans99,Kokkelmans02}.  For a Feshbach
resonance tuned to low energy $\delta$ by a magnetic field, we expect
$\delta \propto (B-B_0)$ near the position of the resonance, where $B_0$ is 
the field at which the resonance is at zero energy.  Determining the precise constant
of proportionality requires a detailed atomic physics analysis that is beyond
the scope of this manuscript.
 However we can approximate $\delta$ by the Zeeman energy difference 
between the closed and open two-atom states.  The latter are approximately dominated
by electronic triplet (open channel) and singlet (closed channel) states, giving 
\be
\delta \approx 2\mubohr (B-B_0), 
\label{eq:approxdetuning}
\ee
with $\mubohr$ the Bohr magneton.  With this approximation, we can extract parameters 
of our model, the most important being the width of the resonance, from current experiments.
Using  Eq.~(\ref{eq:approxdetuning}) inside Eq.~(\ref{eq:adefappendix}), we get 
\be
\as^{-1} = -\frac{2\mubohr\sqrt{m}}{\sqrt{\width}} (B-B_0),
\label{ainversetheory}
\ee
which should be compared with the form~\cite{Timmermans99}
\bea
\as &=& a_{bg}\big( 1- \frac{B_w}{B-B_0}\big),
\label{ainverseexptpre}
\eea
observed near a resonance~\cite{Regal03,Bartenstein05}, with $B_w$ defined to have the same 
sign as $a_{bg}$.  For $B\to B_0$, Eq.~(\ref{ainverseexptpre}) can be written as 
\bea
\as^{-1}&\simeq& -\frac{B-B_0}{a_{bg} B_w},
\label{ainverseexpt}
\eea
and comparison to Eq.~(\ref{ainversetheory}) gives 
\bse
\bea
\sqrt{\width} &=& \frac{2\sqrt{m} \mubohr a_{bg}B_w}{\hbar},
\\
r_0 &=& - \frac{\hbar^2}{m\mubohr a_{bg} B_w},
\eea
\ese
where we have reinserted the correct factors of $\hbar$ so that $\width$ has units of energy 
and $r_0$ has units of length.  Experiments at JILA on $^{40}$K have studied a resonance with ~\cite{Regal03}
 $a_{bg}= 92 $\AA\ and  $B_w = 9.76 G$.
Using $m = 6.64 \times 10^{-26}$Kg gives 
 $r_0 \approx 10$\AA\ and $\sqrt{\width} = 4.06\times10^{-13} \sqrt{J}$.  
Comparing the width $\width$ to the Fermi energy $\ef$ yields  the parameter 
$\gamma \propto \sqrt{\width/\ef}$ [Eq.~(\ref{eq:gammadef})] characterizing
the two-channel model.  We find, for a typical  $\ef = 3\times10^{-33} J$ 
(See Ref.~\onlinecite{Greiner05}), $\gamma \approx 6600$ and $1/\kf |r_0| \approx 2600$.  
Clearly, this Feshbach resonance is quite broad.  

 For the $830 G$  $^6$Li resonance studied in  Ref.~\onlinecite{Bartenstein05}, we have
$|a_{bg}| = 744 $\AA\ and  $|B_w| = 300 G$, that together with 
 $m = 9.96\times10^{-27}$Kg, gives 
$\sqrt{\width} = 3.93\times10^{-11} \sqrt{J}$
and $|r_0| = 0.5 $\AA.
 A typical value for the Fermi energy may be taken from Zwierlein et al~\cite{Zwierlein}, who quote
$\kf^{-1} \simeq 2000 a_0$, with $a_0 = 0.529$ \AA\ the Bohr radius.  This yields $\kf = 9.4 \times10^6 m^{-1}$,
$\ef = \hbar^2 \kf^2/2m = 4.9\times10^{-29} J$, $\gamma \simeq 5000$ and $1/\kf |r_0| \approx 1982$, also a very 
broad resonance.

Finally, we consider the narrow $^6$Li resonance near $543 G$ studied, e.g., in Ref.~\onlinecite{Strecker}.  
Although to our knowledge fermionic superfluidity has not been observed near this resonance, in the near future 
it is quite likely. Estimating $|a_{bg}|  = 100 a_0$ and $|B_w| \approx 0.1 G$ yields (using $T_F \approx 1.4\mu$K
from Ref.~\onlinecite{Strecker}) $\gamma \simeq 18$ and $1/\kf |r_0| \approx 7.3$.

\subsection{Bound states and resonances of scattering amplitude}

The poles of the scattering amplitude determine the positions of 
 bound-states and resonances~\cite{LandauQM}.
Since $r_0$ is negative~\cite{LandauQM,agr,Gurarie}, we write $f_0$ as 
\be
f_0(k) =\frac{1}{-\as^{-1} - |r_0| k^2/2 - ik},
\ee
with the poles given by the quadratic equation:
\be
\label{eq:poles}
k_p = \frac{i \pm i \sqrt{1+ 2|r_0|/\as}}{-|r_0|}.
\ee
The only subtlety is that these solutions do not always correspond to physical bound
states or resonances.  Purely imaginary poles are physical bound states only if
$\imag \,k_p >0$, such that the corresponding 
wavefunction decays at large radius.  Complex poles correspond to a resonance only if the real
part of the energy $\real\, E>0$ {\it and\/}  $\imag\, E<0$.

For the BEC regime $a_s>0$, we can identify the correct pole as the $-$ of Eq.~(\ref{eq:poles}), 
\be
k_p = \frac{i - i \sqrt{1+ 2|r_0|/\as}}{-|r_0|},
\label{eq:pole}
\ee
since it is the one that in the $r_0\to 0$ limit yields the  correct pole at $k_p = i \as^{-1}$. 
This pole is at $k_p=i\kappa$ with $\kappa$ a real and {\it positive\/} (as required)
wavevector (since $\sqrt{1+ 2|r_0|/\as}>1$) and thus corresponds to a 
bound state~\cite{LandauQM,Baym}
at energy $E_p = -\kappa^2/2m_r = -\kappa^2/m$, with $m_r = m/2$ the reduced mass.
  Equation~(\ref{eq:pole}) then 
yields 
\bea
E_p &=& -\frac{2}{m r_0^2} \big[1+\frac{|r_0|}{\as} - \sqrt{1+\frac{2|r_0|}{\as}}
\big],\\
&=& -\frac{\width}{2} \big[1+\frac{2|\delta|}{\width} - \sqrt{1+\frac{4|\delta|}{\width}}
\big],
\eea
where in the final equality we used Eqs.~(\ref{eq:adefappendix}) and ~(\ref{eq:rdefappendix}).
For $|r_0|\ll \as$ or $|\delta|\ll \width$, 
$\kappa \approx 1/\as$, 
a regime that is referred to as \lq\lq universal\rq\rq.  The bound-state energy is
$E_p \approx -\frac{1}{m\as^2} = -\frac{\delta^2}{\width}$.  In the opposite limit of 
  $|r_0|\gg \as$ or $|\delta| \gg \width$,  
$\kappa \approx \sqrt{\frac{2}{\as|r_0|}}$, corresponding to the 
bound-state energy $E \approx -(|r_0| \as m)^{-1} = \delta$.  Thus, in this regime
the bound state simply follows the detuning.

On the BCS side $\as<0$, for $2|r_0|<|\as|$ both the $+$ and $-$ of 
Eq.~(\ref{eq:poles}) apparently give solutions of the form 
$k_p=i\kappa$ for $\kappa$ real. But since, for both, 
$\kappa$ is  negative, they do not correspond to a physical bound state since
the corresponding wavefunction is an
exponentially growing solution of the radial Schr\"odinger equation~\cite{LandauQM,Baym}. 
On the other hand, for
 $2|r_0|>|\as|$, the pole $k_p$ is complex.  Choosing the correct sign of the square root, 
we have the physical pole at 
\be
k_p = \frac{1}{|r_0|}\sqrt{\frac{2|r_0|}{|\as|} -1} -\frac{i}{|r_0|}.
\label{eq:pole2}
\ee
This pole  corresponds to a physical resonance (by definition)
only  when the real part of the energy of the pole is {\it positive\/} while the imaginary part 
is negative.
The resonance energy is 
\bea
E = \frac{k_p^2}{m} = \frac{2}{mr_0^2}\Big(
\frac{|r_0|}{|\as|} - 1 - i \sqrt{2|r_0|/|\as| -1}
\Big),
\eea
which can be written in terms of  the real part of the pole location $E_r$ and width $\Gamma$ as
\bea
E &=& E_r -i\Gamma ,
\\
E_r &\equiv& \frac{2}{mr_0^2} \big(\frac{|r_0|}{|\as|} -1\big) = \delta - \frac{1}{2}\width,
\\
\Gamma&\equiv& \frac{2}{mr_0^2}\sqrt{2|r_0|/|\as| -1}
= \frac{\width}{2}\sqrt{\frac{4\delta}{\width} -1},
\eea
where again we used Eqs.~(\ref{eq:adefappendix})  and (\ref{eq:rdefappendix}).
Thus, there is a true resonance for $1/|\as|> 1/|r_0|$ (or, $\delta>\width/2$), with a width $\Gamma>\width/2$.
We note that $E_r\to 0$ as $\delta\to \width/2$, but $\Gamma(\delta)$ remains finite at this point with 
$\Gamma(\delta=\width/2) = \width/2)$.

The positive-energy resonance in the two-channel model 
 can also be seen in the s-wave partial cross section $\sigma_0 = \frac{4\pi}{k^2} \sin^2 \phaseshift$.
Using Eqs.~(\ref{eq:scattamp}) and  ~(\ref{eq:unitarity}), the phase shift $\phaseshift$ 
satisfies
\bea
{\rm e}^{2i\phaseshift} &=& \frac{g_0 + ik}{g_0 -ik},
\\
g_0 &=& -\as^{-1} + r_0 k^2/2,
\label{eq:gnoughtexpansion}
\eea
so that $\sigma_0$ is given by
\be
\sigma_0 = \frac{4 \pi}{g_0^2+k^2}.
\ee
Using Eq.~(\ref{eq:gnoughtexpansion}) and $E= k^2/m$, we find for $\as<0$ 
after straightforward algebra:
\be
\sigma_0 = \frac{16\pi}{r_0^2 m^2} \frac{1}{(E-E_r)^2 + \Gamma^2},
\ee
the Lorentzian structure expected for a resonance.  
%%%%%%%%%%%%%%%%%%%%%%%%%%%%%%%%%%%%%%%%%%%%%%%%%%%%%%%%%%%%%%%%%%%%%%%%%%%%%%%%%%%%%%%%%%%%%%%%%%%%%%%%%%%%%%

%%%%%%%%%%%%%%%%%%%%%%%%%%%%%%%%%%%%%%%%%%%%%%%%%%%%%%%%%%%%%%%%%%%%%%%%%%%%%%%%%%%%%%%%%%%%%%%%%%%%%%%%%%%%%%
\section{Ground-state energy of two-channel model}
\label{SEC:gsetwochannel}
In this Appendix, we give details of the derivation  of the mean-field  ground-state energy for the two-channel model, Eq.~(\ref{eq:havg}),
presented in Sec.~\ref{SEC:MFT}.  There, we expressed the effective fermion action in terms of the fermion Green function 
$G(\bk,\omega)$, defined in Eqs.~(\ref{eq:green}) and ~(\ref{eq:green2}).  Now, we must simply compute the expectation 
values of $H_K$ and $H_F$, defined in Eqs.~(\ref{eq:onekin}) and ~(\ref{eq:onefbr}).

\subsection{Computation of $\langle H_K \rangle$}

We start by noting that Eq.~(\ref{eq:mudefs}) implies that $H_K$ can be written in the form [this also follows from 
Eq.~(\ref{eq:meanfieldhamiltonianpre})]
\be
 \langle H_K \rangle = \sum_{\bk,\sigma} \xi_k  \langle \ch_{\bk\sigma}^{\dagger} \ch_{\bk\sigma}^{\phdag} \rangle- 
h \Delta N,
\label{eq:hkint}
\ee
with $\Delta N$ the population difference:
\be
\Delta N = \sum_\bk \big( \langle
 \ch_{\bk\uparrow}^{\dagger} \ch_{\bk\uparrow}^{\phdag} \rangle - \langle \ch_{\bk\downarrow}^{\dagger} 
\ch_{\bk\downarrow}^{\phdag} \rangle \big).
\ee
We now compute the two terms of Eq.~(\ref{eq:hkint}) in turn.  The first is given by
\bea
\label{eq:numintermediate}
&&\hspace{-.5cm}\sum_{\bk,\sigma} \xi_k  \langle \ch_{\bk\sigma}^{\dagger} \ch_{\bk\sigma}^{\phdag} \rangle 
 \\
&&\hspace{-.7cm}\quad=
\sum_\bk \xi_{-\bk +\frac{\bQ}{2}} 
T\sum_{\omega_n}\frac{i\omega_n+\xi_{\bk+\frac{\bQ}{2}\downarrow}}{(i\omega_n - \xi_{\bk-\frac{\bQ}{2}\uparrow})
(i\omega_n +\xi_{\bk+\frac{\bQ}{2}\downarrow}) -|\Delta_\bQ|^2}
\nonumber 
\\
&&\hspace{-.7cm}\quad +\sum_\bk \xi_{\bk +\frac{\bQ}{2}} T\sum_{\omega_n}
\frac{i\omega_n+\xi_{\bk-\frac{\bQ}{2}\uparrow}}{(i\omega_n + \xi_{\bk-\frac{\bQ}{2}\uparrow})(i\omega_n 
- \xi_{\bk+\frac{\bQ}{2}\downarrow}) - |\Delta_\bQ|^2 },
\nonumber
\eea
where we have defined $\xi_{k\sigma} \equiv \epsilon_k - \mu_\sigma$.
The first 
frequency sum in Eq.~(\ref{eq:numintermediate}) is easily evaluated after factorizing the denominator: 
\bea\label{eq:numintermediate2}
&&
\hspace{-1cm}T\sum_{\omega_n}\frac{i\omega_n+\xi_{\bk+\frac{\bQ}{2}\downarrow}}{(i\omega_n - \xi_{\bk-\frac{\bQ}{2}\uparrow})
(i\omega_n +\xi_{\bk+\frac{\bQ}{2}\downarrow}) -|\Delta_\bQ|^2}
\\
&&=T\sum_{\omega_n} \frac{i\omega_n + \xi_{\bk+\bQ/2\downarrow}}{(i\omega_n - E_{\bk\uparrow})
(i\omega_n + E_{\bk\downarrow})}\Big],
\nonumber
\\
&& = \frac{1}{2E_\bk} \Big[
(E_{\bk\downarrow} + \xi_{\bk+\frac{\bQ}{2} \downarrow}) n_F(E_{\bk\downarrow})
\nonumber\\
&&\qquad \qquad 
+ 
(E_{\bk\uparrow} - \xi_{\bk+\frac{\bQ}{2} \downarrow}) n_F(E_{\bk\uparrow})
\Big],\nonumber
\eea
with $n_F(x)$ the Fermi function and 
where we used Eqs.~(\ref{eq:energydefs}).
Taking the $T\to 0$ limit, in which $n_F(x) \to \Theta(-x)$, we have 
\bea
\label{eq:numintermediate3}
&&
T\sum_{\omega_n}\frac{i\omega_n+\xi_{\bk+\frac{\bQ}{2}\downarrow}}{(i\omega_n - \xi_{\bk-\frac{\bQ}{2}\uparrow})
(i\omega_n +\xi_{\bk+\frac{\bQ}{2}\downarrow}) -|\Delta_\bQ|^2}
\\
&& = \frac{1}{2} \big[\Theta(-E_{\bk\uparrow}) + \Theta(E_{\bk\downarrow})\big]
+ \frac{\txi_k}{2\te_k}  \big[\Theta(-E_{\bk\uparrow}) - \Theta(E_{\bk\downarrow})\big].
\nonumber
\eea
A similar result for the second frequency sum in Eq.~(\ref{eq:numintermediate}) 
may be obtained by taking $\bQ\to -\bQ$ and $h\to -h$ in Eq.~(\ref{eq:numintermediate3})
(since this operation interchanges $\xi_{\bk+\frac{\bQ}{2}\downarrow}$ and
$\xi_{\bk-\frac{\bQ}{2}\uparrow}$):
\bea
\label{eq:numintermediate4}
&&
T\sum_{\omega_n}\frac{i\omega_n+\xi_{\bk-\frac{\bQ}{2}\uparrow}}{(i\omega_n + \xi_{\bk-\frac{\bQ}{2}\uparrow})
(i\omega_n -\xi_{\bk+\frac{\bQ}{2}\downarrow}) -|\Delta_\bQ|^2}
\\
&& = \frac{1}{2} \big[\Theta(-E_{\bk\downarrow}) + \Theta(E_{\bk\uparrow})\big]
+ \frac{\txi_k}{2\te_k}  \big[\Theta(-E_{\bk\downarrow}) - \Theta(E_{\bk\uparrow})\big],
\nonumber
\\
&&
= 1 -  \frac{1}{2} \big[\Theta(-E_{\bk\uparrow}) + \Theta(E_{\bk\downarrow}) \big]
\nonumber\\
&&
\qquad 
+ \frac{\txi_k}{2\te_k}  \big[\Theta(-E_{\bk\uparrow}) - \Theta(E_{\bk\downarrow}) \big],
\nonumber
\eea
where in the last line we used $\Theta(x) = 1- \Theta(-x)$.  Inserting these sums into 
Eq.~(\ref{eq:numintermediate}), we have after a straightforward rearrangement: 
\bea
&&\hspace{-.5cm}\sum_{\bk,\sigma} \xi_k  \langle \ch_{\bk\sigma}^{\dagger} \ch_{\bk\sigma}^{\phdag} \rangle 
= \sum_\bk \txi_k + \sum_\bk \frac{\txi_k^2}{\te_k} 
\big[\Theta(-E_{\bk\uparrow}) - \Theta(E_{\bk\downarrow}) \big]
\nonumber \\
&&\hspace{-.5cm}\qquad\qquad \qquad
- \sum_\bk \frac{\bk \cdot \bQ}{2m} \big[\Theta(-E_{\bk\uparrow}) + \Theta(E_{\bk\downarrow}) \big].
\label{eq:hkinavgpre}
\eea
Following the same procedure as above, $\Delta N$ is given by 
\bea
&&\hspace{-1cm}\Delta N =  \sum_\bk  T\sum_{\omega_n}\Big[\frac{i\omega_n+\xi_{k+\frac{\bQ}{2}\downarrow}}
{(i\omega_n - \xi_{\bk-\frac{\bQ}{2}\uparrow})(i\omega_n +\xi_{\bk+\frac{\bQ}{2}\downarrow}) -|\Delta_\bQ|^2}
\nonumber
\\
&&\hspace{-1cm}\qquad \qquad 
-\frac{i\omega_n+\xi_{\bk-\frac{\bQ}{2}\uparrow}}
{(i\omega_n + \xi_{\bk-\frac{\bQ}{2}\uparrow})(i\omega_n -\xi_{\bk+\frac{\bQ}{2}\downarrow}) -|\Delta_\bQ|^2},
\\
&&\hspace{-.3cm}= \sum_\bk \Big(-1 + \Theta(-E_{\bk\uparrow})+
\Theta(E_{\bk\downarrow})\Big),
\eea
which can be combined with Eq.~(\ref{eq:hkinavgpre}) to yield (using $\sum_\bk \bk \cdot \bQ = 0$)
\bea
&& \hspace{-1cm}\langle H_K \rangle 
= \sum_\bk \txi_k + \sum_\bk \frac{\txi_k^2}{\te_k} 
\big[\Theta(-E_{\bk\uparrow}) - \Theta(E_{\bk\downarrow}) \big]
\nonumber \\
&&\hspace{-1cm}\qquad
+ \sum_\bk \big(
\frac{\bk \cdot \bQ}{2m}+h
\big)
\big[1-\Theta(-E_{\bk\uparrow}) - \Theta(E_{\bk\downarrow}) \big],
\label{eq:hkinavgappendix}
\eea
used in the main text.

\subsection{Computation of $\langle H_F \rangle$}
The computation of $\langle H_F \rangle$ follows straightforwardly along the lines of 
the calculation of $\langle H_K \rangle$ presented above.  The two terms
comprising $\langle H_F \rangle$ are identical, and yield 
\bea
&&\hspace{-.5cm}\langle H_F\rangle =   \sum_\bk T\sum_{\omega_n} 
\frac{2 |\Delta_\bQ|^2}{(i\omega_n - \xi_{\bk-\frac{\bQ}{2}\uparrow})
(i\omega_n +\xi_{\bk+\frac{\bQ}{2}\downarrow} ) -|\Delta_\bQ|^2},
\nonumber \\
&&\hspace{-.3cm}\qquad  = \sum_k \frac{|\Delta_\bQ|^2}{\te_k} 
\big[\Theta(-E_{\bk\uparrow}) - \Theta(E_{\bk\downarrow}) \big],
\eea
the result used in the main text.

\subsection{Ground state energy at $\Delta_\bQ=0$}

One simple check on our expression for $E_G$, Eq.~(\ref{eq:havgpre}), 
is the limit $\Delta_\bQ \to 0$, required to 
reproduce (at arbitrary $\bQ$)
the ground-state energy for a normal Fermi 
gas under a finite chemical potential difference $h$, a quantity that we also use in the main text.  
Taking this limit, we have 
\bea
&&\hspace{-.5cm}E_G = 
\sum_\bk ( \txi_k - |\txi_k| ) 
  +\sum_k   |\txi_k|\Big(1+   \Theta(-E_{\bk\uparrow})  -
\Theta(E_{\bk\downarrow})
\Big)
\nonumber \\
&& \hspace{-.5cm}\qquad + \sum_\bk  \big(\frac{\bk \cdot \bQ}{2m} + h\big)
\Big(1-   \Theta(-E_{\bk\uparrow})  -
\Theta(E_{\bk\downarrow})
\Big),
\label{eq:havgnormal1}
\eea
where now, at $\Delta_\bQ =0$, $\Theta(-E_{\bk\uparrow})$ and  $\Theta(E_{\bk\downarrow})$ are given by
\bea
\Theta(-E_{\bk\uparrow}) &=& \Theta\big[\frac{\bk \cdot \bQ}{2m} +h - |\txi_k|\big],\label{eq:theta1defnormal}
\\
&=& \Theta(h-\xi_{\bk-\frac{\bQ}{2}})\Theta(\txi_k) +  \Theta(\xi_{\bk+\frac{\bQ}{2}}+h)\Theta(-\txi_k),
\nonumber
\\
\Theta(E_{\bk\downarrow})
&=& \Theta\big[\frac{\bk \cdot \bQ}{2m} +h + |\txi_k| \big],
\label{eq:theta2defnormal}
\\
&=& \Theta(\xi_{\bk+\frac{\bQ}{2}}+h)\Theta(\txi_k) +  \Theta(h-\xi_{\bk-\frac{\bQ}{2}})\Theta(-\txi_k).
\nonumber
\eea
The second lines of Eqs.~(\ref{eq:theta1defnormal}) and ~(\ref{eq:theta2defnormal})
can each be verified by considering the first lines for $\txi_k>0$ and  $\txi_k<0$. With these 
expressions in hand, Eq.~(\ref{eq:havgnormal1}) may be considerably simplified.  Inserting
them into Eq.~(\ref{eq:havgnormal1}), and combining all the momentum sums, we have 
\bea\label{eq:havgnormal2}
&&\hspace{-.7cm}E_G = 
\sum_\bk \Big[
\txi_k\big(1
+ \Theta(h-\xi_{\bk-\frac{\bQ}{2}})- \Theta(\xi_{\bk+\frac{\bQ}{2}}+h)\big)
 \\
&&
\hspace{-.7cm}+  \big(\frac{\bk \cdot \bQ}{2m} + h\big)\big(
1 - \big(\Theta(h-\xi_{\bk-\frac{\bQ}{2}})- \Theta(\xi_{\bk+\frac{\bQ}{2}}+h)\big)
\big)\Big].
\nonumber
\eea
Next, using the identity $\Theta(x) = 1-\Theta(-x)$, Eq.~(\ref{eq:havgnormal2}) 
further simplifies to
\bea
&&\hspace{-.5cm}E_G = 
\sum_\bk \big(\txi_k - \frac{\bk \cdot \bQ}{2m} - h\big)\Theta(h-\xi_{\bk-\frac{\bQ}{2}})
\nonumber \\
&&\hspace{-.5cm}\qquad+ 
\sum_\bk \big(\txi_k + \frac{\bk \cdot \bQ}{2m} + h\big)\Theta(-h-\xi_{\bk+\frac{\bQ}{2}}),
\label{eq:havgnormal3}
\eea
where we note that, since the sums over $\bk$ are restricted to small $\bk$ 
by  the step functions, the sums are each convergent at large $\bk$.  Therefore, it is valid to shift 
$\bk \to \bk + \bQ/2$
and $\bk \to \bk - \bQ/2$ in the first and second terms, respectively, which
yields
\bea
&&\hspace{-1cm}E_G\! =\! 
\sum_\bk \Big[\big(\xi_k\! -\! h\big)\Theta(h\! -\! \xi_k)\!
+ \! \big(\xi_k\! +\! h\big)\Theta(-h\! -\! \xi_k)\Big],
\label{eq:havgnormal4}
\eea
the correct result for the ground-state energy of a normal
Fermi gas under an applied chemical potential difference.  
%%%%%%%%%%%%%%%%%%%%%%%%%%%%%%%%%%%%%%%%%%%%%%%%%%%%%%%%%%%%%%%%%%%%%%%%%%%%%%%%%%%%%%%%%%%%%%%%%%%%%%%%%%%%%%

%%%%%%%%%%%%%%%%%%%%%%%%%%%%%%%%%%%%%%%%%%%%%%%%%%%%%%%%%%%%%%%%%%%%%%%%%%%%%%%%%%%%%%%%%%%%%%%%%%%%%%%%%%%%%%
\section{BEC-BCS crossover at $h = 0$}
\label{SEC:BECBCSreview}
In this Appendix, we review 
the BEC-BCS crossover~\cite{Eagles,Leggett,Nozieres,sademelo,Timmermans01,Holland,Ohashi,agr,Stajic} 
exhibited by $H$, Eq.~(\ref{eq:meanfieldhamiltonianpre}),
at zero chemical potential difference $h = 0$.  This will set the stage for our subsequent treatment 
at finite $h$ of interest to us and will 
 also serve to establish notation.  The mean-field ground-state energy $E_G$ associated with $H$ is Eq.~(\ref{eq:havg}).
After taking $h=0$ in the Eq.~(\ref{eq:havg}),
to be completely general one must then minimize over $\bQ$ and $B_\bQ$.
However, in the absence of any chemical potential difference it is clear that $E_G$ is minimized by
$\bQ=0$ as there is no energetic gain (only cost) of FFLO-type states.
 Thus, at the outset we  set $\bQ=0$ along
with $h=0$ in Eq.~(\ref{eq:havg}), obtaining (writing $\Delta_0$ as $\Delta$ for notational simplicity):
\be
E_G 
= (\delta-2\mu) \frac{\Delta^2}{g^2} + 
\sum_k ( \xi_k - E_k +\frac{\Delta^2}{2\epsilon_k}),
\label{eq:gsefirst}
\ee
where we have defined
\be
\xi_k \equiv \epsilon_k - \mu = \frac{k^2}{2m}-\mu,
\ee
and used Eq.~(\ref{eq:deltaintro}) for the renormalized detuning $\delta$.  
We first compute $E_G$ in the normal state $\Delta = 0$:
\bse
\bea
E_G(\Delta = 0)&=& \sum_k (\xi_k - |\xi_k|),
\label{eq:normalstategsefirst}
\\
&=&  -\frac{8}{15} c\mu^{5/2}\Theta(\mu),
\label{eq:normalstategsefirstp}
\eea
\ese
where we converted the sum to an integral and used the three-dimensional 
density of states $N(E) = c \sqrt{E}$ with 
\be
\label{eq:cdefappendix}
c\equiv \frac{m^{3/2}}{\sqrt{2}{\pi^2}}.
\ee
  Combining 
this with Eq.~(\ref{eq:gsefirst}) then gives:
\bea
E_G &=& (\delta-2\mu) \frac{\Delta^2}{g^2}
- \frac{8}{15}c\mu^{5/2}\Theta(\mu) + I(\mu,\Delta),
\label{eq:cfive}
\eea
where 
\bea
I(\mu,\Delta) &\equiv&
\int \frac{d^3 k}{(2\pi)^3} \big(|\xi_k|-E_k + \frac{\Delta^2}{2\epsilon_k}\big),
\label{eq:gseintsecond}
\eea
where we have converted the momentum sum to an integral.  

%
%-----------------------------
%
% fig%54
%
\begin{figure}[bth]
\vspace{1.4cm}
\centering
\setlength{\unitlength}{1mm}
\begin{picture}(40,40)(0,0)
\put(-50,0){\begin{picture}(0,0)(0,0)
\includegraphics{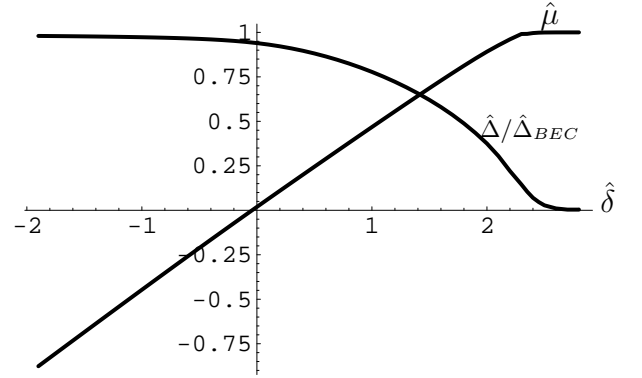}
\end{picture}}
\put(51,48) {\large{$\muh$}}
\put(59,23.0) {\large{$\deltah$}}
\put(43,33) {{$\Deltah/\Deltah_{BEC}$}}
\end{picture}
\vspace{-.5cm}
\caption{(Color online) Plot of $\Deltah/\Deltah_{BEC}$ (i.e., the gap normalized to its asymptotic value 
$\Deltah_{BEC} = \sqrt{2\gamma/3}$) in the BEC regime) and $\muh$, as a function of normalized detuning
$\deltah = \delta/\ef$ for $\gamma = 0.1$.}
\label{BECBCSplot}
\end{figure}
%------------------------------

The standard BEC-BCS crossover follows from finding the minimum of $E_G$ which satisfies
the gap equation 
\bse
\label{eq:eomzeroh}
\be
0 = \frac{\partial E_G}{\partial \Delta},
\ee
while satisfying the number constraint
\be
N = -\frac{\partial E_G}{\partial \mu},
\label{eq:numzeroh}
\ee
\ese
which we evaluate numerically in Fig.~\ref{BECBCSplot}.

For a narrow Feshbach resonance ($\gamma \ll 1$),
we can find accurate analytic approximations to $E_G$ in Eq.~(\ref{eq:cfive})
in all relevant regimes.  The first step is finding an appropriate approximation
to Eq.~(\ref{eq:gseintsecond}), which has drastically 
different
properties depending on whether  $\mu>0$ (so that the low-energy states are near the Fermi
surface) or $\mu<0$ (so that there is no Fermi surface and excitations are gapped with
energy bounded from below by $|\mu|$). We proceed by first evaluating 
the derivative $\frac{\partial I}{\partial \Delta}$ and then integrating 
the expression with constant of integration $I(\mu,0)=0$.
\bse
\bea
&&\hspace{-1cm}\frac{\partial I}{\partial\Delta} = -\Delta  \int \frac{d^3 k}{(2\pi)^3} \big(
\frac{1}{E_k} - \frac{1}{\epsilon_k}
\big),
\\
 &&\hspace{-1cm}\quad\,\,\,\,\,= -c\Delta  \int_0^{\infty} \sqrt{\epsilon} d\epsilon
 \big(
\frac{1}{\sqrt{(\epsilon-\mu)^2+\Delta^2}} - \frac{1}{\epsilon}
\big),
\\
&&\hspace{-1cm}\quad\,\,\,\,\,\simeq -2  N(\mu) \Delta\ln \frac{8{\rm e}^{-2} \mu}{\Delta}, 
\hskip 0.5cm \mu>0; \mu \gg \Delta,
\label{eq:iderbcs}
\\
&&\hspace{-1cm}\quad\,\,\,\,\,\simeq   N(\mu)\Delta  
\Big[ \pi + \frac{\pi}{16} \big(\frac{\Delta}{\mu}\big)^2\Big], \hskip 0.5cm \mu<0;
 |\mu| \gg \Delta,
\label{eq:ibec}
\eea
\ese
Equation~(\ref{eq:ibec}) may be obtained by Taylor expanding the integrand in $\Delta\ll |\mu|$ and 
integrating term by term, with details of the derivation of Eq.~(\ref{eq:iderbcs}) appearing in 
Ref.~\onlinecite{Gurarie}.
Integrating with respect to $\Delta$,  we thus have
\bse
\label{eq:i}
\bea
&&\hspace{-1cm}I \simeq 
\begin{cases}
 - N(\mu)\big(\frac{\Delta^2}{2} + \Delta^2 \ln \frac{8{\rm e}^{-2} \mu}{\Delta}\big)
 & \text{$\mu>0$; $\mu \gg \Delta$}, 
\cr
    N(\mu) \frac{\Delta^2}{2} 
\Big[ \pi + \frac{\pi}{32} \big(\frac{\Delta}{\mu}\big)^2\Big] 
&\text{$\mu<0$; $|\mu| \gg \Delta$},
\label{eq:itwo}
\cr
\end{cases}
\eea
\ese
Having computed $E_G(\mu,\Delta)$ in the regimes of interest, the phase diagram is easily deduced
by finding $\Delta$ that minimizes $E_G(\mu,\Delta)$, subject to the total atom number constraint
Eq.~(\ref{eq:numzeroh}).

\subsection{BCS regime}

The BCS regime is defined by $\delta \gg 2\ef$, where $\Delta \ll \mu$ and $\mu\simeq \ef>0$,
with pairing taking place in a thin shell around the well-formed Fermi surface.
In this regime, $E_G$ is given by 
\bea
&&E_{G} \simeq -c\frac{\sqrt{\mu}}{2}\Delta^2 +\frac{\Delta^2}{g^2} (\delta - 2\mu)  + 
c\sqrt{\mu} \Delta^2 \ln\frac{\Delta} {8 {\rm e}^{-2} \mu}
\nonumber \\
&& \qquad \qquad  -\frac{8}{15} c\mu^{5/2}.
\label{eq:eg0fin}
\eea

It is convenient to work with the dimensionless variables defined in Eq.~(\ref{eq:dimensionlessvariables}).
The normalized ground-state energy $e_G$ in the BEC regime is then given by 
\bea
&&e_G\equiv \frac{E_{G}}{c\ef^{5/2}} \simeq -\frac{\sqrt{\muh}}{2}\Deltah^2 +\Deltah^2 (\deltah - 2\muh) \gamma^{-1}
\nonumber \\
&& \qquad +
\sqrt{\muh} \Deltah^2 \ln\frac{\Deltah} {8 {\rm e}^{-2} \muh} - \frac{8}{15} \muh^{5/2},
\eea
where, $\gamma$, defined in Eq.~(\ref{eq:gammadef}), is a dimensionless measure of the Feshbach resonance
width $\width$ to the Fermi energy.  With this, 
Eqs.~(\ref{eq:eomzeroh}) become 
\bse
\bea
&&\hspace{-1.2cm}0 = \frac{\partial e_G}{\partial\Deltah},
\\
&&\hspace{-1.2cm}\quad\simeq   2\Deltah (\deltah - 2\muh )\gamma^{-1} + 
2\sqrt{\muh} \Deltah \ln \frac{\Deltah} {8 {\rm e}^{-2} \muh},
\label{eq:gapzeroh2}
\\
&&\hspace{-1.2cm}\frac{4}{3} = -\frac{\partial e_G}{\partial\muh},
\\
&&\hspace{-1.2cm}\quad\simeq\frac{5}{4} \frac{\Deltah^2}{\sqrt{\muh}} + \frac{4}{3} \muh^{3/2} + 2\Deltah^2 \gamma^{-1}
- \frac{\Deltah^2}{2\sqrt{\muh}}\ln \frac{\Deltah}{8{\rm e}^{-2} \muh},
\label{eq:numzeroh2}
\eea
\ese
that admits the normal state ($\Deltah = 0$, $\muh = 1$) and the BCS SF state 
\bse
\bea
\label{eq:baredeltanought}
\deltah&\simeq&\deltahbcs(\muh) \equiv 8{\rm e}^{-2} \muh{\rm e}^{-\gamma^{-1}(\deltah - 2\muh)/\sqrt{\muh}},
\\
\frac{4}{3} &\simeq&  \frac{4}{3} \muh^{3/2} + 2\Deltah^2 \gamma^{-1}.
\label{eq:weakcouplingbcsnum}
\eea
\ese
where in the second line we approximately neglected the first term on the right side of Eq.~(\ref{eq:numzeroh2}),
valid since $\deltahbcs \ll 1$ (and $\gamma \ll 1$).
It is easy to show that the BCS solution is always a minimum of $E_G$.

The meaning of the two terms on the right side of Eq.~(\ref{eq:weakcouplingbcsnum})
 is clear once we recall its form in terms of dimensionful quantities:
\be
n \simeq \frac{4}{3}c\mu^{3/2} + 2|b|^2,
\ee
i.e., the  first term simply represents the total unpaired atom density, reduced below $n$ since
$\mu<\ef$, while the second term represents the density of atoms bound into molecules,
i.e., twice the molecular density $|b|^2$.  Qualitatively, we see that at large $\deltah$, 
$\Deltah \ll 1$, implying from the number equation that $\muh \alt 1$.

\subsection{BEC regime}

We next consider the BEC regime defined by  $\delta < 0$.  As we shall see, in this regime
$\mu < 0$ and $|\mu| \gg \Delta$, so that Eq.~(\ref{eq:i}), $I(\mu,\Delta)$, applies.
This yields, for the  normalized ground-state energy, 
\be
e_{G} \simeq (\deltah - 2\muh ) \Deltah^2 \gamma^{-1} 
 + \sqrt{|\muh|} \frac{\Deltah^2}{2}  \Big[ \pi + \frac{\pi}{32} \big(\frac{\Deltah}{\muh}\big)^2\Big],
\ee
and, for the gap and number equations (dividing by an overall factor of $\Deltah$ in the former)
\bse
\bea
\label{becgap}
0 &\simeq& 2\gamma^{-1} (\deltah - 2\muh ) + \sqrt{|\muh|}\Big[ \pi + \frac{\pi}{16} \big(\frac{\Deltah}{\muh}\big)^2\Big],
\\
\frac{4}{3}  &\simeq& 2\gamma^{-1} \Deltah^2 + \frac{\Deltah^2 \pi }{ 4\sqrt{|\muh|}}.
\label{becnum}
\eea
\ese
As noted in Ref.~\onlinecite{sademelo}, in the BEC regime the roles of the two equations are reversed,
with $\muh$ approximately determined by the gap equation and $\Deltah$ approximately determined by the 
number equation.
Thus, $\muh$ is well-approximated by neglecting the term proportional to $\Deltah^2$  in Eq.~(\ref{becgap}),
giving
\bea
\muh &\approx& \frac{\deltah}{2} \Big[
\sqrt{1+ 
\frac{\gamma^2 \pi^2}{32 |\deltah|}} - 
\frac{\gamma \pi}{\sqrt{32 |\deltah|}}
\Big]^2.
\label{eq:eomzeroh2p}
\eea
At large negative detuning, $|\deltah| \gg 1$, where it is valid
in the BEC regime, Eq.~(\ref{eq:eomzeroh2p})  reduces to $\mu\approx \deltah/2$,
with
the chemical potential tracking the detuning.

Inserting Eq.~(\ref{eq:eomzeroh2p}) into Eq.~(\ref{becnum}) yields
\be
\label{eq:deltazeroh}
\Deltah^2 = \frac{2\gamma}{3}\Big[
1- \frac{\gamma \pi}{\sqrt{(\gamma \pi)^2 + 32|\deltah|}}
\Big].
\ee
Using $\Deltah = \Delta/\ef$ and the relation $\Delta^2 = g^2n_m$ between $\Delta$ and
the molecular density, we have 
\bea
n_m &=& \frac{3}{4} \gamma^{-1} \Deltah^2 n,
\\
&\simeq&\frac{n}{2}\Big[
1- \frac{\gamma \pi}{\sqrt{(\gamma \pi)^2 + 32|\deltah|}}
\Big],
\eea
which, as expected (given the fermions are nearly absent for $\mu<0$)
 simply yields $n_m \approx n/2$ in the asymptotic (large $|\deltah|$) BEC regime.

%%%%%%%%%%%%%%%%%%%%%%%%%%%%%%%%%%%%%%%%%%%%%%%%%%%%%%%%%%%%%
%%%%%%%%%%%%%%%%%%%%%%%%%%%%%%%%%%%%%%%%%%%%%%%%%%%%%%%%%%%%%

\section{Derivation of Eq.~(\ref{eq:havgBCS2})}
\label{SEC:appsums}
In the present appendix, we provide the steps leading from Eq.~(\ref{eq:havgBCSp}) to Eq.~(\ref{eq:havgBCS2}).
To this end we need to evaluate
\bea
S(h) &=& S_1(h) + S_2(h) ,
\\
S_1(h) &\equiv& E_k\sum_k \Theta(h - E_k),
\\
S_2(h) &\equiv& -h\sum_k \Theta(h - E_k).
\eea
We shall derive Eq.~(\ref{eq:havgBCS2}) by showing that $\partial S/\partial h = -\mg(h)$,  with $\mg(h)$ given by 
Eq.~(\ref{eq:mbcs}).  Since $S(0) = 0$ (recall $E_k>0$), this is sufficient.  Firstly, 
using $\Theta'(x) = \delta(x)$ with $\delta(x)$ the Dirac delta function, we have 
\bea
\frac{\partial S_1}{\partial h} &=& \sum_k E_k \delta(h - E_k) =  h\sum_k  \delta(h - E_k),
\label{eq:sonediff}
\\
\frac{\partial S_2}{\partial h} &=& -  \sum_k \Theta(h - E_k) -h\sum_k \delta(h - E_k).
\eea
Adding these, we obtain
\be
\frac{\partial S}{\partial h} = -\sum_k \Theta(h - E_k) = -\vol \mg(h),
\ee
which provides the connection between Eq.~(\ref{eq:havgBCSp}) and Eq.~(\ref{eq:havgBCS2}).

%%%%%%%%%%%%%%%%%%%%%%%%%%%%%%%%%%%%%%%%%%%%%%%%%%%%%%%%%%%%%%%%%%%%%%%%%%%%%%%%%%%%%%%%%%%%%%%%%%%%%%%%%%%%%%
\section{Derivation of leading-order contribution to t-matrix}
\label{tmatrix}

The leading-order contribution to the T-matrix is given by the Feynman diagram in Fig.~\ref{fig:bubble}  (corresponding
to molecular scattering), with external momenta
and frequencies set equal to zero.   We find it easiest to compute this diagram by starting at finite temperature $T$ 
before taking the $T\to 0$ limit.  Standard analysis gives 
\be
T_m = g^4 T \sum_\omega \int \frac{d^3p}{(2\pi)^3} \frac{1}{(i\omega - \xi_{p\uparrow})^2 }
\frac{1}{(-i\omega - \xi_{p\downarrow})^2},
\ee
with $\xi_{p\sigma} = \epsilon_p - \mu_\sigma$ as in the main text. The fermionic Matsubara frequency
sum can be straightforwardly evaluated using standard techniques:~\cite{Mahan}
\bea
S(A,B) &\equiv & T\sum_\omega \frac{1}{(i\omega - A)^2} \frac{1}{(i\omega - B)^2} ,
\\
 &=&  \frac{d}{dA}\frac{d}{dB}  T\sum_\omega \frac{1}{i\omega - A} \frac{1}{i\omega - B} ,
\\
 &=&   \frac{d}{dA}\frac{d}{dB} \frac{n_F (A) - n_F(B)}{A-B}.
\eea
Taking the $T\to 0$ limit, in which $n_F(x) \to \Theta(-x)$ and evaluating the derivatives,
we have 
\bea
&&S(A,B)  =  \frac{-2}{(A-B)^3} \big[ \Theta(-A) - \Theta(-B)
\big] 
\nonumber \\
&& \qquad \qquad  \qquad 
- \frac{\delta(A)}{B^2} - \frac{\delta(B)}{A^2}.
\label{eq:expressionsab}
\eea
   Using Eq.~(\ref{eq:expressionsab}) for  $S(A,B)$, 
with $A = \xi_{p\uparrow}$ and $B = -  \xi_{p\downarrow}$,  the molecular T-matrix is  (using the density of states 
$N(\epsilon) = c\sqrt{\epsilon}$) given by:
\bea
&&T_m = - \frac{c g^4}{4} \int_0^{\infty} d\epsilon \sqrt{\epsilon} \frac{1}{(\epsilon - \mu)^3} \big[
 \Theta(-\epsilon + \mu_\uparrow) - \Theta(\epsilon - \mu_\downarrow)
\big]
\nonumber \\ 
&&\qquad \qquad  - c g^4 \int_0^{\infty} d\epsilon\sqrt{\epsilon} \Big[ 
\frac{\delta(\epsilon - \mu_\uparrow)}{(\epsilon - \mu_\downarrow)^2}
+ \frac{\delta(\epsilon - \mu_\downarrow)}{(\epsilon - \mu_\uparrow)^2}\big].
\label{eq:intvertex}
\eea
Since we are in the BEC regime, $\mu<0$.  Taking (without loss of generality) $h>0$, we have $\mu_\downarrow <0$ 
always and the second step function in the first line of Eq.~(\ref{eq:intvertex}) is always unity.  Moreover,
for the same reason,
the second delta function in the second line 
is always zero.  However, $\mu_\uparrow$ does change signs with increasing $h$, so that the first delta function in the second line
can contribute.  Evaluating the remaining integrals in Eq.~(\ref{eq:intvertex}) then yields
\be
T_m =   \frac{c g^4 \pi}{32 |\mu|^{3/2}} F_4(h/|\mu|),
\ee
with $F_4(x)$ defined in Eq.~(\ref{eq:f}), the result used in the main text.

%%%%%%%%%%%%%%%%%%%%%%%%%%%%%%%%%%%%%%%%%%%%%%%%%%%%%%%%%%%%%%%%%%%%%%%%%%%%%%%%%%%%%%%%%%%%%%%%%%%%%%%%%%%%%%
\section{Computation of excluded sums}
\label{app:ex}
In this Appendix, we provide details for the computation of the \lq\lq excluded sums\rq\rq 
appearing in the ground-state energy Eq.~(\ref{eq:havgpre2})~\cite{ff}.  The first such 
sum is 
\be
S_1  \equiv \sum_k \te_k \Big(1+ \Theta(-E_{\bk\uparrow}) - 
\Theta(E_{\bk\downarrow})\Big).
\label{eq:firstone}
\ee
Using Eqs.~(\ref{ekuparrow}) and ~(\ref{ekdownarrow}) for $E_{\bk \sigma}$, we see 
the first step function gives unity for (with $\theta$ the angle between $k$ 
and $Q$, and in this section dropping the subscript $\bQ$ on $\Delta_\bQ$ for simplicity)
\be
\label{condpre1}
\frac{k Q \cos\theta}{2m} +h> \sqrt{\txi_k^2+\Delta^2},
\ee
while the second gives unity for 
\be\label{condpre2}
\frac{k Q \cos\theta}{2m} +h> -\sqrt{\txi_k^2+\Delta^2},
\ee
and each vanishes otherwise.
Clearly, if Eq.~(\ref{condpre1}) is satisfied, then so is Eq.~(\ref{condpre2}).
Thus, out of four possibilities only two nonzero contributions
to Eq.~(\ref{eq:firstone}) occur, when either both inequalities are satisfied or when both are violated.

The conditions Eqs.~(\ref{condpre1}) and (\ref{condpre2}) restrict the momentum sum in 
Eq.~(\ref{eq:firstone}) to the immediate vicinity of the Fermi surface.
Thus, we shall replace $k\to \tkf$ on the left side of 
Eqs.~(\ref{condpre1}) and (\ref{condpre2}), where $\tkf$ is the Fermi
wavevector associated with the adjusted chemical potential 
$\tilde{\mu} \equiv \mu - Q^2/8m$, i.e., $\tkf^2/2m = \tilde{\mu}$.
Following Ref.~\onlinecite{ff}, we now determine where these conditions intersect
the Fermi surface where $\txi_k \to 0$.  In this limit, Eqs.~(\ref{condpre1}) and (\ref{condpre2})
can be written as
\bea
\label{costhetamin}
&&\cos\theta > \cos\theta_{min} \equiv \frac{1 - \dbmu}{\Qb},
\\
&&\cos\theta > \cos\theta_{max} \equiv -\frac{1 + \dbmu}{\Qb},
\label{costhetamax}
\eea
where (as in the main text) we defined the rescaled momenta $\Qb \equiv \frac{\tkf Q}{2m \Delta}$ and 
the rescaled chemical potential difference $\dbmu \equiv \frac{h}{\Delta}$. 

Now, $S_1$ can be straightforwardly computed (converting sums to integrals via
 $\sum_k \to \frac{N(\tilde{\mu})}{2}\int d\varepsilon \int d\cos \theta$)
\bea
&&\hspace{-1cm}S_1 =  \Delta N(\tilde{\mu}) \int_{cos \theta_{min}}^{1} 
d\cos\theta \int_0^{\varepsilon_+} d\varepsilon \sqrt{1+ \frac{\varepsilon^2}{\Delta^2}}
\nonumber \\
 &&\hspace{-1cm}\qquad+ \Delta N(\tilde{\mu}) \int_{-1}^{cos \theta_{max}} 
d\cos\theta \int_0^{\varepsilon_+} d\varepsilon \sqrt{1+ \frac{\varepsilon^2}{\Delta^2}},
\label{eq:exclsumint}
\eea
where 
\be
\varepsilon_+ \equiv  \Delta \sqrt{(\Qb\cos\theta + \dbmu)^2-1}, 
\ee
is the maximum energy at a particular $\theta$ where the contribution to $S_1$ is finite.
The two lines of Eq.~(\ref{eq:exclsumint}) correspond to the cases when both Eq.~(\ref{condpre1}) and 
 Eq.~(\ref{condpre2}) are satisfied, or both violated, respectively.

Clearly, if $\cos \theta_{max} <-1$ or $\cos \theta_{min} > 1$ (as defined by Eq.~(\ref{costhetamin}) and 
Eq.~(\ref{costhetamax}))  then these integrals simply vanish. These limits are manifested in
the final expression by step ($\Theta$) functions in the final result.  Furthermore, we have implicitly
assumed in Eqs.~(\ref{costhetamin}) and (\ref{costhetamax}) that  $\frac{1 - \dbmu}{\Qb}<1$, 
and  $-\frac{1 + \dbmu}{\Qb}>-1$.  If either assumption is not valid, then the 
corresponding integration range over $\cos\theta$ in Eq.~(\ref{eq:exclsumint}) becomes 
$\int_{-1}^1 d\cos \theta$.  Taking this into account, the integrals in Eq.~(\ref{eq:exclsumint})
yield for $S_1$:
\bea
&&\hspace{-.6cm}S_1 =
\frac{N(\tilde{\mu}) \Delta^2}{2\Qb} \big[
H(\Qb+\dbmu) + H(\Qb-\dbmu) - H(\dbmu-\Qb)
\big],
\\
 &&\hspace{-.6cm}H(x) \equiv \Big[\frac{1}{3} (x^2-1)^{3/2}\! +\! x\cosh^{-1}\!x - \sqrt{x^2-1}\Big]\Theta(x-1).
\nonumber 
\eea
Following the same procedure for $S_2$, the final line of Eq.~(\ref{eq:havgpre2}), gives
\bea  
&&\hspace{-1.2cm}S_2 \equiv \sum_k \big(\frac{\bk\cdot \bQ}{2m} + h
\big)  \big(1-\Theta(-E_{\bk\uparrow}) - \Theta(E_{\bk\downarrow}) \big),
\\
&&\hspace{-1.2cm}\quad\,=
 -\frac{N(\tilde{\mu}) \Delta^2}{3\Qb} \big[ J(\Qb+\dbmu)\!
+ \! J(\Qb-\dbmu)\! -\!  J(\dbmu - \Qb) \big],
\label{eq:sumweneed}
\eea
with
\be
J(x) \equiv (x^2 - 1)^{3/2}\Theta(x-1), 
\ee
which is the result used in the main text.

%%%%%%%%%%%%%%%%%%%%%%%%%%%%%%%%%%%%%%%%%%%%%%%%%%%%%%%%%%%%%%%%%%%%%%%%%%%%%%%%%%%%%%%%%%%%%%%%%%%%%%%%%%%%%%

%%%%%%%%%%%%%%%%%%%%%%%%%%%%%%%%%%%%%%%%%%%%%%%%%%%%%%%%%%%%%%%%%%%%%%%%%%%%%%%%%%%%%%%%%%%%%%%%%%%%%%%%%%%%%%
\section{Free expansion}
\label{app:free}
In this Appendix, for completeness we review the free expansion dynamics of a trapped Bose gas, recalling
how it yields information about the initial boson momentum 
distribution~\cite{Cornellexp,Ketterleexp,Hollandetal,read}.
This is of interest here as free expansion is a direct probe able to distinguish and identify the 
phases discussed in this paper. Its application to the FFLO state requires an additional first step of
sweeping the resonance in the usual manner~\cite{Regal} to project the FFLO state onto a finite momentum
molecular condensate.

We take the initial state of the Bose system to be a condensate characterized by a single-particle
wavefunction $B_0(\br)$.  After time $t$ of free expansion, $B_0(\br)$ evolves into $b(\br,t)$
given by
\bea
\label{eq:psiexp}
b(\br,t) &=& \int \frac{d^3k}{(2\pi)^3} \tilde{B}_0(\bk)
{\rm e}^{i \bk \cdot \br}  {\rm e}^{-i \hbar t k^2/2m},
\\
\tilde{B}_0(\bk) &=&\int d^3r  B_0(\br) {\rm e}^{-i \bk \cdot \br},
\label{eq:psitilde}
\eea
governed by  the free-particle Schr\"odinger equation.
The corresponding spatial boson density distribution at time $t$ is
\be
\label{eq:bosondensity}
n(\br,t) = \langle \bh^{\dagger}(\br,t)\bh(\br,t) \rangle \simeq b^*(\br,t)b(\br,t).
\ee
Inserting Eq.~(\ref{eq:psiexp}) into Eq.~(\ref{eq:bosondensity}), using Eq.~(\ref{eq:psitilde}),
and shifting $\bk \to \bk + m\br/\hbar t$, we have 
\bea
\label{eq:nint}
&&
\hspace{-1.1cm}
n(\br,t)\!\simeq\!\!\int\! d^3r_1 d^3r_2 f(\br_1,\br_2) {\rm e}^{i\frac{m}{\hbar t}\br\cdot (\br_1 - \br_2)}
 \!B_0^*(\br_1) \!B_0^\phstar(\br_2),
\\
&&
\hspace{-1cm}
f(\br_1,\br_2) \!\equiv\! \int \frac{d^3 k}{(2\pi)^3} \frac{d^3 k'}{(2\pi)^3}  
{\rm e}^{i(\bk \cdot \br_1 -\bk' \cdot \br_2)}
{\rm e}^{i  \frac{\hbar t}{2m} (k^2-k'^2)},
\eea
The function $f(\br_1,\br_2)$ can be evaluated by changing variables to $\bk,\bk' = \bp \pm \bq/2$, giving
\bea
f(\br_1,\br_2)\!\! &=&\!\!\int \frac{d^3 p}{(2\pi)^3} \frac{d^3 q}{(2\pi)^3}  
{\rm e}^{i\bp \cdot (\br_1-\br_2)}{\rm e}^{i\bq\cdot(\br_1+ \br_2)/2}
{\rm e}^{i (\hbar t/m)\bp\cdot\bq}, \nonumber
\\
&=&\Big(\frac{m}{2\pi\hbar t}
\Big)^3
{\rm e}^{im(\br_1 +\br_2)(\br_1 - \br_2)/2\hbar t},
\label{eq:ffin}
\eea
Inserting Eq.~(\ref{eq:ffin}) into Eq.~(\ref{eq:nint}), we find 
\bea
&&
\nonumber
n(\br,t)
\simeq
\Big(\frac{m}{2\pi\hbar t}
\Big)^3\int d^3 r_1 d^3r_2\,
 {\rm e}^{i\frac{m}{\hbar t}(\br_1-\br_2)\cdot (\br +\frac{1}{2}[\br_1+\br_2])}
\\
&&\qquad\qquad\qquad\times
 B_0^*(\br_1) B_0^\phstar(\br_2).
\label{eq:nfin}
\eea
Noting that the initial cloud is small compared to the expanded one, we may neglect $\br_1$ and 
$\br_2$ compared to $\br$ in the exponential in the first line of Eq.~(\ref{eq:nfin}) since
$\br$ is measured in the expanded cloud while $\br_1$ and $\br_2$ are confined to the initial cloud.
This reduces $n(\br,t)$ to
\bea
n(\br,t)
&\simeq&
\Big(\frac{m}{2\pi\hbar t}
\Big)^3
\tilde{B}_0^*\big(\frac{m}{\hbar t}\br\big) \tilde{B}_0^\phstar\big(\frac{m}{\hbar t}\br\big) ,
\\
&\simeq&
\Big(\frac{m}{2\pi\hbar t}
\Big)^3
n_{\bk = \frac{m}{\hbar t} \br},
\label{eq:fedd}
\eea
where $n_\bk$ is the momentum distribution function.
Thus, as advertised, the density profile $n(\br,t)$ in the expanded cloud probes 
the  {\it initial\/} momentum distribution.  For the simplest trapped FFLO-type state 
$B_0(\br) = B_\bQ(\br) {\rm e}^{i\bQ\cdot \br}$, Eq.~(\ref{eq:fedd}), with $B_\bQ(\br)$
the shape of the amplitude envelope determined by the trap.  Taking it (for concreteness
and simplicity) to be a Gaussian $B_\bQ(\br) \propto {\rm e}^{-r^2/2R_0^2}$ and
using Eq.~(\ref{eq:fedd}) we find 
\be
n(\br,t) \propto \exp\Big[
-(\br - \frac{\hbar t}{m} \bQ)^2/(\frac{\hbar t}{m} R_0^{-1})^2
\Big],
\label{eq:nofrgauss}
\ee
i.e., a Gaussian peaked at $\br =  \frac{\hbar t \bQ}{m}$ thus probing the FFLO wavevector $\bQ$.
Requiring the peak location be much larger than the Gaussian 
width $\hbar t/mR_0$ thus implies that $QR_0\gg 1$ is necessary to observe the FFLO state in this manner.

%%%%%%%%%%%%%%%%%%%%%%%%%%%%%%%%%%%%%%%%%%%%%%%%%%%%%%%%%%%%%%%%%%%%%%%%%%%%%%%%%%%%%%%%%%%%%%%%%%%%%%%%%%%%%%

%%%%%%%%%%%%%%%%%%%%%%%%%%%%%%%%%%%%%%%%%%%%%%%%%%%%%%%%%%%%%%%%%%%%%%%%%%%%%%%%%%%%%%%%%
%%%%%%%%%%%%%%%%%%%%%%%%%%%%%%%%%%%%%%%%%%%%%%%%%%%%%%%%%%%%%%%%%%%%%%%%%%%%%%%%%%%%%%%%%
%%%%%%%%%%%%%%%%%%%%%%%%%%%%%%%%%%%%%%%%%%%%%%%%%%%%%%%%%%%%%%%%%%%%%%%%%%%%%%%%%%%%%%%%%


\begin{thebibliography}{10}
%--------------------------------
%--------------------------------
\bibitem{Regal}
C.A. Regal, M. Greiner, and D.S. Jin,
%
Phys. Rev. Lett. {\bf 92}, 040403 (2004).
%
% Observation of resonance condensation of Fermionic atom pairs
%
%--------------------------------
%--------------------------------
\bibitem{Zwierlein}
M.W. Zwierlein, C.A. Stan, C.H. Schunck, S.M.F. Raupach, A.J. Kerman, and W. Ketterle,
%
Phys. Rev. Lett. {\bf 92}, 120403 (2004).
%
% Condensation of pairs of fermionic atoms near a Feshbach resonance
%--------------------------------
%--------------------------------
\bibitem{Kinast}
J. Kinast, S.L. Hemmer, M.E. Gehm, A. Turlapov, and J.E. Thomas,
 Phys. Rev. Lett. {\bf 92}, 150402 (2004).
%
% Evidence for Superfluidity in a Resonantly Interacting Fermi Gas
%--------------------------------
%--------------------------------
\bibitem{Bartenstein}
    M. Bartenstein, A. Altmeyer, S. Riedl, S. Jochim, C. Chin, J.H. Denschlag, and R. Grimm,
 Phys. Rev. Lett. {\bf 92}, 203201 (2004).
%
% Collective Excitations of a Degenerate Gas at the BEC-BCS Crossover
%-------------------------------- 
%-------------------------------- 
\bibitem{Bourdel04}
T. Bourdel, L. Khaykovich, J. Cubizolles, J. Zhang, F. Chevy, M. Teichmann, L. Tarruell, 
S.J.J.M.F. Kokkelmans, and C. Salomon,
Phys. Rev. Lett. {\bf 93}, 050401 (2004).
%
%Experimental Study of the BEC-BCS Crossover Region in Lithium 6
%--------------------------------
%--------------------------------
\bibitem{Chin}
C. Chin, M. Bartenstein A. Altmeyer, S. Riedl, S. Jochim, J.H. Denschlag, and 
R. Grimm, Science {\bf 305}, 1128 (2004).
%
%Observation of the Pairing Gap in a Strongly Interacting Fermi Gas
%--------------------------------
%--------------------------------
\bibitem{Greiner}
M. Greiner, C.A. Regal, and D.S. Jin, 
 Phys. Rev. Lett. {\bf 94}, 070403 (2005).
%
%Probing the Excitation Spectrum of a Fermi Gas in the BCS-BEC Crossover Regime
%--------------------------------
%--------------------------------
\bibitem{Partridge05p}
G.B. Partridge, K.E. Strecker, R.I. Kamar, M.W. Jack, and R.G. Hulet, 
Phys. Rev. Lett. {\bf 95\/}, 020404 (2005).
%
% Molecular Probe of Pairing in the BEC-BCS Crossover
%--------------------------------
%--------------------------------
\bibitem{Zwierlein05p}
M.W. Zwierlein, J.R. Abo-Shaeer, A. Schirotzek, C.H. Schunck, and W. Ketterle, Nature {\bf 435\/}, 1047 (2005).
%
% Vortices and Superfluidity in a Strongly Interacting Fermi Gas
%--------------------------------
%--------------------------------
%---------%** early and review papers on FR -- 
%-------------------------------- 
%%-------------------------------- 
\bibitem{Timmermans99}
E. Timmermans, P. Tommasini, M. Hussein, and A. Kerman, 
Phys. Rep. {\bf 315\/}, 199 (1999).
%
% Feshbach resonances in atomic Bose-Einstein condensates
%
%---------------------------
%-------------------------------- 
\bibitem{Dieckmann}
K. Dieckmann, C. A. Stan, S. Gupta, Z. Hadzibabic, C.H. Schunck, and W. Ketterle,
Phys. Rev. Lett. {\bf 89\/}, 203201 (2002).
%
%Decay of an ultracold fermionic lithium gas near a Feshbach resonance
%-------------------------------- 
%-------------------------------- 
\bibitem{Ohara}
K.M. O'Hara, S.L. Hemmer, M.E. Gehm, S.R. Granade, and J.E. Thomas,
Science {\bf 298}, 2179 (2002). 
%
% Observation of a Strongly-Interacting Degenerate Fermi Gas of Atoms
%------------------------------------------------------- 
%--------------------------------
\bibitem{Regal03}
C.A. Regal and D.S. Jin, Phys. Rev. Lett. {\bf 90}, 230404 (2003).
%
% C.A. Regal and D.S. Jin, 
% Measurement of positive and negative scattering lengths in a Fermi Gas of atoms  
%-------------------------------- 
%--------------------------------
\bibitem{Bourdel}
T. Bourdel, J. Cubizolles, L. Khaykovich, K.M.F. Mag\~alhaes,  
S.J.J.M.F. Kokkelmans, G.V. Shlyapnikov, and C. Salomon,
Phys. Rev. Lett {\bf 91}, 020402 (2003). 
%
%Measurement of the Interaction Energy near a Feshbach Resonance in a 6Li Fermi Gas
%-------------------------------- 
%--------------------------------
\bibitem{Strecker}
K.E. Strecker, G.B. Partridge, and R.G. Hulet,
 Phys. Rev. Lett. {\bf 91}, 080406 (2003). 
%
% Phys. Rev. Lett. 91, 080406 (2003) 
%-------------------------------- 
%--------------------------------
\bibitem{Duine04}
R.A. Duine and H.T.C. Stoof,
Phys. Rep. {\bf 396\/}, 115 (2004).
%
% Atom-molecule coherence in Bose gases
% 
%--------------------------------
%-------------------------------- 
\bibitem{Bartenstein05}
M. Bartenstein,  A. Altmeyer, S. Riedl, R. Geursen, S. Jochim, C. Chin, J.H. Denschlag, R. Grimm,
A. Simoni, E. Tiesinga, C.J. Williams, and P.S. Julienne,
Phys. Rev. Lett. {\bf 94}, 103201 (2005). 
%
% Precise determination of $^6$Li Cold Collision Parameters by Radio-Frequency
% Spectroscopy on Weakly Bound Molecules
%-------------------------------- 
%--------------------------------
\bibitem{Kohler}
T. K\"{o}hler, K. G{\'o}ral, and P.S. Julienne,
cond-mat/0601420.
%
% Production of cold molecules via magnetically
%    tunable Fesh\-bach resonances
%--------------------------------
%--------------------------------
%--------------------------------
%
%  BEC-BCS crossover theory 
%
%
%--------------------------------
%--------------------------------
\bibitem{Eagles}
D.M. Eagles, Phys. Rev. {\bf 186\/}, 456 (1969).
%
% Possible Pairing without Superconductivity at Low Carrier 
% Concentrations in Bulk and Thin-Film Superconducting Semiconductors
%--------------------------------
\bibitem{Leggett}
A.J. Leggett, 
 in {\it Modern Trends in the Theory of Condensed Matter\/}, 
edited by A. Pekalski and R. Przystawa, (Springer-Verlag, Berlin, 1980).
%--------------------------------
%--------------------------------
\bibitem{Nozieres}
 P. Nozi\`eres and S. Schmitt-Rink, J. Low Temp. Phys. {\bf 59}, 195 (1985).
%--------------------------------
%--------------------------------
\bibitem{sademelo}
C.A.R. S\'a de Melo, M. Randeria, and J.R. Engelbrecht,
Phys. Rev. Lett. {\bf 71}, 3202 (1993).
%
% Crossover from BCS to Bose superconductivity: Transition temperature and time-dependent Ginzburg Landau
% Theory
%--------------------------------
%-------------------------------- 
\bibitem{Timmermans01}
E. Timmermans, K. Furuya, P.W. Milonni, and A.K. Kerman,
Physics Letters A {\bf 285}, 228 (2001).
% Prospect of creating a composite Fermi- Bose superfluid
%--------------------------------
%--------------------------------
\bibitem{Holland}
M. Holland, S.J.J.M.F. Kokkelmans, M.L. Chiofalo, and R. Walser,
 Phys. Rev. Lett. {\bf 87}, 120406 (2001).
% 
% Resonance Superfluidity in a Quantum Degenerate Fermi Gas
%--------------------------------
%--------------------------------
\bibitem{Ohashi}
Y. Ohashi and A. Griffin, 
Phys. Rev. Lett. {\bf 89}, 130402 (2002);
%
% BEC-BCS crossover in a Gas of Fermi Atoms with a Feshbach Resonance
%
Phys. Rev. A {\bf 67}, 033603 (2003).
%
% Superfluid transition temperature in a trapped gas of Fermi atoms with a Feshbach resonance
%
%--------------------------------
%--------------------------------
\bibitem{agr}
A.V. Andreev, V. Gurarie, and L. Radzihovsky,
Phys. Rev. Lett. {\bf 93}, 130402 (2004).
% 
% Nonequilibrium dynamics and thermodynamics of a degenerate Fermi gas across a Feshbach resonance
%--------------------------------
%--------------------------------
\bibitem{Stajic}
J. Stajic, J.N. Milstein, Q. Chen, M.L. Chiofalo, M.J. Holland, and K. Levin,  
Phys. Rev. A {\bf 69}, 063610 (2004).
%
% Nature of superfluidity in ultracold Fermi gases near Feshbach resonances
%--------------------------------
%---------------------------
\bibitem{Levinsen}
J. Levinsen and V. Gurarie,
 Phys. Rev. A {\bf 73\/}, 053607 (2006). 
%
% Properties of the BCS-BEC condensate in the BEC regime 
%---------------------------
%--------------------------------
\bibitem{Gurarie}
V. Gurarie and L. Radzihovsky, cond-mat/0611022.
%--------------------------------
%--------------------------------
\bibitem{Zwierlein05}
M.W. Zwierlein, A. Schirotzek, C.H. Schunck, and 
W. Ketterle, Science {\bf 311}, 492 (2006).
%
% Fermionic superfluidity with imbalanced spin populations and the quantum phase
% transition to the normal state
%-------------------------------- 
%--------------------------------
\bibitem{Partridge05} 
G.B. Partridge, W. Li, R.I. Kamar, Y. Liao, and R.G. Hulet,
Science {\bf 311\/}, 503 (2006).
%
%Pairing and Phase Separation in a Polarized Fermi Gas
%--------------------------------
%--------------------------------
\bibitem{Zwierlein06}
M.W. Zwierlein and  W. Ketterle,
cond-mat/0603489.
% Comment on "Pairing and Phase Separation in a Polarized Fermi Gas" 
%by G. B. Partridge, W. Li, R. I. Kamar, Y. Liao, R. G. Hulet, 
%Science 311, 503 (2006)
%--------------------------------
%--------------------------------
\bibitem{PartridgeComment}
G.B. Partridge, W. Li, R.I. Kamar, Y. Liao, and R.G. Hulet,
cond-mat/0605581.
%
%Response to Comment on "Pairing and Phase Separation in a Polarized Fermi Gas"
%--------------------------------
%--------------------------------
 \bibitem{Zwierlein06p}
M.W. Zwierlein, C.H. Schunck, A. Schirotzek, and  W. Ketterle,
cond-mat/0605258.
%
%Direct Observation of the Superfluid Phase Transition in Ultracold Fermi Gases
%--------------------------------
%--------------------------------
\bibitem{Shin}
Y. Shin, M.W. Zwierlein, C.H. Schunck, A. Schirotzek, and W. Ketterle, Phys. Rev. Lett. {\bf 97}, 030401 (2006).
%
%Observation of Phase Separation in a Strongly-Interacting Imbalanced Fermi Gas
%--------------------------------
%--------------------------------
\bibitem{shortpaper}
D.E. Sheehy and L. Radzihovsky, Phys. Rev. Lett. {\bf 96}, 060401 (2006).
%
% BEC-BCS crossover in "magnetized" Feshbach-resonantly paired superfluids
%--------------------------------
%--------------------------------
%
% *** Next: Recent theory papers on asymmetric fermi superfluids 
%
%--------------------------------
%--------------------------------
\bibitem{Combescot01}
R. Combescot, Europhys. Lett. {\bf 55}, 150 (2001).
%
% BCS superfluidity in ultracold gases with unequal atomic populations
%--------------------------------
%--------------------------------
\bibitem{Liu}
W.V. Liu and F. Wilczek,
Phys. Rev. Lett. {\bf 90}, 047002 (2003).
% Interior gap superfluidity
%-------------------------------
%--------------------------------
\bibitem{Bedaque}
P.F. Bedaque, H. Caldas, and G. Rupak,
Phys. Rev. Lett. {\bf 91}, 247002 (2003).
% 
% Phase separation in Asymmetrical Fermion superfluids
%-------------------------------
%-------------------------------
\bibitem{Caldas}
H. Caldas,
 Phys. Rev. A {\bf 69}, 063602 (2004).
%
%Cold asymmetrical fermion superfluids
%--------------------------------
%--------------------------------
\bibitem{Mizushima}
T. Mizushima, K. Machida, and M. Ichioka,
Phys. Rev. Lett. {\bf 94}, 060404 (2005).
%
% Direct Imaging of Spatially Modulated Superfluid Phases in Atomic Fermion Systems.
%--------------------------------
%--------------------------------
\bibitem{Carlson}
J. Carlson and S. Reddy, 
Phys. Rev. Lett. {\bf 95}, 060401 (2005).
%
% Asymmetric Two-Component Fermion Systems in Strong Coupling
%--------------------------------
%--------------------------------
\bibitem{Cohen}
T.D. Cohen, Phys. Rev. Lett. {\bf 95}, 120403 (2005).
%
% Phase Separation and an Upper Bound for a Generalized Superfluid Gap for Cold Fermi Fluids in the Unitary Regime
%--------------------------------
%--------------------------------
\bibitem{Castorina}
P. Castorina, M. Grasso, M. Oertel, M. Urban and D. Zappal\`a,
Phys. Rev. A {\bf 72}, 025601 (2005).
%
% Nonstandard pairing in asymmetric trapped Fermi gases
% 
%--------------------------------
%--------------------------------
\bibitem{Sedrakian}
A. Sedrakian, J. Mur-Petit, A. Polls, and H. M\"uther,
Phys. Rev. A {\bf 72},  013613 (2005).
%
%Pairing in a two-component ultracold Fermi gas: phases with broken
%space symmetries
%
%--------------------------------
%--------------------------------
\bibitem{Yang1}
K. Yang,  Phys. Rev. Lett. {\bf 95}, 218903 (2005).
%
% Realization and Detection of Fulde-Ferrell-Larkin-Ovchinnikov Superfluid Phases
% in Trapped Atomic Fermion Systems
%--------------------------------
%--------------------------------
\bibitem{Pao}  C.-H. Pao, S.-T. Wu, and S.-K. Yip, Phys. Rev. B {\bf 73}, 132506 (2006).
%--------------------------------
%--------------------------------
\bibitem{Son}
D.T. Son and M.A. Stephanov, Phys. Rev. A {\bf 74}, 013614 (2006).
%
% Phase Diagram of Cold Polarized Fermi Gas 
%--------------------------------
%--------------------------------
\bibitem{Yang2}
K. Yang, cond-mat/0508484.
%
% Quantum Liquid Crystal Phases in Fermionic Superfluids with Pairing between Fermion
% Species of Unequal Densities
%--------------------------------
%--------------------------------
\bibitem{Dukelsky}
J. Dukelsky, G. Ortiz, and S.M.A. Rombouts, and K. van Houcke,
Phys. Rev. Lett. {\bf 96}, 180404 (2006).
%
% Integrable models for asymmetric Fermi superfluids: Emergence of a new exotic phase
%-------------------------------
%---------------------------------
\bibitem{YangSachdev}
K. Yang and S. Sachdev, 
Phys. Rev. Lett. {\bf 96}, 187001 (2006).
%
% Quantum criticality of a Fermi gas with a spherical dispersion minimum
%--------------------------------
%--------------------------------
\bibitem{Pieri}
 P. Pieri and G.C. Strinati, Phys. Rev. Lett.
 {\bf 96}, 150404 (2006).
%
% Trapped fermions with density imbalance in the BEC limit
%--------------------------------
%--------------------------------
\bibitem{Torma}
 J. Kinnunen, L. M. Jensen, and P. T\"orm\"a,
Phys. Rev. Lett. {\bf 96}, 110403 (2006).
%
% Strongly interacting Fermi gases with density imbalance,
%--------------------------------
%--------------------------------
\bibitem{Yi}
 W. Yi and L.-M. Duan, Phys. Rev. A {\bf 73}, 031604 (2006).
%
%Trapped Fermions across a Feshbach resonance with population imbalance
%--------------------------------
%--------------------------------
\bibitem{Chevy}
F. Chevy, Phys. Rev. Lett. {\bf 96}, 130401 (2006).
%
% Density Profile of a Trapped Strongly Interacting Fermi Gas with Unbalanced Spin Populations
%--------------------------------
%--------------------------------
\bibitem{He}
L. He, M. Jin, and P. Zhuang, Phys. Rev. B {\bf 73}, 214527 (2006).
%
% LOFF pairing vs. Breached Pairing in Asymmetric Fermion Superfluids
%--------------------------------
%--------------------------------
\bibitem{Caldas06}
H. Caldas, cond-mat/0601148.
% Finite size effects in cold asymmetrical Fermion superfluids
%--------------------------------
%--------------------------------
\bibitem{DeSilva}
T.N. De Silva and E.J. Mueller,
Phys. Rev. A {\bf 73}, 051602 (2006).
%
% Density profiles of an imbalance trapped  Fermi gas near a Feshbach resonance
%
%-------------------------------- 
\bibitem{Haque}
M. Haque and H.T.C. Stoof, Phys. Rev. A {\bf 74}, 011602 (2006). 
%
% Pairing of a trapped resonantly-interacting fermion mixture with unequal spin populations
%
%--------------------------------
%---------------------------------
\bibitem{SachdevYang}
S. Sachdev and K. Yang, Phys. Rev. B {\bf 73}, 174504  (2006).
%
%Fermi surfaces and Luttinger's theorem in paired fermion systems
%--------------------------------
%--------------------------------
\bibitem{Bulgac06}
A. Bulgac, M.M. Forbes, and  A. Schwenk, Phys. Rev. Lett. {\bf 97}, 020402 (2006).
%
% Induced P-wave Superfluidity in Asymmetric Fermi Gases
%--------------------------------
%--------------------------------
\bibitem{HoZhai}
T.-L. Ho and H. Zhai, cond-mat/0602568.
%
% Homogeneous fermion superfluid with unequal spin populations
%--------------------------------
%--------------------------------
\bibitem{LiuHu}
X.-J. Liu and H. Hu, 
Europhys. Lett. {\bf 75}, 364 (2006).
%
%BCS-BEC crossover in an asymmetric two-component Fermi gas
%--------------------------------
%--------------------------------
\bibitem{Gu}
Z.-C. Gu, G. Warner, and F. Zhou,
cond-mat/0603091. 
% Fermion pairing with population imbalance: energy landscape and phase 
% separation in a constrained Hilbert subspace
%--------------------------------
%--------------------------------
\bibitem{Yang06}
 K. Yang, cond-mat/0603190.
%
% Realization, Characterization, and Detection of Novel Superfluid Phases with Pairing between Unbalanced Fermion Species
%--------------------------------
%--------------------------------
\bibitem{HuLiu}
H. Hu and  X.-J. Liu, Phys. Rev. A {\bf 73}, 051603  (2006).
%Hui Hu, Xia-Ji Liu, 
%cond-mat/0603332.
%
% Mean field phase diagrams of imbalanced Fermi gases near a Feshbach resonance
%
%--------------------------------
%--------------------------------
\bibitem{Iskin}
M. Iskin and C.A.R. S\'a de Melo, cond-mat/0604184.
%
% Two-species fermion mixtures with population imbalance
%
%--------------------------------
%--------------------------------
\bibitem{Imambekov}
A. Imambekov, C.J. Bolech, M. Lukin, and  E. Demler, cond-mat/0604423.
%
% Breakdown of the Local Density Approximation in Interacting Systems of Cold Fermions in Strongly Anisotropic Traps
%
%--------------------------------
%--------------------------------
\bibitem{Jensen}
L.M. Jensen, J. Kinnunen, and P. T\"orm\"a, cond-mat/0604424.
%
%Non-BCS superfluidity in trapped ultracold Fermi gases,
%
%--------------------------------
%--------------------------------
\bibitem{Mannarelli}
M. Mannarelli, G. Nardulli, and  M. Ruggieri, cond-mat/0604579.
%Evaluating the phase diagram of superconductors with asymmetric spin populations
% 
%--------------------------------
%--------------------------------
\bibitem{YiDuan}
W. Yi and  L.-M. Duan, Phys. Rev. A {\bf 74}, 013610 (2006).
%
%Phase diagram of a polarized Fermi gas across a Feshbach resonance in a potential trap
%
%--------------------------------
%--------------------------------
\bibitem{PaoYip}
C.-H. Pao and  S.-K. Yip, J. Phys. Cond. Mat. {\bf 18}, 5567 (2006).
%Asymmetric Fermi superfluid in a harmonic trap
%--------------------------------
%--------------------------------
\bibitem{DeSilva06}
T.N. De Silva and  E.J. Mueller, cond-mat/0604638.
%
%Surface tension in population imbalanced unitary Fermi gases
%--------------------------------
%--------------------------------
\bibitem{Caldas06p}
H. Caldas, cond-mat/0605005.
% Phase Transition in Imbalanced Fermion Superfluids
%--------------------------------
%--------------------------------
\bibitem{Chien}
C.-C. Chien, Q. Chen, Y. He, and  K. Levin, cond-mat/0605039.
%
%Intermediate temperature superfluidity in an atomic Fermi gas 
%with population imbalance
%--------------------------------
%--------------------------------
\bibitem{Koponen}
T. Koponen, J. Kinnunen, J.-P. Martikainen, L.M. Jensen, and P. T\"orm\"a,
cond-mat/0605169.
%
%Fermion pairing with spin-density imbalance in an optical lattice
%--------------------------------
%--------------------------------
\bibitem{Sedrakian06}
A. Sedrakian, H. Muther, and A. Polls,
cond-mat/0605085.
%
%Specific heat jump in a two-component ultracold Fermi gas
%
%--------------------------------
%--------------------------------
\bibitem{Parish}
M.M. Parish, F.M. Marchetti, A. Lamacraft, and B.D. Simons,
cond-mat/0605744.
%
%Finite temperature phase diagram of a polarised Fermi condensate
%--------------------------------
%--------------------------------
\bibitem{Chevy2}
F. Chevy, cond-mat/0605751.
%
% Universal phase diagram of a strongly interacting Fermi gas with unbalanced spin populations
%--------------------------------
%--------------------------------
\bibitem{Recati}
A. Recati, I. Carusotto, C. Lobo, and S. Stringari, cond-mat/0605754.
%
% Spin polarizability of a trapped superfluid Fermi gas
%--------------------------------
%--------------------------------
\bibitem{He06}
L. He, M. Jin and P. Zhuang, cond-mat/0606322.
%
% Finite temperature phase diagram of a two-component Fermi gas
% with density imbalance
%--------------------------------
%--------------------------------
\bibitem{Gubbels}
K.B. Gubbels, M.W.J. Romans, and H.T.C. Stoof,
cond-mat/0606330.
%
% Sarma phase in trapped unbalanced Fermi gases
%--------------------------------
%--------------------------------
\bibitem{Martikainen}
J.-P. Martikainen, Phys. Rev. A {\bf 74\/}, 013602 (2006). 
%
% Ultracold polarized Fermi gas at intermediate temperatures
%
%--------------------------------
%--------------------------------
\bibitem{Clogston}
A.M. Clogston, Phys. Rev. Lett. {\bf 9}, 266 (1962).
%
% Upper limit for the critical field in hard superconductors
%
%--------------------------------
%--------------------------------
\bibitem{Sarma}
G. Sarma, J. Phys. Chem. Solids {\bf 24}, 1029 (1963).
%
% On the influence of a uniform exchange field acting on the spins of
% the conduction electrons in a superconductor
%--------------------------------
%---------------------------
\bibitem{ff}
P. Fulde and R.A. Ferrell,
Phys. Rev. {\bf 135}, A550 (1964).
% Superconductivity in a strong spin exchange field 
%---------------------------
%---------------------------
\bibitem{lo}
A.I. Larkin and Yu.N. Ovchinnikov,
Zh. Eksp. Teor. Fiz. {\bf 47}, 1136 (1964) 
[Sov. Phys. JETP {\bf 20}, 762 (1965)].
%-------------------------
%-------------------------
\bibitem{Aslamazov}
L.G. Aslamazov, 
Zh. Eksp. Teor. Fiz. {\bf 55}, 1477 (1968) 
Sov. Phys. JETP {\bf 28}, 773 (1969). 
%-------------------------
%---------------------------
\bibitem{Takada}
S. Takada and T. Izuyama,
Prog. Theor. Phys. {\bf 41}, 635 (1969).
%
% Superconductivity in a molecular field I. 
%---------------------------
%--------------------------------
\bibitem{Matsuo}
S. Matsuo, S. Higashitani, Y. Nagato, and 
K. Nagai,
J. Phys. Soc. Jpn. {\bf 67},  280 (1998).
%
% Phase diagram of the Fulde-Ferrell-Larkin-Ovchinnikov State in a Three-Dimensional
% Superconductor
%--------------------------------
%--------------------------------
\bibitem{Agterberg}
D.F. Agterberg and K. Yang, 
J. Phys. Cond. Mat. {\bf 13}, 9259 (2001).
%
% The effect of impurities on Fulde-Ferrell-Larkin-Ovchinnikov superconductors
%--------------------------------
%--------------------------------
\bibitem{Alford}
M. Alford, J.A. Bowers, and K. Rajagopal,
Phys. Rev. D {\bf 63}, 074016 (2001).
%
% Crystalline color superconductivity  
%--------------------------------
%--------------------------------
\bibitem{Combescot}
R. Combescot and C. Mora,
Europhys. Lett. {\bf 68}, 79 (2004).
%
% The low temperature Fulde-Ferrell-Larkin-Ovchinnikov phases in 3 dimensions
%--------------------------------
%--------------------------------
\bibitem{Bowers}
J.A. Bowers and K. Rajagopal,
Phys. Rev. D {\bf 66}, 065002 (2002).
%
% Crystallography of color superconductivity
%--------------------------------
%--------------------------------
\bibitem{CeCoIn5}
H.A. Radovan, N.A. Fortune, T.P. Murphy, S.T. Hannahs, E.C. Palm, S.W. Tozer, and D. Hall,
Nature {\bf 425\/}, 51 (2003);
%
% Magnetic enhancement of superconductivity from electron spin domains
%--------------------------------
%--------------------------------
%
A. Bianchi, R. Movshovich, C. Capan, P.G. Pagliuso, and J.L. Sarrao, 
 Phys. Rev. Lett. {\bf 91}, 187004 (2003).
%
% Possible Fulde-Ferrell-Larkin-Ovchinnikov Superconducting State in CeCoIn5
%--------------------------------
%--------------------------------
\bibitem{AndreevLifshitz}
A.F. Andreev and I.M. Lifshitz, Zh. Eksp. Teor. Fiz. {\bf 56}, 2057 (1969) 
[Sov. Phys. JETP {\bf 29}, 1107 (1969)]. 
%--------------------------------
%--------------------------------
\bibitem{Chester}
G.V. Chester, Phys. Rev. A {\bf 2}, 256 (1970).
%--------------------------------
%--------------------------------
\bibitem{Leggett70}
A.J. Leggett, Phys. Rev. Lett. {\bf 25}, 1543 (1970).
%--------------------------------
%--------------------------------
\bibitem{KimChan}
E. Kim and M.H.W. Chan, Nature {\bf 427}, 225 (2004); Science {\bf 305}, 1941 (2004).
%--------------------------------
%--------------------------------
\bibitem{Footnote:FFLO}
To show this, first calculate the thermodynamic critical field by equating the energy-density
{\it penalty\/}  $H^2/8\pi$ for expelling a magnetic field with the condensation-energy density
{\it gain\/} for pairing $N(\ef) \Delta^2/2$.  They are equal at the thermodynamic critical
field $H_c$:
%
\be
\nonumber
H_c = \sqrt{4\pi N(\ef)} \Delta.
%\label{eq:hcbcssc}
\ee
%
We define $\Hcz$ as the critical Zeeman field at which superconductivity would be destroyed by the 
purely Zeeman effect. Under an applied field a normal Fermi gas acquires a 
magnetization $M = \chip H$ due to the Zeeman effect,
with the Pauli susceptibility $\chip = 2N(\ef) \mubohr^2$ (see, e.g., Ref.~\onlinecite{Negele}). 
At $\Hcz$, the energy gained by magnetizing is equal to the BCS condensation energy:
%
\be
\nonumber
%\label{eq:hzbcssc}
\frac{\chip (\Hcz)^2}{2} = \frac{N(\ef)\Delta^2}{2},
\ee
%
or $\Hcz = \Delta\sqrt{N(\ef)/\chi_P} = \Delta/\sqrt{2}\mubohr$, a field known as the Clogston upper limit~\cite{Clogston}
for the critical field of a superconductor. The ratio $H_c/\Hcz$ therefore
satisfies
%
\be
\nonumber
\frac{H_c^2}{(\Hcz)^2} = \Big( \frac{\sqrt{3}}{\pi^2} \Big)^{\frac{2}{3}} \frac{e^2n^{1/3}}{mc^2},
\ee
with $n$ the electron density.
  Since the ratio on the right side
is the Coulomb energy divided by the electron rest energy, clearly it is $\ll 1$.  Hence,
in conventional superconductors, superconductivity is destroyed by orbital effects 
long before the Zeeman effect comes into play. 
Models that yield the FFLO state essentially assume the opposite.
%--------------------------------
%--------------------------------
\bibitem{Footnote:situation}
The situation in type-II superconductors is even less hopeful as there at the lower critical field
$H_{c1} < H_c$, the vortex lattice appears, further obscuring effects of the Zeeman field.
%--------------------------------
%--------------------------------
\bibitem{SingleQ}
One may ask whether more complicated FFLO-type pairing states, containing more
Fourier components, are more stable.  
The simplest such generalization, referred to as the Larkin-Ovchinnikov (LO) state~\cite{lo}, takes the form 
$\Delta(\br) = \Delta_\bQ \cos \bQ \cdot \br$.  
Using a finite temperature Ginzburg-Landau analysis, LO in fact showed that the LO order parameter has 
slightly lower energy than the single plane wave ansatz, which we refer to as the 
Fulde-Ferrell (FF) state~\cite{ff}.  
In fact, subsequent work~\cite{Bowers} has found that states containing even more $Q$ vectors are slightly more favorable.
Fortunately, such more complicated FFLO-type states are
generally found~\cite{Combescot,Matsuo} to be only marginally more stable than the FF state we consider here (see, however,
Ref.~\onlinecite{Bowers,RajagopalComment}).
This implies that although the true ground state is very likely more complicated, possessing a 
richer real-space structure, the region of the phase diagram where we find the FFLO state
should be  {\it quantitatively
accurate\/}. Of course, the difficulty of finding the true FFLO-type ground state (unsolved even in  
the simpler case of an infinite 
homogeneous three dimensional system), is further complicated in the setting of atomic physics experiments that are
typically conducted a harmonic trap.~\cite{Mizushima,Torma}
%--------------------------------
%--------------------------------
\bibitem{RajagopalComment}
 In Ref.~\onlinecite{Bowers} Bowers and Rajagopal (BR) studied a large class of generalized FFLO states,
finding that  states 
having more Fourier components $\bQ$ (forming a FCC crystal structure) are much lower 
in energy than the single plane-wave state studied here.  Based on  this
instability BR  argued that such generalized crystalline superfluid FFLO states may be stable
over a far larger range of Zeeman fields than the single-$Q$ FFLO state considered
here. Although intriguing, we do not find these arguments persuasive enough
to overturn the predictions of our detailed controlled calculation,
that is consistent with a general wisdom on this issue. Firstly, Ref.~\onlinecite{Bowers} is
based on a Landau [Taylor expansion $E(\Delta_\bQ)$ in $\Delta_\bQ$] expansion
at $T=0$ for a strongly first-order transition. As a result there is no
small parameter that controls or validates predictions that follow from
this uncontrolled Taylor expansion.  Consistent with this concern, BR find the energy
of these states to be {\it unbounded from below\/}.  Furthermore, as discussed above~\cite{SingleQ},
these results of BR directly contradict other work~\cite{lo,Combescot,Matsuo}  on this issue that 
find such more complicated FFLO states to be only {\it marginally\/} more stable. 
%
%--------------------------------
%--------------------------------
\bibitem{noteBroad}
Since
$\gamma$ is inversely proportional to $\sqrt{\ef}$, i.e., the proper measure
of the width of the resonance is in its relation to the Fermi energy,
it is quite conceivable that even presently studied Feshbach
resonances can be made to be narrow in this dimensionless sense, by
considerably increasing the size of the atomic cloud.
%--------------------------------
%--------------------------------
\bibitem{MFTcomment} The FFLO$\to$N transition is only
continuous within mean-field theory, and can be argued to be
generically driven first order by
fluctuations.~\cite{Landau,Brazovskii,Alexander}
%--------------------------------
%--------------------------------
\bibitem{Landau}
L.D. Landau, Phys. Z. Sowejetunion {\bf 11}, 26 (1937); reprinted in {\it Collected papers of L.D. Landau\/},
D. ter Haar, ed.
(Permagon, New York, 1965).
%--------------------------------
%--------------------------------
\bibitem{Brazovskii}
S.A. Brazovskii, Zh. Eksp. Teor. Fiz. {\bf 68}, 42 (1975) [Sov. Phys. JETP {\bf 41}, 85 (1975)].
%--------------------------------
%--------------------------------
\bibitem{Alexander}
S. Alexander and J. McTague, Phys. Rev. Lett. {\bf 41}, 702 (1978). 
%
% Should All Crystals Be bcc? Landau Theory of Solidification and Crystal Nucleation
%--------------------------------
%--------------------------------
\bibitem{GreinerOL}
M. Greiner, O. Mandel, T. Esslinger, T.W. H\"ansch, and I. Bloch, Nature {\bf 415}, 39 (2002).
%
% Quantum phase transition from a superfluid to a Mott insulator in a gas of ultracold atoms
%---------------------------
%--------------------------------
\bibitem{Altman}
E. Altman, E. Demler, and M.D. Lukin,
Phys. Rev. A {\bf 70}, 013603 (2004).
%
%Probing many-body states of ultra-cold atoms via noise correlations
%--------------------------------
%--------------------------------
\bibitem{Greiner05}
M. Greiner, C.A. Regal, J.T. Stewart, and D.S. Jin,
Phys. Rev. Lett. {\bf 94}, 110401 (2005). 
%
%
% Probing Pair-Correlated Fermionic Atoms through Correlations in Atom Shot Noise 
%--------------------------------
%--------------------------------
\bibitem{Lamacraft}
A. Lamacraft, Phys. Rev. A {\bf 73}, 011602 (2006).
% 
% Particle correlations in a fermi superfluid
%--------------------------------
%--------------------------------
\bibitem{Rumer}
Y.B. Rumer, JETP {\bf 10}, 409 (1960).
%--------------------------------
%---------------------------
\bibitem{LandauQM}
L.D. Landau and E.M. Lifshitz,
 {\it Quantum Mechanics: non-relativistic theory\/}, Permagon, New York, 1977.
%--------------------------------
\bibitem{commentPhysicalBoundState}
For small $|\delta| < \width$ the negative energy
pole $E_p=-\delta^2/\width$ corresponds to a bound state  only for
$\delta < 0$. For $\delta > 0$ this pole is a so-called virtual bound
state~\cite{LandauQM}, corresponding to $\imag(k) < 0$ and therefore an
exponentially growing rather than a decaying eigenstate and therefore is
not a physical bound state.
%---------------------------
\bibitem{Gurariepwave}
V. Gurarie, L. Radzihovsky and A.V. Andreev,
Phys. Rev. Lett. {\bf 94\/}, 230403 (2005).
%
% Quantum phase transitions across p-wave Feshbach resonance
%--------------------------------
%--------------------------------
\bibitem{Negele}
J.W. Negele and H. Orland, {\it Quantum many-particle systems}, Addison Wesley, 1988.
%--------------------------------
%--------------------------------
\bibitem{Schrieffer}
J.R. Schrieffer, \textit{Theory of Superconductivity}, Perseus, 1999.
%---------------------------
%---------------------------
\bibitem{deGennes} 
P.-G. de Gennes, 
\textit{Superconductivity of Metals and Alloys}, Benjamin, New York, 1966.
%-------------------------
%-------------------------------
\bibitem{Tinkham}
M. Tinkham, \textit{Introduction to Superconductivity},
McGraw-Hill, 1996.
%--------------------------------
%--------------------------------
\bibitem{Paonote}
Recently, a number of papers~\cite{Pao,Iskin} have appeared in the literature with a mean-field phase diagram
for resonantly interacting fermions that 
disagrees qualitatively with our mean-field phase diagram reported in Ref.~\onlinecite{shortpaper}
(particularly in the negative-detuning BEC regime).
These Refs.~\onlinecite{Pao,Iskin} also disagree with other recent work~\cite{Gu,Chien,Parish} on the 
one and two-channel models that are in agreement with our results.~\cite{shortpaper}
We believe the origin of this discrepancy is that the authors of Refs.~\onlinecite{Pao,Iskin} 
erroneously study solutions of the gap equation that are maxima or saddle points (rather than
minima) and which do not correspond to any physical ground state of the system; this thereby leads to 
incorrect results (see also Ref.~\cite{Comment})..
  The authors of 
Ref.~\onlinecite{Sedrakian} (and references therein) erroneously report a polarization-dependent gap of the BCS state
[Eq.(1) of Ref.~\onlinecite{Sedrakian}] that is actually the polarization dependence of the unstable Sarma solution
[a local maximum of $E_G(\Delta)$],
%
\be
\nonumber 
\frac{\Delta_{\rm Sarma}}{\deltabcs} = \sqrt{1-\frac{4}{3} \frac{\Delta N}{N} \frac{\ef}{\deltabcs}},
\ee
%
obtained by combining Eqs.~(\ref{eq:approxmagsarma}), (\ref{eq:sarmastate}) and (\ref{magrelation2})
(approximating $\muh \simeq 1$ for simplicity). Such problems stem from the existence of first-order
transitions at finite $h$ (where $E_G$ exhibits maxima and saddle points), but do not arise at 
$h=0$.
%--------------------------------
%--------------------------------
\bibitem{Comment}
D.E. Sheehy and L. Radzihovsky, preprint cond-mat/0608172.
%--------------------------------
%--------------------------------
\bibitem{CommentNote}
Subsequent to the initial appearence of the preprint of this manuscript, as well as our Comment (Ref.~\cite{Comment})
on Ref.~\cite{Pao}, the authors
of Ref.~\cite{Iskin} altered their previously incorrect phase diagram (appearing in versions 1,2,3 of cond-mat/0604184).  
The published version now appears to be in agreement with our work, at least in the equal-mass case considered here.
%--------------------------------

%--------------------------------
\bibitem{footnoteexact} 
Note that here we used a more general expression for $e_{G,SF}$ than is necessary for the computation
of $\deltahbcs$.  This is necessary for the computation of $\hcone$ and $\hctwo$ in Sec.~\ref{SEC:firstorderBCS}.
%---------------------------
%---------------------------
 \bibitem{CommentsONn} Since we are studying a homogeneous \lq\lq box\rq\rq\
trap, fixed $N$ corresponds to fixed uniform density. We will generalize
this to a fixed $N$ in the experimentally-relevant case of a
harmonic trap in Sec.~\ref{SEC:LDA}.
%---------------------------
%---------------------------
%---------------------------
%
%--------------------------------
\bibitem{Viverit}
L. Viverit, C.J. Pethick, and H. Smith,
Phys. Rev. A {\bf 61}, 053605 (2000).
%
% Zero-temperature phase diagram of binary boson-fermion mixtures
%--------------------------------
%--------------------------------
\bibitem{Fetter}
A.L. Fetter and J.D. Walecka, {\it Quantum theory of many-particle systems}, McGraw Hill, New York, 1971. 
%--------------------------------
%---------------------------
\bibitem{commentLG}
While our calculation of $a_m(h)$ is confined to the Born approximation, we expect
the qualitative finding of a vanishing of $a_m(h)$ at a critical $h$ to remain 
in a full calculation.~\cite{Levinsen}
%--------------------------------
%--------------------------------
\bibitem{ChaikinLubensky}
P. Chaikin and T.C. Lubensky, {\it Principles of Condensed Matter Physics},
Cambridge University Press, Cambridge (1995).
%--------------------------------
%--------------------------------
\bibitem{sounddiscussion}
In actuality, the first-order transition slightly precedes the vanishing of the scattering length
(and concomitant vanishing of the sound velocity).  This follows from the fact that, prior to
the  first-order transition, the \sfm state is in a locally stable minimum of $E_G(B)$ 
with a finite Bogoliubov sound velocity.
%---------------------------
%--------------------------------
\bibitem{Kleinert}
H. Kleinert, Fortschr. Phys. {\bf 26}, 565 (1978).
%---------------------------
%---------------------------
\bibitem{Uzunov}
D.I. Uzunov, Phys. Lett. {\bf 87}A, 11 (1981).
%--------------------------------
%--------------------------------
\bibitem{Sachdev}
S. Sachdev, {\it Quantum Phase Transitions}, Cambridge University Press, 
Cambridge, 1999. 
%--------------------------------
%--------------------------------
\bibitem{Petrov}
D.S. Petrov, C. Salomon, and G.V. Shlyapnikov,
Phys. Rev. Lett {\bf 93}, 090404 (2004).
% Weakly bound dimers of fermionic atoms
%---------------------------
%---------------------------
\bibitem{Brodsky}
I.V. Brodsky, M.Yu. Kagan, A.V. Klaptsov, R. Combescot, and 
X. Leyronas,
cond-mat/0507240.
% Bound states of three and four resonantly interacting particles
%---------------------------
%---------------------------
\bibitem{LDAvalidity}  Clearly, LDA breaks down at the edge of the cloud 
as well as near $r_c$, corresponding to the first-order transition where LDA 
unphysically predicts an infinitely sharp jump discontinuity in the spatial 
profile, that will be cut off by the kinetic energy.
%--------------------------------
%--------------------------------
%---------------------------
\bibitem{Butts} D.A. Butts and D.S. Rokhsar,
 Phys. Rev. A {\bf 55}, 4346 (1997)
%---------------------------
%---------------------------
\bibitem{Gubankova}
E. Gubankova, W.V. Liu, and F. Wilczek,
Phys. Rev. Lett. {\bf 91}, 032001 (2003).
%
% Breached pairing superfluidity: Possible realization in QCD
%--------------------------------
%--------------------------------
\bibitem{Forbes}
M.M. Forbes, E. Gubankova, W.V. Liu, and F. Wilczek,
Phys. Rev. Lett. {\bf 94}, 017001 (2005).
%
% Stability Criteria for Breached Pair Superfluidity
%--------------------------------
%%--------------------------------
%\bibitem{commentError}Unfortunately, a number of other
%works\cite{Iskin} have also reproduced the erroneous
%phase boundary on BEC side, for reasons similar to Pao, et
%al.\cite{Pao} 
%%--------------------------------
%--------------------------------
\bibitem{comment:AtfiniteT} At finite $T$ it is easy to see that the \sfm is no longer a
phase distinct from the SF since they both have a finite magnetization and have the same symmetry.
%--------------------------------
%--------------------------------
\bibitem{agd}
A.A. Abrikosov, L.P. Gorkov, and I.E. Dzyaloshinskii,
{\it Methods of Quantum Field Theory in Statistical Physics\/},
Dover, 1975.
%-------------------------------
%--------------------------------
\bibitem{bragg}
J. Steinhauer, N. Katz, R. Ozeri, N. Davidson, C. Tozzo, and F. Dalfovo,
Phys. Rev. Lett. {\bf 90}, 060404 (2003); 
%
% 
%
S. Richard, F. Gerbier, J.H. Thywissen, M. Hugbart, P. Bouyer, and A. Aspect,
Phys. Rev. Lett. {\bf 91}, 010405 (2003).
%
%--------------------------------
%--------------------------------
\bibitem{Kokkelmans02}
S.J.J.M.F. Kokkelmans and  M.J. Holland, Phys. Rev. Lett. {\bf 89\/}, 180401 (2002).
%
% Ramsey fringes in a Bose-Einstein condensate between atoms and molecules
%--------------------------------
%--------------------------------
\bibitem{Baym}
G. Baym, {\it Lecture notes in quantum mechanics\/}, Addison-Wesley, 1969.
%---------------------------
%-------------------------------
\bibitem{Mahan}
G.D. Mahan, {\it Many Particle Physics\/}, Plenum, New York, 1990.
%---------------------------
%---------------------------
\bibitem{Cornellexp}
 M.H. Anderson, J.R. Ensher, M. R. Matthews, 
C.E. Wieman, and E.A. Cornell, Science {\bf 269\/}, 198 (1995).
%--------------------------------
%--------------------------------
\bibitem{Ketterleexp}
K.B. Davis, M.-O. Mewes, M.R. Andrews, N.J. van Druten, D.S. Durfee, 
D.M. Kurn, and W. Ketterle,  Phys. Rev. Lett. {\bf 75}, 3969 (1995).
%--------------------------------
%--------------------------------
\bibitem{Hollandetal}
M. Holland and J. Cooper, Phys. Rev. A {\bf 53}, 1954 (1996).
%--------------------------------
%---------------------------
\bibitem{read}
N. Read and N.R. Cooper, Phys. Rev. A {\bf 68\/}, 035601 (2003).
% Free expansion of lowest Landau level states of trapped atoms: 
% a wavefunction microscope 
%---------------------------
%---------------------------
\end{thebibliography}
\end{document}